\PassOptionsToPackage{svgnames}{xcolor}
\documentclass[twocolumn]{aastex62}
\usepackage{fixfoot}
\usepackage[clockwise, figuresright]{rotating}
\usepackage{amsmath}
\usepackage{gensymb}
\usepackage{color,soul}
\usepackage{multirow}
\usepackage{IEEEtrantools}
\usepackage{bm}
\usepackage{float}
\usepackage{textcomp}
\usepackage{booktabs}
\usepackage{longtable}
\usepackage{csquotes}
\usepackage{morefloats}
\usepackage{appendix}
\received{2019 January 15}
\revised{2019 March 4}
\accepted{2019 March 11, by The Astrophysical Journal}

\shorttitle{$M_{\rm BH}$--$M_{\rm *,sph}$ and $M_{\rm BH}$--$M_{\rm *,gal}$ for ETGs}
\shortauthors{Sahu, Graham, and Davis}

\begin{document}

\title{Black Hole Mass Scaling Relations for Early-Type Galaxies: $M_{BH}$--$M_{*,sph}$ and $M_{BH}$--$M_{*,gal}$ }

\correspondingauthor{Nandini Sahu}
\email{nsahu@swin.edu.au}

\author[0000-0003-0234-6585]{Nandini Sahu}
\affil{OzGrav-Swinburne, Centre for Astrophysics and Supercomputing, Swinburne University of Technology, Hawthorn, VIC 3122, Australia}
\affil{Centre for Astrophysics and Supercomputing, Swinburne University of Technology, Hawthorn, VIC 3122, Australia}

\author[0000-0002-6496-9414]{Alister W. Graham}
\affil{Centre for Astrophysics and Supercomputing, Swinburne University of Technology, Hawthorn, VIC 3122, Australia}

\author[0000-0002-4306-5950]{Benjamin L. Davis}
\affil{Centre for Astrophysics and Supercomputing, Swinburne University of Technology, Hawthorn, VIC 3122, Australia}

\keywords{black hole physics --- galaxies: bulges --- galaxies: evolution --- galaxies: photometry  --- galaxies: elliptical and lenticular, cD  --- galaxies: structure}

\begin{abstract}
Analyzing a sample of 84 early-type galaxies with directly-measured super-massive black hole masses---nearly doubling the sample size of such galaxies with multi-component decompositions---a symmetric linear regression on the reduced (merger-free) sample of 76 galaxies reveals $M_{BH}\propto M_{*,sph}^{1.27\pm 0.07}$ with a total scatter of $\Delta_{rms}=$ 0.52~dex in the $\log(M_{BH})$ direction. However, and importantly, we discover that the ES/S0-type galaxies with disks are offset from the E-type galaxies by more than a factor of ten in their $M_{BH}/M_{*,sph}$ ratio, with ramifications for formation theories, simulations, and some virial factor measurements used to convert AGN virial masses into $M_{BH}$. Separately, each population follows a steeper relation with slopes of $1.86\pm0.20$ and $1.90\pm0.20$, respectively. The offset mass ratio is mainly due to the exclusion of the disk mass, with the two populations offset by only a factor of two in their $M_{BH}/M_{*,gal}$ ratio in the $M_{BH}$--$M_{*,gal}$ diagram where $M_{BH}\propto M_{*,gal}^{1.8\pm 0.2}$ and $\Delta_{rms}=0.6\pm 0.1$~dex depending on the sample. For $M_{BH} \gtrsim 10^7 M_{\odot}$, we detect no significant bend nor offset in either the $M_{BH}$--$M_{*,sph}$ or $M_{BH}$--$M_{*,gal}$ relations due to barred versus non-barred, or core-S\'ersic versus S\'ersic, early-type galaxies. For reference, the ensemble of late-type galaxies (which invariably are S\'ersic galaxies) follow $M_{BH}$--$M_{*,sph}$ and $M_{BH}$--$M_{*,gal}$ relations with slopes equal to $2.16\pm 0.32$ and $3.05\pm 0.70$, respectively.  Finally, we provide some useful conversion coefficients, $\upsilon$, accounting for the different stellar mass-to-light ratios used in the literature, and we report the discovery of a local, compact massive spheroid in NGC~5252. 

\end{abstract}

\section{Introduction}

There is growing evidence suggesting that black holes exist in a continuum of masses, from stellar mass black holes \citep[a few\,$M_\odot$ to $\approx 100\,M_\odot$;][]{Belczynski:2010, Abbott:GW150914:2016} to super-massive black holes 
\citep[$10^5 \, M_\odot - 10^{10}\, M_\odot $;][]{Lynden:1969, Wolfe:1970, Lynden-Bell:Rees:1971, Natarajan:2009, Inayoshi:2016}. In between these two mass ranges lie the intermediate-mass black holes \citep[][and references therein]{Miller:2003, Mapelli:2016, Mezcua:2017, Graham:Soria:Davis:2018}. A galaxy may contain several thousand \citep{Hailey:2018} to millions \citep{Elbert:2018} of stellar mass black holes, but typically only one central Super-Massive Black Hole (SMBH) for which there are many theories \citep{Miller:2003, Mayer:2007:Merger:SMBH, Hirano:2017:Primordial:SMBH, Morganti:2017:AGN:FEEDBACK}.

In order to obtain insight for these theories, for the last three decades, astronomers have been investigating the underlying relations between SMBHs and various properties of the host galaxies \citep[see the review in][and references therein]{Graham:Review:2016}. Based on \citet{Dressler:1989}, and various black hole formation scenarios and feedback models, most astronomers have come to envision a fundamental scaling relation existing between the mass of an SMBH and that of the spheroidal stellar component of the host galaxy.

Building on some of the previous estimates of black hole masses, \citet{Dressler:Richstone:1988} predicted an upper limit of $10^{9} \,M_\odot$ for the central SMBH mass of the galaxies with the largest spheroids. Their prediction was based on the central black hole mass ($M_{BH}$) and spheroid stellar mass ($M_{sph}$ or $ M_{bulge}$) ratios in the two neighboring galaxies M31 and M32. \citet{Dressler:1989} directly, and \citet{Yee:1992} indirectly, suggested a linear relationship between the black hole mass and bulge mass of a galaxy. \citet{Kormendy:Richstone:1995} and \citet{Magorrian:1998} subsequently observed a linear relation between $M_{BH}$ and $M_{bulge}$. 

Using larger samples of galaxies and updated black hole masses, most astronomers continued to report a near-linear $M_{BH}$--$M_{bulge}$ relation for nearly two decades \citep[e.g.][]{Ho:1999, Ferrarese:Ford:2005, Graham:2007:Mbh, Gultekin:Richstone:2009, Sani:2011}. However, during the same period, some astronomers \citep{Laor:1998, Wandel:1999} found a steeper relation due to the addition of low-mass galaxies in their datasets. \citet{Salucci:2000} reported that spiral galaxies have a steeper $M_{BH}-M_{bulge}$ slope than massive elliptical galaxies. Further, \citet{Laor:2001} reported $M_{BH} \propto M^{1.53\pm 0.14}_{bulge}$ from his work on an updated sample of 40 quasars. 

\citet{Graham:2012} observed two different slopes in the $M_{BH}$--$L_{bulge}$ diagram for galaxies with S{\'e}rsic or core-S{\'e}rsic spheroids \citep{Alister:2003:CS}. He found a near-linear $M_{BH}-L_{bulge}$ relation for the massive core-S{\'e}rsic galaxies (all of which were early-type galaxies), and a \enquote{super-quadratic}\footnote{The phrase \enquote{super-quadratic} was used to describe a power-law with a slope greater than 2 but not as steep as 3.} relation for the low-mass S{\'e}rsic galaxies (most of which were late-type galaxies). Further, \citet{Graham:Scott:2013} and \citet{Scott:Graham:2013}, with their work on a bigger sample of galaxies, recovered this bent relation and \citet{Graham:Scott:2015} showed that the so-called pseudobulges \citep{Gadotti:2009, Kormendy:Bender:2011} also complied with the non-linear (super-quadratic) arm of the bent relation. The bent relation strongly suggested the need to re-visit various theories and implications based on the previously assumed linear relation. For example, if there is evolution along the $M_{BH}$--$M_{sph}$ relation, then the steeper relation reveals that the fractional growth of a black hole's mass is faster than that of low-mass spheroids (S{\'e}rsic galaxies), consistent with many other works \citep[e.g.][]{Diamond:Rieke:2012, Seymour:2012, LaMassa:2013, Drouart:2014}.

These $M_{BH}$ scaling relations will help us understand the rate at which the black hole mass grows relative to the star formation rate in the host galaxy, which further aids formation and evolution theories of black holes and the galaxies which encase them \citep[e.g.][]{Shankar:2009}. It also provides insight into the understanding of AGN feedback models between an SMBH and its host galaxy \citep[e.g.][]{Hopkins:Hernquist:2006}. In the past, some simulations have reported steeper (at the low-mass end) and bent $M_{BH}$--$M_{*,sph}$ relations \citep{Cirasuolo:2005, Fontanot:2006, Dubois:2012, Khandai:2012, Bonoli:2014, Neistein:2014, Angles-Alcazar:2017}, which partly supports our findings. 

\citet{Gadotti:Kauffmann:2009} reported discrepancies between the black hole mass estimated from the $M_{\rm BH}$--$\sigma$ relation and the single linear $M_{BH}$--$M_{*,sph}$ relation for all type of (elliptical, lenticular and spiral) galaxies. There are in fact many influential works which have based their predictions on a single linear $M_{BH}$--$M_{*,sph}$ relation, for all type of galaxies \citep{Fabian:1999, Wyithe:Loeb:2003, Marconi:Risaliti:2004, Springel:2005, Begelman:Nath:2005, Croton:2006, DiMatteo:2008, Natarajan:2012}. This can affect the inferred science; hence, we recommend that these simulations be revisited using the new scaling relations.

Numerous investigations of the $M_{BH}$--$M_{sph}$ relation were based on the belief that there is a large possibility of black hole mass correlating better with its host bulge stellar mass, rather than with its host galaxy (or total) stellar mass, reflected by the smaller scatter in the $M_{BH}$--$M_{sph}$ relation. However, \citet{Lasker:2014}\footnote{\citet{Lasker:2014} had only 4 late-type galaxies in their sample} with their (early-type galaxy)-dominated sample of 35 galaxies claimed that black hole mass correlates with total galaxy luminosity equally well as it does with the bulge luminosity. Additionally, there have been several detections of bulge-less galaxies which harbor massive black holes at their center \citep[e.g.][]{Reines:2011, Secrest:2012, Schramm:2013, Simmons:2013, Satyapal:2014}. This suggests the possibility of the black hole mass correlating directly with the galaxy mass ($M_{gal}$), whether this be the stellar, baryonic, or total mass \citep{Ferrarese:2002, Baes:2003, Sabra:DM:2015, Davis:2018:b}.

The recent work by \citet{Savorgnan:2016:Slopes} used a larger sample of 66 galaxies---consisting of 47 early-type galaxies (ETGs) and 19 late-type galaxies (LTGs)---and reported that black hole mass correlates equally well with bulge luminosity and total galaxy luminosity only for ETGs, not for LTGs (see their Figures 1 and 2). They also suggested a different idea for the bend in the $M_{BH}$--$M_{sph}$ relation that was not detected by \citet{Lasker:2014}. For the core-S{\'e}rsic and S{\'e}rsic galaxies in \citet{Savorgnan:2016:Slopes},  they found $ M_{BH}\propto M_{*,sph}^{1.19 \pm 0.23}$ and $ M_{BH}\propto M_{*,sph}^{1.48 \pm  0.20}$, respectively. These slopes for the two populations have overlapping uncertainties (within the $1\sigma$ level) and unlike in \citet{Scott:Graham:2013}, which estimated the bulge masses using a morphologically-dependent bulge-to-total ratio for 75 late-type and early-type galaxies, there was no clear bend. Furthermore, \citet{Savorgnan:2016:Slopes} found different trends for their early-type and late-type galaxies, which they referred to as a \enquote{red sequence} and a \enquote{blue sequence}, respectively, although color information was not shown in that diagram.

Our work on the hitherto largest dataset of 84 early-type galaxies, with directly-measured black hole masses, builds on \citet{Savorgnan:Graham:2016} and nearly doubles their number of ETGs with multi-component decompositions. ETGs consist of ellipticals (E), elliculars\footnote{ETGs with intermediate stellar disks \citep{Liller:1966, Graham:Ciambur:Savorgnan:2016}} (ES), and lenticulars (S0), where the latter two types have disks. Ellicular and lenticular galaxies often contain bars, bar-lenses, inner disks, rings, and ansae in addition to the bulge and disk. ETGs are often misclassified, as many catalogs, e.g., Third Reference Catalogue of Bright Galaxies (RC3), \citet{RC3:1991}, failed to identify disks from a visual inspection of the images. For our set of ETGs, we perform multi-component decompositions to identify disks, and bars, and separate the bulge luminosity from the total galaxy luminosity. We intend to refine how the black hole mass correlates with its host spheroid stellar mass, and determine how it correlates with the host galaxy stellar mass. We investigate whether or not the core-S{\'e}rsic and S{\'e}rsic galaxies cause the bend in $M_{BH}$--$M_{sph}$ relation. Also, we combine our work on ETGs with the study of LTGs by \citet{Davis:2018:a, Davis:2018:b} to further explore the reason behind the bend in the $M_{BH}$--$M_{sph}$ relation. We additionally explore the possibility of different $M_{BH}$--$M_{sph}$ relations depending on the ETG sub-morphology, i.e., for galaxies with and without a disk, and galaxies with and without a bar. In all the cases, we also investigate the prospect of a better or equally likely correlation of black hole mass with total galaxy stellar mass.

In the following Sections, we describe our imaging dataset and primary data reduction techniques. Section \ref{Modeling and decomposing} illustrates the galaxy modeling and multi-component decomposition of the galaxy light. This section also presents a detailed discussion of the stellar mass-to-light ratios that we applied to the luminosity to determine the stellar masses. We compare the masses of the galaxies calculated using different (color-dependent) stellar mass-to-light ratios, and we provide a conversion coefficient which can be applied to bring them into agreement with alternate prescriptions for the mass-to-light ratio. In Section \ref{Results}, we present the black hole scaling relations for our ETG sample, along with an extensive discussion of the nature of the $M_{BH}$--$M_{*,sph}$ and $M_{BH}$--$M_{*,gal}$ relations for various cases: S{\'e}rsic and core-S{\'e}rsic galaxies; galaxies with and without a disk; galaxies with and without a bar; and ETGs versus LTGs. Finally, in Section \ref{conclusions}, we summarize our work and present the main implications.
Henceforth, we will be using the terms spheroid and bulge of a galaxy interchangeably.

\section{Imaging Data}
\label{Imaging}

We have compiled an exhaustive sample of all 84 ETGs currently with a directly measured SMBH mass. We use the black hole masses measured from direct methods, i.e., modeling of stellar and gas dynamics. Gas-dynamical modeling is fundamentally simpler, as gases being viscous, easily settle down and rotate in a circular disk-like structure, while stellar dynamical modeling is complex and computationally expensive \citep{Walsh:2013}. Although both have their pros and cons, we prefer to use the black hole masses measured from stellar dynamics, as stars are influenced only by gravitational forces, while gas dynamics are more prone to non-gravitational forces. In order to know more about the above primary methods of black hole mass measurement, readers are directed to the review by \citet{Ferrarese:Ford:2005}. 

Out of a total of 84 ETGs, we obtain SMBH masses, distances, and light profile component parameters for 40 galaxies from \citet{Savorgnan:Graham:2016}. For NGC~1271 and NGC~1277, we directly used the SMBH masses, and the bulge and total galaxy stellar masses, from the work on their H- and V- band Hubble Space Telescope (HST) Images retrieved and reduced by \citet{Graham:Ciambur:Savorgnan:2016} and \citet{Graham:Durr:Savorgnan:2016}, respectively. The remaining 42 galaxies were modeled by us, which also includes seven galaxies (A3565~BCG, NGC~524, NGC~2787, NGC~1374, NGC~4026, NGC~5845, and NGC~7052) from the dataset of \citet{Savorgnan:Graham:2016} that we remodeled. About 80\% of the galaxy images used in this work are Spitzer Space Telescope (SST) $3.6\,\mu$m images, taken with the Infra-Red Array Camera (IRAC). The remaining few images are Sloan Digital Sky Survey \citep[SDSS,][]{York:2000} $r^{\prime}$-band images and Two Micron All Sky Survey \citep[2MASS,][]{Jarrett:2003} $K_s$-band images.

\subsection{Image Sources}

IRAC $3.6\,\mu$m images (IRAC1) are unaffected by dust absorption, have large fields-of-view, and are sufficiently spatially resolved to enable us to visually identify the primary galaxy components, thereby increasing the accuracy of disassembling galaxy images. Hence, for our analysis, we preferred to use IRAC $3.6\,\mu$m images. However, for some galaxies whose Spitzer images are not available, we used images from the SDSS archive and 2MASS catalog.

The 42 galaxy images (including seven remodeled) that we modeled were comprised of 33 images in the $3.6\,\mu$m band, out of which five images are downloaded from the Spitzer Survey of Stellar Structure in Galaxies \citep[$S^4G$:][]{Sheth:2010, Mateos:2013, Querejeta:2015} pipeline-1, and 28 images are obtained from the Spitzer Heritage Archive \citep[SHA:][]{Levine:2009, Wu:2010, Capak:2013}. Of the remaining 9 galaxies, six $K_s$-band images are obtained from 2MASS \citep{Jarrett:2003} and three $r'$-band images are from the SDSS Data Release-8 \citep{SDSS:DR8:2011}.

Images from the $S^4G$ pipeline-1 (P1)\footnote{\url{http://irsa.ipac.caltech.edu/data/SPITZER/S4G/docs/pipelines_readme.html}} are science-ready, calibrated images formed by mosaicking individual Basic Calibrated Data (BCD) frames. The $S^4G$ survey is limited to galaxies with a maximum distance of 40 Mpc, brighter than a B-band apparent magnitude of 15.5 mag, and a size limit $ D_{25} > 1\arcmin$ \citep{Sheth:2010}. Hence, we obtained $3.6\,\mu$m images of galaxies not fitting this criteria from SHA, which are level-2, post-Basic Calibrated Data (pBCD)\footnote{\url{https://irsa.ipac.caltech.edu/data/SPITZER/docs/dataanalysistools/cookbook/6/}} images. The pBCD images are a mosaicked form of level-1 corrected Basic Calibrated Data (cBCD) frames. Level-1 cBCD frames have already undergone dark current subtraction, flat-field correction, various instrument artifact corrections, and flux calibration.

The $r^{\prime}$-band images of three galaxies (NGC~6086, NGC~307, NGC~4486B) from the SDSS catalog are also basic corrected and calibrated. Although optical-band images suffer from dust extinction, we justify our choice of SDSS images, as they have a large field-of-view and sufficient resolution to help us identify galaxy components.
For the remaining six galaxies (A1836~BCG, MRK~1216, NGC~1550, NGC~4751, NGC~5328, NGC~5516,), we used flux calibrated\footnote{\url{https://www.ipac.caltech.edu/2mass/releases/allsky/doc/sec4_1.html}, \url{https://www.ipac.caltech.edu/2mass/releases/allsky/doc/sec4_2.html}} $K_{s}$-band images from the 2MASS catalog.

About $95\%$ of the images in our total galaxy sample of 84 are in either the $3.6\,\mu$m (roughly L-band) or the $2.17\,\mu$m ($K_s$-band), which helps us obtain a more reliable distribution and measurement of luminosity and stellar mass, due in part to a stable stellar mass-to-light ratio in these bands (described in Section \ref{M_L ratio}). Table \ref{Photometric_para} lists the flux calibration zero points, image pixel scale, stellar mass-to-light ratios used in this work, and solar absolute magnitude in different image pass-bands. 

\begin{deluxetable}{ccccc}
\tablecolumns{5}
\tablecaption{Photometric Parameters}
\tabletypesize{\scriptsize}
\tablehead{ 
\colhead{Image Source} & \colhead{Zero-Point} & \colhead{Pixel Scale} & \colhead{$\Upsilon_*$} & \colhead{$MAG_{\odot}$} \\ 
\colhead{} & \colhead{( mag\tablenotemark{a} )}  & \colhead{( \arcsec )} & \colhead{$M_\odot/L_\odot$} & \colhead{mag}
}

\startdata
 S4G & 21.097\tablenotemark{b} & 0.75 & 0.6\tablenotemark{f}& 6.02\\ 
 SHA & 21.581\tablenotemark{c} & 0.6 & 0.6\tablenotemark{f}& 6.02\\
  2MASS & Image specific\tablenotemark{d} & 1 & 0.7\tablenotemark{g}& 5.08\\
  SDSS & 22.5\tablenotemark{e} & 0.4 & 2.8\tablenotemark{h}& 4.65\\
 \enddata
 \tablecomments{Columns:
(1) Image Source.
(2) Photo-metric zero-points of images in AB magnitude.
(3) Pixel size of images.
(4) Stellar mass-to-light ratios used to convert measured luminosities into stellar masses.
(5) Absolute magnitude of sun in AB magnitude system.
}
 \tablenotetext{a}{AB magnitude system.}
 \tablenotetext{b}{\citet[][their Equation-13]{Salo:2015}.}
 \tablenotetext{c}{\citet[][their Equation-1]{Mateos:2016}.}
 \tablenotetext{d}{Zero-points specified in image headers were converted from Vega magnitude to AB magnitude using equation (5) from \citet{Blanton:2005}.}
 \tablenotetext{e}{ \citet[][their Equation-4]{Blanton:2005}.}
 \tablenotetext{f}{Taken from \citet{Meidt:2014} for $3.6\,\mu$m band.}
 \tablenotetext{g}{Using $\Upsilon_*^{3.6}$ in the equation $\Upsilon^{3.6\mu m}_* = 0.92\times \Upsilon^{K_s}_* - 0.05$ from \citet{Oh:2008}.}
 \tablenotetext{h}{Calibrated using $\Upsilon^{r^{\prime}}_*= \Upsilon^{K_s}_* \times L_{K_s}/L_{r^{\prime}}$ with $\Upsilon^{K_s}_* = 0.7$.}
 
\label{Photometric_para}
 \end{deluxetable}

\subsection{Image Reduction and Analysis}
All the images obtained from the various telescope pipelines described above have already undergone dark current subtraction, flat fielding, bad pixel and cosmic ray correction, sky-subtraction (except for $S^4G$ and 2MASS images), and flux calibration. The automated routines in the telescope pipelines either over or under-estimated the sky-background intensity, which we observed for most of our galaxies. 
Hence, we started our image analysis by measuring the sky-background intensities, then generating the image masks and calculating the telescope's point spread function.

\subsubsection{Sky Backgrounds}
Sky-background level subtraction is one of the crucial steps to measure a galaxy's luminosity accurately. As our target galaxy images are extended over a large number of pixels in the CCD images that we are using, an error in sky background intensity subtraction will lead to a systematic error in the surface brightness profile, especially at the larger radii and result in an erroneous measurement of the galaxy component at large radii, and in turn the inner components and the galaxy luminosity. The wide-field images that we obtained from the SHA and SDSS pipelines have already undergone sky subtraction, but as we analyzed the intensity distribution of the images, we found that the peak of the sky-background level was offset from zero for almost all of the images. Hence, it was necessary to calculate the correction in order to tune the sky level of these images to zero.

\begin{figure}
\begin{center}
\includegraphics[clip=true,trim= 4mm 4mm 0mm 2mm,width=\columnwidth]{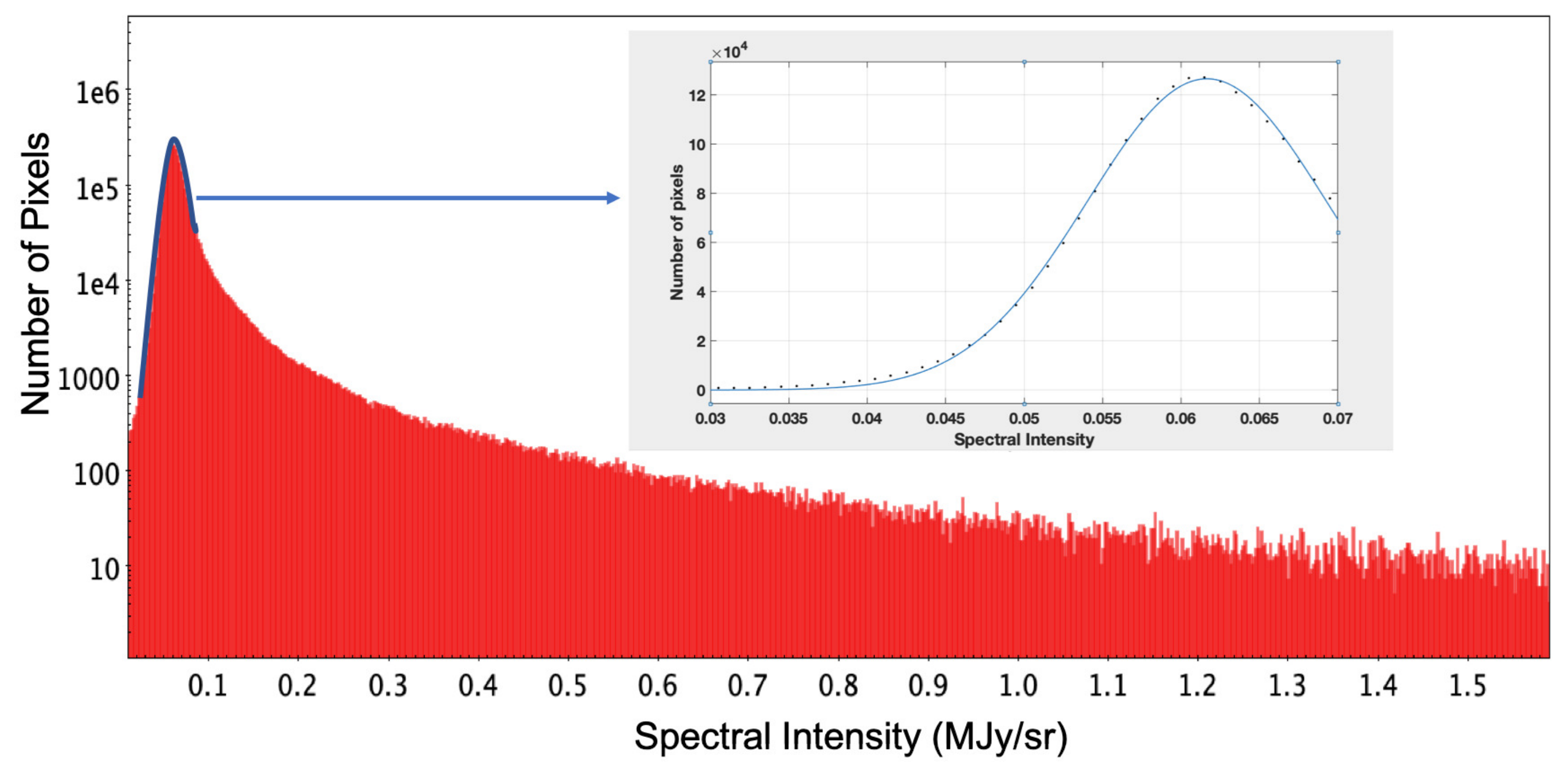}
\caption{Gaussian fit to the sky-background intensity of the \enquote{level-2 corrected}, $3.6\,\mu$m -band image of NGC 1600 from SHA, which has already undergone sky subtraction, but the sky level peaking at a non-zero value indicates that it still requires adjustment. The \textcolor{red}{red} distribution shows the faint (sky-dominated) end of the intensity histogram (number of pixels at each intensity value) from the CCD image of NGC 1600. The inset plot shows a Gaussian fit (\textcolor{Blue}{blue} curve) to the sky values in the range of 0.03 to 0.07 MJy/sr, peaking at 0.062 MJy/sr. The intensity distribution following the peak includes the intensity of our target galaxy and other radiating sources (added with the sky value).}
\label{skyGaussfit}
\end{center}
\end{figure}

To calculate the sky-background intensity level, we follow a similar procedure as explained in \citet{Almoznino:Loinger:1993}. 
The intensity distribution of the sky-background photons incident on a CCD image ideally follows a Poisson distribution when the only source of systematic error is random emission from the radiating object, in this case, the \enquote{sky-background}. However, many other systematic errors are introduced in a CCD image when it undergoes telescope pipelining. In that case, a Gaussian distribution (normal distribution) can be a better approximation for the intensity distribution of the \enquote{sky-background}.
We constructed the intensity function (pixel number of given intensity versus intensity histogram) of the entire image frame (not just a few portions of the sky that appear free of sources) and fit a Gaussian to the portion of the histogram dominated by the sky (the peak at lower-intensity values), as shown in Figure~\ref{skyGaussfit}. Intensity values of the pixels occupied by other radiating sources, including our target galaxy, produce the long tail towards higher intensities. The Gaussian fit gives us an optimally accurate mean sky value and the standard deviation (rms error) in any one pixel.

\subsubsection{Masking}
Images for our galaxy sample have large fields-of-view. Apart from our target galaxy, these images also contain other radiating sources around and overlapping with the target galaxy. Major contaminating sources are background quasars and foreground stars that overlap the pixel area occupied by the galaxy of interest. Hence, for an accurate measurement of the galaxy luminosity, we eliminate the contribution of these contaminating sources by generating a mask file. A mask is either a .fits or .pl file marking (with their pixel coordinates and pixel size) the areas and sources to be discarded during the analysis. 

We used the task \textsc{mskregions} in the Image Reduction and Analysis Facility \textsc{(IRAF)} software to read a list of user-specified regions to be masked in our image. The task then generates a mask file (.pl or .fits file) using our galaxy image as a reference for the size of the mask file. The list of contaminating objects and subsequent masks are generated in two parts by us:

\begin{enumerate}
\item \textsc{Source Extractor} \citep{Bertin:Arnouts:1996}: It uses a threshold background value to automatically identify all the objects present in an image and makes a catalog of them, designating each object by its physical coordinates in the image. We can identify and remove our target galaxy from this list (knowing its physical coordinates) and generate a mask file using this catalog using the task \textsc{mskregions}.
 
\item \textsc{Manual masking}: \textsc{Source-Extractor} cannot identify the background and foreground objects overlapping with the pixel area of our target galaxy. However, it is important to mask them in order to avoid biasing the image decomposition; therefore, we need to mask them manually. We carefully find the overlapping sources by observing our galaxy at different brightness (contrast) levels. For this purpose, we use the astronomical imaging and data visualization application \textsc{SAOImage DS9}. We generate the second mask file of contaminating objects with the \textsc{mskregion} task.
\end{enumerate} 

We combine the above two mask files using the \textsc{imarith} task in IRAF and further use the final mask as a reference for avoiding the contaminated pixels during extraction and modeling of the target galaxy light. Extra care was taken to manually mask dust in the three SDSS $r^{\prime}$-band images.

\vspace{4mm}
 
\subsubsection{PSF determination}
\label{PSF determination}
 
The spatial resolution of an image is limited by the telescope's aperture size, the wavelength of observation, the pixel size of its instrument, and the atmospheric blurring for ground-based observations. 
A distant star is a point source, whose light profile is ideally described by a delta function, but due to the collective resolution limitations, it is imaged as an extended object, and its light profile becomes a function with a non-zero width. Hence, the Full Width at Half Maximum (FWHM) of the light profile of a star in an image is a measure of the total \textit{seeing effect}, which is quantified by the Point Spread Function (PSF) of the telescope. 

The image of an object obtained by a telescope can be mathematically described as a convolution of its actual profile with the telescope's PSF. Hence, in order to measure the parameters of the actual light (or surface brightness) profile of a galaxy and its components, we need our fitting functions to be convolved with the telescope's PSF.

\citet{Moffat:1969} describes how the wings of the seeing profile (PSF) of a telescope is represented better by a Moffat function rather than a Gaussian function. A \enquote{Moffat function} has the mathematical form 
\begin{equation}
I(R)=I_0 \left(1+\left(\frac{R}{\alpha}\right)^2\right)^{-\beta},
\end{equation}
where $\alpha$ is the width parameter and $\beta$ controls the spread in the wings of the seeing profile \citep[see Figure 3 in][]{Moffat:1969}. The parameters $\alpha$ and $\beta$ are related to the FWHM of the profile through the equation
FWHM $=2\alpha\sqrt{2^{\frac{1}{\beta}}-1}$. 
The value of $\alpha$ and $\beta$ increases with poor seeing (e.g., higher atmospheric turbulence) and gradually, the profile that they describe approaches a Gaussian.
We used the IRAF task \textsc{imexamine} to determine the PSF of our images. The \textsc{imexamine} task fits the radial profile of selected stars with a Moffat function and provides the required parameters: FWHM and $\beta$.

\section{Modeling and decomposing the galaxy light}
\label{Modeling and decomposing}
The luminosity of a galaxy is modeled by fitting quasi-elliptical isophotes\footnote{A curve which connects the points of equal brightness} at each radius along the semi-major axis ($R_{maj}$). \citet{Ciambur:2016:Profiler}, in his introduction section, and \citet{Savorgnan:Graham:2016}, in their Section 4.1, employ both 1D (one-dimensional) and 2D (two-dimensional) modeling and provide a critical comparison of the two techniques. \citet{Savorgnan:Graham:2016} had more success modeling the galaxies as a set of 1D profiles; hence we also prefer to use 1D profile modeling, which takes into account the radial variation in all of the isophotal parameters such as ellipticity ($\epsilon$), position angle (PA), and the irregularity in an isophote's shape across the whole $2\pi$ azimuthal range as quantified using Fourier harmonic coefficients. Therefore, 1D modeling should not be confused with the light profile obtained only from a one-dimensional cut of a galaxy image. 

Early-type galaxies are commonly ill-considered to be featureless (no sub-components) and are expected to have regular elliptical isophotes, a scenario which is only valid for purely elliptical galaxies.
Early-type galaxies can be morphologically sub-classified as ellipticals (E) consisting of an extended spheroid, elliculars (ES) consisting of an extended spheroid with an intermediate-scale disk \citep[e.g.,][]{Graham:Ciambur:Savorgnan:2016}, and lenticulars (S0) comprised of a spheroid and an extended large-scale disk. Apart from these standard components, ETGs may also contain nuclear disks, inner rings, bars, bar-lenses \citep{Sandage:1961, Laurikainen:2009, SAHA:GRAHAM:2018}, outer rings, and ansae \citep{SAHA:GRAHAM:2018, Martinez-Valpuesta:2007}, which can cause non-elliptical or irregular isophotes in a galaxy.

\subsection{One-dimensional Representation of the Galaxy Light}

We use the new \textsc{IRAF} tasks \textsc{Isofit} and \textsc{Cmodel} \citep{Ciambur:2015} to extract the 1D light profile and associated parameter profiles (e.g., ellipticity, PA, etc.), and create a 2D model of each galaxy. \textsc{Isofit} and \textsc{Cmodel} are upgraded versions of the \textsc{IRAF} tasks \textsc{Ellipse} and \textsc{Bmodel} \citep{Jedrzejewski1:1987, Jedrzejewski2:1987}, respectively. 

In order to extract a galaxy light profile, \textsc{Isofit} reads a 2D image of a galaxy, the associated mask file, and fits quasi-elliptical isophotes at each radius of the galaxy, starting from its photometric center to its apparent edge, thus including every part of the galaxy. Further, \textsc{Isofit} uniformly samples each isophote across the whole azimuthal range, using a natural angular coordinate for ellipses, known as the \enquote{Eccentric Anomaly} \citep[$\psi$, for more details see Section 3 of][]{Ciambur:2015:Ellipse}, and provides average intensity and associated parameters of the isophotes as a function of semi-major axis radii. The isophotal intensity can be expressed in terms of the average intensity $\langle I_{ell} \rangle$ and Fourier perturbations such that 
\begin{equation}
 I(\psi) = \langle I_{ell} \rangle + \sum_n \big[A_n sin(n\psi) + B_n cos(n\psi) \big]  
\end{equation}
where, $A_n$ and $B_n$ are nth order Fourier harmonic coefficients. 

As explained by \citet{Ciambur:2015:Ellipse}, while fitting each isophote, \textsc{Isofit} calculates $A_{n}$ and $B_{n}$, these Fourier coefficients when added together, account for the irregular isophotal shapes and give a near-perfect fit. \citet{Ciambur:2015:Ellipse} also mentions that the value of $A_{n}$ and $B_{n}$ decreases with increasing order (n); therefore, we calculate a sufficient number of even harmonic coefficients, up to a maximum of $n=10$. Apart from the $n=3$ harmonic, odd-ordered Fourier harmonic coefficients ($n=5, 7, 9,$etc.) appear to provide almost no refinement in an isophote's shape; thus we can obtain a very good light profile and galaxy model, without them. Also, for the light profile along the major axis ($\psi = 0$), the sine terms are zero; hence we corrected our major-axis intensity values only for the cosine perturbations ($B_{n}$).

\begin{figure*}
\begin{center}
\includegraphics[clip=true,trim= 1mm 0mm 0mm 1mm,width=\textwidth]{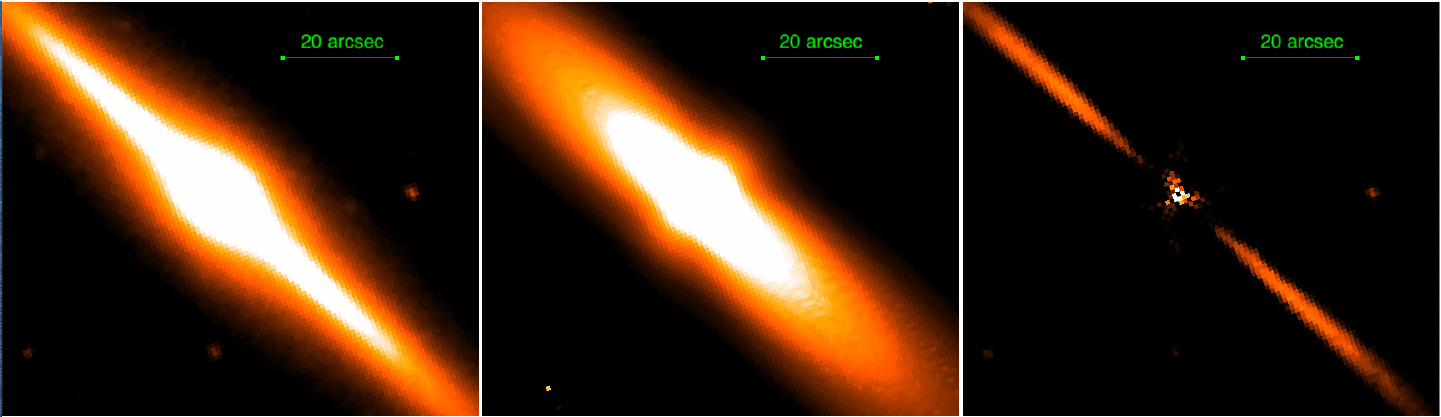}
\includegraphics[clip=true,trim= 1mm 0mm 0mm 1mm,width=\textwidth]{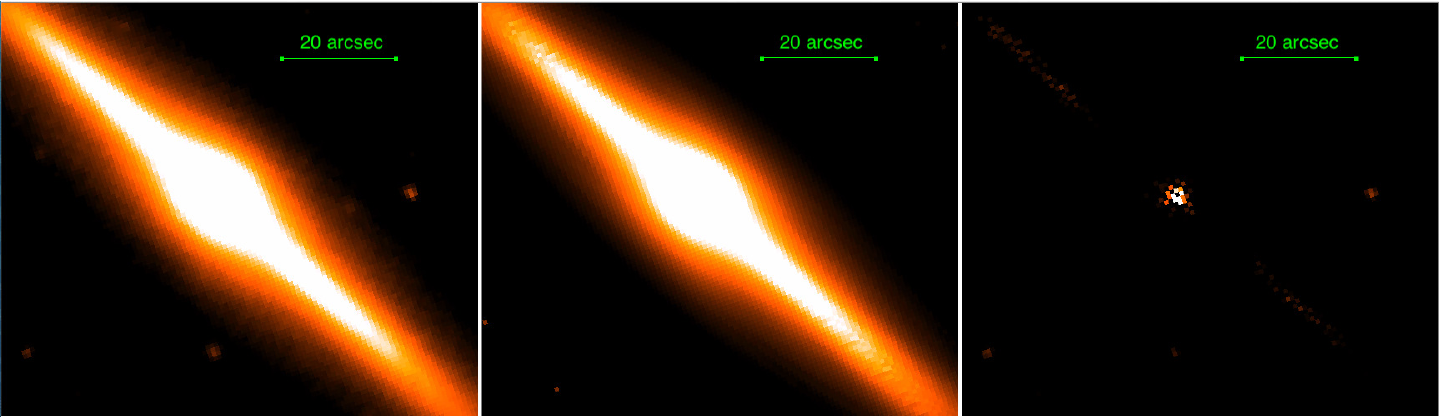}
\caption{Comparison of models and residual images for NGC~4762. First row of images are the galaxy image, model, and the residual image generated using the \textsc{Ellipse} and \textsc{Bmodel} tasks in \textsc{IRAF}. The second-row of images are the galaxy image, model, and the residual image generated using the \textsc{Isofit} and \textsc{Cmodel} tasks \citep{Ciambur:2015}.}
\label{ellipse_isofit}
\end{center}
\end{figure*}

The original \textsc{Ellipse} task is limited to only work well for face-on galaxies with almost purely elliptical isophotes (with few or no additional components), as it does not properly utilize the higher-order harmonics to fit and quantify irregularities in the isophotal shapes.
Figure \ref{ellipse_isofit} provides a comparison of models obtained for NGC~4762  using the \textsc{Ellipse} and \textsc{Isofit} tasks.

Various isophotal parameters ($\epsilon$, PA, $A_{n}$ and $B_{n}$) obtained from the \textsc{Isofit} task, are sufficient to generate an excellent 2D model of a galaxy using the \textsc{Cmodel} task. The galaxy model can be further subtracted from the galaxy image to obtain a residual image, which is useful to study various foreground and background sources overlapping with the galaxy pixels. The quality of the residual image depends on how accurately the isophotal model emulates the galaxy. The quality of the model generated using the \textsc{Isofit} and \textsc{Cmodel} tasks can be appreciated in Figure \ref{ellipse_isofit}. 

It is evident in Figure \ref{ellipse_isofit} that the \textsc{Ellipse} task could not construct a very good fit to the irregular isophotes of NGC~4762 due to the high inclination of the galaxy and its (peanut shell)-shaped bulge associated with the bar (as seen in the light profile, Figure \ref{profiler}). The \textsc{Ellipse} task fails to properly model the galaxy light along the disk, leaving behind the bright stripes in the residual image.

\subsection{Disassembling the Galaxy Image }

The isophotal table, obtained from \textsc{Isofit}, is used by the software \textsc{Profiler} \citep{Ciambur:2016:Profiler} to plot and fit the 1D radial surface brightness profile of a galaxy, with respect to both its semi-major axis radius ($R_{maj}$) and the equivalent axis ($R_{eq}$). $R_{eq}$ is the geometric mean of $R_{maj}$ and $R_{min}$. It is the radius of an imaginary circular isophote equivalent in area to the elliptical isophote with major- and minor-axis radius $R_{maj}$ and $R_{min}$, conserving the total surface brightness of the elliptical isophote. This gives $R_{eq}=\sqrt{R_{maj} R_{min}} = R_{maj}\sqrt{1-\epsilon }$, where $\epsilon $ is the ellipticity of the isophote.  
Along with the surface brightness profile, \textsc{Profiler} also plots the radial profiles of the isophote's ellipticity, position angle, and some of the higher-order Fourier harmonic coefficients (B4, B6, B8).

To decompose the galaxy light into its components, we use a wide variety of parametric analytical functions available in \textsc{Profiler}. For example, \citet{Sersic:1963} and Core-S{\'e}rsic \citep{Alister:2003:CS} functions for galactic bulges; exponential, truncated/anti-truncated exponential, and inclined-disk models for various types and orientations of disks; \citet{Ferrers:1877} function for bars; S{\'e}rsic for bar-lenses/pseudobulges, Gaussian for rings, and ansae (centered at the ring/anase radius); and PSFs for nuclear point sources. Table \ref{profiler_functions} presents the mathematical formulae for the radial surface brightness profiles of these functions and the corresponding expressions to determine the apparent magnitudes from the fit parameters. More details about the surface brightness profiles of the various fitting functions can be found in Section 3 of \citet{Ciambur:2016:Profiler}. 
\pagebreak
\movetabledown=2.2in
\begin{rotatetable*}
\begin{deluxetable*}{|c|c|c|c|}
\tabletypesize{\scriptsize}
\tablecolumns{4}
\tablecaption{Fitting Functions}
\tablehead{ 
\colhead{Function} & \colhead{Radial Surface Brightness\tablenotemark{a}, $\mu(R) $} & \colhead{Apparent Magnitude\tablenotemark{b}, $m$} & \colhead{Profile Parameters} \\ 
  \colhead{} & \colhead{(mag arcsec$^{-2}$)}  & \colhead{(mag)} & \colhead{}
}
\startdata
S{\'e}rsic\tablenotemark{c}        &  $\mu_e + \left(2.5*b_n/ \ln{10} \right)[(R/R_e)^{1/n}-1]$    & $\mu_e -5\log{R_e}-2.5\log[2\pi n \left(\exp{b_n}/(b_n)^{2n}\right)\Gamma{(2n)}] $   & $\mu_e,n,R_e$ \\
\hline
Core-S{\'e}rsic\tablenotemark{d}      &   $\mu^\prime-2.5 \gamma/\alpha\log[1+(R_b/R)^\alpha] + 2.5/\ln(10)[b_n \left((R^\alpha+R_b^\alpha)/R_e^\alpha \right)^{1/n\alpha}] $ & $\mu_b -2.5\log2\pi[R_b^2/(2-\gamma)+ne^{(b_n(R_b/R_e)^{1/n})}(\Gamma(2n)-\gamma(2n,(R_b/R_e)^{1/n} ))] $   & $\mu^\prime,R_b,R_e,n,\alpha,\gamma$ \\
\hline
Exponential\tablenotemark{e}     &   $ \mu_0 + (2.5/\ln(10))(R/h)$   & $\mu_0 - 2.5\log[2\pi h^2] $   & $\mu_0 , h$\\
\hline
Truncated\tablenotemark{f} &  $ \mu_0 + (2.5/ln(10)) (R/h_1)$ (for $R \leq R_b)$& $\mu_0 -2.5\log2\pi[h_1^2 + e^{-R_b/h_1}(h_2-h_1)(h_2+h_1+R_b)]$ & $\mu_0,h_1,h_2,R_b$ \\
         disk                         &  $ \mu_b + (2.5/\ln(10))((R-R_b)/h_2)$ (for $R > R_b)$  &      & \\
\hline
Inclined disk\tablenotemark{g} & $\mu_0 -2.5\log[(R/h_r) K_1(R/h_r)]$&Integrated Numerically &$\mu_0,h_r$ \\
\hline
Ferrer\tablenotemark{h} & $\mu_0 -2.5\alpha\log[1-(R/R_{out})^{2-\beta}] $& $ \mu_0 -2.5\log[\pi R_{out}^2*hyp2F1(-\alpha, 2/(2-\beta), (4-\beta)/(2-\beta),1)]$ & $\mu_0,R_{out},\alpha,\beta$ \\
\hline
Gaussian\tablenotemark{i} & $\mu_r + (2.5/\ln(10))((R-R_r)^2/2\sigma^2) $ &$\mu_r - 2.5\log2\pi[\sigma^2 e^{-R_r^2/2\sigma^2} + \sigma R_r \sqrt{\pi/2} (1+erf(R_r/\sigma\sqrt{2}))] $ & $\mu_r,R_r,\sigma$ 
\label{profiler_functions}
\enddata
\tablenotetext{a}{The radial surface brightness profile was obtained from the intensity profile, using $\mu(R)=-2.5 \log(I(R))$+ zero-point (see Table \ref{Photometric_para}).}

\tablenotetext{b}{$m=-2.5 \log(L)$, where luminosity (L)$=\int 2\pi R \,I(R)dR$, $I(R)$ is the radial intensity profile.}

\tablenotetext{c}{From \citet{Ciotti:1991} and \citet{Graham:Driver:2005}, the quantity $b_n$ is calculated by solving $\Gamma(2n)=2\gamma(2n,b_n)$.}

\tablenotetext{d}{Equation 5 from \citet{Alister:2003:CS}, $\mu^\prime$ and $\mu_b$ are related through Equation 6 from \citet{Alister:2003:CS}. The expression for the apparent magnitude is deduced under the approximation, $\alpha\rightarrow\infty$ \citep[Equation A20 from][]{Trujillo:2004:A20}.}

\tablenotetext{e}{Equation 14 from \citet{Graham:Driver:2005}, for n=1.}

\tablenotetext{f}{Equation 10 from \citet{Ciambur:2016:Profiler}.}

\tablenotetext{g}{Equation 12 from \citet{Ciambur:2016:Profiler} along the major axis, and $K_1(R/h_r)$ is the modified Bessel function of the second kind.}

\tablenotetext{h}{From \citet{Ferrers:1877}; $hyp2F1$ in the apparent magnitude expression represents the hyper-geometric function.}

\tablenotetext{i}{The parameter $\mu_r$ is the peak value of the Gaussian surface brightness profile at the \enquote{peak radius} r, and $\sigma$ is the width of the Gaussian.}

\end{deluxetable*}
\end{rotatetable*}

We disassemble the galaxy light into its components by fitting various features present in the galaxy light profile, using the functions mentioned in Table \ref{profiler_functions}. To help identify the components that are present in a galaxy, we visually inspect the galaxy image at various contrast levels using \textsc{DS9}, and we also inspect various features present in the ellipticity, position angle, $B4$, and $B6$ profiles (if required), which is beneficial in discerning galaxy components. Apart from that, we went through the literature, reviewing previous structural and kinematical studies of our galaxies, which gave us clues about the components present, their relative intensity (or surface brightness) levels, and their radial extents (sizes). In order to distinguish the components, like an inner disk, inner ring, nuclear star cluster, and most importantly, to identify the deficit of light at the center of a galaxy (core-S{\'e}rsic), we consulted previous works with highly resolved Hubble Space Telescope images \citep[e.g.,][]{Dullo:2014}.

Having obtained a fit for the light profile---based on real physical structure/components---for the major-axis, we map it to the equivalent-axis ($R_{eq}$), ensuring the central (R=0) surface brightness of each component remains roughly constant. The equivalent-axis parameters for each component of a galaxy are required so that \textsc{Profiler} can use the circular symmetry of the equivalent-axis to integrate the surface brightness profiles and calculate the apparent magnitudes for all the components and the whole galaxy itself. 

\begin{figure*}
\begin{center}
\includegraphics[clip=true,trim= 2mm 0mm 0mm 2mm, width=1.5\columnwidth]{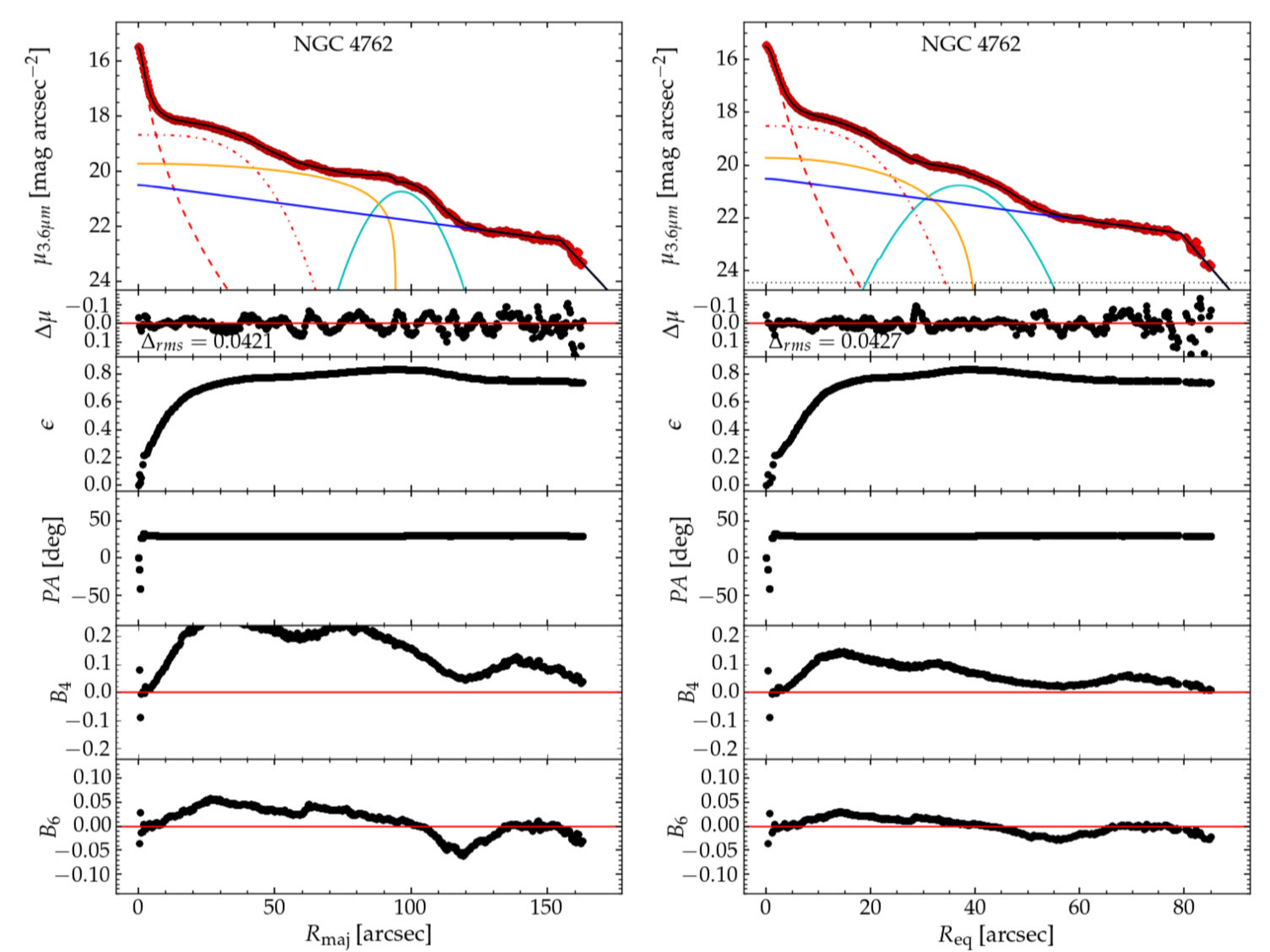}
\caption{$3.6\,\mu$m surface brightness profile of NGC~4762, plotted and fit using \textsc{PROFILER}. The left panel shows the profile along the major-axis with $\Delta_{rms}=0.0421$ mag\,arcsec$^{-2}$, and the right panel shows the profile along the equivalent-axis with $\Delta_{rms}=0.0427$ mag\,arcsec$^{-2}$. Physical sizes can be derived using a scale of $11$~pc/$\arcsec$ based on a distance of 22.6~Mpc. NGC 4762 is a barred lenticular galaxy with its multi-component fit comprised of a S{\'e}rsic function for the bulge (\textcolor{red}{- - -}), a low index S{\'e}rsic function for the bar-lens/pseudobulge (\textcolor{red}{ - $\cdot$ - $\cdot$ -}), a Ferrers function for the bar (\textcolor{orange}{---}), a Gaussian for the ansae (\textcolor{cyan}{---}), and a truncated exponential model for the extended warped disk (\textcolor{blue}{---}).}
\label{profiler}
\end{center}
\end{figure*}

Figure \ref{profiler} shows the multi-component fit to the surface brightness profile of NGC~4762, for both the major- and equivalent-axes. It is a barred-lenticular galaxy with a small bulge, an (oval-shaped) bar-lens, a bar, ansae, and a truncated disk. 
\citet{Laurikainen:2005, Laurikainen:2007, Laurikainen:2011} observed that many S0 galaxies contain bars and \enquote{ovals} (also known as \enquote{lenses} or \enquote{bar-lenses}), with the inner regions of vertically-heated bars appearing as boxy/(peanut shell)-shaped structures referred to by some as pseudobulges \citep[see][]{Combes:Sanders:1981, Athanassoula:2002, Athanassoula:2005}.
The bumps in the light profile of NGC~4762, as well as the ellipticity, $B_4$, and $B_6$ profiles at $R_{maj} \approx 30\arcsec$ and $R_{maj} \approx 80\arcsec$ correspond to the perturbation of the isophotes due to the bar-lens/pseudobulge and the bar, respectively.
As shown in the simulations by \citet[][their Figure 7]{SAHA:GRAHAM:2018}, the adjacent bump ($R_{maj} \approx 80\arcsec$) and dip ($R_{maj} \approx 120\arcsec$)  in the $B_6$ profile suggest the presence of an ansae at $R_{maj} \approx 100\arcsec$, at the end of the bar.

We also note that the decomposition results from
\citet[][e.g., their Figure 11; see also NGC~4026 and NGC~4371 in our Appendix]{SAHA:GRAHAM:2018} support the truncated disk model\footnote{A truncated disk model has a change in slope beyond the truncation radius} in NGC 4762. Also, according to \citet{Kormendy:Bender:4762:2012}, the warped disk at the outer edge is possibly due to some ongoing tidal encounter. Table \ref{NGC_4762_para} lists the fit parameters for the components in NGC 4762. Light profile fits for all other galaxies can be found in the Appendix.

\begin{deluxetable*}{cccc}
\label{NGC_4762_para}
\tablecolumns{4}
\tablecaption{Model parameters for the NGC 4762 light profile}
\tablehead{ 
\colhead{Component} & \colhead{Function} & \colhead{Major-axis parameters} & \colhead{Equivalent-axis parameters}
}
\startdata
 Bulge & S{\'e}rsic & $\mu_e=17.89, n=2.36, R_e=4.39 $ &  $\mu_e=17.09, n=1.85, R_e=2.24 $\\ 
 Barlens & S{\'e}rsic & $\mu_e=18.98, n=0.28, R_e=28.81 $ &  $\mu_e=18.89, n=0.31, R_e=14.4 $\\
 Bar & Ferrers & $\mu_0=19.72, R_{out}=94.56 ,\alpha=1.65,\beta=0.01 $ &  $\mu_0=19.72, R_{out}=40.66, \alpha=3.81 ,\beta=0.01 $ \\
 Ansae & Gaussian & $\mu_r=20.74, R_r=96.45 , FWHM=21.30 $ &  $\mu_r= 20.77, R_r=37.06, FWHM=15.89 $ \\
 Disk & Truncated Exponential & $\mu_0=20.48, R_b=155.07,  h1=82.62, h2=10.23 $ & $\mu_0=20.48, R_b=79.36, h1=40.92, h2=4.72$\\
\enddata
\tablecomments{Scale size parameters ($R_e, R_{out}, R_r, h1$, and $ h2$) are in units of arcseconds, and surface brightnesses ($\mu_e, \mu_0$, and $\mu_r$) pertains to the $3.6\,\mu$m-band (AB mag). FWHM of the Gaussian can be related to its standard deviation ($\sigma$) by, $ FWHM= 2\sigma \sqrt{2\ln2}$. Equivalent-axis is also known as the \enquote{geometric mean} axis, given by the square root of the product of major- and minor-axis.}
\end{deluxetable*}



\subsection{Stellar Mass Calculation }
\label{M_L ratio}

We calculate the absolute magnitudes for all the galaxies, and their spheroids, using their apparent magnitudes measured using \textsc{Profiler}, and the distances in Table \ref{data_table}. 
These absolute magnitudes, after applying the small corrective term for cosmological dimming\footnote{A magnitude of $10\log(1+z)$ is subtracted to account for the dimming of the observed magnitudes due to the expansion of the Universe, where z is redshift based on the galaxy distance. Red-shift was calculated assuming the latest cosmological parameters $H_0=67.4$, $\Omega_m=0.315$, $\Omega_{vacuum} = 0.685$ \citep{Planck:Collaboration:2018}.} \citep{Tolman:1930} are used to calculate the corresponding intrinsic luminosities. The intrinsic luminosity is derived in terms of the solar luminosity in each band (see Table \ref{Photometric_para}), and these luminosity values are then converted into stellar masses by multiplying them with the stellar mass-to-light ratio ($\Upsilon_*$) for each band.  

Stellar mass-to-light ratios depend on many factors, such as the Initial Mass Function (IMF) of stars in a galaxy, star formation history, metallicity, age, and they can be biased due to attenuation from dust in a galaxy. The interdependence of these factors and their effect on the stellar mass-to-light ratio is not very well known. Therefore, the mass-to-light ratio dependence on these properties has large uncertainties associated with it.
\citet{Meidt:2014} suggest a constant, optimal, stellar mass-to-light ratio of $\Upsilon_* = 0.6$ for the $3.6\,\mu$m band, based on the \citet{Chabrier:2003} IMF, which is consistent with the age-metallicity relation and can be used for both old, metal-rich and young, metal-poor stellar populations. The emission at $3.6\,\mu$m and $2.2\,\mu$m is largely unaffected by the luminosity bias due to young stars, and also it undergoes minimal dust extinction \citep{Querejeta:2015}, enabling us a somewhat stable mass-to-light ratio. Using $\Upsilon_*^{3.6\mu m} = 0.6$ in the following equation from \citet{Oh:2008}:
\begin{equation}
\label{oh}
\Upsilon^{3.6\mu m}_* = 0.92\times \Upsilon^{K_s}_* - 0.05,   
\end{equation}
which relates the stellar mass-to-light ratio at $3.6\,\mu$m and that of the $K_s$-band, we obtained a constant stellar mass-to-light ratio of $\Upsilon^{K_s}_* = 0.7$ for the $K_s$-band images. The latest relation: $\Upsilon^{3.6\mu m}_* = 1.03\times \Upsilon^{K_s}_* - 0.16 $ (J.Schombert, private communication), which is based on a larger $K_s-3.6\,\mu$m dataset, also revealed a consistent value for $\Upsilon^{K_s}_*$. 

For our three $r^{\prime}$-band data, we used an average stellar mass-to-light ratio of $\Upsilon_{*}^{r^{\prime}} \equiv M_*/L_{r^{\prime}} = 2.8$ to obtain the corresponding stellar masses. $\Upsilon_{*}^{r^{\prime}}$ was calibrated using
\begin{equation}
\frac{M_*}{L_{r^{\prime}}} = \left(\frac{L_{K_s}}{L_{r^{\prime}}}\right) \left(\frac{M_*}{L_{K_s}}\right),   
\end{equation}
ensuring that the galaxy stellar masses are consistent with the masses obtained using $K_s$-band magnitudes (obtained from 2MASS imaging of these galaxies), and a stellar mass-to-light ratio of $\Upsilon_{*}^{K_s} = 0.7$. We present the spheroid and total galaxy stellar masses for our galaxies in Table \ref{data_table}.

\subsection{Comparison of Stellar Masses}
\label{Comparison of Masses}

\begin{figure}
\begin{center}
\includegraphics[clip=true,trim= 12mm 3mm 16mm 14mm,width=\columnwidth]{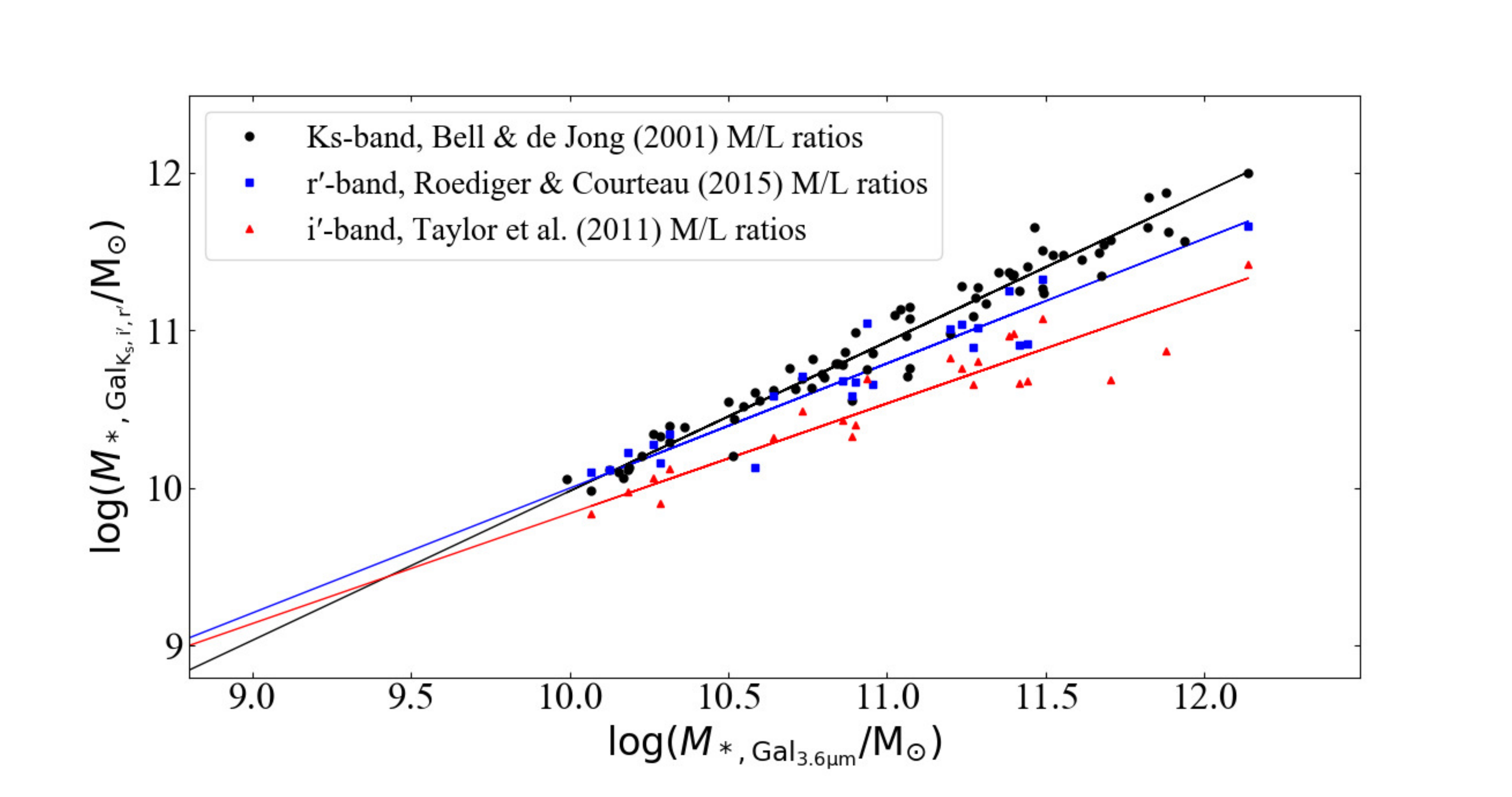}
\caption{Comparison of the galaxy stellar masses for our sample. The masses on the horizontal axis are calculated from $3.6\,\mu$m imaging with $\Upsilon_{*}^{3.6\,\mu m}=0.6$, while the ($K_s$-, $r^{\prime}$-, and $i^{\prime}$- band)-derived masses are shown on the vertical axis. The \textbf{black} dots represent the total galaxy stellar masses of 71 galaxies based on improved $K_s$-band magnitudes and ($B-K_s$ color-dependent) $K_s$-band stellar mass-to-light ratios from \citet{Bell:deJong:2001}. \textcolor{blue}{Blue} squares show the total galaxy stellar masses of 23 galaxies obtained using $r^{\prime}$-band magnitudes and $g^{\prime}-r^{\prime}$ color-dependent mass-to-light ratios from \citet{Roediger:Courteau:2015}, and the \textcolor{red}{red} triangles mark the total galaxy stellar masses of the same 23 galaxies calculated using $i^{\prime}$-band magnitudes and $g^{\prime}-i^{\prime}$ color-dependent mass-to-light ratios from \citet{Taylor:2011}.
Black, blue, and red lines are the least-square regression lines defining a relation between these masses. 
}
\label{Stellar_mass_comparison}
\end{center}
\end{figure}

Here we compare the galaxy stellar masses measured using the $3.6\,\mu$m-band images (calculated as described above) with the galaxy stellar masses calculated using (already available) $K_s$, $i^{\prime}$, and $r^{\prime}$-band magnitudes and three different formula for the corresponding stellar mass-to-light ratios. The comparison and the best fit lines are shown in Figure \ref{Stellar_mass_comparison}, where the horizontal-axis designates the ($3.6\,\mu$m-band)-derived masses, labeled $\log(M_{*,Gal_{3.6\mu m}}/M_\odot)$, and the vertical-axis depicts the masses based on the $K_s$, $i^{\prime}$ and $r^{\prime}$ band magnitudes, labeled $\log(M_{*,Gal_{K_s,i^\prime,r^\prime}}/M_\odot)$. 

\begin{figure}
\begin{center}
\includegraphics[clip=true,trim= 3mm 3mm 14mm 14mm,width=\columnwidth]{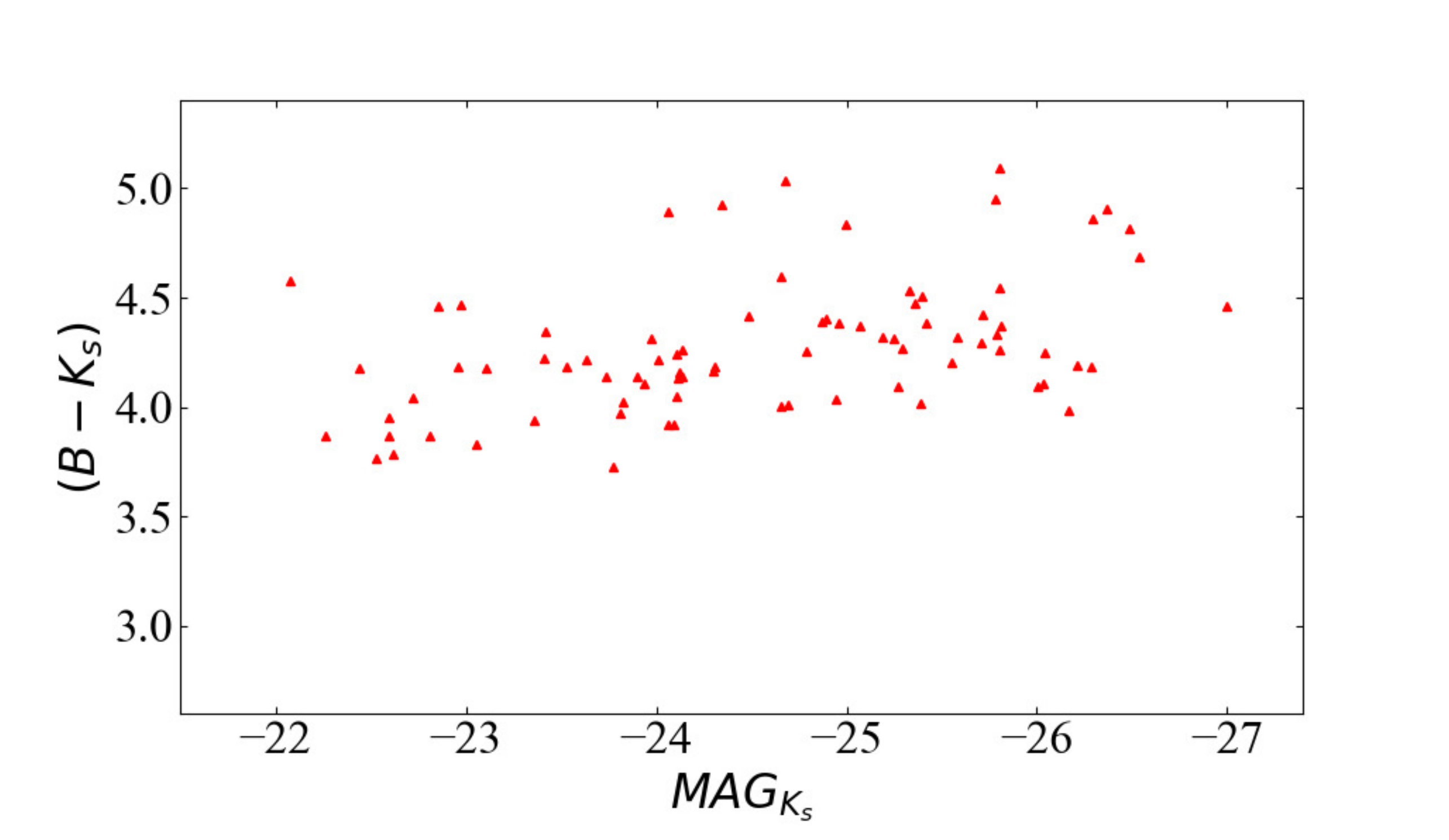}
\caption{($B-K_s$)-color versus $K_s$-band absolute magnitude (in Vega system) diagram for 82 ETGs. Most of our sample resides along the relatively flat arm (for $MAG_{K_s} < -22 \,$mag) of the color-magnitude diagram presented by \citet{Graham:Soria:2018}.
}
\label{color_mag}
\end{center}
\end{figure}

The black dots in Figure \ref{Stellar_mass_comparison} show the masses of 71 galaxies calculated here using $K_s$-band magnitudes and ($B-K_s$ color-dependent) $K_s$-band stellar mass-to-light ratios from \citet[][their Table 1]{Bell:deJong:2001}, placed with respect to our ($3.6\,\mu$m-band) stellar masses. The $K_s$ and $B$-band magnitudes were obtained from the 2MASS catalog \citep{Jarrett:2003} and the Third Reference Catalogue (RC3) of Bright Galaxies \citep{RC3:1991}, respectively. The $K_s$-band magnitudes obtained from the 2MASS data reduction pipelines are usually underestimated \citep{Schombert:Smith:2012}, therefore we used Equation 1 from \citet{Scott:Graham:2013} to correct for this. The size of this correction was $< 0.35$ mag. The $K_s$-band stellar mass-to-light ratios were brought to a Chabrier IMF, from the scaled/diet Salpeter IMF used by \citet{Bell:deJong:2001}, by subtracting an IMF dependent constant of 0.093 dex \citep{Taylor:2011, Mitchell:2013}. In Figure \ref{color_mag}, we also present the ($B-K_s$)-color versus the $K_s$-band magnitude for our sample, which is consistent with the color-magnitude diagram presented by \citet[][their Figure 11]{Graham:Soria:2018}, implying that our galaxies belong to the red-sequence, which flattens ($B-K_s \approx 4$ ) at bright magnitudes ($MAG_{K_s} < -22 \,$mag). 

The red triangles in Figure \ref{Stellar_mass_comparison} are the masses of 23 galaxies calculated using $i^{\prime}$-band magnitudes and ($g^{\prime}-i^{\prime}$ color-dependent) $i^{\prime}$-band stellar mass-to-light ratios (based on a Chabrier IMF) from \citet[][their Equation 7]{Taylor:2011}. 

The blue squares represent the masses of 23 galaxies calculated using $r^{\prime}$-band magnitudes and ($g^{\prime}-r^{\prime}$ color-dependent) $r^{\prime}$-band stellar mass-to-light ratios from \citet{Roediger:Courteau:2015}, which are based on the Stellar Population Synthesis (SPS) model by \citet{Conroy:Gunn:White:2009}. The apparent galaxy magnitudes in the $g^{\prime}$, $r^{\prime}$, and $i^{\prime}$-bands were obtained from the SDSS data release 6 \citep{SDSS:DR6:2008}. 

The black, blue, and red lines in Figure \ref{Stellar_mass_comparison} represent the least-squares fits to the three corresponding types of data points. We found that there is almost a linear one-to-one relationship between the ($K_s$-band)-derived masses (black line) and our ($3.6\,\mu$m)-derived masses. The galaxy stellar masses based on $r^{\prime}$- and $i^{\prime}$-band magnitudes (blue line and red line, respectively) are systematically offset. Although the offset is small, it systematically increases at higher galaxy masses. Such an offset has been noticed in a few other studies  \citep[e.g.][]{Taylor:2011, Graham:Soria:Davis:2018}. The systematic offset between the above three lines can be attributed mainly to the initial mass functions, star formation rates, and the stellar evolutionary histories assumed to derive the mass-to-light ratios, and possibly some systematic uncertainties introduced in the apparent magnitudes by various telescope pipeline processes. 

Figure \ref{Stellar_mass_comparison} mainly serves to depict that the use of different stellar mass-to-light ratio prescriptions for luminosities (magnitudes) obtained in different bands can produce different stellar masses for a galaxy and its components \citep[see][for a detailed comparison of masses calculated using different methods]{Kannappan:Gawiser:2007}. 
In passing, we note that we will explore if this may be a factor contributing to the offset observed by \citep{Shankar:2016} between galaxies with directly measured black hole masses and the population at large.

Differences in estimated stellar mass will lead to different estimates of a galaxy's black hole mass when using the black hole mass scaling relations presented here and elsewhere. Hence, in our forth-coming equations for the $M_{BH}$--$M_{*,sph}$ and $M_{BH}$--$M_{*,gal}$ relations, we are including a conversion or correcting coefficient, $\upsilon$ (lower case upsilon), for the stellar masses \citep[see][]{Davis:2018:a}. This stellar mass correcting coefficient accounts for the difference in stellar mass of a galaxy due to either the difference in the stellar mass-to-light ratio ($\Upsilon_*$) used for the same passband, or due to a different passband magnitude as well as a different mass-to-light ratio applied to it. If $\Upsilon_{*}^{IRAC1}$ is a user-preferred Spitzer $3.6\,\mu$m-band stellar mass-to-light ratio, the correction coefficient $\upsilon_{*,IRAC1}$ is given by, 
\begin{equation}
\upsilon_{*,IRAC1}= \frac{\Upsilon_{*}^{IRAC1}}{0.6},  
\end{equation}
where 0.6 is the stellar mass-to-light ratio for the IRAC1 ($3.6\,\mu$m) passband used in this work, adopted from \citet{Meidt:2014}.

The correcting coefficient ($\upsilon$), for the masses ($M_{*,K_s}$, $M_{*,r^{\prime}}$, $M_{*,i^{\prime}}$) derived using the $K_s$-, $i^{\prime}$-, and $r^{\prime}$-band magnitudes with the three stellar mass-to-light ratio trends shown in Figure \ref{Stellar_mass_comparison}, can be expressed as follows:
\begin{equation}
\label{Ks}
\log\upsilon_{*,K_s} = -0.06 \log\left( \frac{ M_{*,K_s}}{10^{10} M_{\odot}} \right) - 0.06,    
\end{equation}

\begin{equation}
\label{r}
\log\upsilon_{*,r^{\prime}} = -0.26 \log\left(\frac{M_{*,r^{\prime}}}{10^{10} M_{\odot}}\right) + 0.03,     
\end{equation}

\begin{equation}
\label{i}
\log\upsilon_{*,i^{\prime}} = -0.43 \log\left(\frac{M_{*,i^{\prime}}}{10^{10} M_{\odot} }\right) - 0.21.     
\end{equation}

These equations are obtained by calculating the offset of the three lines shown in Figure \ref{Stellar_mass_comparison} from our ($3.6\,\mu$m)-derived galaxy masses calculated in Section \ref{M_L ratio}.

\subsection{Error Analysis}
\label{error}
Our spheroid and galaxy stellar masses depend on three main independent quantities, which are: the stellar mass-to-light ratio ($\Upsilon_*$); distance ($D$); and the apparent magnitude ($m$). We have estimated the error in the above three quantities and added them in quadrature.

 Our galaxy sample, dominated by near-infrared imaging, enables us to apply a relatively stable stellar mass-to-light ratio adopted from \citet{Meidt:2014} and \citet{Querejeta:2015}. \citet{Meidt:2014} recommend the use of a more liberal $15\%$ uncertainty on the $3.6\,\mu$m stellar mass-to-light ratio, accounting for an atypical evolutionary history or non-stellar emissions (which are dominant in red colors). As $\Upsilon^{r^{\prime}}_{*}$ for our $r^{\prime}$-band images are calibrated against 2MASS imaging and $\Upsilon^{K_s}_{*}$, and $\Upsilon^{K_s}_{*}$ in turn is derived from $\Upsilon^{3.6\mu m}_{*}$, as described in Section \ref{M_L ratio}, we assign a constant uncertainty of $15\%$ to the stellar mass-to-light ratios for all the galaxies.

For most of the 42 galaxies (Table \ref{data_table}) that we modeled, we obtained the error in their distances from the publication which presented their directly measured SMBH mass. For the rest of the galaxies (including the galaxies from \citet{Savorgnan:Graham:2016}), we are using a constant error of $7\%$ in their distances, which is a typical percentage error in the ($Virgo + GA + Shapley$)-corrected Hubble flow distances, obtained from NASA/IPAC Extragalactic Database. 

Some of the sources of error in the apparent magnitudes are imprecise sky subtraction; error in the telescope's PSF size measurement; and error in the decomposition of the galaxy light. The decomposition error can include an error due to neglecting a component of the galaxy; misinterpreting a component's size or position; error in the calibrated zero-point magnitude; misinterpreting nuclear components or being unable to resolve it; etc. It is nearly impossible to quantify all these errors. 

If we assume that we have used an accurate method to measure the sky level and the telescope's PSF, and trust various telescope pipelines (where we downloaded our images) for their zero-point flux calibration, then our main source of error in magnitude will be the error in the galaxy light decomposition process. Although, \textsc{Profiler} provides the formal random error for each fit parameter of the various components of a galaxy, which is the rms error obtained by least square minimization between data and the fitting function, it is very small. To better quantify the uncertainty in the decomposition, we have followed the (light profile fit-quality) grading scheme described by \citet[][in their section-4.2.1]{Savorgnan:Graham:2016}, except that we have assigned a symmetric error of 0.2 mag, 0.6 mag, and 0.8 mag to the spheroidal component of our grade-1, grade-2, and grade-3 galaxies, respectively.

As we are dealing with the stellar masses in log, we calculate these errors in log (dex). An error of $\delta m$ mag in apparent magnitude, a $\delta D$ error in distance, and a $\delta\Upsilon_*$ error in the stellar mass-to-light ratio, added in quadrature, give us the error in the stellar mass (in dex), as
\begin{equation}
\footnotesize{\scriptsize}
 \delta \log M = \sqrt{\left(\frac{\delta m}{2.5}\right)^2 + \left(2\frac{\delta D}{D\ln(10)}\right)^2 + \left(\frac{\delta\Upsilon_*}{\Upsilon_* \ln(10)}\right)^2}. 
 \label{quad_error}
\end{equation}

We assign a constant error of $0.12$ dex to the galaxy masses, which is equivalent to the total quadrature error (calculated using Equation \ref{quad_error}) assigned to the spheroid masses of our grade-1 galaxies, which are mostly single component galaxies. 


\startlongtable
\begin{deluxetable*}{lclllcrrr}
\tablecolumns{9}
\tablecaption{Galaxy Sample}
\tabletypesize{\scriptsize}
\tablehead{
\colhead{Galaxy}     & \colhead{Type} & \colhead{Core} & \colhead{Distance}  & \colhead{$\log\left(M_{BH}/M_{\odot}\right)$}  & \colhead{$MAG_{sph}$} & \colhead{$MAG_{gal}$}     & \colhead{$\log\left(M_{*,sph}/M_{\odot}\right)$} & \colhead{$\log\left(M_{*,gal}/M_{\odot}\right) $}\\
 \colhead{}          &  \colhead{}      &  \colhead{}     &   \colhead{(Mpc)}   &  \colhead{ }   & \colhead{(mag)} & \colhead{(mag)} &   \colhead{ }        &  \colhead{ }    \\
     \colhead{(1)}      &     \colhead{(2)}        &    \colhead{(3)}   &     \colhead{(4)}  &  \colhead{(5)}  & \colhead{(6)}   &     \colhead{(7)}      &      \colhead{(8)}  & \colhead{(9)}   
}
\startdata
A1836 BCG\tablenotemark{a}   & E1-2      & yes   & 158.00$\pm$11.06     & 9.59$\pm$0.06[5a,G]    &  -24.56$\pm$0.20  & -24.56$\pm$0.20     & 11.70$\pm$0.12  
& 11.70$\pm$0.12  \\
A3565 BCG     & E1        & no    & 40.70 $\pm$2.90[4a]      & 9.04$\pm$0.09[5a,G]              &  -23.22$\pm$0.6    & -23.26$\pm$0.20       & 11.47$\pm$ 0.26               & 11.49$\pm$0.12  \\
NGC 0307\tablenotemark{b}    & SAB0      & no    & 52.80$\pm$3.70      & 8.34$\pm$0.13[5c,S]      &  -20.31$\pm$0.80    & -21.14$\pm$0.20      & 10.43$\pm$0.33                    
& 10.76$\pm$0.12    \\
NGC 0404    & S0        & no    & 3.06$\pm$0.37      & 4.85$\pm$0.13[5d,S]                &  -14.43$\pm$0.60    &  -17.33$\pm$0.20      & 7.96$\pm$0.27                     & 9.12$\pm$0.12   \\
NGC 0524    & SA0(rs)   & yes   & 23.30$\pm$1.63      & 8.92$\pm$0.10[5e,S]                &   -20.97$\pm$0.60  & -22.21$\pm$0.20     & 10.57$\pm$0.26                    & 11.07$\pm$0.12  \\
NGC 1194    & S0        & no    & 53.20$\pm$3.70      & 7.81$\pm$0.04[5f,M]                &    -21.31$\pm$0.80   & -21.87$\pm$0.20       & 10.71$\pm$0.33                    & 10.94$\pm$0.12  \\
NGC 1275    & E      & no    & 72.9$\pm$5.10[4a]      & 8.90$\pm$0.20[5g,G]                &   -24.14$\pm$0.60    & -24.23$\pm$0.20       & 11.84$\pm$0.26              & 11.88$\pm$0.12    \\
NGC 1374    & S0        & no?   & 19.20$\pm$1.34      & 8.76$\pm$0.05[5h,S]                &    -20.09$\pm$0.60   &  -20.83$\pm$0.20     & 10.22$\pm$0.26                    & 10.52$\pm$0.12  \\
NGC 1407    & E         & yes   & 28.05$\pm$3.37      & 9.65$\pm$0.08[5h,S]                &  -23.19$\pm$0.60  &   -23.34$\pm$0.02  & 11.46$\pm$0.27                        & 11.52$\pm$0.12  \\
NGC 1550\tablenotemark{a}    & E1        & yes   & 51.57$\pm$3.61      & 9.57$\pm$0.06[5h,S]     &  -23.14$\pm$0.20   & -23.14$\pm$0.20    & 11.13$\pm$0.12           & 11.13$\pm$0.12  \\
NGC 1600    & E3        & yes   & 64.00$\pm$4.48      & 10.23$\pm$0.05[5i,S]                &  -24.09$\pm$0.20  &  -24.09$\pm$0.20   & 11.82$\pm$0.12                       & 11.82$\pm$0.12  \\
NGC 2787    & SB0(r)    & no    & 7.30$\pm$0.51      & 7.60$\pm$0.06[5j,G]                &    -17.35$\pm$0.60   &  -19.51$\pm$0.20     & 9.13$\pm$ 0.26                      & 9.99$\pm$0.12   \\
NGC 3665    & S0        & no    & 34.70$\pm$2.43      & 8.76$\pm$0.10[5k,G]                &    -22.12$\pm$0.60  & -22.74$\pm$0.20       & 11.03$\pm$0.26                      & 11.28$\pm$0.12  \\
NGC 3923    & E4        & yes   & 20.88$\pm$2.70      & 9.45$\pm$0.13[5l,S]                &   -23.02$\pm$0.20 & -23.02$\pm$0.20     & 11.40$\pm$0.15                        & 11.40$\pm$0.12  \\
NGC 4026    & SB0       & no    & 13.20$\pm$0.92      & 8.26$\pm$0.11[5m,S]                &   -19.82$\pm$0.80  & -20.44$\pm$0.20     & 10.11$\pm$0.33                      & 10.36$\pm$0.12  \\
NGC 4339    & S0        & no    & 16.00$\pm$1.33      & 7.63$\pm$0.33[5n,S]                &  -18.72$\pm$0.60  & -19.96$\pm$0.20     & 9.67$\pm$0.26                      & 10.17$\pm$0.12  \\
NGC 4342    & ES/S0     & no    & 23.00$\pm$1.00      & 8.65$\pm$0.18[5o,S]                &  -19.38$\pm$0.60   &  -20.20$\pm$0.20   & 9.94$\pm$0.25            & 10.26$\pm$0.12  \\
NGC 4350    & EBS       & no    & 16.80$\pm$1.18      & 8.86$\pm$0.41[5p,SG]    &  -20.22$\pm$0.60   & -20.90$\pm$0.20   &  10.28$\pm$0.26    & 10.55$\pm$0.12  \\
NGC 4371    & SB(r)0    & no    & 16.90$\pm$1.48      & 6.84$\pm$0.08[5l,S]                &  -19.27$\pm$0.60  & -21.03$\pm$0.20   & 9.89$\pm$0.26                          & 10.60$\pm$0.12  \\
NGC 4429    & SB(r)0    & no    & 16.50$\pm$1.60      & 8.18$\pm$0.09[5q,G]                &   -20.69$\pm$0.60   &  -21.79$\pm$0.20      & 10.46$\pm$ 0.26             & 10.90$\pm$0.12  \\
NGC 4434    & S0        & no    & 22.40$\pm$1.57      & 7.84$\pm$0.17[5n,S]                &   -19.32$\pm$0.60  & -20.00$\pm$0.20     & 9.91$\pm$0.26                      & 10.18$\pm$0.12  \\
NGC 4486B\tablenotemark{b}   & E1        & no    & 15.30$\pm$0.32      & 8.76$\pm$0.24[5r,S]    &  -17.90$\pm$0.80  &  -17.90$\pm$0.20   & 9.46$\pm$ 0.33                      & 9.46$\pm$0.12   \\
NGC 4526    & S0        & no    & 16.90$\pm$1.69      & 8.67$\pm$0.04[5s,G]                &  -21.27$\pm$0.60    & -22.14$\pm$0.20       & 10.70$\pm$ 0.26                    & 11.04$\pm$0.12   \\
NGC 4552    & E         & no    & 14.90$\pm$0.95      & 8.67$\pm$0.05[5t,S]                &  -21.75$\pm$0.60   & -21.92$\pm$0.20    & 10.88$\pm$ 0.25                          & 10.95$\pm$0.12   \\
NGC 4578    & S0(r)    & no    & 16.30$\pm$1.14      & 7.28$\pm$0.35[5n,S]                &  -18.97$\pm$0.60   & -20.10$\pm$0.20    & 9.77$\pm$ 0.26                     & 10.23$\pm$0.12   \\
NGC 4649    & E2      & yes   & 16.40$\pm$1.10      & 9.67$\pm$0.10[5u,S]                &  -23.14$\pm$0.20   & -23.14$\pm$0.20    & 11.44$\pm$ 0.12                & 11.44$\pm$0.12   \\
NGC 4742    & S0        & no    & 15.50$\pm$1.15      & 7.15$\pm$0.18[5v,S]                &  -19.21$\pm$0.60  & -19.92$\pm$0.20     & 9.87$\pm$ 0.26                         & 10.15$\pm$0.12   \\
NGC 4751\tablenotemark{a}    & S0        & yes?  & 26.92$\pm$1.88      & 9.15$\pm$0.05[5h,S]      &   -21.53$\pm$0.60 & -22.11$\pm$0.20  & 10.49$\pm$ 0.26                  & 10.72$\pm$0.12   \\
NGC 4762    & SB0       & no    & 22.60$\pm$3.39      & 7.36$\pm$0.15[5n,S]                &  -19.45$\pm$0.60   & -22.19$\pm$0.20    & 9.97$\pm$ 0.28                     & 11.06$\pm$0.12   \\
NGC 5018    & S0        & no    & 40.55$\pm$4.87      & 8.02$\pm$0.09[5l,S]                &  -21.97$\pm$0.60  & -22.91$\pm$0.20    & 10.98$\pm$ 0.27                       & 11.35$\pm$0.12   \\
NGC 5252    & S0        & no    & 96.80$\pm$6.78      & 9.00$\pm$0.40[5w,G]                &   -21.67$\pm$0.60   &  -23.00$\pm$0.20      & 10.85$\pm$ 0.26                  & 11.38$\pm$0.12   \\
NGC 5328\tablenotemark{a}    & E1        & yes   & 64.10$\pm$4.49      & 9.67$\pm$0.15[5h,S]      &  -24.03$\pm$0.20   &  -24.03$\pm$0.20   & 11.49$\pm$ 0.12                        & 11.49$\pm$0.12   \\
NGC 5419    & E2-3      & yes   & 56.20$\pm$3.93      & 9.86$\pm$0.14[5x,S]                &   -23.15$\pm$0.20  & -23.15$\pm$0.20    & 11.44$\pm$ 0.12                     & 11.44$\pm$0.12   \\
NGC 5516\tablenotemark{a}    & E1-2      & yes?  & 58.44$\pm$4.09      & 9.52$\pm$0.06[5h,S]                 &  -23.91$\pm$0.20  & -23.91$\pm$0.20   & 11.44$\pm$ 0.12                        & 11.44$\pm$0.12   \\
NGC 5813    & S0        & yes   & 31.30$\pm$2.60      & 8.83$\pm$0.06[5y,S]                &  -21.68$\pm$0.60  & -22.62$\pm$0.20    & 10.86$\pm$ 0.26                           & 11.23$\pm$0.12   \\
NGC 5845    & ES        & no    & 25.20$\pm$1.76      & 8.41$\pm$0.22[5z,S]                &  -19.83$\pm$0.60   &  -20.32$\pm$0.20   & 10.12$\pm$ 0.26                     & 10.32$\pm$0.12   \\
NGC 6086\tablenotemark{b}    & E        & yes   & 138.00$\pm$9.66      & 9.57$\pm$0.16[5aa,S]   &  -23.03$\pm$0.60   & -23.03$\pm$0.20    & 11.52$\pm$ 0.26                   & 11.52$\pm$0.12   \\
NGC 6861    & ES        & no    & 27.30$\pm$4.49      & 9.30$\pm$0.08[5h,S]                 &   -21.88$\pm$0.60  & -22.10$\pm$0.20   & 10.94$\pm$0.29                         & 11.02$\pm$0.12   \\
NGC 7052    & E4        & yes   & 66.40$\pm$4.65[4a]      & 8.57$\pm$0.23[5ab,G]                &  -23.19$\pm$0.20     &  -23.19$\pm$0.20     & 11.46$\pm$0.12               & 11.46$\pm$0.12   \\
NGC 7332    & SB0(pec)  & no    & 24.89$\pm$2.49      & 7.11$\pm$0.20[5ac,S]                &  -20.08$\pm$0.80   &  -21.63$\pm$0.20  & 10.22$\pm$0.34                    & 10.84$\pm$0.12   \\
NGC 7457    & S0        & no    & 14.00$\pm$0.98      & 7.00$\pm$0.30[5ad,S]                &   -18.04$\pm$0.60   & -20.00$\pm$0.20       & 9.40$\pm$0.26           & 10.19$\pm$0.12       
\label{data_table}
\enddata
\tablecomments{Columns:
(1) Galaxy name.
(2) Morphology, based on our decompositions.
(3) Presence of partially depleted core.
(4) Distance, primarily from the corresponding paper presenting the measured SMBH mass ($M_{BH}$). For some galaxies which did not have any error associated with these, we assigned an error of $7\%$ (see Section \ref{error}).
(5) Directly measured super-massive black hole mass, reference, and method used (S: Stellar dynamics, G: Gas dynamics, M: $H_2O$ Megamaser). The error in $M_{BH}$, obtained from the corresponding papers, was added in quadrature with the distance error. 
(6) Spheroid absolute magnitude at $3.6 \,\mu$m, unless otherwise noted in Column 1 (AB mag system).
(7) Total galaxy absolute magnitude at $3.6 \,\mu$m, unless otherwise noted in Column 1 (AB mag system).
(8) Spheroidal mass measured in this work, see Section \ref{M_L ratio}.
(9) Galaxy mass measured in this work.\\
References: 4a=NED (Virgo + GA + Shapley)-corrected Hubble flow distances; 5a=\citet{DallaBonta:2009}; 5b=\citet{Walsh:2017}; 5c=\citet{Erwin:2018}; 5d=\citet{Nguyen:2017}; 5e=\citet{Krajnovic:2009}; 5f =\citet{Kuo:2011}; 5g=\citet{Scharwachter:2013}; 5h=\citet{Rusli:2013}; 5i=\citet{Thomas:2016}; 5j=\citet{Sarzi:2001}; 5k=\citet{Onishi:2017}; 5l=\citet{Saglia:2016}; 5m=\citet{Gultekin:2009}; 5n=\citet{Krajnovic:2018}; 5o=\citet{Cretton:VanDenBosch:1999}; 5p=\citet{Pignatelli:2001}; 5q=\citet{TimothyDavis:2018}; 5r=\citet{Kormendy:1996}; 5s=\citet{Gould:2013}; 5t=\citet{Hu:2008}; 5u=\citet{Shen:Gebhardt:2010}; 5v=\citet{Tremaine:ngc4742:2002}; 5w=\citet{Capetti:2005}; 5x=\citet{Mazzalay:2016}; 5y=\citet{Hu:2008}; 5z=\citet{Gebhardt:2003}; 5aa=\citet{McConnell:2011}; 5ab=\citet{Marel:Bosch:1998}; 5ac=\citet{Batcheldor:2013}; 5ad=\citet{Schulze:Gebhardt:2011}. 
}
\tablenotetext{a}{\,\,2MASS $K_s$-band galaxy images}
\tablenotetext{b}{\,\,SDSS $r^{\prime}$-band galaxy images}
\end{deluxetable*}

\pagebreak

\section{Results and discussion}
\label{Results}
We performed a Bivariate Correlated Errors and Intrinsic Scatter (\textsc{BCES}) regression \citep{Akritas:Bershady:1996} between the SMBH masses and both the spheroid masses and the total galaxy masses of our sample. \textsc{BCES} is simply an extension of Ordinary Least Squares (OLS) estimator permitting dependent measurement errors in both the variables. We use the bisector line obtained by the \textsc{BCES}\footnote{To perform the \textsc{BCES} regression, we used the \textsc{PYTHON} script (available at \url{https://github.com/rsnemmen/BCES}) written by \citet{Nemmen:2012}, we modified it to calculate the intrinsic scatter \citep[Equation 1 from][]{Graham:Driver:2007}.} regression; this line symmetrically bisects the regression lines obtained using $\textsc{BCES(X\textbar Y)}\footnote{Minimizes scatter in the X-direction.}$ and $\textsc{BCES(Y\textbar X)}\footnote{Minimizes scatter in the Y-direction.}$. The bisector regression line offers equal treatment to the measurement errors in both the coordinates, and allows for intrinsic scatter. In addition to the \textsc{BCES} routine, we also used the modified \textsc{FITEXY} routine \citep{Press:1992, Tremaine:ngc4742:2002} to perform a regression on our data for the $M_{BH}$--$M_{*,sph}$ and $M_{BH}$--$M_{*,sph}$ relations. We found results highly consistent with that of the \textsc{BCES} regression, within the $1\sigma$ bounds.

In our analysis, we have excluded eight galaxies (MRK~1216, NGC~404, NGC~1277, NGC~1316, NGC~2787,  NGC~4342, NGC~4486B, and NGC~5128), which leaves us with a reduced dataset of 76 ETGs. In all our plots hereafter, these galaxies are shown by a black star (except for MRK 1216). We excluded MRK~1216 from our regression analysis because we did not obtain a suitably resolved and deep image to determine the spheroidal component of this galaxy.

NGC 1316 (Fornax-A) and NGC 5128 (Cen A) are galaxy mergers in progress. According to \citet{Kormendy:Ho:2013}, these two galaxies have much higher bulge masses compared to their central supermassive black hole masses, which can make them stand out in the black hole mass scaling relations. 

NGC 404 has the lowest SMBH mass in our sample. \citet{Nguyen:2017} provide a measured black hole mass of $7^{+1.5}_{-2.0} \times 10^4 M_\odot$, using Jeans Anisotropic Modeling (JAM) of stellar orbits, along with a $3\sigma$ upper limit of $1.5 \times 10^5 M_\odot$ in $M_{BH}$. Although, NGC 404 does not appear to be an outlier in our dataset, as it follows the regression lines at the low-mass end, we still exclude it as it would anchor the low-mass end of the relationship and we do not want our regression lines to be biased by any individual galaxy.

We also exclude NGC~4342  and NGC~4486B because they have been tidally stripped due to the gravitational pull of their nearby massive companion galaxies, NGC~4365 \citep{Blom:Forbes:2014} and NGC~4486 \citep{Batcheldor:2010}, respectively. NGC~4342 and NGC~4486B are left with a significantly reduced galaxy mass and can be seen clearly offset in our $M_{BH}$--$M_{*,sph}$ and $M_{BH}$--$M_{*,gal}$ diagrams (towards the low-mass side of the $M_{*,sph}$ and $M_{*,gal}$ coordinate axes). NGC 221 (M32) is another, similar, well known offset galaxy due to the tidal stripping from the massive companion galaxy M31 \citep[e.g.,][]{Graham:CE:2002}. Such compact elliptical galaxies are relatively rare among the general population and are recommended to be excluded from $M_{BH}$--$M_{*,gal}$ scaling relations \citep[see][]{Graham:Soria:2018}.

NGC~1277 (peculiar morphology) and NGC~2787 are two disk galaxies which are potential outliers at the high- and low-mass end of our relations, respectively.  They have a torquing effect on our regression lines, especially for the sub-category of galaxies with a disk (ES/S0).  We have therefore excluded these galaxies from our regressions to avoid biasing the slope of our scaling relations. Furthermore, the stellar mass for NGC~1277 is measured from V-band imaging \citep{Graham:Durr:Savorgnan:2016} and a stellar mass-to-light ratio based on an unusual bottom heavy IMF \citep{Martin:2015}. According to \citet[][their Figure 8]{Courteau:2014}, stellar mass-to-light ratios based on a bottom heavy IMF can be a factor $\sim$6  higher than stellar mass-to-light ratios based on the Chabrier IMF that we have adopted, which is likely to be the principal reason for NGC~1277 outstanding at the high-mass end of our relations.

The above galaxies remain excluded in all the regressions presented in this paper. In Figures \ref{CS_S}-\ref{ETG_LTG}, we identify an additional five galaxies with a peculiar morphology, to investigate if they might be outliers, but they are included in the regressions. 

In our search for the underlying relation between super-massive black hole mass and host galaxy property, we explored various possibilities for the scaling relations by dividing the galaxy sample into different categories. Specifically: S{\'e}rsic and core-S{\'e}rsic galaxies; galaxies with and without a disk; and galaxies with and without a bar. We will analyze and discuss the scaling relations for these categories in the following sections.

\subsection{S{\'e}rsic and Core-S{\'e}rsic Galaxies}

Core-S{\'e}rsic galaxies are massive ETGs with a central supermassive black hole that likely formed from the merging of the central black holes of two or more galaxies \citep{Begelman:1980, Graham:2004, Merritt:2006}. They occupy the high-mass end of the black hole mass scaling relations. The discovery of the bent   $M_{BH}$--$L_{sph}$ ($M_{*,sph}$) relation for S{\'e}rsic and core-S{\'e}rsic galaxies was based on a mixed sample of elliptical, lenticular, and spiral galaxies \citep{Graham:2012, Graham:Scott:2013, Scott:Graham:2013}. In our work, we investigated the nature of the above relation based on a larger sample of only early-type galaxies. 

\begin{figure*}
\begin{center}
\includegraphics[clip=true,trim= 9mm 2mm 12mm 10mm,height=6cm,width=\textwidth]{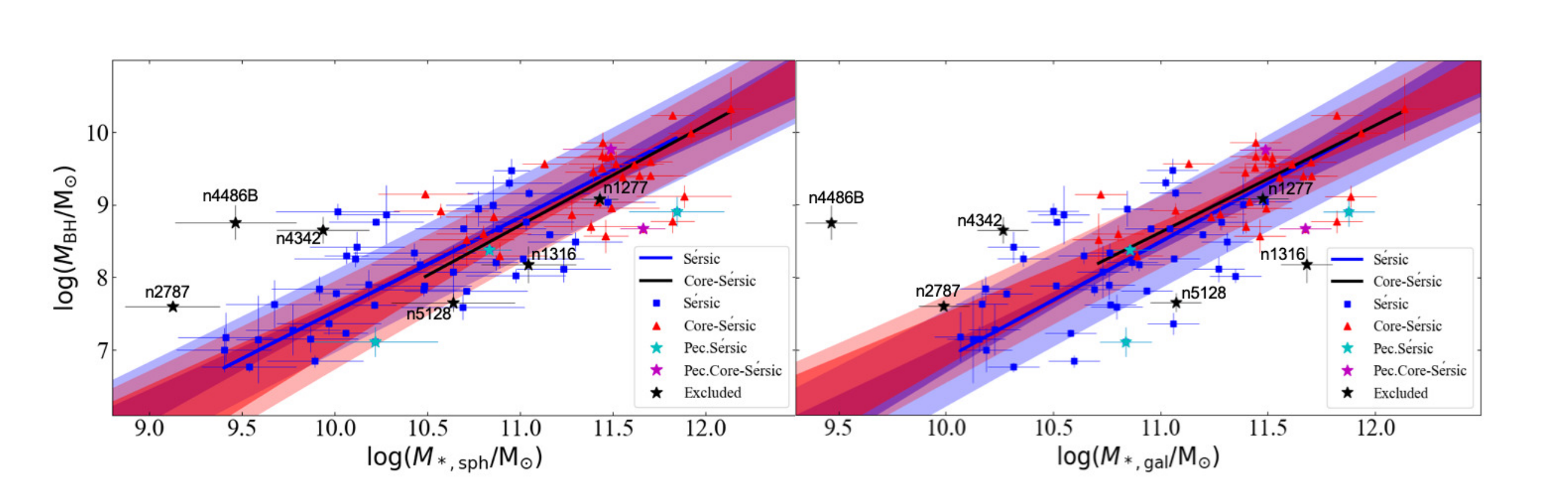}
\caption{Black hole mass versus spheroid stellar mass (left) and total galaxy stellar mass (right). Over-plotted are S{\'e}rsic galaxies (\textcolor{blue}{blue} squares) and core-S{\'e}rsic galaxies (\textcolor{red}{red} triangles). The \textcolor{blue}{blue} and \textbf{black} lines represent the corresponding bisector regression lines of S{\'e}rsic and core-S{\'e}rsic galaxies, and the dark \textcolor{blue}{blue} and dark \textcolor{red}{red} bands display the $\pm 1\sigma$ uncertainty on the slope and intercept of the lines. The light \textcolor{blue}{blue} and light \textcolor{red}{red} regions show the $\pm 1 \sigma$ rms scatter of the data about the \textcolor{blue}{blue} and \textbf{black} regression lines for S{\'e}rsic and core-S{\'e}rsic galaxies, respectively. Peculiar S{\'e}rsic (three \textcolor{cyan}{cyan} stars) and peculiar core-S{\'e}rsic (two \textcolor{magenta}{magenta} stars) galaxies are depicted with a different symbol but they were  included in the regressions. The six \textbf{black} stars are galaxies excluded from the regression:  NGC~1316 and NGC~5128 are mergers; NGC~4486B and NGC~4342 are  stripped  galaxies; and NGC~1277 and NGC~2787 are potential outliers at the extremities of the spheroid mass range which may bias the regression line. Their relative position remains the same from Figures~\ref{CS_S} to \ref{BNB1}. We do not show the remaining two excluded galaxies: NGC~404 lies at low mass end of the diagrams (see Figure \ref{ETG_LTG}) and for MRK~1216, we could not properly measure its spheroid and total galaxy stellar masses due to the lack of a good image. It is evident that both populations overlie with each other, leading us to the conclusion that there is no \enquote{bend} in the $M_{BH}$--$M_{*,sph}$ nor $M_{BH}$--$M_{*,gal}$ relations for ETGs with $M_{BH} \gtrsim 10^7\, M_{\odot}$ due to S{\'e}rsic or core-S{\'e}rsic galaxies \citep[see also][]{Savorgnan:2016:Slopes}.}
\label{CS_S}
\end{center}
\end{figure*}

We categorized S{\'e}rsic and core-S{\'e}rsic galaxies based on their central light profiles, as determined from previous studies of high-resolution images \citep{Ferrarese:2006, Richings:2011, Dullo:2014}. Figure \ref{CS_S} presents two regressions performed on the two categories (S{\'e}rsic and core-S{\'e}rsic) for the SMBH mass versus both the spheroid stellar mass (left panel) and the total galaxy stellar mass (right panel) relations.

The \textsc{BCES} bisector regression of our 45 S{\'e}rsic and 31 core-S{\'e}rsic galaxies revealed $M_{BH} \propto M_{*,sph}^{1.30\pm 0.14}$ and $M_{BH} \propto M_{*,sph}^{1.38\pm 0.21}$, respectively. For the black hole mass versus total galaxy mass diagram we obtained $M_{BH} \propto M_{*,gal}^{1.61\pm 0.18}$ and $M_{BH} \propto M_{*,gal}^{1.47\pm 0.18}$ for S{\'e}rsic and core-S{\'e}rsic galaxies, respectively. For both the $M_{BH}$--$M_{sph}$ and $M_{BH}$--$M_{gal}$ relations, the slopes and intercepts of the regression lines for the S{\'e}rsic (blue line) and core-S{\'e}rsic (red line) ETGs are consistent within the $1\sigma$ confidence interval. Slopes and intercepts for the \textsc{BCES} bisector, as well as \textsc{BCES($Y|X$)} and \textsc{BCES($X|Y$)}, regression lines for the S{\'e}rsic and core-S{\'e}rsic galaxies, for both the $M_{BH}$--$M_{sph}$ and $M_{BH}$--$M_{gal}$ relations, can be found in Table \ref{fit parameters}.

Our findings are unlike the relations $M_{BH}\propto M_{*,sph}^{(2.22\pm 0.58)}$ and $M_{BH}\propto M_{*,sph}^{(0.94\pm 0.14)}$ obtained by \citet{Scott:Graham:2013} for their S{\'e}rsic and core-S{\'e}rsic galaxies, respectively. It appears that they may have found the break in the $M_{BH}$--$M_{*,sph}$ relation due to the inclusion of spiral galaxies, which steepened the $M_{BH}$--$M_{sph}$ relation for for their S{\'e}rsic galaxies (see Section \ref{ETGs&LTGs}).
 
\begin{figure*}
\begin{center}
\includegraphics[clip=true,trim= 8mm 0mm 12mm 9mm,height=6cm,width=\textwidth]{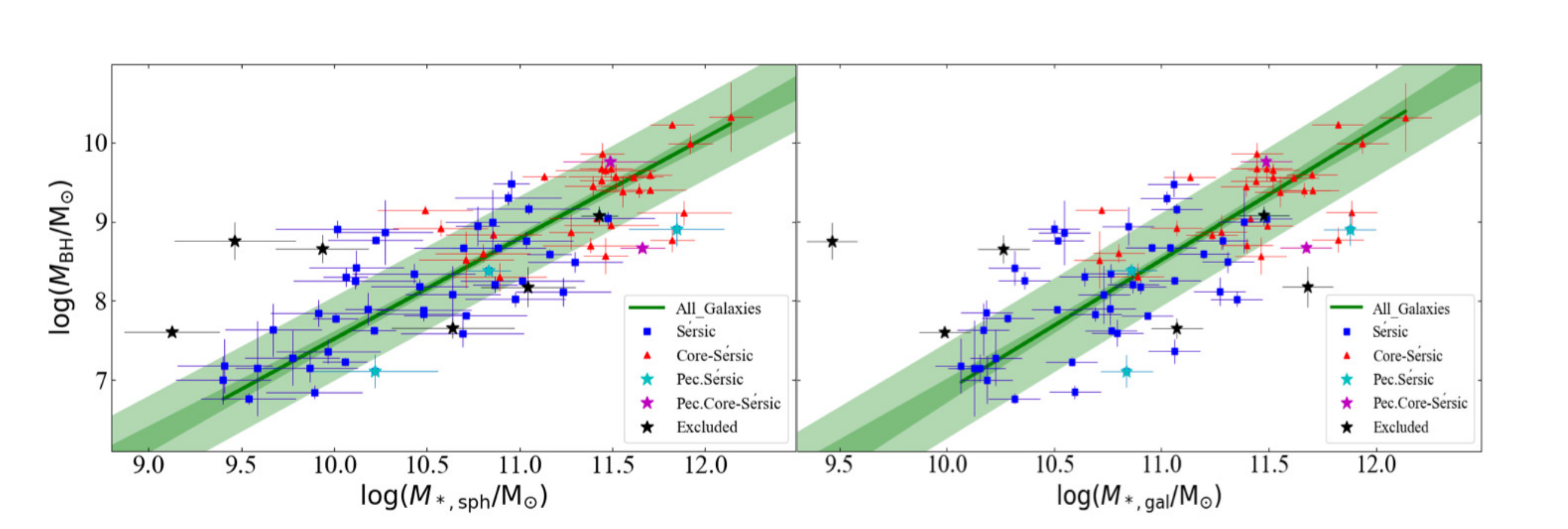}
\caption{Similar to Figure \ref{CS_S}. The \textcolor{green}{green} lines represent the single bisector regression lines for the sample of (84-8=) 76 ETGs with $M_{BH} \gtrsim 10^7\, M_{\odot}$. Both diagrams depict S{\'e}rsic and core-S{\'e}rsic ETGs following a unique relation in both the $M_{BH}$--$M_{*,sph}$ and $M_{BH}$--$M_{*,gal}$ diagrams. Such that, $M_{BH}\propto M^{1.27 \pm 0.07}_{*,sph}$ and $M_{BH}\propto M^{1.65 \pm 0.11}_{*,gal}$ with an rms scatter of 0.52 dex and 0.58 dex (in the $\log M_{BH}$ direction), respectively.}
\label{single_reg_scs}
\end{center}
\end{figure*}

The consistency of the regression lines for the S{\'e}rsic and core-S{\'e}rsic ETGs suggest that all the early-type galaxies (whether S{\'e}rsic or core-S{\'e}rsic) may follow single log-linear relations in the $M_{BH}$--$M_{*,sph}$ and $M_{BH}$--$M_{*,gal}$ diagrams.
Fitting single \textsc{BCES} bisector regression lines, for the $M_{BH}$--$M_{*,sph}$ and $M_{BH}$--$M_{*,gal}$ relations over our total (reduced) sample of 76 ETGs (Figure \ref{single_reg_scs}), revealed two tight relations, which can be expressed as,   
\begin{IEEEeqnarray}{rCl}
\label{Mbh_Msph}
\log(M_{BH}/M_\odot) &=& (1.27\pm 0.07)\log\left(\frac{M_{*,sph}}{\upsilon (5\times10^{10}\,M_\odot)}\right) \nonumber \\
&& +\> (8.41\pm 0.06),
\end{IEEEeqnarray}
and
\begin{IEEEeqnarray}{rCl}
\label{Mbh_Mgal}
\log(M_{BH}/M_\odot) &=& (1.65\pm 0.11)\log\left(\frac{M_{*,gal}}{\upsilon (5\times10^{10}\,M_\odot)}\right) \nonumber \\
&& +\> (8.02\pm 0.08),
\end{IEEEeqnarray}
with total rms scatters, in $\log(M_{BH})$, of 0.52 dex and 0.58 dex, respectively. 

The dark green line in both panels of Figure \ref{single_reg_scs} represents the \textsc{BCES} bisector regression line for our sample of 76 ETGs, which is surrounded by a dark green shade showing the $\pm 1 \sigma$ uncertainty in the slope and the intercept of the line. The light green shade represents the $\pm 1\sigma$ rms scatter of the data about the regression line. 

The similarity in the scatter about both relations (Equations \ref{Mbh_Msph} and \ref{Mbh_Mgal}) suggests that the black hole mass correlates nearly as well with galaxy stellar mass (or luminosity) as it does with spheroid stellar mass (or luminosity) for ETGs. This partly supports the claim of \citet{Lasker:2014}, albeit qualified by the restriction to ETGs, as was noted by \citet{Savorgnan:2016:Slopes}. Hence, with knowledge of the galaxy stellar mass, it would appear (at this stage of the analysis) that one can use the $M_{BH}$--$M_{*,gal}$ relation to estimate the black hole mass of an ETG nearly as accurately as if estimated using the $M_{BH}$--$M_{*,sph}$ relation. Additionally, it should be remembered that a poor bulge/disk decomposition may introduce an error of noticeably more than 0.1 dex to the bulge stellar mass, and thus the $M_{BH}$--$M_{*,gal}$ relation may in many instances be preferable.

For our total galaxy stellar masses, we used a constant uncertainty of 0.12 dex (see Section \ref{M_L ratio}) in all the regressions. However, we also derived the $M_{BH}$--$M_{*,gal}$ relation using a range of different uncertainties (0.10 dex, 0.12 dex, 0.15 dex, 0.20 dex) on $\log M_{*,gal}$, and found that the slope and intercept of equation \ref{Mbh_Mgal} remained within the $\pm 1\sigma$ bound.

Our scaling relations are based on the use of a different constant stellar mass-to-light ratio for each passband (see Table \ref{Photometric_para} and Section \ref{M_L ratio}). However, we checked the robustness of our $M_{BH}$--$M_{*,sph}$ and $M_{BH}$--$M_{*,gal}$ relations, using the color-dependent stellar mass-to-light ratios to calculate galaxy and spheroid stellar masses for our galaxies. As explained in Section \ref{Comparison of Masses}, we calculated $B-K_s$ color-dependent $K_s$-band stellar mass-to-light ratios ($\Upsilon^{K_s}_*$) for all our galaxies, using the equation $\log\left(\Upsilon^{K_s}_* \right) = 0.2119\times(B-K_s) - 0.9586$ from \citet{Bell:deJong:2001}. Further, we used this $\Upsilon^{K_s}_*$ in the formulae from \citet{Oh:2008}, (Equation \ref{oh}) to obtain color-dependent $\Upsilon^{3.6\mu m}_*$. For the remaining two\footnote{NGC 4486B, which is excluded from our regressions, is one of the three galaxies for which we used SDSS $r^{\prime}$-band images.} SDSS $r^{\prime}$-band images we used $\Upsilon^{r^{\prime}}_* = 2.8$, calibrated against 2MASS imaging as described in Section \ref{M_L ratio}. The use of color-dependent stellar mass-to-light ratios for the spheroid and galaxy stellar masses of our sample resulted in $M_{BH}\propto M_{*,sph}^{1.20\pm 0.07}$ and $M_{BH}\propto M_{*,gal}^{1.52\pm 0.10}$. These relations are consistent within the $\pm 1 \sigma$ bound of our previous relations (Equations \ref{Mbh_Msph} and \ref{Mbh_Mgal}), obtained using the masses based on the constant stellar mass-to-light ratios described in Section \ref{M_L ratio}.

\subsection{Galaxies With a Disk (ES/S0) and Without a Disk (E)}
\label{EESS0}

\begin{figure*}
\begin{center}
\includegraphics[clip=true,trim= 10mm 1mm 15mm 9mm,height=6cm,width=\textwidth]{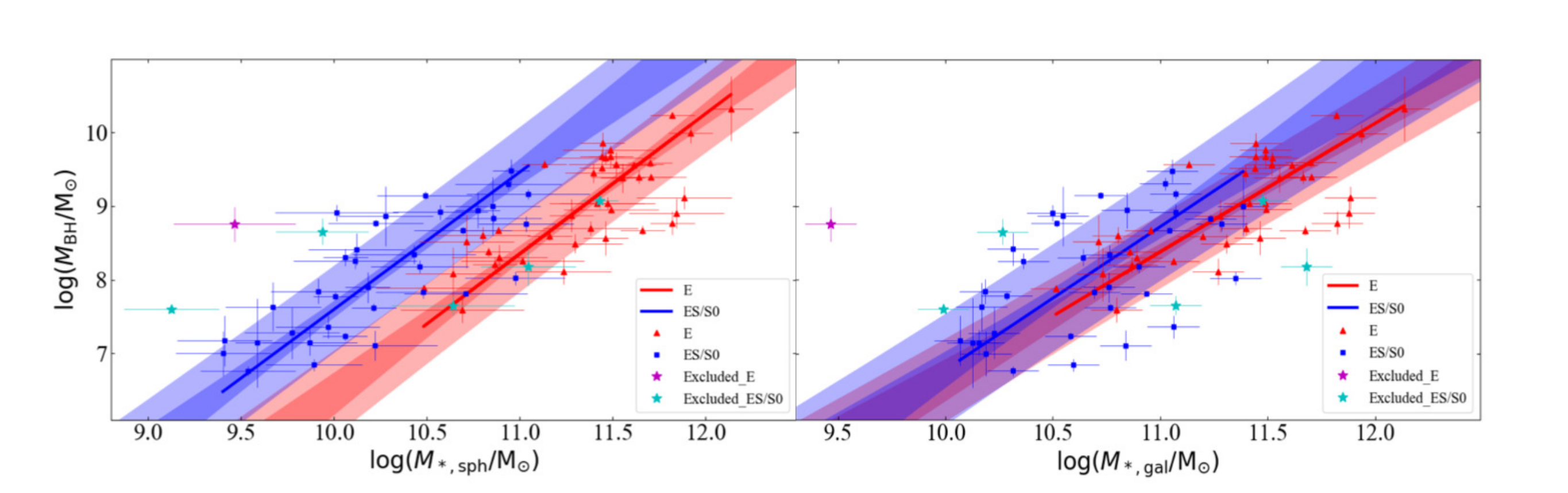}
\caption{Similar to Figure \ref{CS_S}, but now showing ETGs with (ES/S0) and without (E) a disk. In the $M_{BH}$--$M_{*,sph}$ diagram, the \textcolor{blue}{blue} regression line for galaxies with a disk (\textcolor{blue}{blue} squares) is offset from the \textcolor{red}{red} regression line for galaxies without a disk (\textcolor{red}{red} triangles) by more than an order of magnitude. This offset reveals two different scaling relations (Equation \ref{Mbh_Msph_ESS0} and \ref{Mbh_Msph_E}) for the two sub-morphological types (ES/S0 and E) with rms scatters in the $\log(M_{BH})$ direction of 0.57 dex and 0.50 dex, respectively. In the $M_{BH}$--$M_{*,gal}$ diagram, both the regression lines (Equation \ref{Mbh_Mgal_ESS0} and \ref{Mbh_Mgal_E}) are consistent with each other, suggesting a single relation (Equation \ref{Mbh_Mgal}) for galaxies with and without a disk.}
\label{E_ESS0}
\end{center}
\end{figure*}

We divided our ETG sample into those with an intermediate or extended disk (ES- and S0-type) and those without a disk (E-type), and performed separate \textsc{BCES} bisector regressions on each category. 
Figure \ref{E_ESS0} reveals separate relations for galaxies with a disk and galaxies without a disk in the $M_{BH}$--$M_{*,sph}$ diagram. The two relations are: 
\begin{IEEEeqnarray}{rCl}
\label{Mbh_Msph_ESS0}
\log(M_{BH}/M_\odot) &=& (1.86\pm 0.20)\log\left(\frac{M_{*,sph}}{\upsilon (5\times10^{10} M_\odot)}\right) \nonumber \\
&& +\> (8.90\pm 0.13),
\end{IEEEeqnarray}
for 36 galaxies with a disk, and
\begin{IEEEeqnarray}{rCl}
\label{Mbh_Msph_E}
\log(M_{BH}/M_\odot) &=& (1.90\pm 0.20)\log\left(\frac{M_{*,sph}}{\upsilon (5\times10^{10} M_\odot)}\right) \nonumber \\
&& +\> (7.78\pm 0.15),
\end{IEEEeqnarray}
for 40 galaxies without a disk, with an rms scatter of 0.57 dex and 0.50 dex, respectively. While the slopes are consistent, the intercepts, are different by 1.12 dex (more than an order of magnitude). Therefore, to estimate the black hole mass using the spheroid stellar mass of an ETG, it is beneficial to know if the galaxy has a disk (ES/S0) or not (E).

In the $M_{BH}$--$M_{*,gal}$ diagram (Figure \ref{E_ESS0}, right panel), the slopes of the regression lines for galaxies with (Equation \ref{Mbh_Mgal_ESS0}) and without (Equation \ref{Mbh_Mgal_E}) a disk are again consistent. However, the intercepts of each relation now only differ by a factor of 2, rather than 13 (i.e, 1.12~dex), in black hole mass. While the $1\sigma$ uncertainty on these two intercepts does not quite overlap, we derive a single $M_{BH}$--$M_{*,gal}$ relation for ES/S0 and E-type galaxies. Given that one may not know if their ETG of interest contains a disk, to estimate black hole mass using the total galaxy stellar mass, one may prefer the relation obtained by performing the single regression (Equation \ref{Mbh_Mgal}) on the whole ETGs sample. The bisector regression line for the 36 ETGs with a disk is
\begin{IEEEeqnarray}{rCl}
\label{Mbh_Mgal_ESS0}
\log(M_{BH}/M_\odot) &=& (1.94\pm 0.21)\log\left(\frac{M_{*,gal}}{\upsilon (5\times10^{10} M_\odot)}\right) \nonumber \\
&& +\> (8.14\pm 0.12),
\end{IEEEeqnarray}
with an rms scatter of 0.71 dex, and for the 40 galaxies without a disk we obtained
\begin{IEEEeqnarray}{rCl}
\label{Mbh_Mgal_E}
\log(M_{BH}/M_\odot) &=& (1.74\pm 0.16)\log\left(\frac{M_{*,gal}}{\upsilon (5\times10^{10} M_\odot)}\right) \nonumber \\
&& +\> (7.85\pm 0.12),
\end{IEEEeqnarray}
with an rms scatter of 0.48 dex.

The above results agree with the fact that most \textit{elliptical} galaxies primarily consist of an extended spheroid; hence their total galaxy mass is nearly equal to their spheroid mass. Thus, in both the $M_{BH}$--$M_{*,sph}$ and $M_{BH}$--$M_{*,gal}$ diagrams, elliptical galaxies reside at the same place, usually at the high-mass end. The ellicular (ES) and lenticular (S0) galaxies have their total galaxy stellar mass distributed in their spheroid, disk, and sometimes other components. Therefore, their spheroid stellar mass can be significantly less than the galaxy stellar mass, and in the $M_{BH}$--$M_{*,sph}$ diagram they reside at the low-mass (left) side creating an offset from the galaxies without a disk. We also performed \textsc{BCES($Y|X$)} and \textsc{BCES($X|Y$)} regressions for the above cases and the best fit parameters can be found in Table \ref{fit parameters}.

\subsection{Barred and Non-barred Galaxies}

The $M_{BH}-\sigma$ relation is often reported to be the most fundamental relationship between the super-massive black hole mass and any galaxy property, where $\sigma$ is the velocity dispersion of the host galaxy's spheroid \citep{Ferrarese:Merritt:2000, Gebhardt:2000}. However, previous studies have found that barred galaxies are offset towards higher $\sigma$ values in the $M_{BH}-\sigma$ diagram \citep{Graham:2007, Graham:2008, Graham:2011}. This offset can be accounted for in one of two ways: either the velocity dispersion of barred galaxies is systematically higher than non-barred galaxies \citep{Hartmann:2014}, or their central super-massive black hole mass is under-estimated.  

\begin{figure*}[h]
\begin{center}
\includegraphics[clip=true,trim= 9mm 3mm 14mm 8mm,height=6cm,width=\textwidth]{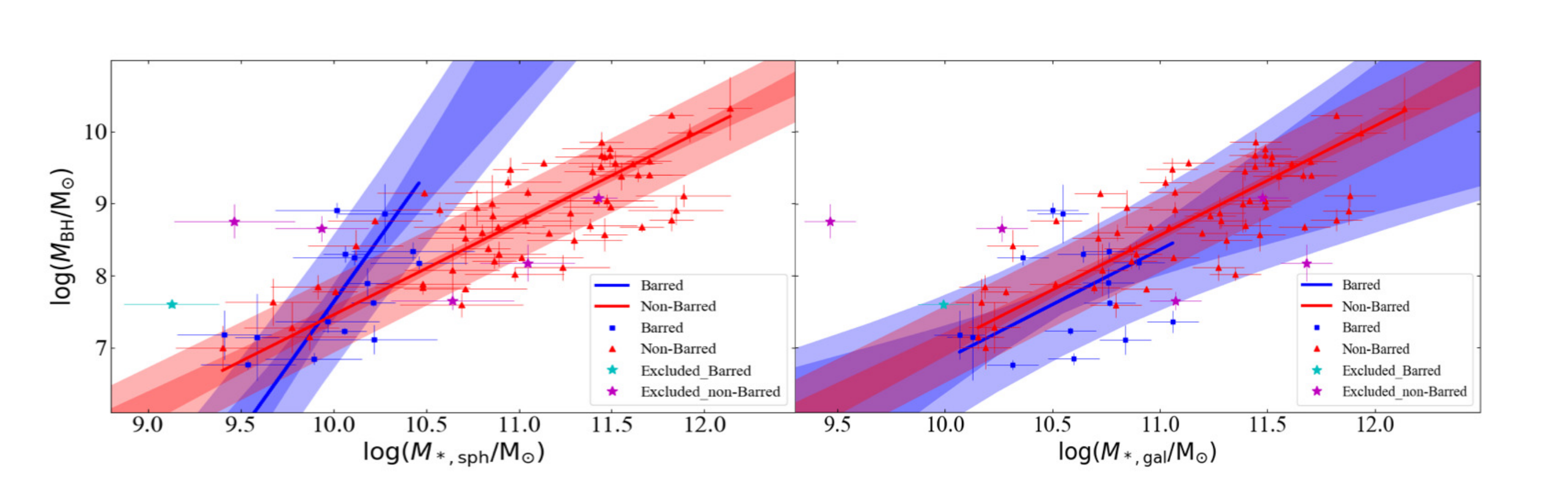}
\caption{Similar to Figure \ref{CS_S}, but now showing galaxies with a bar (15 \textcolor{blue}{blue} squares) and without a bar (61 \textcolor{red}{red} triangles). Upon performing separate regressions for barred (\textcolor{blue}{blue} line) and non-barred (\textcolor{red}{red} line) galaxies, we found that the slopes of the two lines in the $M_{BH}$--$M_{*,gal}$ diagram are consistent (see Table \ref{fit parameters}), suggesting a single slope for barred and non-barred ETGs (see Figure \ref{BNB1}). However, we require a larger dataset of barred galaxies to draw a firm conclusion on whether or not barred galaxies create an offset in the $M_{BH}$--$M_{*,sph}$ relation.}
\label{BNB}
\end{center}
\end{figure*}

In an attempt to solve this problem, we performed separate regressions for the barred and non-barred galaxies in the $M_{BH}$--$M_{*,sph}$ and $M_{BH}$--$M_{*,gal}$ diagrams (see Figure \ref{BNB}). Our reduced sample of 76 ETGs consists of 15 barred galaxies (red squares) and 61 non-barred galaxies (blue triangles). The slope of the $M_{BH}$--$M_{*,gal}$ relation for barred and non-barred ETGs are consistent with each other. However, with only 15 barred ETGs in our sample, the uncertainty on the slope of the $M_{BH}$--$M_{*,sph}$ relation for the barred galaxies is large (see Table \ref{fit parameters}) and makes it problematic to determine at what mass to compare the intercepts. From a visual inspection of Figure \ref{BNB}, we feel that it would be premature to draw any firm conclusion until more barred ETGs are in the sample. 

The parameters of the \textsc{BCES} bisector, along with \textsc{BCES($Y|X$)} and \textsc{BCES($Y|X$)}, regression lines for our dataset of 15 barred and 61 non-barred ETGs can be found in Table \ref{fit parameters}. 

\begin{figure*}[h]
\begin{center}
\includegraphics[clip=true,trim= 7mm 1mm 12mm 9mm,height=6cm,width=\textwidth]{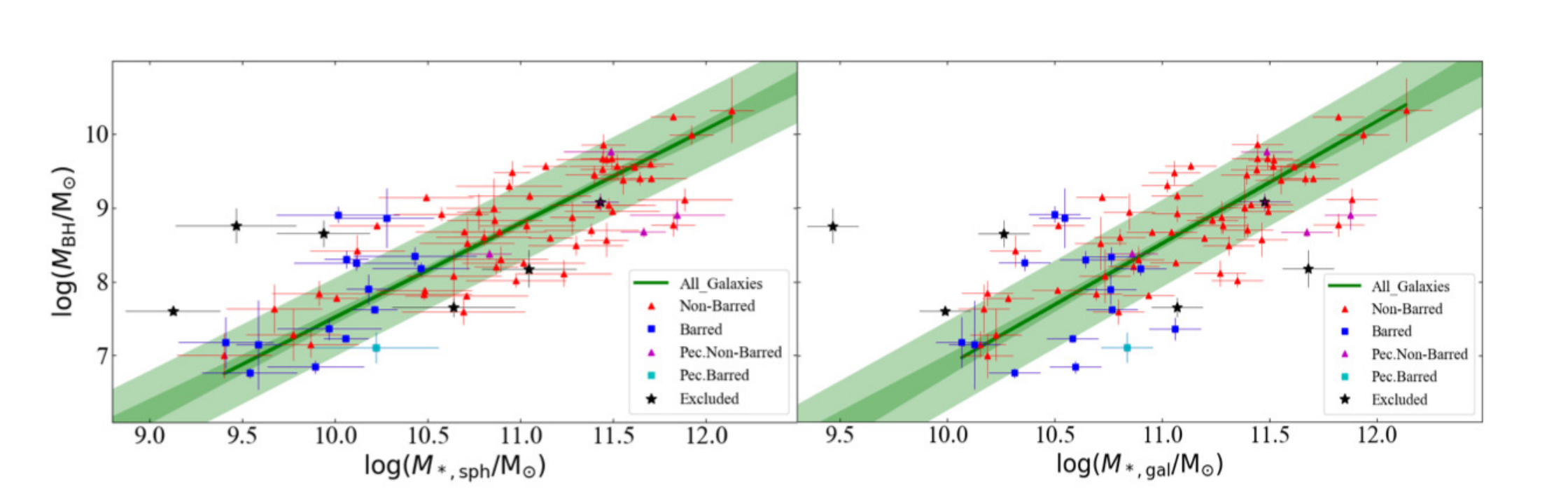}
\caption{Similar to Figure \ref{single_reg_scs}, but showing which galaxies are barred.
}
\label{BNB1}
\end{center}
\end{figure*}

In Figure \ref{BNB1}, we have again shown the single ETG regression line for both the $M_{BH}$--$M_{*,sph}$ and the $M_{BH}$--$M_{*,gal}$ relations (as in Figure \ref{single_reg_scs}), but here we identify the barred (blue squares) and non-barred (red triangles) galaxies with different symbols. The barred galaxies are not offset in the $M_{BH}$--$M_{*,gal}$ diagram, and there is no clear evidence for an offset to lower black hole masses in the $M_{BH}$--$M_{*,sph}$ diagram, implying that the barred galaxies likely have a higher velocity dispersion relative to the non-barred galaxies thereby creating the offset in the $M_{BH}-\sigma$ diagram. 

\subsection{Early-type Galaxies and Late-type Galaxies}
\label{ETGs&LTGs}
\begin{figure*}[h]
\begin{center}
\includegraphics[clip=true,trim= 8mm 1mm 12mm 8mm,height=6cm,width=\textwidth]{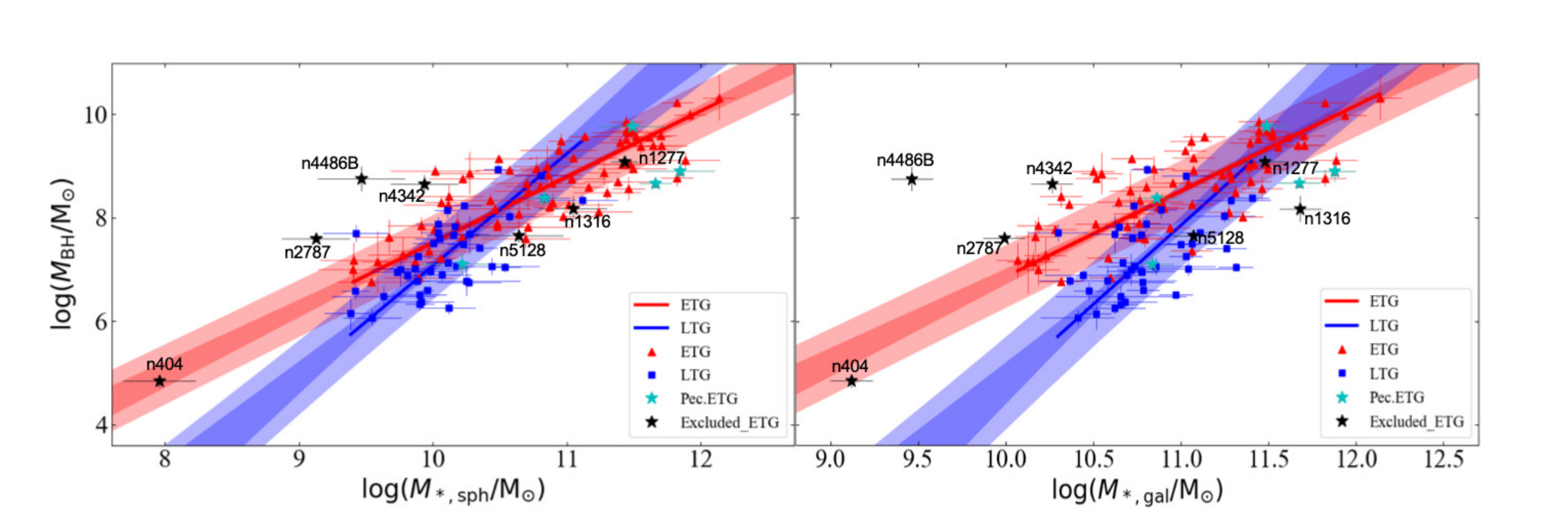}
\caption{$M_{BH}$--$M_{*,sph}$ and $M_{BH}$--$M_{*,gal}$ relations for ETGs (\textcolor{red}{red} triangles) and LTGs (\textcolor{blue}{blue} squares). Data for the late-type galaxies is taken from \citet{Davis:2018:a}. In both panels, the \textcolor{red}{red} and \textcolor{blue}{blue} lines represent the bisector regression lines for ETGs and LTGs, respectively. In the $M_{BH}$--$M_{*,sph}$ diagram, $M_{BH} \propto M_{*,sph}^{1.27\pm 0.07}$ for ETGs and $M_{BH} \propto M_{*,Sph}^{2.17\pm 0.32}$ for LTGs. In the $M_{BH}$--$M_{*,gal}$ diagram, $M_{BH} \propto M_{*,gal}^{1.65\pm 0.11}$ for ETGs and $M_{BH} \propto M_{*,Gal}^{3.05\pm 0.70}$ for LTGs. Although, the ETG NGC 404 ($\log M_{BH}/M\odot= 4.84$) is excluded from the regressions, it follows the regression lines for ETGs.  NGC 4486B, which has the second lowest galaxy stellar mass in our sample is a stripped compact elliptical galaxy.
} 
\label{ETG_LTG}
\end{center}
\end{figure*}

We have combined our ETG data with the recent work on the largest sample of late-type galaxies (LTGs, i.e. spirals) by \citet{Davis:2018:a}. We found that the regression lines followed by these two populations, ETGs and LTGs\footnote{We have taken the BCES bisector regression line from \citet{Davis:2018:b}}, in the $M_{BH}$--$M_{*,sph}$ and $M_{BH}$--$M_{*,gal}$ diagrams are not consistent with each other (see Figure-\ref{ETG_LTG}). 


In the black hole mass versus spheroid mass diagram, the regression line for the reduced sample of 40 LTGs from \citet[][accepted]{Davis:2018:a} can be expressed as,
\begin{IEEEeqnarray}{rCl}
\label{Mbh_Msph_LTG}
\log(M_{BH}/M_\odot) &=& (2.16\pm 0.32)\log\left(\frac{M_{*,sph}}{\upsilon (5\times10^{10} M_\odot)}\right) \nonumber \\
&& +\> (8.58\pm 0.22),
\end{IEEEeqnarray}
which has a slope approximately twice as steep as that of the ETGs: $ M_{BH}\propto M_{*,sph}^{1.27\pm 0.07}$ (Equation \ref{Mbh_Msph}). 
Similarly, in the black hole mass versus galaxy stellar mass diagram, LTGs define the relation
\begin{IEEEeqnarray}{rCl}
\label{Mbh_Mgal_LTG}
\log(M_{BH}/M_\odot) &=& (3.05\pm 0.70)\log\left(\frac{M_{*,gal}}{\upsilon (5\times10^{10} M_\odot)}\right) \nonumber \\
&& +\> (6.93\pm 0.14),
\end{IEEEeqnarray}
while the ETGs follow the proportionality $ M_{BH}\propto M_{*,gal}^{1.65 \pm 0.11}$ (Equation \ref{Mbh_Mgal}). 

This shallow and steep relation is roughly consistent with the bend observed by \citet{Savorgnan:2016:Slopes}, where they found a near-linear relation, $ M_{BH}\propto M_{*,sph}^{1.04 \pm 0.10}$, for their reduced\footnote{\citet{Savorgnan:2016:Slopes} excluded 2 ETGs and 2 LTGs from their total sample.} sample of 45 ETGs, with an rms scatter of 0.51 dex in the black hole mass, and $ M_{BH}\propto M_{*,sph}^{2-3}$ for their 17 LTGs. They refer to the two correlations as an \textit{early-type sequence} (or \textit{red-type sequence}) and a \textit{late-type sequence} (or \textit{blue-type sequence}). Parameters for our \textsc{BCES($Y|X$)} and \textsc{BCES($X|Y$)} regression lines for LTGs and ETGs can be found in Table \ref{fit parameters}. 

From our work, we infer that the previous papers found a bent $M_{BH}$--$M_{*,sph}$ relation due to S{\'e}rsic and core-S{\'e}rsic galaxies \citep[e.g.][]{Scott:Graham:2013} because most of the S{\'e}rsic galaxies in their sample were LTGs and most of the core-S{\'e}rsic galaxies were ETGs. The bend in their relation was supposedly due to the different formation processes (dry merging versus gaseous growth), as traced by the difference in the central surface brightness profile of the galaxies. However, we find that the bend is due to the two broad morphological classes of galaxies: ETGs (consisting of ellipticals E, elliculars ES, and lenticulars S0) and LTGs (consisting of spirals Sp), supporting the finding in \citet{Savorgnan:2016:Slopes}, which was also later shown by \citet[][see his Figure 2]{van_den_Bosch:2016}.

The situation is, however, a little more complicated than presented above. As explained in \citet{Graham:Soria:2018}, the color-magnitude relation for ETGs had confounded the situation when working with B-band magnitudes. This results in the fainter S{\'e}rsic ETGs following a steep B-band $M_{BH}$--$L_{B,sph}$ relation (and a shallow $L_B$--$\sigma$ relation). Additionally, we have established that the bulges of ETGs follow a steep $M_{BH}$--$M_{*,sph}$ relation if one has a sample consisting of pure E-type or a sample of ES and S0 type. Section \ref{EESS0} reveals a slope of around $1.9\pm 0.2$ for both of these populations, which is not overly dissimilar to the slope of $2.16\pm 0.32$ for bulges in spiral galaxies. 

\begin{figure*}[h]
\begin{center}
\includegraphics[clip=true,trim= 6mm 1mm 12mm 8mm,height=6cm,width=\textwidth]{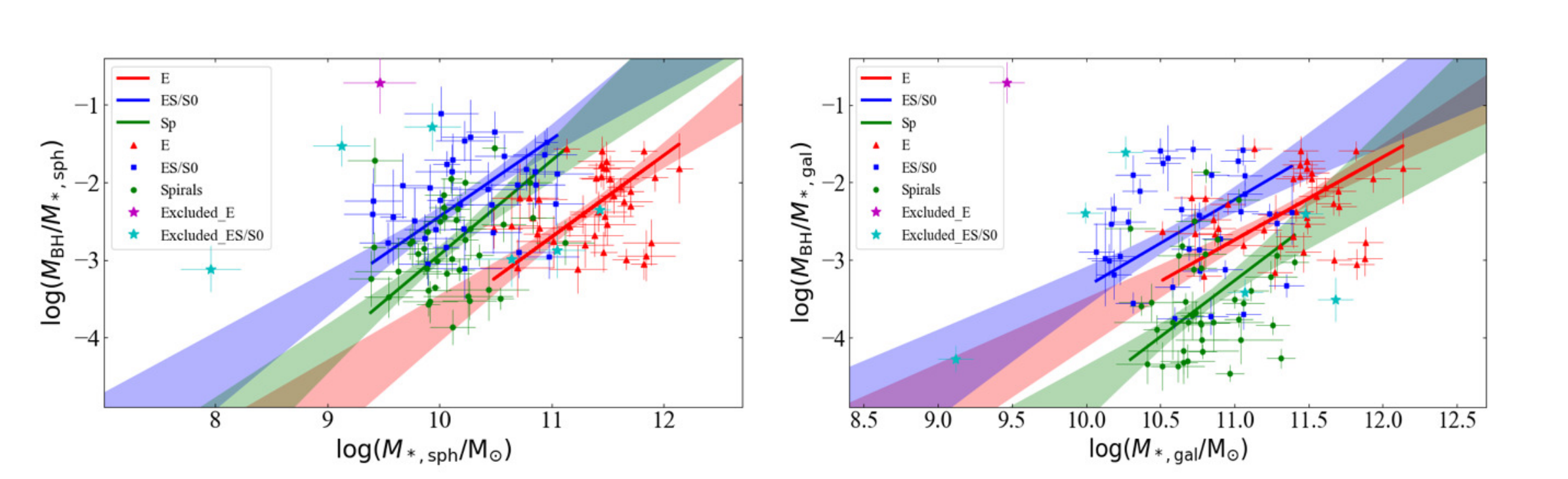}
\caption{$(M_{BH}/M_{*,sph})$--$M_{*,sph}$ and $(M_{BH}/M_{*,gal})$--$M_{*,gal}$ relations for ETGs with a disk (\textcolor{blue}{blue} squares), ETGs without a disk (\textcolor{red}{red} triangles), and LTGs (\textcolor{green}{green} circles). In both the panels, \textcolor{blue}{blue}, \textcolor{red}{red}, and \textcolor{green}{green} lines represent the bisector regression lines for the three sub-populations of ES/S0-, E-, and Sp-type galaxies, respectively. Dark bands around the lines shows the $\pm 1 \sigma$ uncertainty in the corresponding slopes and intercepts. In the $(M_{BH}/M_{*,sph})$--$M_{*,sph}$ diagram, the regression line for ETGs with a disk is offset from the regression line for ETGs without a disk by $1.28\pm 0.17$~dex in their $(M_{BH}/M_{*,sph})$ ratios, which is consistent with the offset observed in the $M_{BH}$--$M_{*,sph}$ diagram within the $1\sigma$ bound. In the $M_{BH}/M_{*,gal}$--$M_{*,gal}$ diagram, spiral galaxies follow steeper relation than ETGs, analogous to the right panel of Figure \ref{ETG_LTG}.
} 
\label{Ratio}
\end{center}
\end{figure*}

Importantly, we find that the $(M_{BH}/M_{*,sph})$--$M_{*,sph}$ and $(M_{BH}/M_{*,gal})$--$M_{*,gal}$ relations (see Figure \ref{Ratio}) are qualitatively and quantitatively consistent with our $M_{BH}$--$M_{*,sph}$ and $M_{BH}$--$M_{*,gal}$ relations for the sub-populations of ETGs (ES/S0 and E) and LTGs (Sp), within $1 \sigma$ bound. Parameters for these regression lines can be found in Table \ref{fit parameters}. Figure \ref{Ratio} also depicts how the $M_{BH}/M_{*,sph}$ and $M_{BH}/M_{*,gal}$ ratios do not have a constant value as was implied by our $M_{BH}$--$M_{*,sph}$ and $M_{BH}$--$M_{*,gal}$ relations.

\subsection{NGC 5252: A Compact Massive Spheroid}

In addition to the above scaling relations, we have discovered a compact massive spheroid in NGC 5252 ($z \approx 0.02$), with a stellar mass of $M_{*,sph}= 7.1^{+5.8}_{-3.2} \times 10^{10} M_{\odot}$ and a half light radius ($R_{e,sph}$) of just 0.672~kpc, adding to the sample of 21 identified by \citet{Graham:CMS:2015}.

\startlongtable
\begin{deluxetable*}{lcrrccccrrrr}
\tabletypesize{\footnotesize}
\tablecolumns{11}
\tablecaption{Linear Regressions\label{fit parameters}}
\tablehead{
\colhead{Regression} & \colhead{Minimization} & \colhead{$\alpha$} & \colhead{$\beta$} & \colhead{$\epsilon$} & \colhead{$\Delta_{rms}$} & \colhead{} & \colhead{$r$} & \colhead{$\log{p}$} & \colhead{$r_s$} & \colhead{$\log{p_s}$} \\
\colhead{} & \colhead{} & \colhead{} & \colhead{(dex)} & \colhead{(dex)} & \colhead{(dex)} & \colhead{} & \colhead{} & \colhead{(dex)} & \colhead{} & \colhead{(dex)} \\
\colhead{(1)} & \colhead{(2)} & \colhead{(3)} & \colhead{(4)} & \colhead{(5)} & \colhead{(6)} & \colhead{} & \colhead{(7)} & \colhead{(8)} & \colhead{(9)} & \colhead{(10)}
}
\startdata
\multicolumn{11}{c}{\textbf{76 Early-Type Galaxies}} \\
\hline
\multicolumn{11}{c}{$\bm{\log(M_{\rm BH}/{\rm M_{\sun}})=\alpha\log(M_{\rm *,sph}/[\upsilon(5\times10^{10}\ {\rm M_{\sun}})])+\beta}$} \\
\hline
\textsc{bces}$(Bisector)$ & \textit{Symmetric} & $1.27\pm0.07$ & $8.41\pm0.06$ & $0.41$ & $0.52$ & \multirow{3}{*}{$\left\}\begin{tabular}{@{}l@{}} \\ \\ \\ \end{tabular}\right.$} & \multirow{3}{*}{$0.82$} & \multirow{3}{*}{$-18.96$} & \multirow{3}{*}{$0.80$} & \multirow{3}{*}{$-17.20$} \\
\textsc{bces}$(M_{\rm BH}|M_{\rm *,sph})$ & $M_{\rm BH}$ & $1.12\pm0.08$ & $8.43\pm0.06$ & $0.40$ & $0.49$ & & & & \\
\textsc{bces}$(M_{\rm *,sph}|M_{\rm BH})$ & $M_{\rm *,sph}$ & $1.45\pm0.09$ & $8.38\pm0.07$ & $0.45$ & $0.57$ & & & & \\
\hline
\multicolumn{11}{c}{$\bm{\log(M_{\rm BH}/{\rm M_{\sun}})=\alpha\log(M_{\rm *,gal}/[\upsilon(5\times10^{10}\ {\rm M_{\sun}})])+\beta}$} \\
\hline
\textsc{bces}$(Bisector)$ & \textit{Symmetric} & $1.65\pm0.11$ & $8.02\pm0.08$ & $0.53$ & $0.58$ & \multirow{3}{*}{$\left\}\begin{tabular}{@{}l@{}} \\ \\ \\ \end{tabular}\right.$} & \multirow{3}{*}{$0.76$} & \multirow{3}{*}{$-15.12$} & \multirow{3}{*}{$0.76$} & \multirow{3}{*}{$-14.71$} \\
\textsc{bces}$(M_{\rm BH}|M_{\rm *,gal})$ & $M_{\rm BH}$ & $1.33\pm0.12$ & $8.13\pm0.08$ & $0.51$ & $0.55$ & & & & \\
\textsc{bces}$(M_{\rm *,gal}|M_{\rm BH})$ & $M_{\rm *,gal}$ & $2.10\pm0.18$ & $7.86\pm0.11$ & $0.63$ & $0.69$ & & & & \\
\hline\\
\hline
\multicolumn{11}{c}{\textbf{S{\'e}rsic and Core-S{\'e}rsic Galaxies}} \\
\hline
\multicolumn{11}{c}{45 S{\'e}rsic Galaxies: $\bm{\log(M_{\rm BH}/{\rm M_{\sun}})=\alpha\log(M_{\rm *,sph}/[\upsilon(5\times10^{10}\ {\rm M_{\sun}})])+\beta}$} \\
\hline
\textsc{bces}$(Bisector)$ & \textit{Symmetric} & $1.30\pm0.14$ & $8.43\pm0.10$ & $0.42$ & $0.55$ & \multirow{3}{*}{$\left\}\begin{tabular}{@{}l@{}} \\ \\ \\ \end{tabular}\right.$} & \multirow{3}{*}{$0.71$} & \multirow{3}{*}{$-7.34$} & \multirow{3}{*}{$0.71$} & \multirow{3}{*}{$-7.23$} \\
\textsc{bces}$(M_{\rm BH}|M_{\rm *,sph})$ & $M_{\rm BH}$ & $1.05\pm0.14$ & $8.37\pm0.09$ & $0.40$ & $0.50$ & & & & \\
\textsc{bces}$(M_{\rm *,sph}|M_{\rm BH})$ & $M_{\rm *,sph}$ & $1.63\pm0.23$ & $8.52\pm0.13$ & $0.49$ & $0.66$ & & & & \\
\hline
\multicolumn{11}{c}{31 Core-S{\'e}rsic Galaxies: $\bm{\log(M_{\rm BH}/{\rm M_{\sun}})=\alpha\log(M_{\rm *,sph}/[\upsilon(5\times10^{10}\ {\rm M_{\sun}})])+\beta}$} \\
\hline
\textsc{bces}$(Bisector)$ & \textit{Symmetric} & $1.38\pm0.21$ & $8.30\pm0.20$ & $0.43$ & $0.50$ & \multirow{3}{*}{$\left\}\begin{tabular}{@{}l@{}} \\ \\ \\ \end{tabular}\right.$} & \multirow{3}{*}{$0.56$} & \multirow{3}{*}{$-2.96$} & \multirow{3}{*}{$0.47$} & \multirow{3}{*}{$-2.11$} \\
\textsc{bces}$(M_{\rm BH}|M_{\rm *,sph})$ & $M_{\rm BH}$ & $0.92\pm0.27$ & $8.62\pm0.20$ & $0.39$ & $0.43$ & & & & \\
\textsc{bces}$(M_{\rm *,sph}|M_{\rm BH})$ & $M_{\rm *,sph}$ & $2.20\pm0.55$ & $7.72\pm0.47$ & $0.58$ & $0.72$ & & & & \\
\hline
\multicolumn{11}{c}{45 S{\'e}rsic Galaxies: $\bm{\log(M_{\rm BH}/{\rm M_{\sun}})=\alpha\log(M_{\rm *,gal}/[\upsilon(5\times10^{10}\ {\rm M_{\sun}})])+\beta}$} \\
\hline
\textsc{bces}$(Bisector)$ & \textit{Symmetric} & $1.61\pm0.18$ & $8.00\pm0.09$ & $0.59$ & $0.63$ & \multirow{3}{*}{$\left\}\begin{tabular}{@{}l@{}} \\ \\ \\ \end{tabular}\right.$} & \multirow{3}{*}{$0.58$} & \multirow{3}{*}{$-4.62$} & \multirow{3}{*}{$0.58$} & \multirow{3}{*}{$-4.52$} \\
\textsc{bces}$(M_{\rm BH}|M_{\rm *,gal})$ & $M_{\rm BH}$ & $1.05\pm0.17$ & $8.04\pm0.09$ & $0.54$ & $0.57$ & & & & \\
\textsc{bces}$(M_{\rm *,gal}|M_{\rm BH})$ & $M_{\rm *,gal}$ & $2.71\pm0.55$ & $7.93\pm0.14$ & $0.86$ & $0.92$ & & & & \\
\hline
\multicolumn{11}{c}{31 Core-S{\'e}rsic Galaxies: $\bm{\log(M_{\rm BH}/{\rm M_{\sun}})=\alpha\log(M_{\rm *,gal}/[\upsilon(5\times10^{10}\ {\rm M_{\sun}})])+\beta}$} \\
\hline
\textsc{bces}$(Bisector)$ & \textit{Symmetric} & $1.47\pm0.18$ & $8.17\pm0.17$ & $0.43$ & $0.46$ & \multirow{3}{*}{$\left\}\begin{tabular}{@{}l@{}} \\ \\ \\ \end{tabular}\right.$} & \multirow{3}{*}{$0.58$} & \multirow{3}{*}{$-3.22$} & \multirow{3}{*}{$0.48$} & \multirow{3}{*}{$-2.21$} \\
\textsc{bces}$(M_{\rm BH}|M_{\rm *,gal})$ & $M_{\rm BH}$ & $0.96\pm0.24$ & $8.56\pm0.18$ & $0.39$ & $0.42$ & & & & \\
\textsc{bces}$(M_{\rm *,gal}|M_{\rm BH})$ & $M_{\rm *,gal}$ & $2.44\pm0.64$ & $7.45\pm0.55$ & $0.62$ & $0.68$ & & & & \\
\hline\\
\hline
\multicolumn{11}{c}{\textbf{Galaxies with a Disk (ES/S0) and Galaxies without a Disk (E)}} \\
\hline
\multicolumn{11}{c}{36 Galaxies with a Disk (ES/S0): $\bm{\log(M_{\rm BH}/{\rm M_{\sun}})=\alpha\log(M_{\rm *,sph}/[\upsilon(5\times10^{10}\ {\rm M_{\sun}})])+\beta}$} \\
\hline
\textsc{bces}$(Bisector)$ & \textit{Symmetric} & $1.86\pm0.20$ & $8.90\pm0.13$ & $0.28$ & $0.57$ & \multirow{3}{*}{$\left\}\begin{tabular}{@{}l@{}} \\ \\ \\ \end{tabular}\right.$} & \multirow{3}{*}{$0.77$} & \multirow{3}{*}{$-7.39$} & \multirow{3}{*}{$0.77$} & \multirow{3}{*}{$-7.49$} \\
\textsc{bces}$(M_{\rm BH}|M_{\rm *,sph})$ & $M_{\rm BH}$ & $1.70\pm0.22$ & $8.83\pm0.14$ & $0.29$ & $0.54$ & & & & \\
\textsc{bces}$(M_{\rm *,sph}|M_{\rm BH})$ & $M_{\rm *,sph}$ & $2.05\pm0.26$ & $8.98\pm0.15$ & $0.29$ & $0.62$ & & & & \\
\hline
\multicolumn{11}{c}{40 Galaxies without a Disk (E): $\bm{\log(M_{\rm BH}/{\rm M_{\sun}})=\alpha\log(M_{\rm *,sph}/[\upsilon(5\times10^{10}\ {\rm M_{\sun}})])+\beta}$} \\
\hline
\textsc{bces}$(Bisector)$ & \textit{Symmetric} & $1.90\pm0.20$ & $7.78\pm0.15$ & $0.36$ & $0.50$ & \multirow{3}{*}{$\left\}\begin{tabular}{@{}l@{}} \\ \\ \\ \end{tabular}\right.$} & \multirow{3}{*}{$0.75$} & \multirow{3}{*}{$-7.63$} & \multirow{3}{*}{$0.70$} & \multirow{3}{*}{$-6.32$} \\
\textsc{bces}$(M_{\rm BH}|M_{\rm *,sph})$ & $M_{\rm BH}$ & $1.68\pm0.24$ & $7.92\pm0.15$ & $0.34$ & $0.46$ & & & & \\
\textsc{bces}$(M_{\rm *,sph}|M_{\rm BH})$ & $M_{\rm *,sph}$ & $2.16\pm0.26$ & $7.60\pm0.21$ & $0.39$ & $0.56$ & & & & \\
\hline
\multicolumn{11}{c}{36 Galaxies with a Disk (ES/S0): $\bm{\log(M_{\rm BH}/{\rm M_{\sun}})=\alpha\log(M_{\rm *,gal}/[\upsilon(5\times10^{10}\ {\rm M_{\sun}})])+\beta}$} \\
\hline
\textsc{bces}$(Bisector)$ & \textit{Symmetric} & $1.94\pm0.21$ & $8.14\pm0.12$ & $0.67$ & $0.71$ & \multirow{3}{*}{$\left\}\begin{tabular}{@{}l@{}} \\ \\ \\ \end{tabular}\right.$} & \multirow{3}{*}{$0.57$} & \multirow{3}{*}{$-3.52$} & \multirow{3}{*}{$0.56$} & \multirow{3}{*}{$-3.47$} \\
\textsc{bces}$(M_{\rm BH}|M_{\rm *,gal})$ & $M_{\rm BH}$ & $1.26\pm0.25$ & $8.12\pm0.11$ & $0.62$ & $0.64$ & & & & \\
\textsc{bces}$(M_{\rm *,gal}|M_{\rm BH})$ & $M_{\rm *,gal}$ & $3.47\pm0.76$ & $8.16\pm0.18$ & $1.01$ & $1.08$ & & & & \\
\hline
\multicolumn{11}{c}{40 Galaxies without a Disk (E): $\bm{\log(M_{\rm BH}/{\rm M_{\sun}})=\alpha\log(M_{\rm *,gal}/[\upsilon(5\times10^{10}\ {\rm M_{\sun}})])+\beta}$} \\
\hline
\textsc{bces}$(Bisector)$ & \textit{Symmetric} & $1.74\pm0.16$ & $7.85\pm0.12$ & $0.42$ & $0.48$ & \multirow{3}{*}{$\left\}\begin{tabular}{@{}l@{}} \\ \\ \\ \end{tabular}\right.$} & \multirow{3}{*}{$0.74$} & \multirow{3}{*}{$-7.28$} & \multirow{3}{*}{$0.70$} & \multirow{3}{*}{$-6.27$} \\
\textsc{bces}$(M_{\rm BH}|M_{\rm *,gal})$ & $M_{\rm BH}$ & $1.38\pm0.18$ & $8.10\pm0.12$ & $0.40$ & $0.45$ & & & & \\
\textsc{bces}$(M_{\rm *,gal}|M_{\rm BH})$ & $M_{\rm *,gal}$ & $2.27\pm0.29$ & $7.50\pm0.24$ & $0.51$ & $0.58$ & & & & \\
\hline\\
\hline
\multicolumn{11}{c}{\textbf{Galaxies with and without a Bar}} \\
\hline
\multicolumn{11}{c}{15 Galaxies with a Bar: $\bm{\log(M_{\rm BH}/{\rm M_{\sun}})=\alpha\log(M_{\rm *,sph}/[\upsilon(5\times10^{10}\ {\rm M_{\sun}})])+\beta}$} \\
\hline
\textsc{bces}$(Bisector)$ & \textit{Symmetric} & $3.59\pm1.79$ & $10.14\pm1.15$ & $0.34$ & $0.86$ & \multirow{3}{*}{$\left\}\begin{tabular}{@{}l@{}} \\ \\ \\ \end{tabular}\right.$} & \multirow{3}{*}{$0.60$} & \multirow{3}{*}{$-1.76$} & \multirow{3}{*}{$0.56$} & \multirow{3}{*}{$-1.53$} \\
\textsc{bces}$(M_{\rm BH}|M_{\rm *,sph})$ & $M_{\rm BH}$ & $3.58\pm2.40$ & $10.13\pm1.55$ & $0.33$ & $0.86$ & & & & \\
\textsc{bces}$(M_{\rm *,sph}|M_{\rm BH})$ & $M_{\rm *,sph}$ & $3.61\pm1.37$ & $10.15\pm0.90$ & $0.34$ & $0.86$ & & & & \\
\hline
\multicolumn{11}{c}{61 Galaxies without a Bar: $\bm{\log(M_{\rm BH}/{\rm M_{\sun}})=\alpha\log(M_{\rm *,sph}/[\upsilon(5\times10^{10}\ {\rm M_{\sun}})])+\beta}$} \\
\hline
\textsc{bces}$(Bisector)$ & \textit{Symmetric} & $1.29\pm0.09$ & $8.36\pm0.07$ & $0.41$ & $0.51$ & \multirow{3}{*}{$\left\}\begin{tabular}{@{}l@{}} \\ \\ \\ \end{tabular}\right.$} & \multirow{3}{*}{$0.78$} & \multirow{3}{*}{$-13.14$} & \multirow{3}{*}{$0.73$} & \multirow{3}{*}{$-10.78$} \\
\textsc{bces}$(M_{\rm BH}|M_{\rm *,sph})$ & $M_{\rm BH}$ & $1.10\pm0.10$ & $8.42\pm0.07$ & $0.39$ & $0.47$ & & & & \\
\textsc{bces}$(M_{\rm *,sph}|M_{\rm BH})$ & $M_{\rm *,sph}$ & $1.52\pm0.13$ & $8.28\pm0.10$ & $0.46$ & $0.58$ & & & & \\
\hline
\multicolumn{11}{c}{15 Galaxies with a Bar: $\bm{\log(M_{\rm BH}/{\rm M_{\sun}})=\alpha\log(M_{\rm *,gal}/[\upsilon(5\times10^{10}\ {\rm M_{\sun}})])+\beta}$} \\
\hline
\textsc{bces}$(Bisector)$ & \textit{Symmetric} & $1.52\pm0.59$ & $7.90\pm0.22$ & $0.73$ & $0.73$ & \multirow{3}{*}{$\left\}\begin{tabular}{@{}l@{}} \\ \\ \\ \end{tabular}\right.$} & \multirow{3}{*}{$0.18$} & \multirow{3}{*}{$-0.29$} & \multirow{3}{*}{$0.14$} & \multirow{3}{*}{$-0.20$} \\
\textsc{bces}$(M_{\rm BH}|M_{\rm *,gal})$ & $M_{\rm BH}$ & $0.53\pm0.56$ & $7.79\pm0.18$ & $0.67$ & $0.67$ & & & & \\
\textsc{bces}$(M_{\rm *,gal}|M_{\rm BH})$ & $M_{\rm *,gal}$ & $13.19\pm16.19$ & $9.19\pm1.56$ & $3.41$ & $3.51$ & & & & \\
\hline
\multicolumn{11}{c}{61 Galaxies without a Bar: $\bm{\log(M_{\rm BH}/{\rm M_{\sun}})=\alpha\log(M_{\rm *,gal}/[\upsilon(5\times10^{10}\ {\rm M_{\sun}})])+\beta}$} \\
\hline
\textsc{bces}$(Bisector)$ & \textit{Symmetric} & $1.52\pm0.10$ & $8.10\pm0.08$ & $0.46$ & $0.50$ & \multirow{3}{*}{$\left\}\begin{tabular}{@{}l@{}} \\ \\ \\ \end{tabular}\right.$} & \multirow{3}{*}{$0.78$} & \multirow{3}{*}{$-12.65$} & \multirow{3}{*}{$0.74$} & \multirow{3}{*}{$-11.05$} \\
\textsc{bces}$(M_{\rm BH}|M_{\rm *,gal})$ & $M_{\rm BH}$ & $1.23\pm0.12$ & $8.23\pm0.08$ & $0.44$ & $0.48$ & & & & \\
\textsc{bces}$(M_{\rm *,gal}|M_{\rm BH})$ & $M_{\rm *,gal}$ & $1.90\pm0.16$ & $7.93\pm0.11$ & $0.54$  $0.59$ & & & & \\
\hline\\
\hline
\multicolumn{11}{c}{\textbf{40 Late-Type Galaxies}} \\
\hline
\multicolumn{11}{c}{$\bm{\log(M_{\rm BH}/{\rm M_{\sun}})=\alpha\log(M_{\rm *,sph}/[\upsilon(5\times10^{10}\ {\rm M_{\sun}})])+\beta}$} \\
\hline
\textsc{bces}$(Bisector)$ & \textit{Symmetric} & $2.16\pm0.32$ & $8.58\pm0.22$ & $0.48$ & $0.64$ & \multirow{3}{*}{$\left\}\begin{tabular}{@{}l@{}} \\ \\ \\ \end{tabular}\right.$} & \multirow{3}{*}{$0.66$} & \multirow{3}{*}{$-5.35$} & \multirow{3}{*}{$0.62$} & \multirow{3}{*}{$-4.62$} \\
\textsc{bces}$(M_{\rm BH}|M_{\rm *,sph})$ & $M_{\rm BH}$ & $1.70\pm0.35$ & $8.30\pm0.22$ & $0.46$ & $0.56$ & & & & \\
\textsc{bces}$(M_{\rm *,sph}|M_{\rm BH})$ & $M_{\rm *,sph}$ & $2.90\pm0.55$ & $9.03\pm0.39$ & $0.59$ & $0.82$ & & & & \\
\hline
\multicolumn{11}{c}{$\bm{\log(M_{\rm BH}/{\rm M_{\sun}})=\alpha\log(M_{\rm *,gal}/[\upsilon(5\times10^{10}\ {\rm M_{\sun}})])+\beta}$} \\
\hline
\textsc{bces}$(Bisector)$ & \textit{Symmetric} & $3.05\pm0.70$ & $6.93\pm0.14$ & $0.70$ & $0.79$ & \multirow{3}{*}{$\left\}\begin{tabular}{@{}l@{}} \\ \\ \\ \end{tabular}\right.$} & \multirow{3}{*}{$0.47$} & \multirow{3}{*}{$-2.70$} & \multirow{3}{*}{$0.53$} & \multirow{3}{*}{$-3.34$} \\
\textsc{bces}$(M_{\rm BH}|M_{\rm *,gal})$ & $M_{\rm BH}$ & $2.04\pm0.72$ & $7.04\pm0.14$ & $0.61$ & $0.66$ & & & & \\
\textsc{bces}$(M_{\rm *,gal}|M_{\rm BH})$ & $M_{\rm *,gal}$ & $5.60\pm1.57$ & $6.66\pm0.22$ & $1.11$ & $1.31$ & & & & \\
\hline\\
\hline
\multicolumn{11}{c}{\textbf{ETGs with a disk (ES/S0), ETGs without a disk (E) and LTGs (Sp)}} \\
\hline
\multicolumn{11}{c}{36 Galaxies with a Disk (ES/S0): $\bm{\log(M_{\rm BH}/M_{\rm *,sph})=\alpha\log(M_{\rm *,sph}/[\upsilon(5\times10^{10}\ {\rm M_{\sun}})])+\beta}$} \\
\hline
\textsc{bces}$(Bisector)$ & \textit{Symmetric} & $1.00\pm0.14$ & $-1.74\pm0.12$ & $0.46$ & $0.60$ &  & $0.25$ & $-0.84$ & $0.31$ & $-1.17$ \\
\hline\\
\hline
\multicolumn{11}{c}{40 Galaxies without a Disk (E): $\bm{\log(M_{\rm BH}/M_{\rm *,sph})=\alpha\log(M_{\rm *,sph}/[\upsilon(5\times10^{10}\ {\rm M_{\sun}})])+\beta}$} \\
\hline
\textsc{bces}$(Bisector)$ & \textit{Symmetric} & $1.05\pm0.11$ & $-3.02\pm0.12$ & $0.45$ & $0.53$ &  & $0.23$ & $-0.82$ & $0.21$ & $-0.69$ \\
\hline\\
\hline
\multicolumn{11}{c}{40 Late-Type Galaxies (Sp):$\bm{\log(M_{\rm BH}/M_{\rm *,sph})=\alpha\log(M_{\rm *,sph}/[\upsilon(5\times10^{10}\ {\rm M_{\sun}})])+\beta}$} \\
\hline
\textsc{bces}$(Bisector)$ & \textit{Symmetric} & $1.22\pm0.21$ & $-2.08\pm0.16$ & $0.56$ & $0.65$ &  & $0.18$ & $-0.56$ & $0.18$ & $-0.59$ \\
\hline\\
\hline
\multicolumn{11}{c}{36 Galaxies with a Disk (ES/S0): $\bm{\log(M_{\rm BH}/M_{\rm *,gal})=\alpha\log(M_{\rm *,gal}/[\upsilon(5\times10^{10}\ {\rm M_{\sun}})])+\beta}$} \\
\hline
\textsc{bces}$(Bisector)$ & \textit{Symmetric} & $1.12\pm0.17$ & $-2.56\pm0.12$ & $0.72$ & $0.74$ &  & $0.10$ & $-0.25$ & $0.12$ & $-0.30$ \\
\hline\\
\hline
\multicolumn{11}{c}{40 Galaxies without a Disk (E): $\bm{\log(M_{\rm BH}/M_{\rm *,gal})=\alpha\log(M_{\rm *,gal}/[\upsilon(5\times10^{10}\ {\rm M_{\sun}})])+\beta}$} \\
\hline
\textsc{bces}$(Bisector)$ & \textit{Symmetric} & $1.07\pm0.08$ & $-3.06\pm0.10$ & $0.50$ & $0.54$ &  &$0.23$ & $-0.83$ & $0.21$ & $-0.72$ \\
\hline\\
\hline
\multicolumn{11}{c}{40 Late-Type Galaxies (Sp):$\bm{\log(M_{\rm BH}/M_{\rm *,gal})=\alpha\log(M_{\rm *,gal}/[\upsilon(5\times10^{10}\ {\rm M_{\sun}})])+\beta}$} \\
\hline
\textsc{bces}$(Bisector)$ & \textit{Symmetric} & $1.45\pm0.66$ & $-3.70\pm0.14$ & $0.67$ & $0.70$ &  & $0.12$ & $-0.32$ & $0.18$ & $-0.56$ 
\enddata
\tablecomments{The data and linear regression for late-type galaxies is taken from \citet{Davis:2018:a}.
Columns:
(1) Regression performed.
(2) The coordinate direction in which the offsets from the regression line is minimized.
(3) Slope of the regression line.
(4) Intercept of the regression line.
(5) Intrinsic scatter in the $M_{\rm BH}$ direction \citep[using Equation 1 from][]{Graham:Driver:2007}.
(6) Root mean square scatter in the $M_{\rm BH}$ direction.
(7) Pearson correlation coefficient.
(8) The Pearson correlation probability value.
(9) Spearman rank-order correlation coefficient.
(10) The Spearman rank-order correlation probability value.
}
\end{deluxetable*}
 
\section{Conclusions and Implications}
\label{conclusions}

Our work, based on the largest sample of ETGs with directly-measured SMBH masses, establishes a robust relation between the black hole mass and both the spheroid and galaxy stellar mass. While the color-magnitude relation for ETGs results in a steep $M_{BH}$--$L_{*,sph}$ relation in the optical bands for $MAG_{K_s}>-22$ mag, i.e., $B-K_s \leq 4.0$ \citep{Graham:Soria:2018}, the slopes at the low- and high-luminosity end of the $M_{BH}$--$L_{*,sph}$ relation based on infrared magnitudes are equal to each other. That is, the $M_{BH}$--$M_{*,sph}$ relation for ETGs appears to be defined by a single log-linear relation. This helps to clarify debate over the existence of a steeper (at the low-mass end) and \enquote{bent} $M_{BH}$--$M_{*,sph}$ relation for ETGs. 

Using our image reduction, profile extraction, and multi-component decomposition techniques, we carefully measured the spheroid and galaxy stellar luminosities and masses. We applied the \textsc{BCES} bisector regression to our dataset, providing a symmetric treatment to both the $M_{BH}$ and $M_{*,sph}$ or $M_{*,gal}$ data (we additionally report the scaling relations obtained from other asymmetric regressions in Table \ref{fit parameters}).

We checked the consistency of our $M_{BH}$--$M_{*,sph}$ and $M_{BH}$--$M_{*,gal}$ scaling relations using stellar masses based on color-dependent stellar mass-to-light ratios and found it to be in agreement with our scaling relations based on the constant stellar mass-to-light ratios. This may in part be because our ETGs have fairly constant, red, colors (Figure \ref{color_mag}). 
Our key results can be summarized as follows:

\begin{itemize}
 
\item{Having performed separate regressions using 45 S{\'e}rsic and 31 core-S{\'e}rsic galaxies, we found that, for ETGs, there is no significant bend in either the $M_{BH}$--$M_{*,sph}$ or $M_{BH}$--$M_{*,gal}$ diagram due to S{\'e}rsic and core-S{\'e}rsic galaxies (Figure-\ref{CS_S}).}

\item{ETGs follow a steep $M_{BH}\propto M^{1.27\pm 0.07}_{*,sph}$ relation, with total rms scatter of 0.52 dex in
the $\log M_{BH}$. The slope of this relation is non-linear at the $3\sigma$ bound, leading us to the conclusion that a steeper than linear $M_{BH}$--$M_{*,sph}$ relation exists for ETGs. This also implies that the $M_{BH}/M_{*,sph}$ ratio is not a constant but varies along the relation. }

\item{The SMBH mass of ETGs follow an even steeper relation with the host \textit{galaxy} stellar mass: $M_{BH}\propto M^{1.65\pm 0.11}_{*,gal}$ with an rms scatter (in the $\log M_{BH}$ direction) of 0.58 dex. The slope of this relation is non-linear at the $5.9\sigma$ level. The similarity in the rms scatter of this relation with that of $M_{BH}$--$M_{*,sph}$ relation suggests that black hole mass correlates almost equally well with galaxy mass (luminosity) as it does with spheroid mass (luminosity) for ETGs (Figure \ref{single_reg_scs}). Hence, for the cases where bulge/disk decomposition is difficult, the $M_{BH}$--$M_{*,gal}$ relation can be used to estimate the black hole mass of an ETG using the total galaxy stellar mass. However, as noted below, this approach is not preferred if one knows whether or not the ETG under study contains a disk.}

\item{We discovered separate relations for ETGs with an intermediate-scale or extended disk (ES or S0) and ETGs without a disk (E), having slopes $1.86\pm0.20$ and $1.90\pm0.20$  in the $M_{BH}$--$M_{*,sph}$ diagram, with an rms scatter in the $\log M_{BH}$ direction of 0.57 dex and 0.50 dex, respectively. Crucially, galaxies with a disk are offset from galaxies without a disk (Figure \ref{E_ESS0}) by more than an order of magnitude (1.12 dex) in their $M_{BH}/M_{*,sph}$ ratio. This is likely due to the exclusion of the disk light, rather than an issue with the black hole mass. To better estimate the black hole mass of an ETG, one should use the corresponding $M_{BH}$--$M_{*,sph}$ relation depending on whether the ETG has a disk or not.}

\item{For the $M_{BH}$--$M_{*,gal}$ relation, the intercepts of the two regression lines (for galaxies with and without a disk) differ only by a factor of 2. Hence, the relation obtained by a single regression (Equation \ref{Mbh_Mgal}) may still prove to be preferable for estimating the black hole mass when uncertain about the presence of a disk in an ETG, or for those without a careful multi-component decomposition.}

\item{We found that the regression line for the barred galaxies (which reside at the lower-mass end of our diagrams) are largely consistent with the regression line for the non-barred galaxies in both the $M_{BH}$--$M_{*,sph}$ and $M_{BH}$--$M_{*,gal}$ diagrams (Figures \ref{BNB} and \ref{BNB1}). However, with only 15 barred galaxies, we restrict our conclusion to noting that the barred galaxies do not appear to have lower SMBH masses than the non-barred galaxies in either the $M_{BH}$--$M_{*,sph}$ diagram or  the $M_{BH}$--$M_{*,gal}$ diagram.}

\item{Combining the 76 ETGs studied here, with the 40 LTGs from \citet{Davis:2018:a}, we observe a difference in the slope of the regression lines for ETGs and LTGs (Figure \ref{ETG_LTG}) in both the $M_{BH}$--$M_{*,sph}$ and $M_{BH}$--$M_{*gal}$ diagrams. The LTGs define steeper relations, such that $M_{BH}\propto M_{*,sph}^{2.17\pm 0.32}$ and $M_{BH}\propto M_{*,gal}^{3.05\pm 0.70}$. These slopes for the LTGs are almost double that of the ETGs. This agrees with the change noticed by \citet{Savorgnan:2016:Slopes} in the $M_{BH}$--$M_{*,sph}$ diagram.}

\item{We also found that the behaviour of three sub-populations of galaxies (E, ES/S0 and, Sp) in the $(M_{BH}/M_{*,sph})$--$M_{*,sph}$ and $(M_{BH}/M_{*,gal})$--$M_{*,gal}$ diagrams agree with the corresponding $M_{BH}$--$M_{*,sph}$ and $M_{BH}$--$M_{*,gal}$ relations (see Figures \ref{E_ESS0}, \ref{ETG_LTG} and \ref{Ratio}), supporting the obvious implication of our non-linear $M_{BH}$ vs $M_{*,sph}$ and $M_{*,gal}$ scaling relations, specifically that the $M_{BH}/M_{*,sph}$ and $M_{BH}/M_{*,sph}$ ratios are not constant.}

\end{itemize}

The existence of substructure within the $M_{BH}$--$M_{*,sph}$ diagram, due to sub-populations of ETGs with and without disks, and spiral galaxy bulges, means that past efforts to calibrate the virial $f$-factor using the $M_{BH}$--$M_{*,sph}$  diagram---used for converting virial masses of active galactic nuclei into black hole masses \citep[e.g.,][]{Bentz:Manne:2018}--- will benefit from revisiting. Calibration of the offset between the ensemble of virial masses for AGN and the ensemble of directly measured black hole mass should be performed separately using the significantly different, non-linear, $M_{BH}$--$M_{*,sph}$ relations for ETGs and LTGs, while taking into account the presence or absence of a disk in the ETGs. A similar situation exists with the $M_{BH}$--$\sigma$ diagram, due to the offset sub-populations of galaxies with and without bars \citep{Graham:2011}. In Sahu et al. (2019, in preparation) we will present an analysis of the $M_{BH}$--$\sigma$ relation based on the various sub-samples of the ETG population used in this paper. We will also do this using our combined sample of 120 ETGs and LTGs.
 
Extending our search for the most fundamental black hole mass scaling relation, we will explore the correlation of black hole mass with the spheroid's S\'ersic index\footnote{The S{\'e}rsic index is a measure of the radial concentration of stellar mass.} (n) and half light radius ($R_e$). We already have these two parameters from our homogeneous bulge/disk decomposition of ETGs and LTGs \citep{Davis:2018:a} . We intend to check for the existence of a fundamental plane rather than a line. However, care needs to be taken given that the $L$--$R_e$ relation is curved \citep[e.g.][Graham 2019, submitted]{Graham:Worley:2008}.

The black hole mass scaling relations presented in this work, based on a local ($z\approx 0$) sample of ETGs, can be used to estimate the black hole masses in other galaxies which do not have their SMBH's gravitational sphere-of-influence spatially resolved.

These scaling relations can be further used to derive the black hole mass function from the galaxy luminosity function, for the first time separating the galaxy population according to their morphological type. We plan to calculate the SMBH mass function by applying the black hole mass scaling relations for ETGs and LTGs to the updated spheroid and galaxy luminosity functions from GAMA data \citep{Driver:GAMA:2009} for which the morphological types are known and bulge/disk decompositions have been performed. 

The SMBH mass function, accompanied with knowledge of the galaxy/SMBH merger rate, can be used to constrain the ground-based detection rate of long-wavelength gravitational waves, which are actively being searched for by the Parkes Pulsar Timing Array \citep[PPTA,][]{Shannon:2015, Hobbs:2017}, the European Pulsar Timing Array \citep[EPTA,][]{Stappers:Kramer:2011}, and the North American Nanohertz Observatory for Gravitational Waves \citep[NANOGrav,][]{Siemens:2019}. Using the forth-coming SMBH mass function, we intend to improve the predictions for the detection of the gravitational waves from PTA and make new predictions for detection from the recently inaugurated MeerKAT telescope \citep{Jonas:2007}. The revised black hole scaling relations can also be used to predict the detection of gravitational waves from future space-based detectors. For example, \citet{Mapelli:2012:GW:Space:detectors} investigate the detection of gravitational waves produced from the merger of SMBHs with stellar mass BHs and neutron stars in the central nuclear star clusters of galaxies \citep{Hartmann:2011}.

\acknowledgements
We thank Edward (Ned) Taylor for his helpful comments on calibrating the stellar mass-to-light ratios for $r^{\prime}$-band images and conversion of IMFs. This research was conducted by the Australian Research Council Centre of Excellence for Gravitational Wave Discovery (OzGrav), through project number CE170100004. AWG was supported under the Australian Research Council’s funding scheme DP17012923. This work has made use of the NASA/IPAC Infrared Science Archive and the NASA/IPAC Extragalactic Database (NED). This research has also made use of the Two Micron All Sky Survey and Sloan Digital Sky Survey database. We also acknowledge the use of the HyperLeda database \url{http://leda.univ-lyon1.fr}.

\bibliography{bibliography}

\appendix


\setcounter{figure}{0}    
\setcounter{table}{0}
\vspace*{-1cm}
\section{Surface Brightness Profiles for Early-type Galaxies }
Here we provide the major-axis and equivalent-axis (i.e. geometric mean axis $=\sqrt{R_{maj} R_{min}}$) surface brightness profiles (AB magnitude system) for the 41 ETGs that we modeled (apart from NGC~4762, Figure \ref{profiler}) . Magnitudes and stellar masses of these galaxies, and their spheroids are presented in Table \ref{data_table} in the main paper. The current paper does not directly use the parameters from our decomposition of these light profiles; however, we intend to use them in our upcoming work, where we will tabulate them there.
\subsection{Light profiles from Spitzer $3.6\mu$m images}
\vspace{-3\baselineskip}
\begin{figure}[H]
\includegraphics[clip=true,trim= 1mm 1mm 1mm 1mm,height=12cm,width=0.49\textwidth]{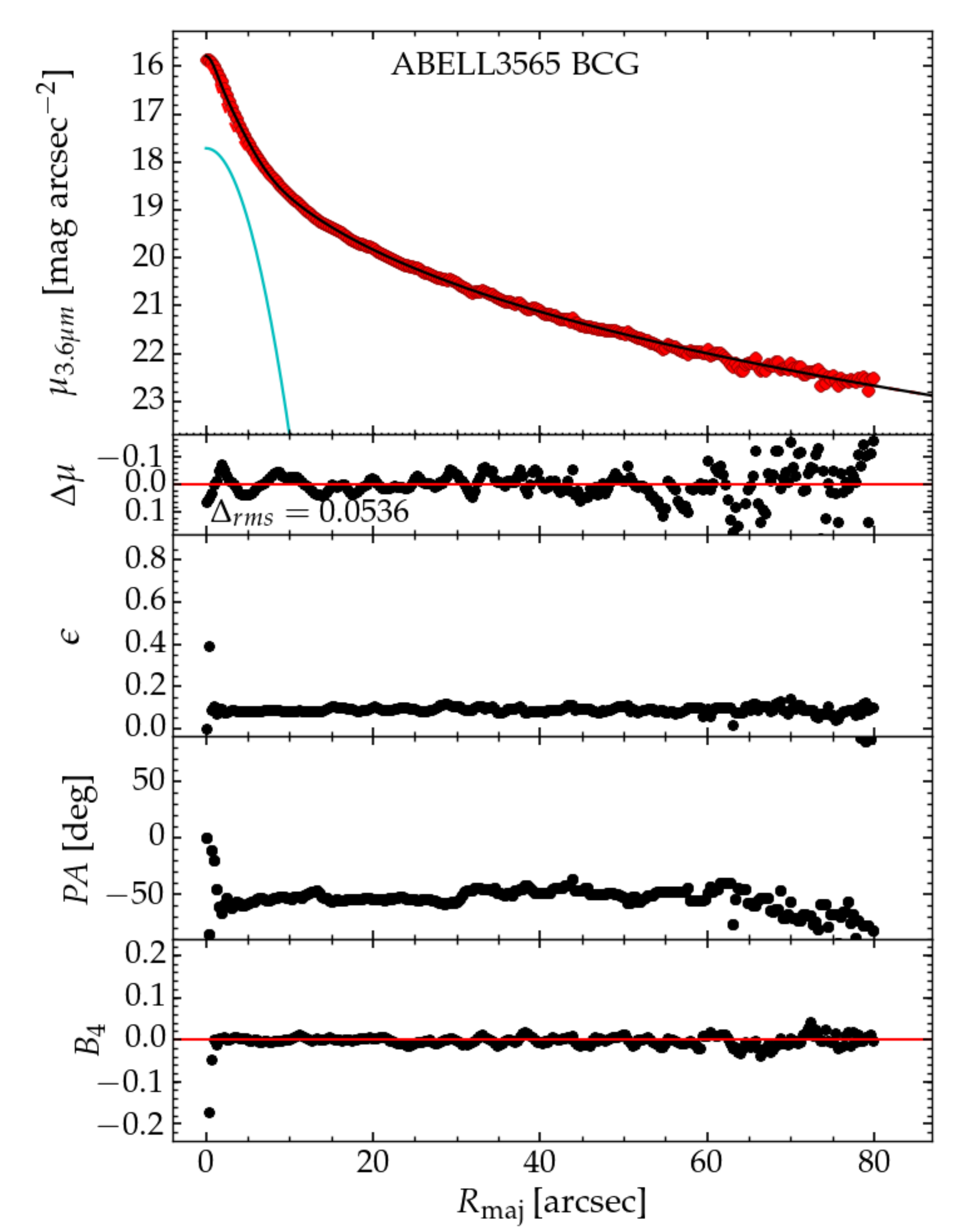}
\includegraphics[clip=true,trim= 1mm 1mm 1mm 1mm,height=12cm,width=0.49\textwidth]{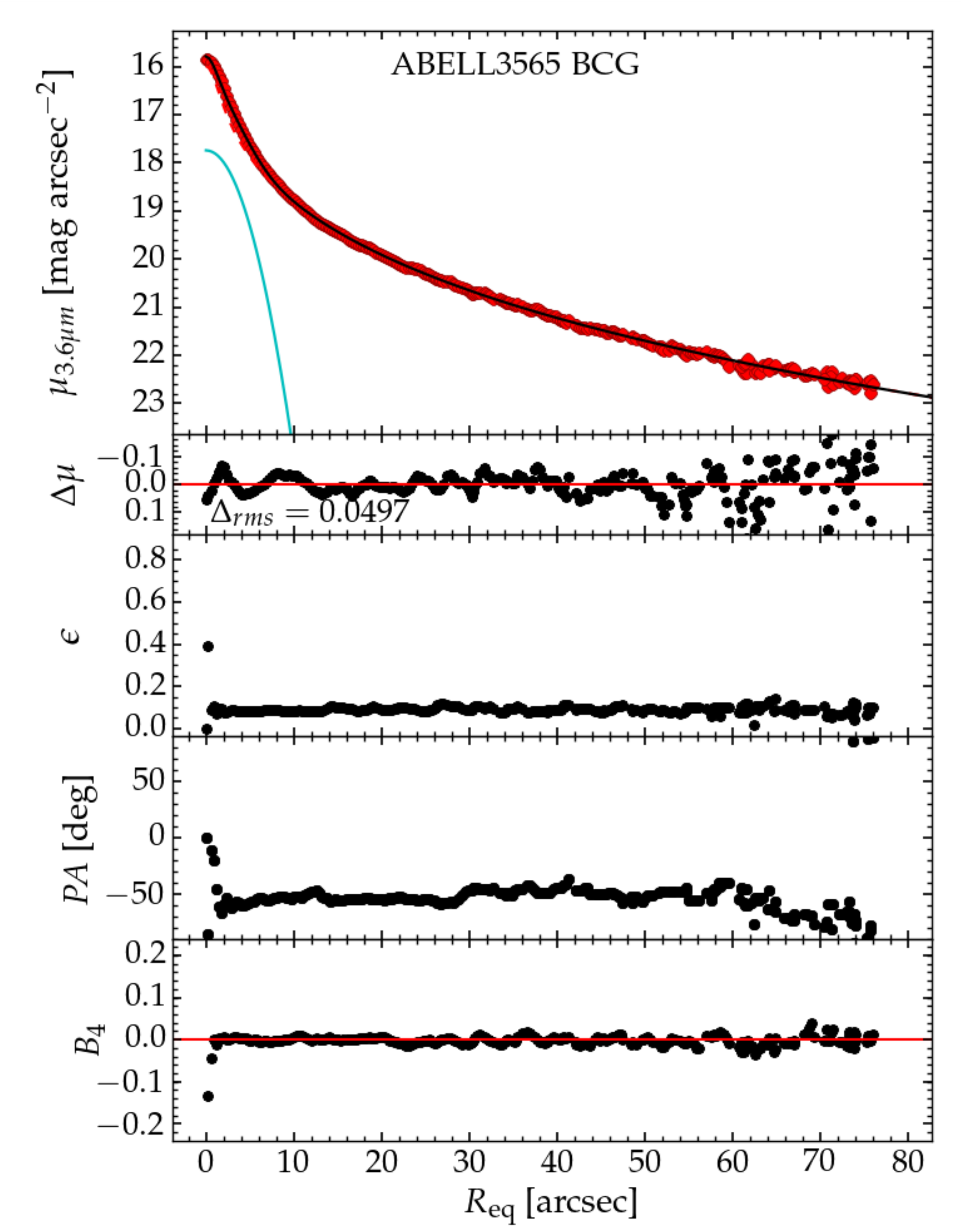}
\caption{ABELL 3565 BCG (IC 4296): elliptical galaxy with an extended spheroid fit using a S{\'e}rsic function (\textcolor{red}{---}) plus a Gaussian (\textcolor{cyan}{---}) accounting for extra light from a central source. IC 4296 has a very high velocity dispersion suggesting it may be a core-S{\'e}rsic galaxy, but we do not have evidence for a deficit of light at its core in the Spitzer data.}
\label{A3565BCG}
\end{figure}

\begin{figure}[H]
\includegraphics[clip=true,trim= 1mm 1mm 1mm 1mm,height=12cm,width=0.49\textwidth]{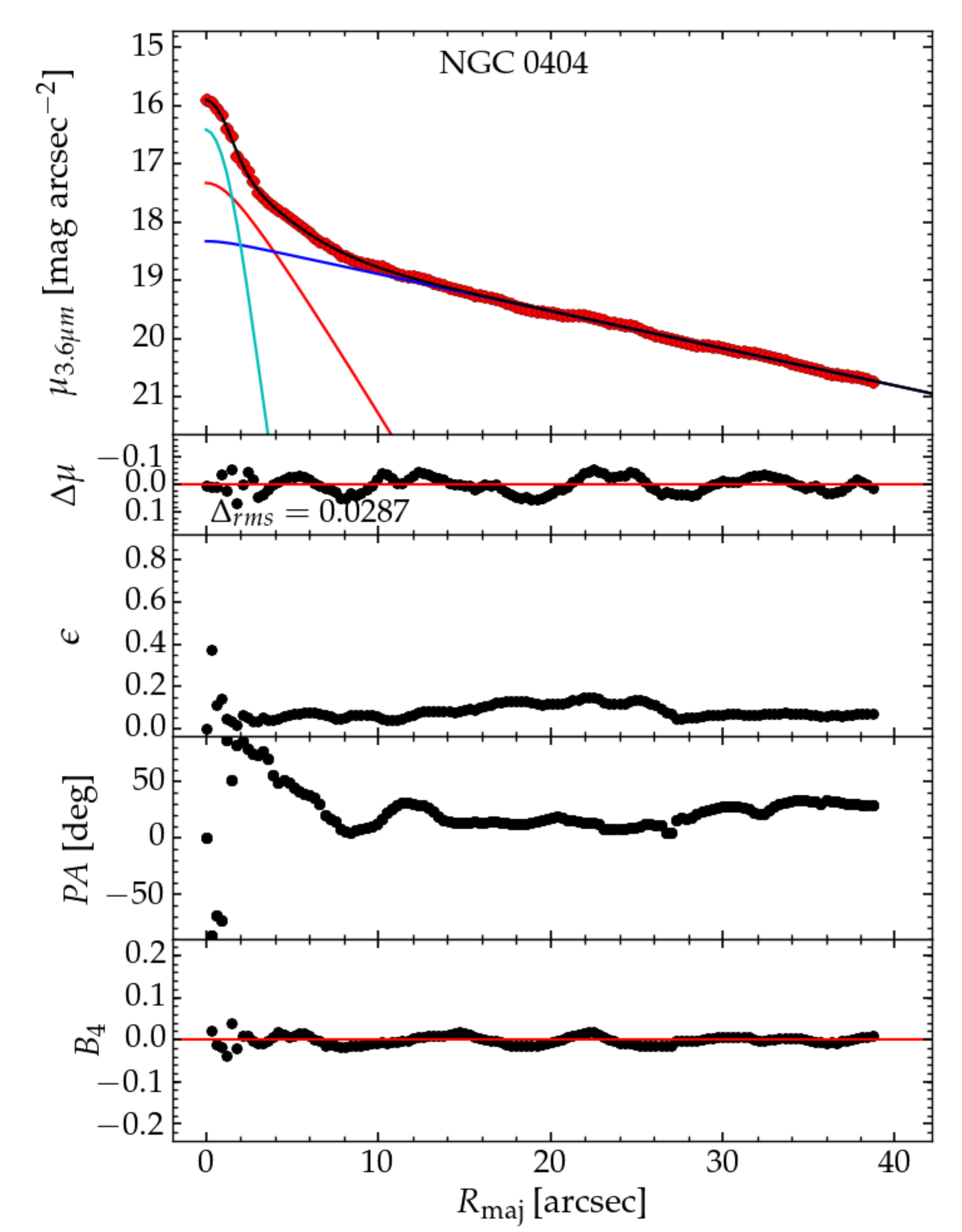}
\includegraphics[clip=true,trim= 1mm 1mm 1mm 1mm,height=12cm,width=0.49\textwidth]{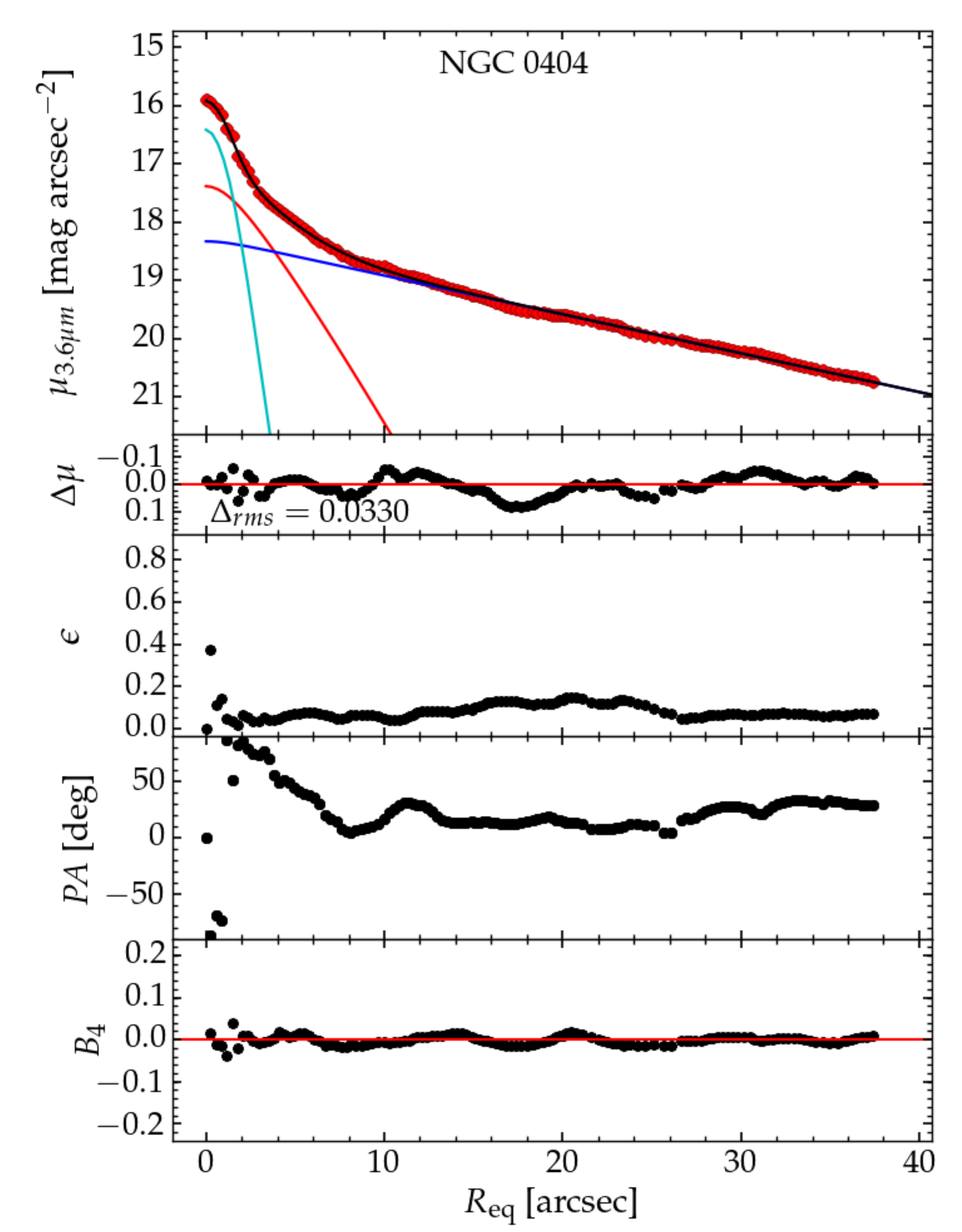}
\caption{NGC 404: a dwarf lenticular galaxy hosting an AGN at its center and a nuclear star cluster \citep{Nguyen:2017}. We fit a S{\'e}rsic function (\textcolor{red}{---}) for its bulge, an exponential for the disk (\textcolor{blue}{---}), and a Gaussian (\textcolor{cyan}{---}) for the central AGN. }
\label{NGC 404}
\end{figure}


\begin{figure}[H]
\includegraphics[clip=true,trim= 1mm 1mm 1mm 1mm,height=12cm,width=0.49\textwidth]{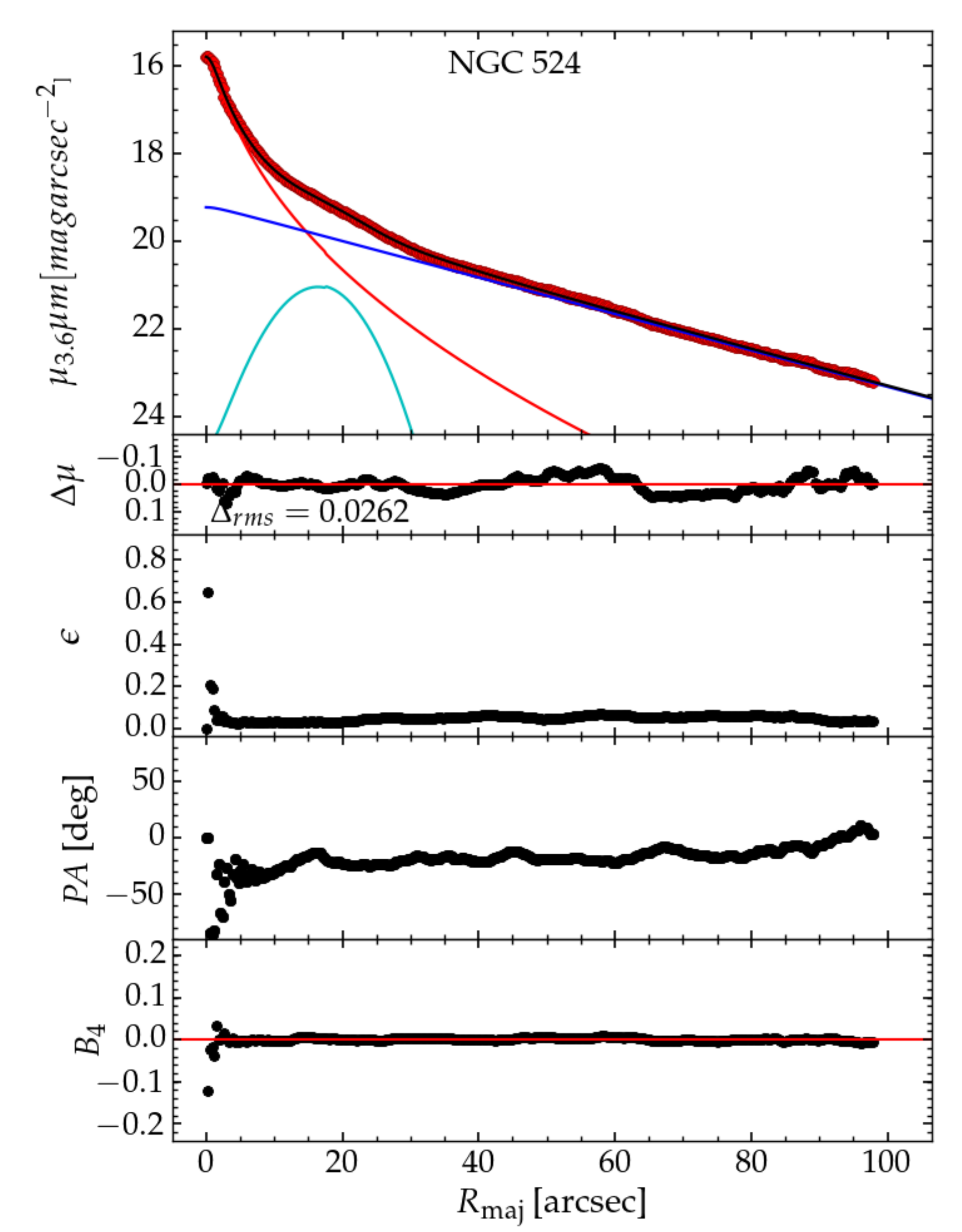}
\includegraphics[clip=true,trim= 1mm 1mm 1mm 1mm,height=12cm,width=0.49\textwidth]{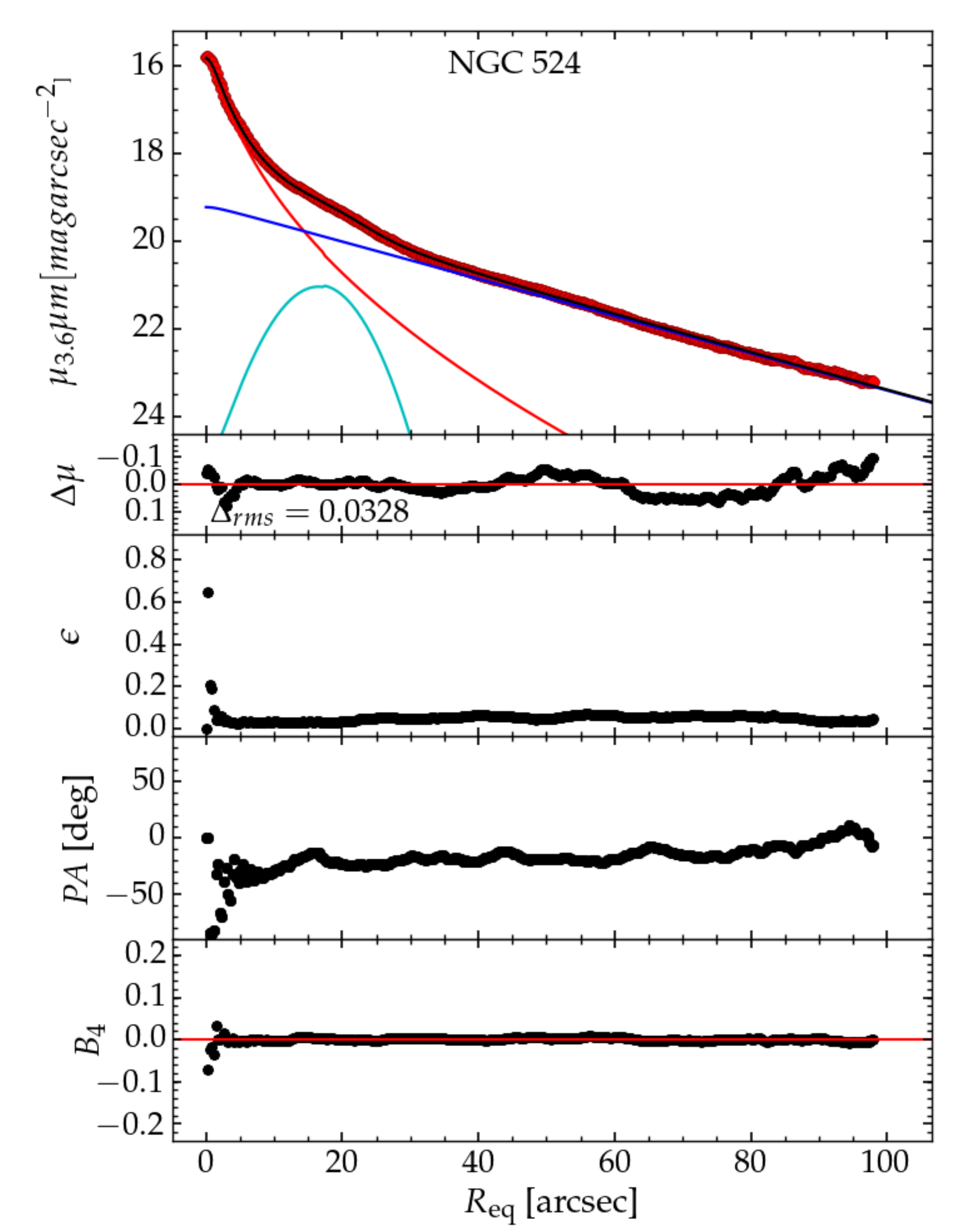}
\caption{NGC 524: a face-on lenticular galaxy with a core-S{\'e}rsic (\textcolor{red}{---}) bulge \citep{Richings:2011}. The galaxy has a faint ring at about $R_{maj}=20\arcsec$ which we fit using a Gaussian (\textcolor{cyan}{---}), and there is an extended exponential disk (\textcolor{blue}{---}). }
\label{NGC 524} 
\end{figure}


\begin{figure}[H]
\includegraphics[clip=true,trim= 1mm 1mm 1mm 1mm,height=12cm,width=0.49\textwidth]{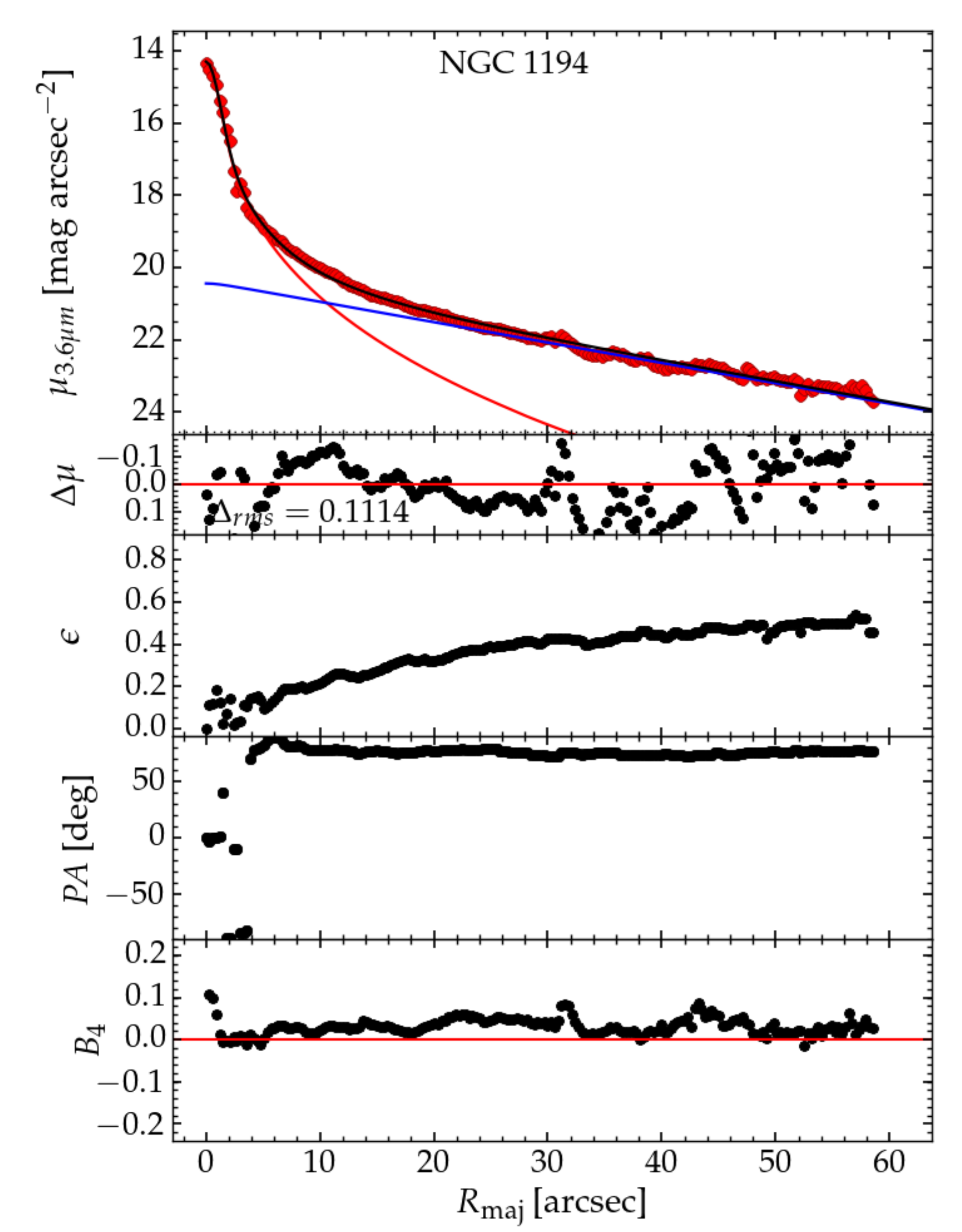}
\includegraphics[clip=true,trim= 1mm 1mm 1mm 1mm,height=12cm,width=0.49\textwidth]{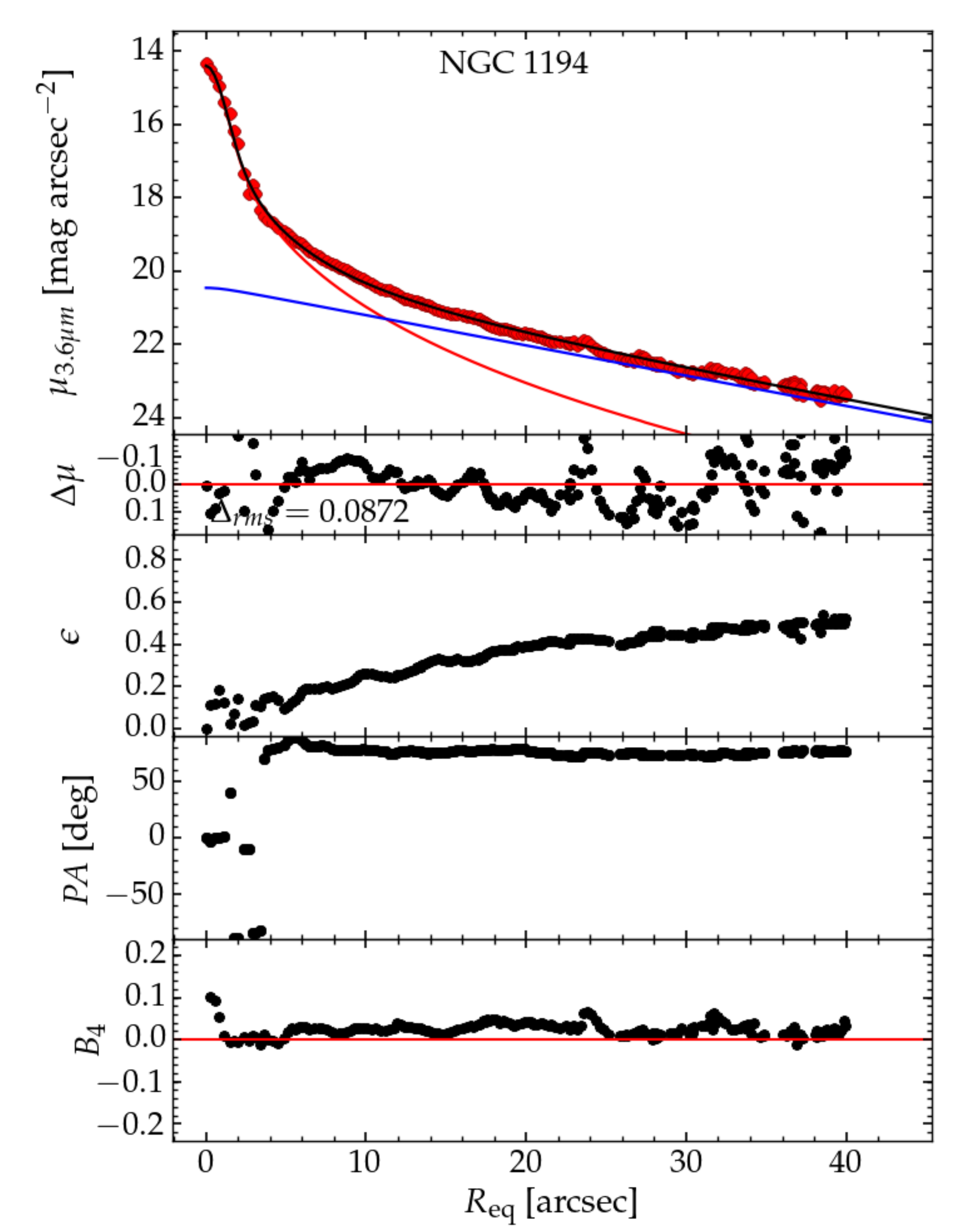}
\caption{NGC 1194: a lenticular, warped disk \citep{Fedorova:2016} galaxy, fit with a S{\'e}rsic bulge (\textcolor{red}{---}) and an extended exponential disk (\textcolor{blue}{---}). It also has a faint debris tail, suggesting it may have undergone a merger, and \citet{Fedorova:2016} also hypothesize that NGC 1194 may harbor two black holes.}
\label{NGC 1194}
\end{figure}

\begin{figure}[H]
\includegraphics[clip=true,trim= 1mm 1mm 1mm 1mm,height=12cm,width=0.49\textwidth]{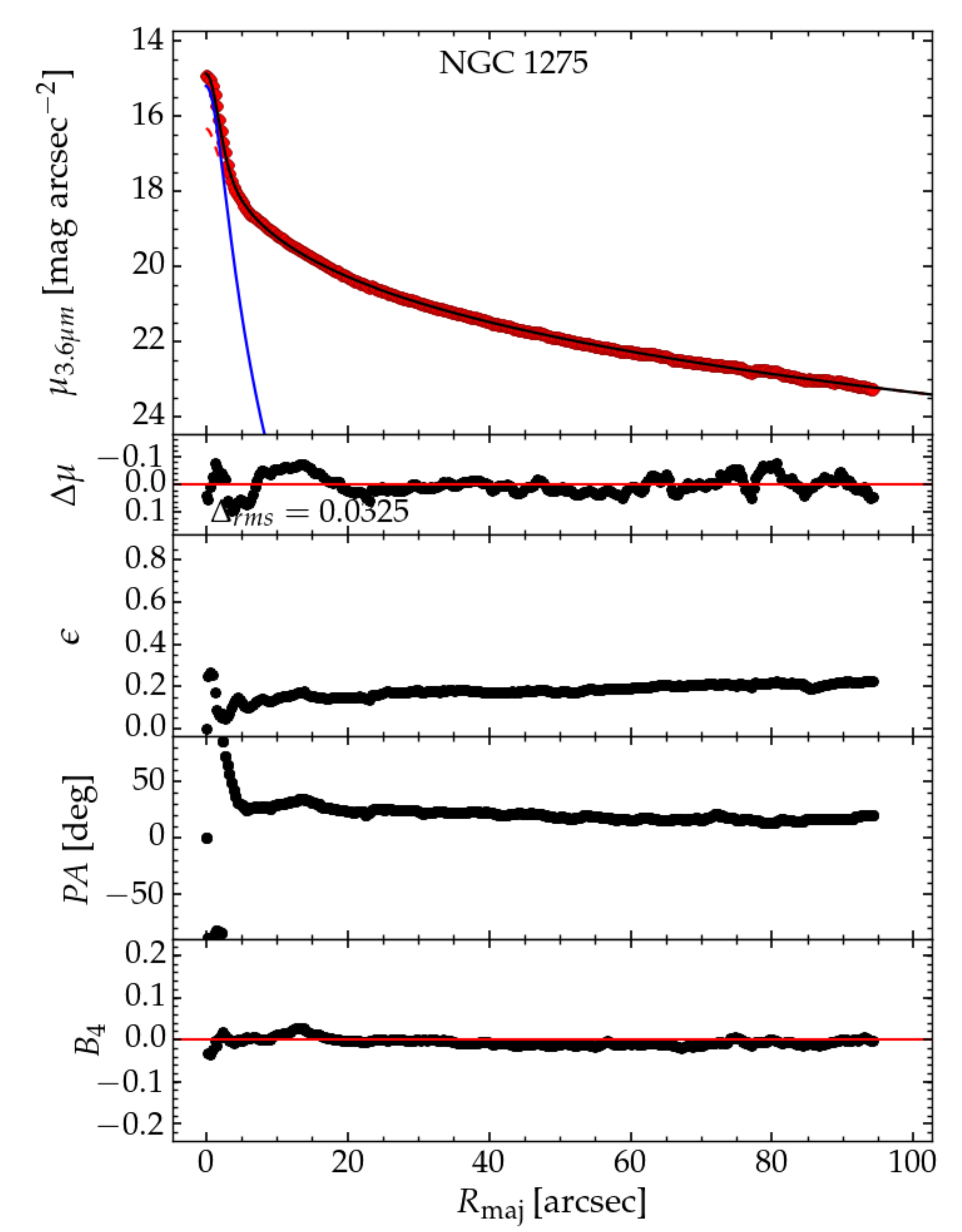}
\includegraphics[clip=true,trim= 1mm 1mm 1mm 1mm,height=12cm,width=0.49\textwidth]{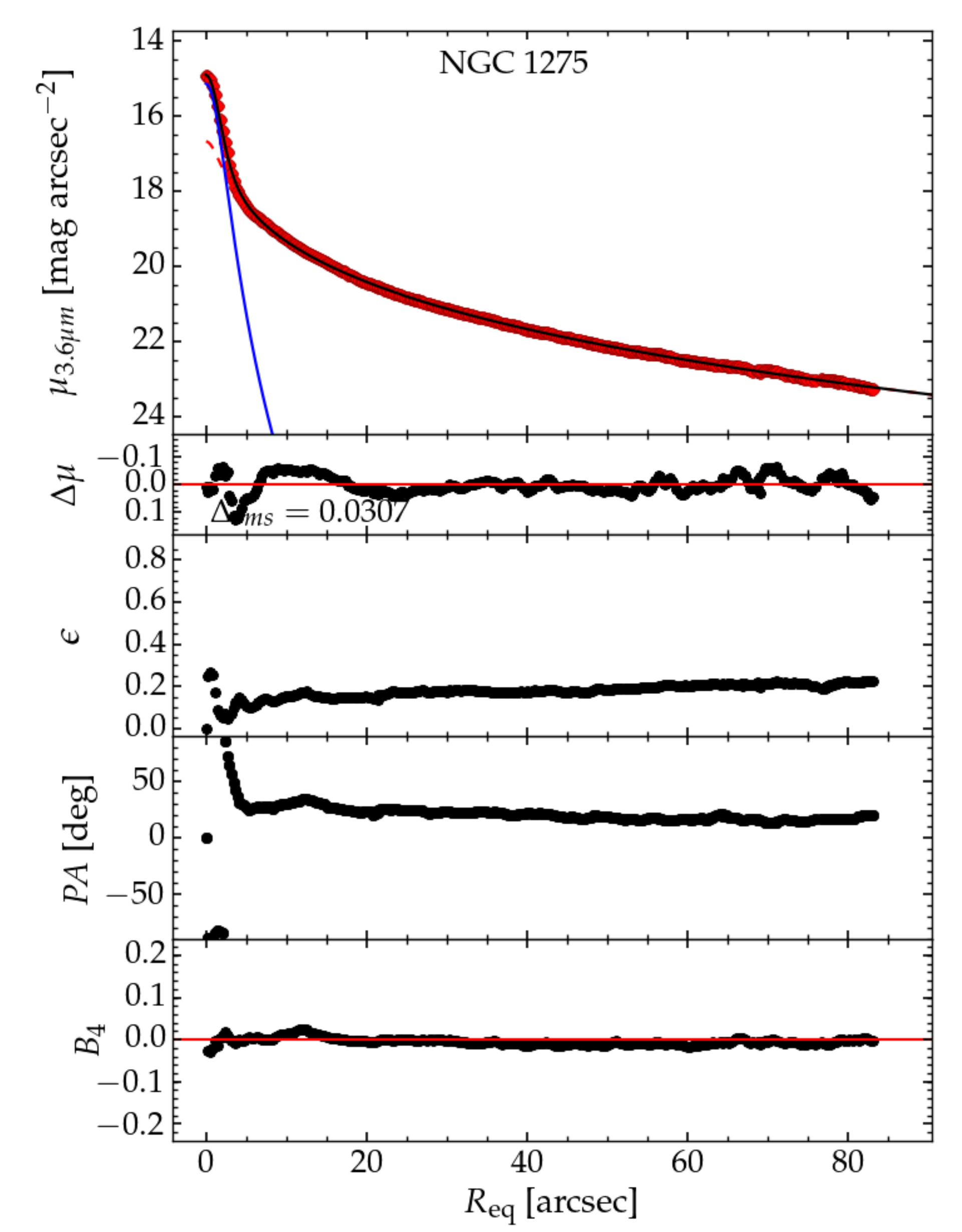}
\caption{NGC 1275: a peculiar elliptical galaxy with an extended bright object at the center (resolved in HST images), fit using an inclined disk (\textcolor{blue}{---}) along with the extended S{\'e}rsic spheroid (\textcolor{red}{---}). The sharp bump in the ellipticity and position profile also hints at the presence of a central disky object.}
\label{NGC 1275}
\end{figure}

\begin{figure}[H]
\includegraphics[clip=true,trim= 1mm 1mm 1mm 1mm,height=12cm,width=0.49\textwidth]{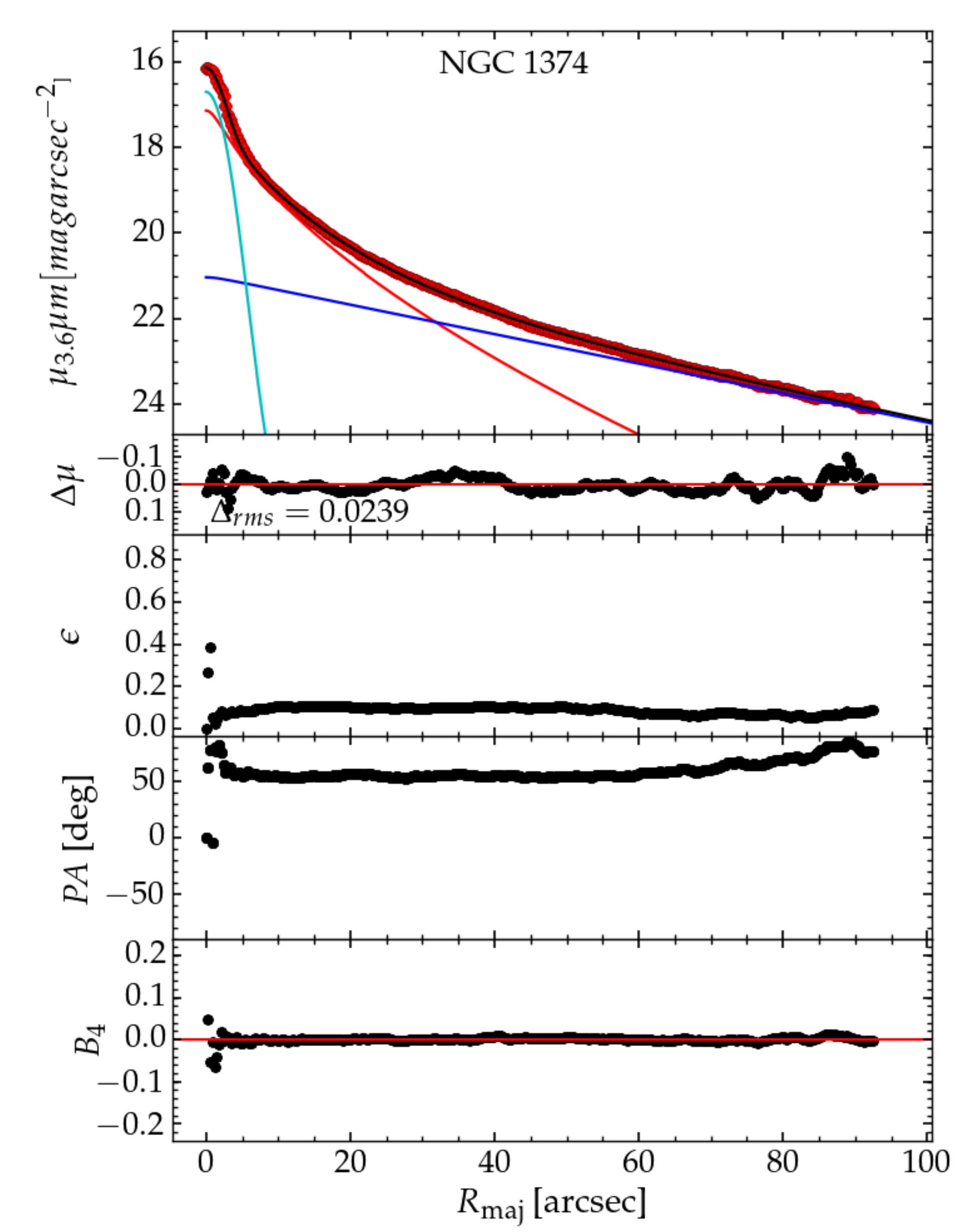}
\includegraphics[clip=true,trim= 1mm 1mm 1mm 1mm,height=12cm,width=0.49\textwidth]{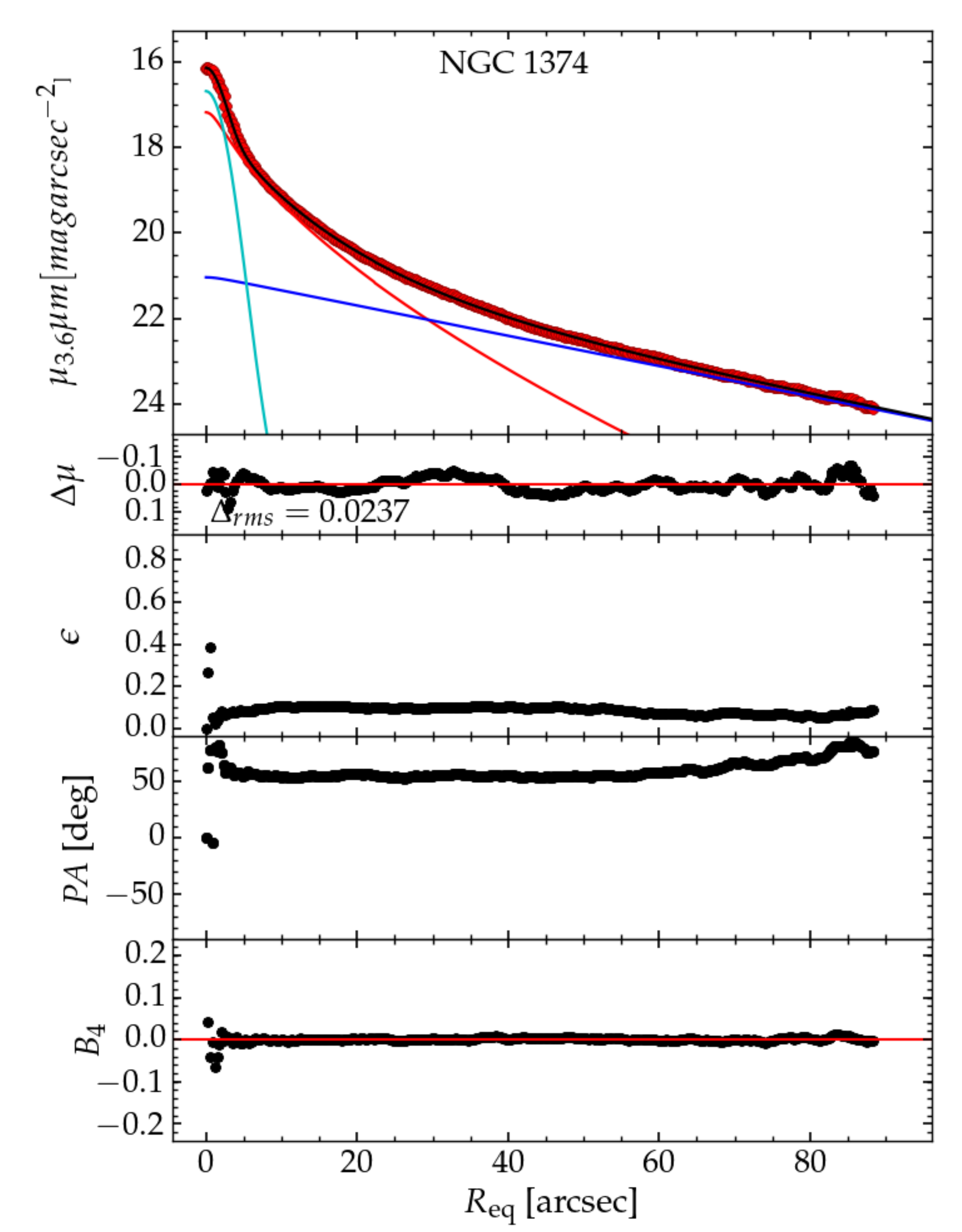}
\caption{NGC 1374:  a face-on lenticular galaxy \citep{Longo:1994, DOnofrio:1995} suspected to have a depleted stellar core \citep{Rusli:core:2013}. Due to the lack of evidence for a depleted core we fit a S{\'e}rsic function (\textcolor{red}{---}) to its bulge plus a Gaussian (\textcolor{cyan}{---}) for a nuclear source possibly related to a peak at $\sim 5\arcsec$ from the center of the rotation curve presented by \citet{Longo:1994}. We also fit an exponential disk (\textcolor{blue}{---}) component, based on the kinematic profile from \citet{DOnofrio:1995}. }
\label{NGC 1374}
\end{figure}

\begin{figure}[H]
\includegraphics[clip=true,trim= 1mm 1mm 1mm 1mm,height=12cm,width=0.49\textwidth]{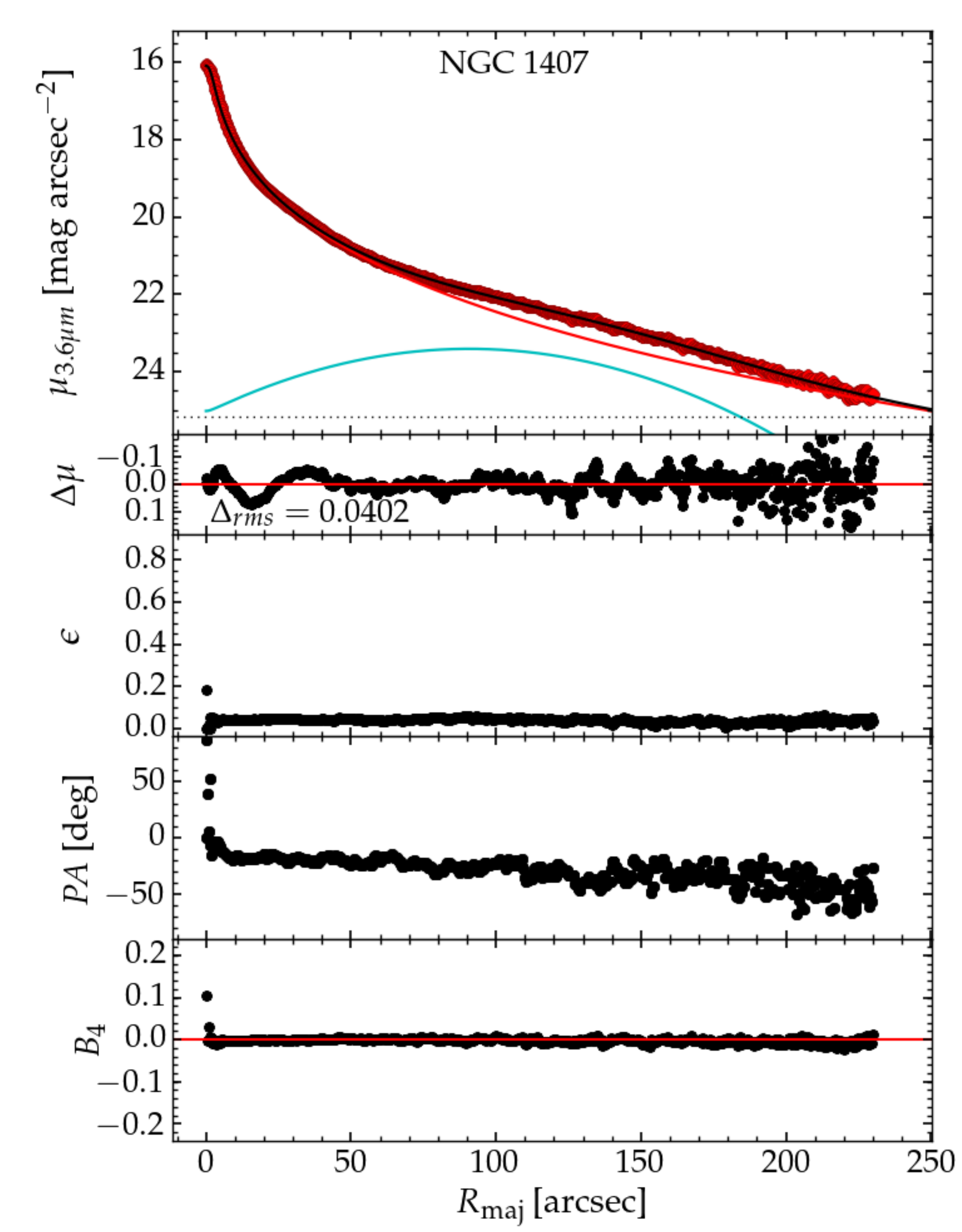}
\includegraphics[clip=true,trim= 1mm 1mm 1mm 1mm,height=12cm,width=0.49\textwidth]{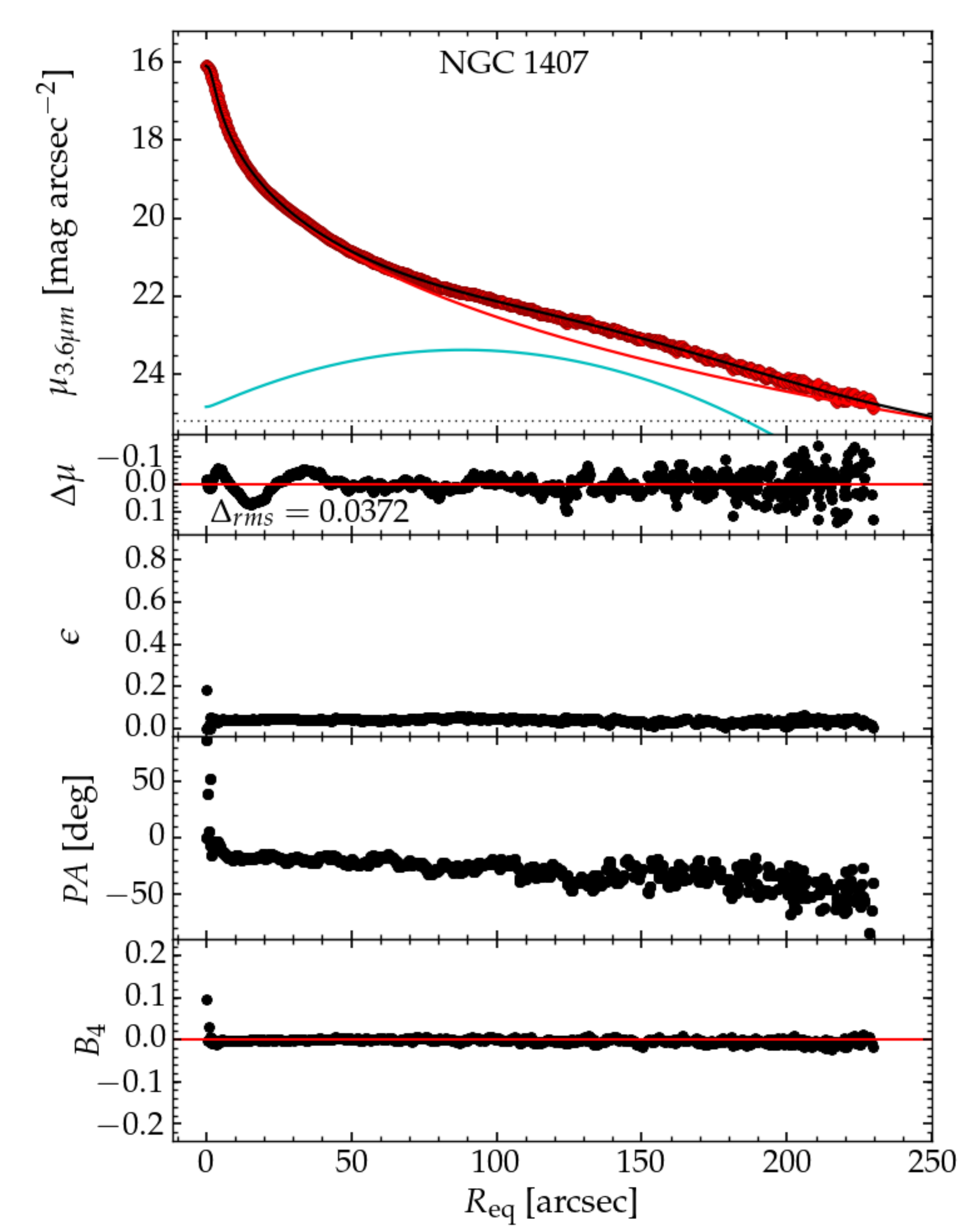}
\caption{NGC 1407: a massive elliptical galaxy with a deficit of light at it core. Its surface brightness profile is fit using a core-S{\'e}rsic function (\textcolor{red}{---})  and a broad Gaussian (\textcolor{cyan}{---}) which accounts well for the bump in the light profile, possibly due to a semi-digested galaxy.}
\label{NGC 1407}
\end{figure}

\begin{figure}[H]
\includegraphics[clip=true,trim= 1mm 1mm 1mm 1mm,height=12cm,width=0.49\textwidth]{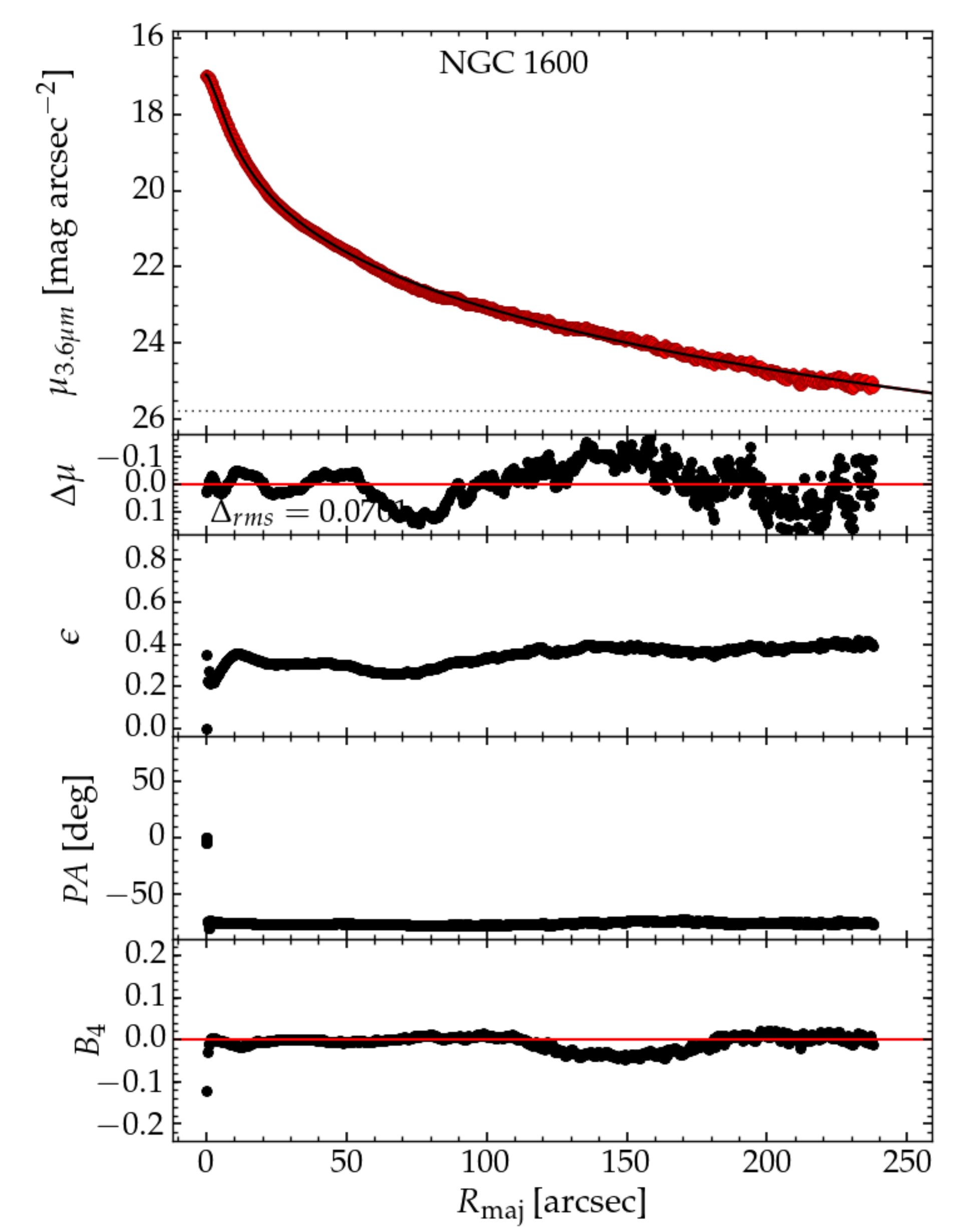}
\includegraphics[clip=true,trim= 1mm 1mm 1mm 1mm,height=12cm,width=0.49\textwidth]{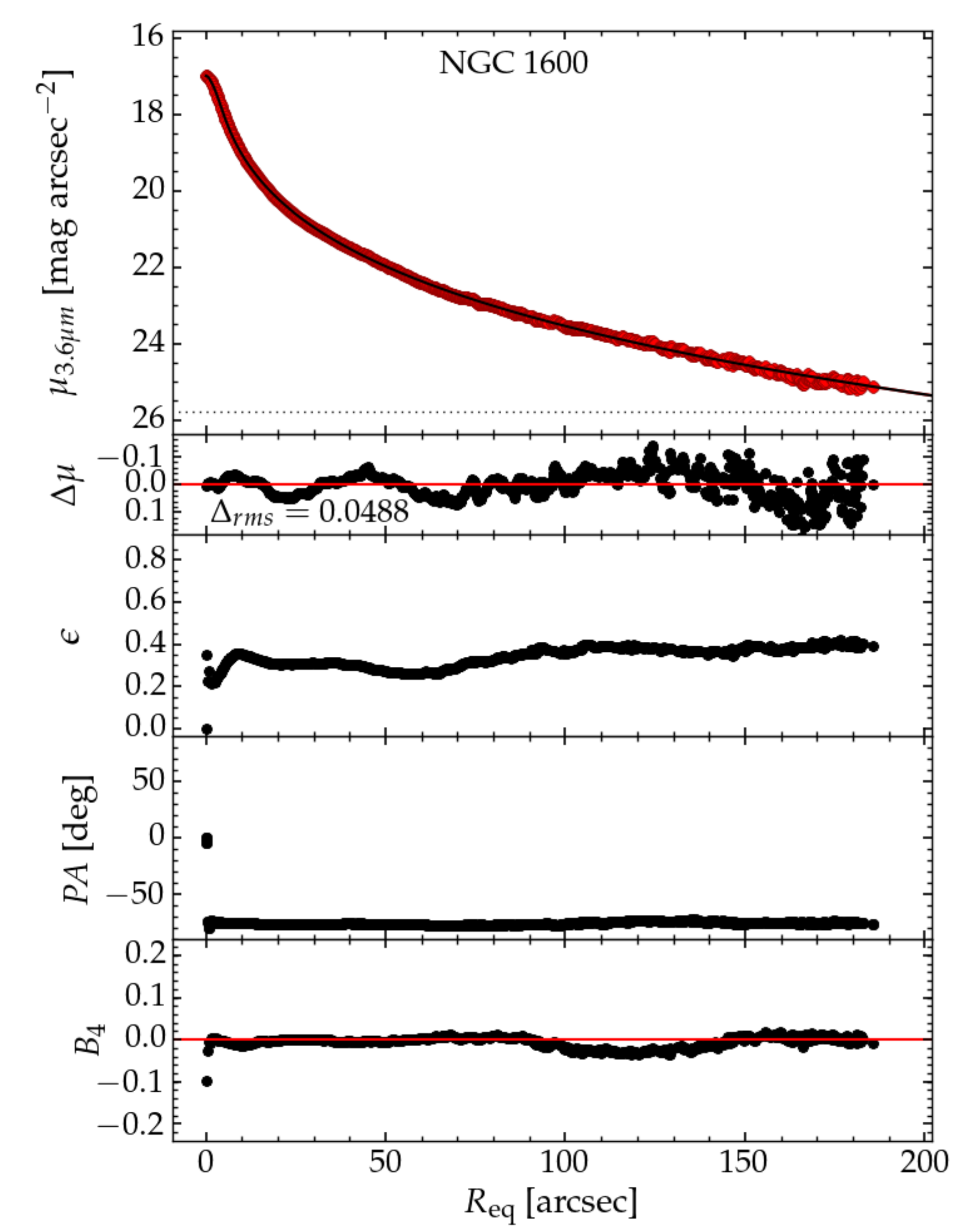}
\caption{NGC 1600: an elliptical galaxy with a depleted core. Its spheroid is fit using a core-S{\'e}rsic function (\textcolor{red}{---}). The shallow dip in the $B_4$ profile is associated with the presence of a tidal debris tail at $\sim 150\arcsec$ along the semi-major axis, which makes the galaxy look boxy (negative $B_4$) at those radii. }
\label{NGC 1600}
\end{figure}
\begin{figure}[H]
\includegraphics[clip=true,trim= 1mm 1mm 1mm 1mm,height=12cm,width=0.49\textwidth]{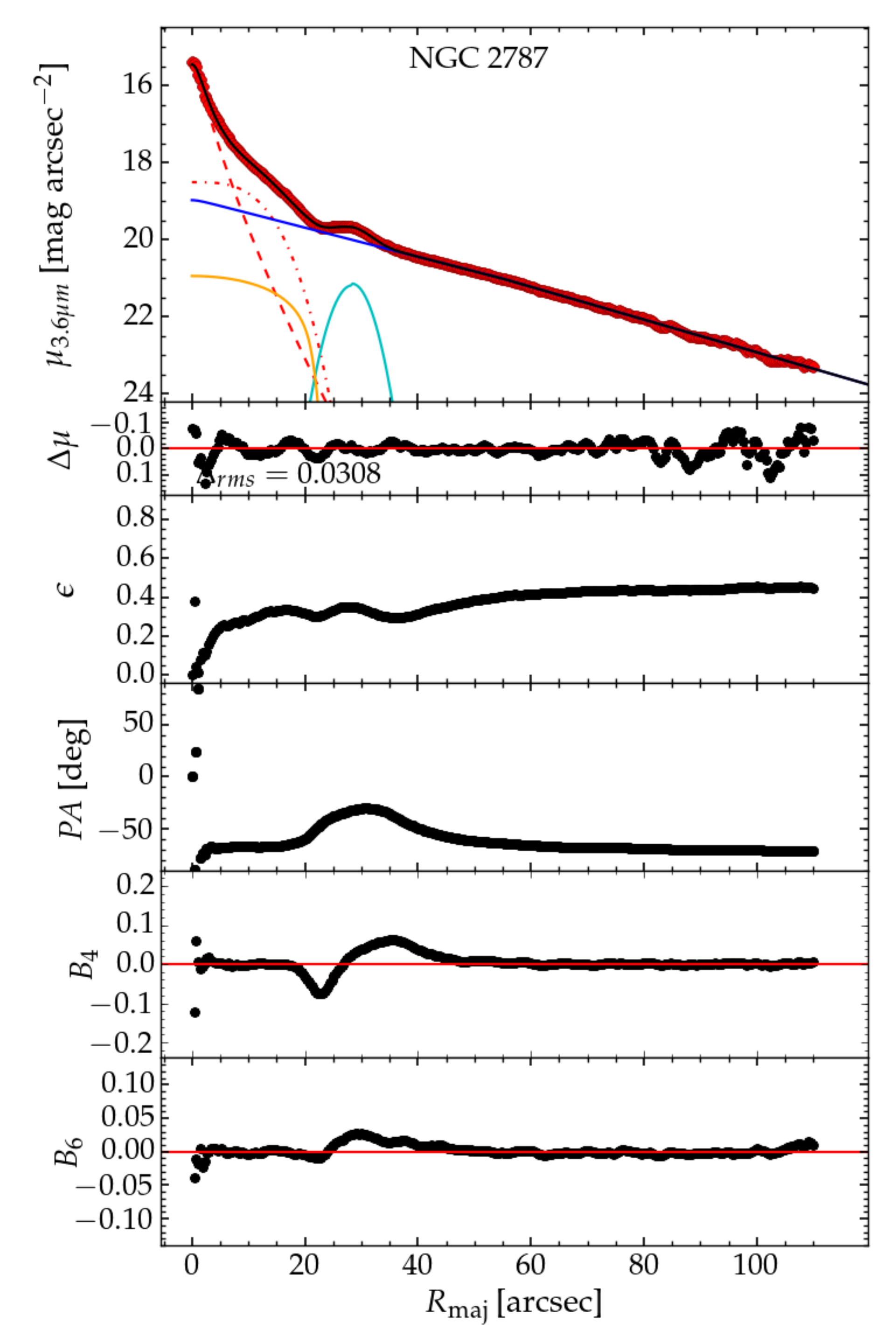}
\includegraphics[clip=true,trim= 1mm 1mm 1mm 1mm,height=12cm,width=0.49\textwidth]{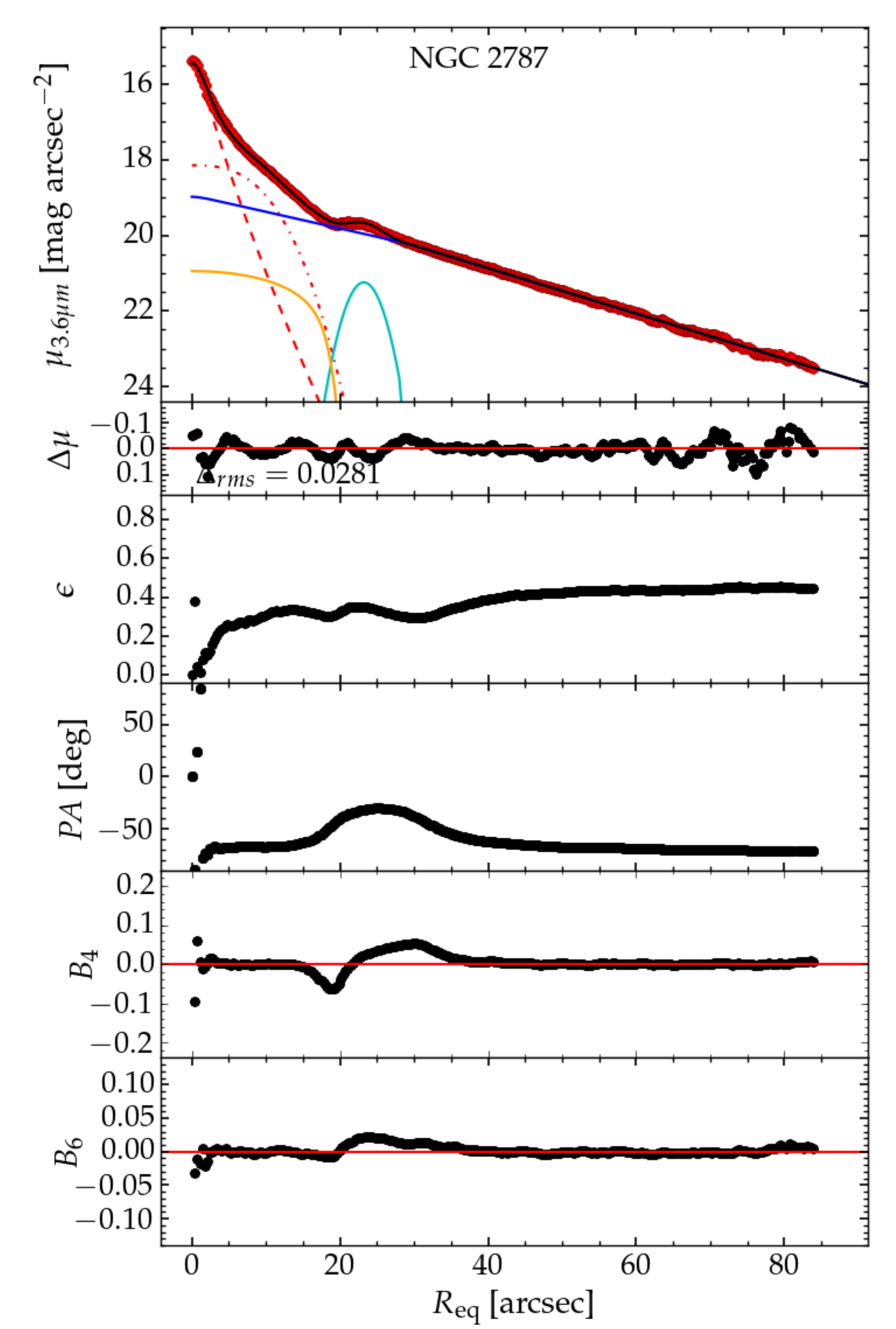}
\caption{NGC 2787: it is a barred lenticular galaxy with its multi-component fit comprised of a S{\'e}rsic function for the bulge (\textcolor{red}{- - -}), a low index S{\'e}rsic function for the prominent bar-lens/pseudobulge (\textcolor{red}{ - $\cdot$ - $\cdot$ -}), a Ferrers function for the bar (\textcolor{orange}{---}), a Gaussian for the ansae (\textcolor{cyan}{---}), and a slightly truncated exponential model for the extended disk (\textcolor{blue}{---}).
The dip in the ellipticity, $B_4$, and $B_6$ profiles at $R_{maj} \approx 22\arcsec$, and the bump in the ellipticity, position angle, $B_4$ and $B_6$ profiles at $\sim 30\arcsec$, corresponds to the perturbation of the isophotes due to the bar/barlens and ansae, respectively.}
\label{NGC 3665}
\end{figure}

\begin{figure}[H]
\includegraphics[clip=true,trim= 1mm 1mm 1mm 1mm,height=12cm,width=0.49\textwidth]{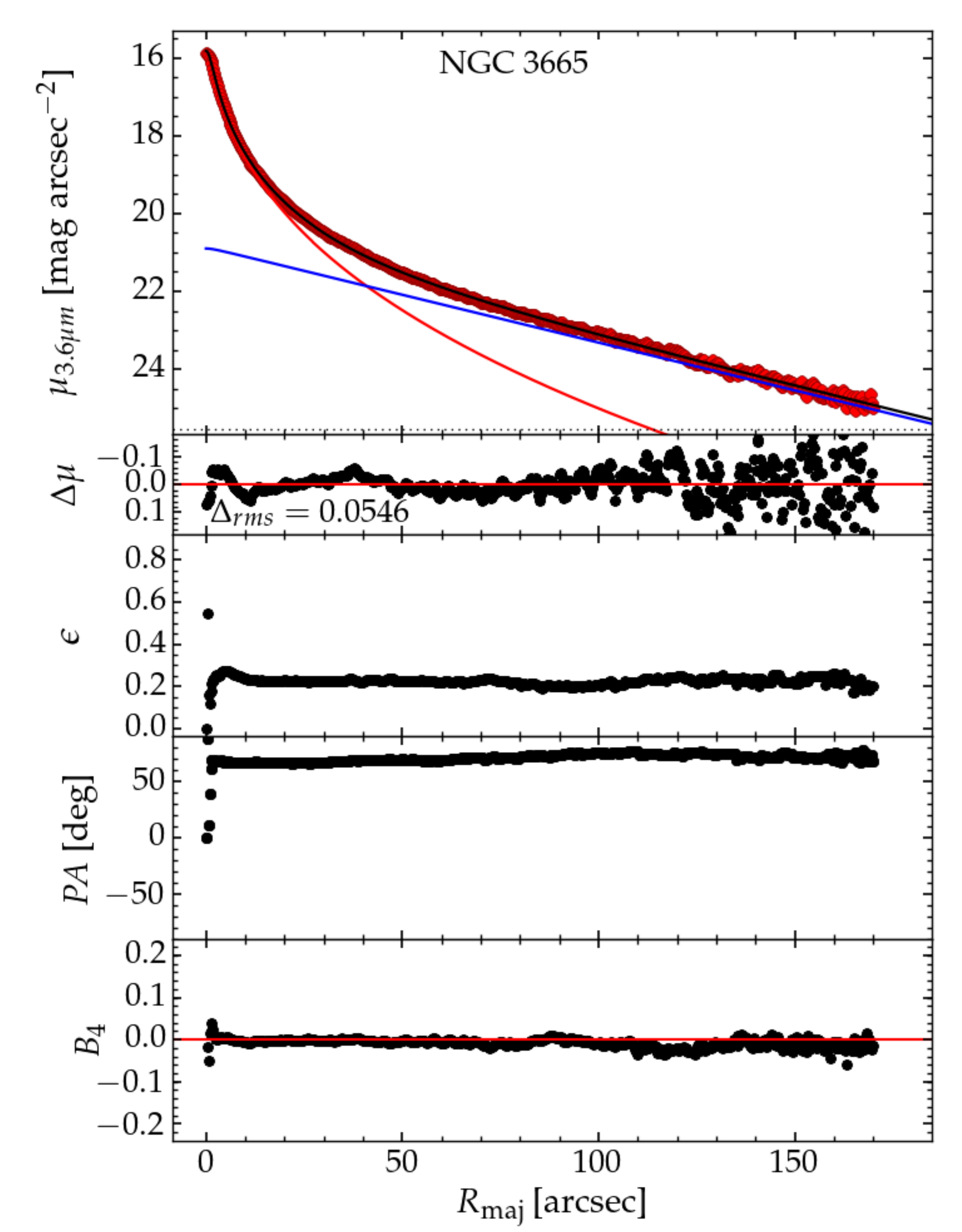}
\includegraphics[clip=true,trim= 1mm 1mm 1mm 1mm,height=12cm,width=0.49\textwidth]{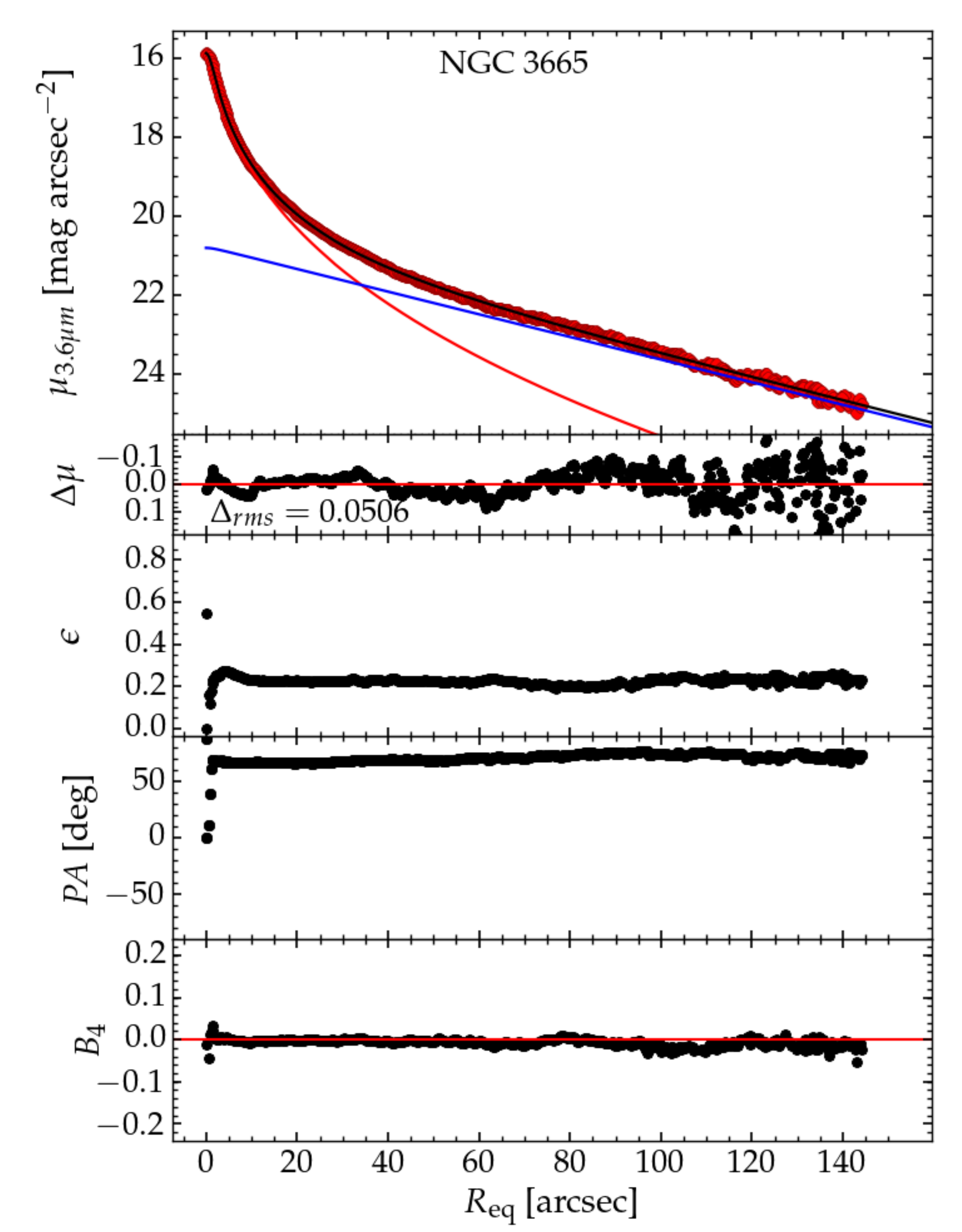}
\caption{NGC 3665: a lenticular galaxy with a S{\'e}rsic bulge (\textcolor{red}{---}) and an extended exponential disk (\textcolor{blue}{---}).}
\label{NGC 3665}
\end{figure}

\begin{figure}[H]
\includegraphics[clip=true,trim= 1mm 1mm 1mm 1mm,height=12cm,width=0.49\textwidth]{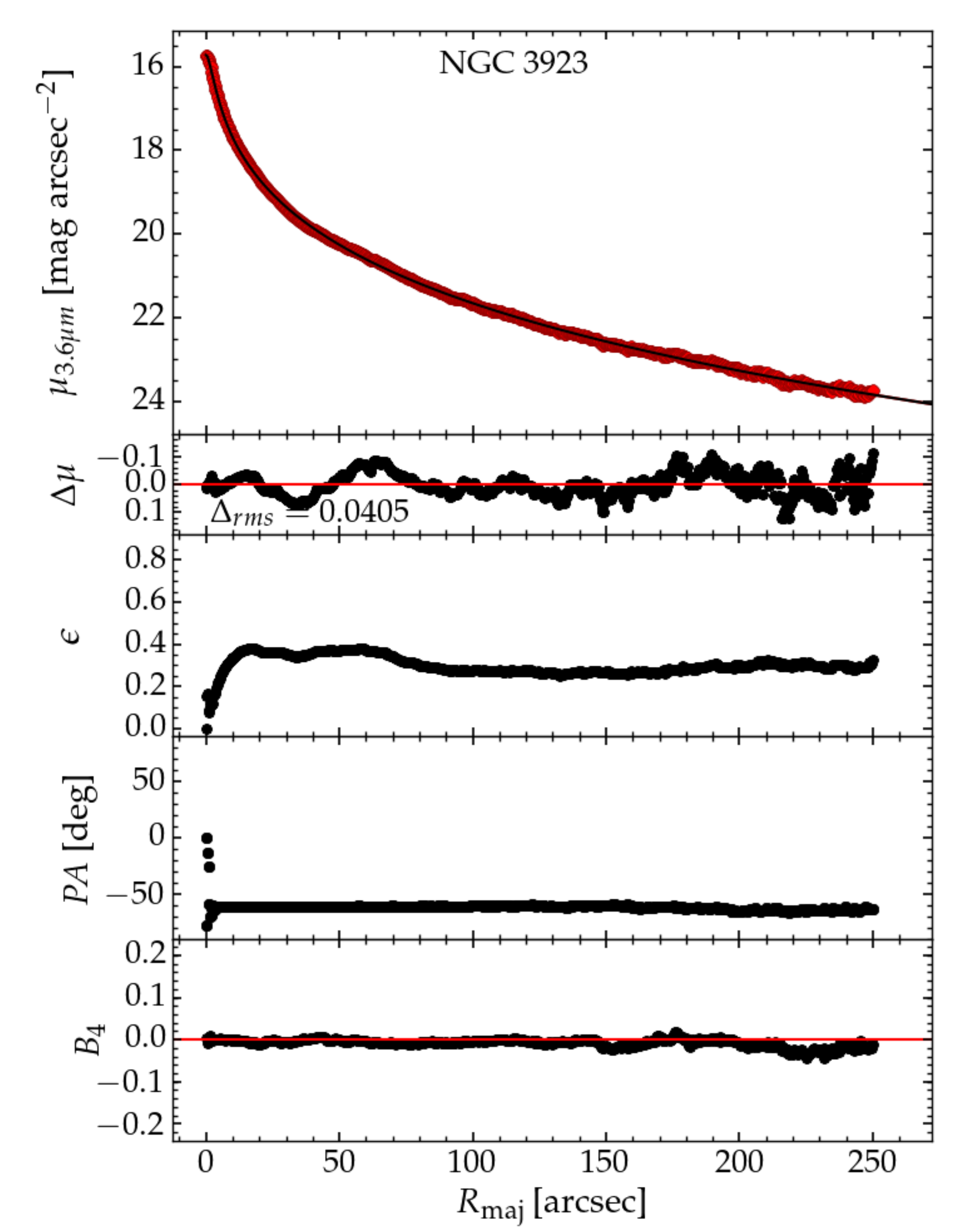}
\includegraphics[clip=true,trim= 1mm 1mm 1mm 1mm,height=12cm,width=0.49\textwidth]{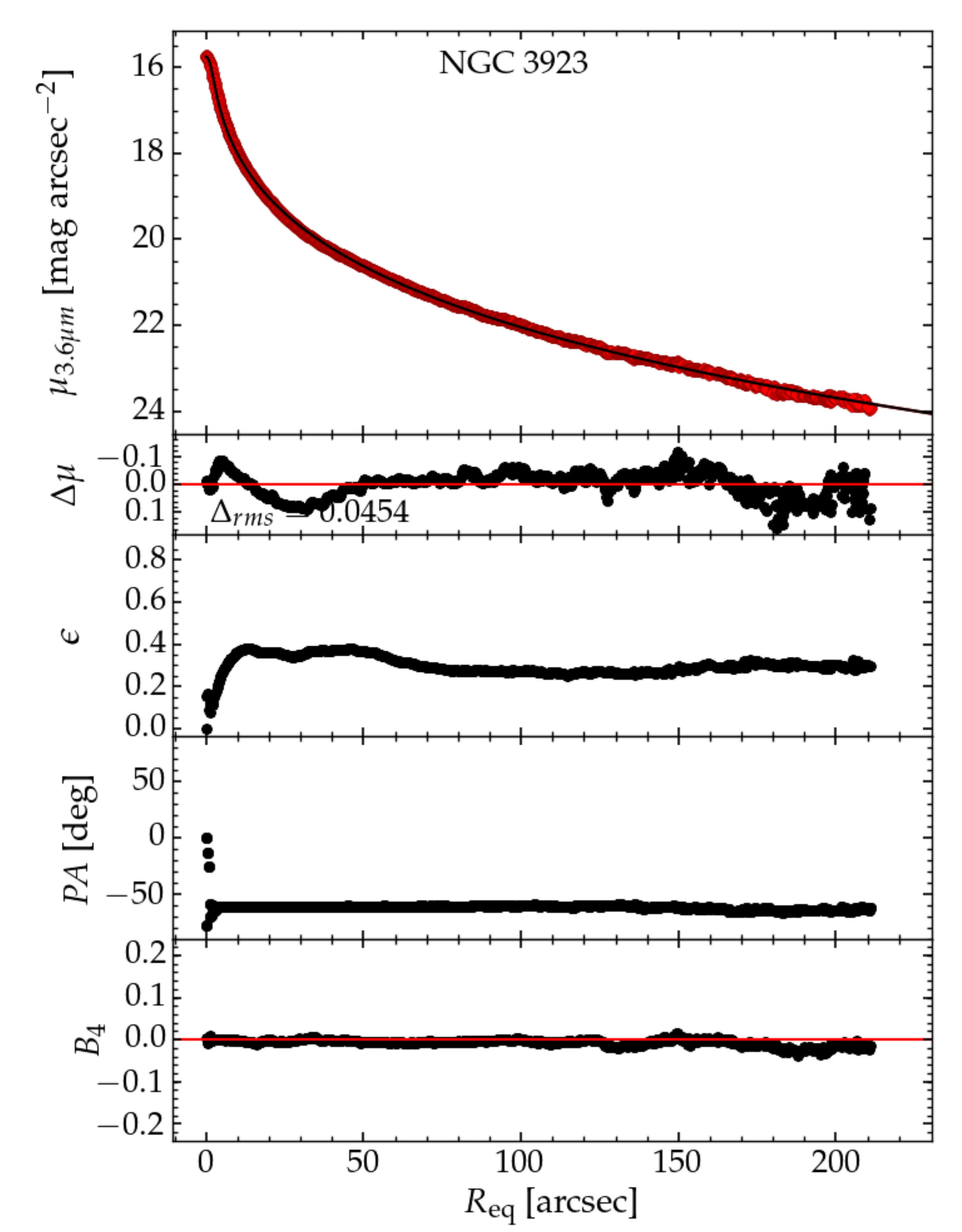}
\caption{NGC 3923: a massive elliptical with a deficit of light in its core, fit using a core-S{\'e}rsic function (\textcolor{red}{---}).}
\label{NGC 3923}
\end{figure}

\begin{figure}[H]
\includegraphics[clip=true,trim= 1mm 1mm 1mm 1mm,height=12cm,width=0.49\textwidth]{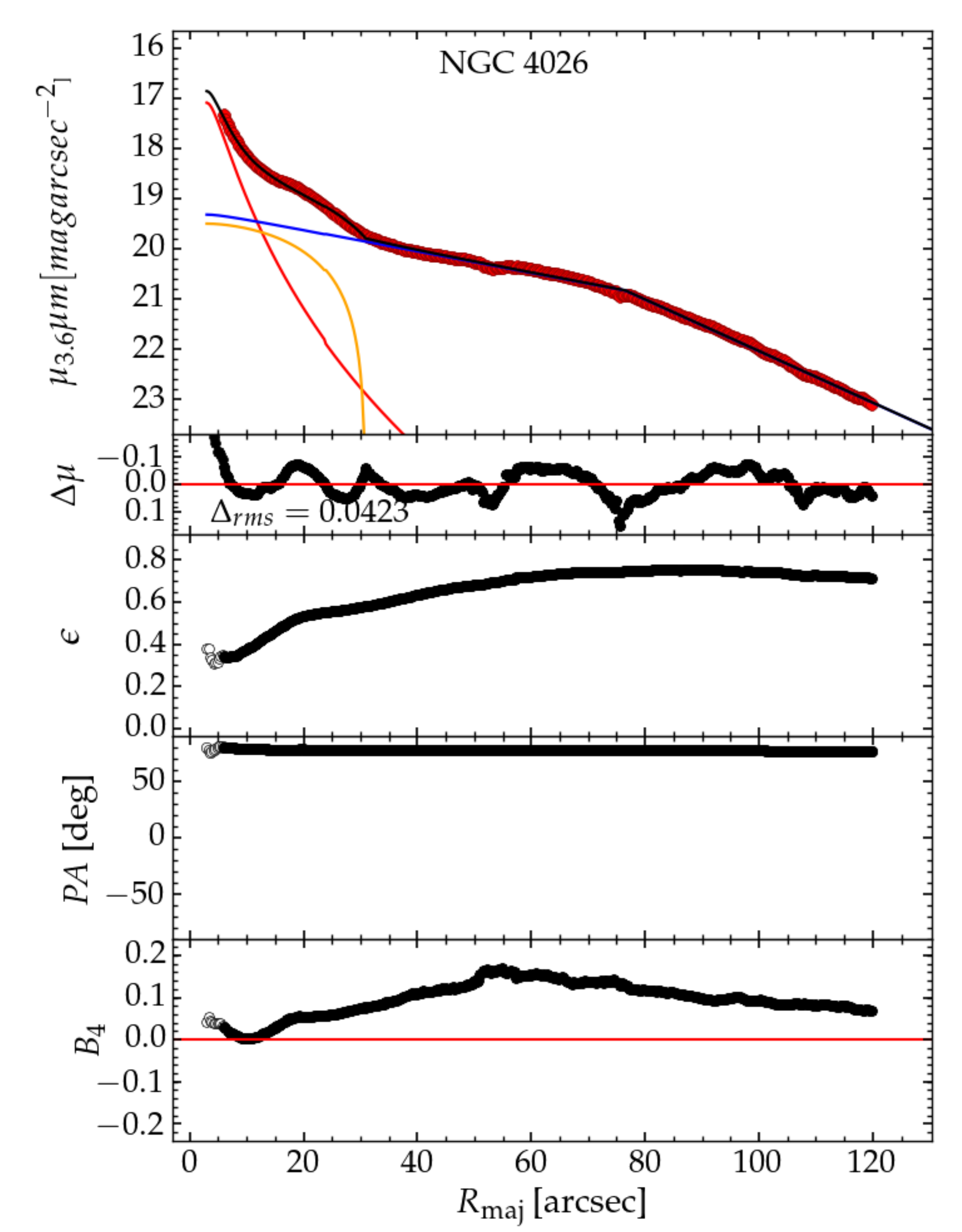}
\includegraphics[clip=true,trim= 1mm 1mm 1mm 1mm,height=12cm,width=0.49\textwidth]{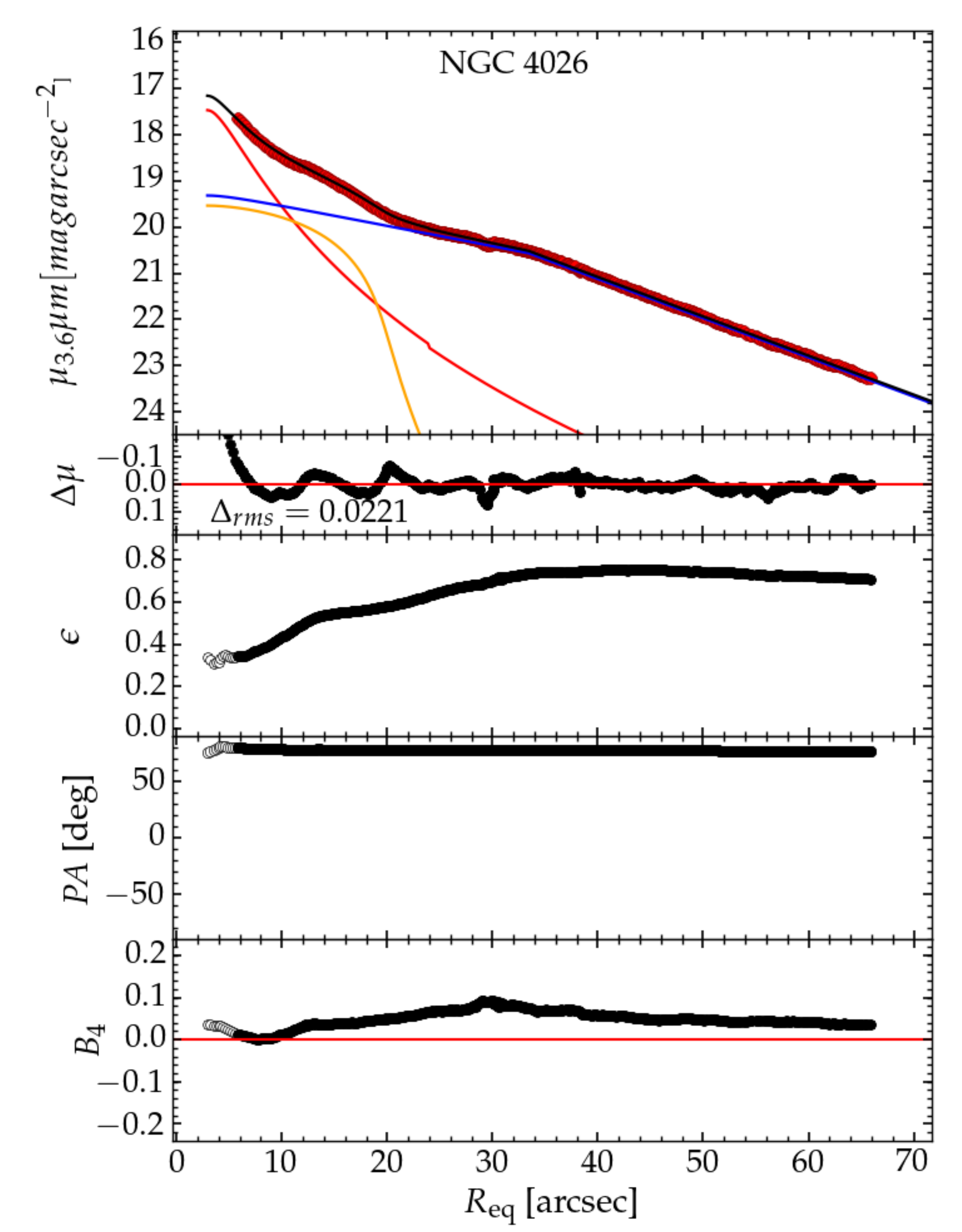}
\caption{NGC 4026: an edge-on lenticular galaxy with a S{\'e}rsic bulge (\textcolor{red}{---}), a faint bar ending at about $R_{maj}=30\arcsec$ and fit using Ferrers (\textcolor{orange}{---}) function, plus a truncated exponential disk (\textcolor{blue}{---}).}
\label{NGC 4026}
\end{figure}

\begin{figure}[H]
\includegraphics[clip=true,trim= 1mm 1mm 1mm 1mm,height=12cm,width=0.49\textwidth]{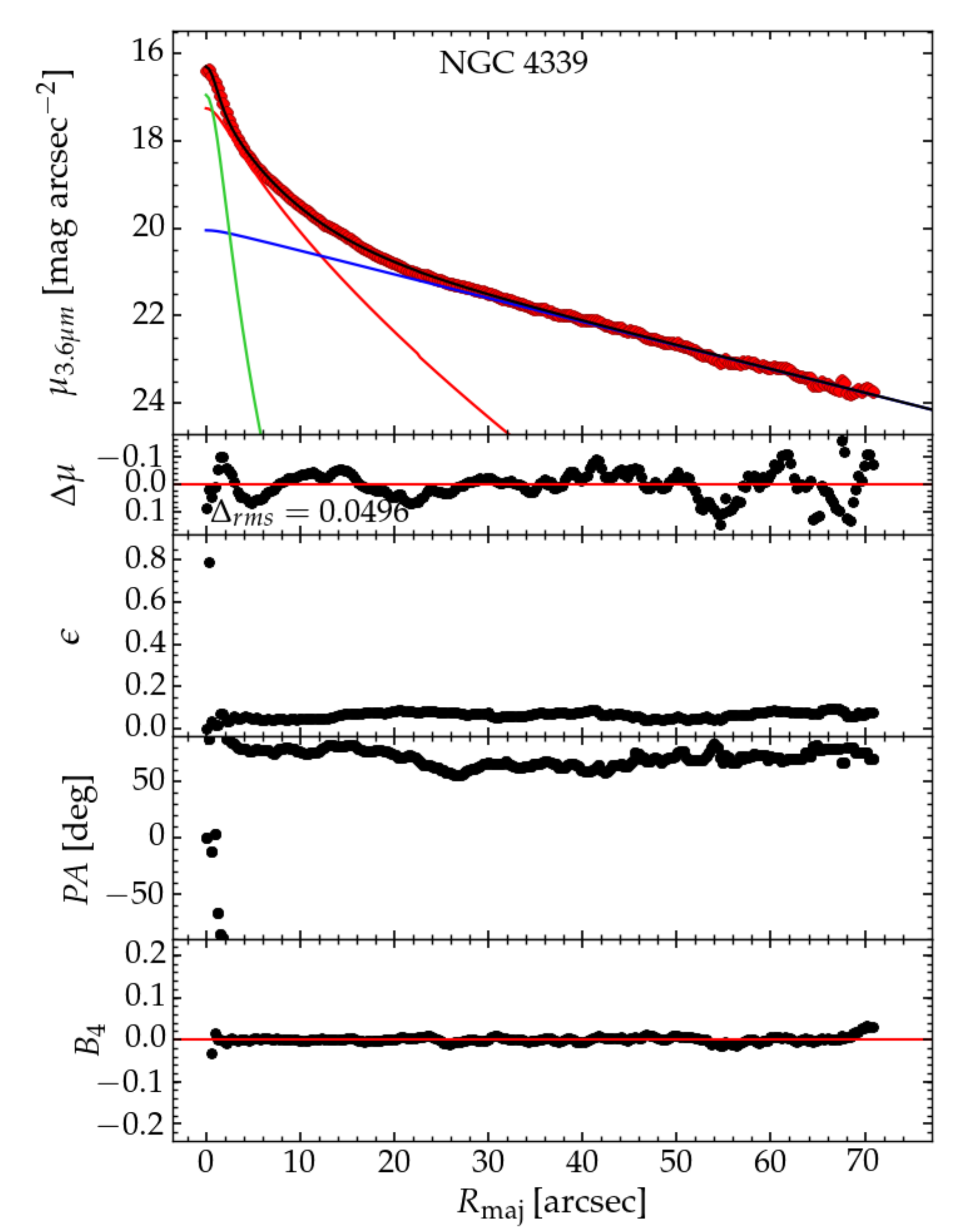}
\includegraphics[clip=true,trim= 1mm 1mm 1mm 1mm,height=12cm,width=0.49\textwidth]{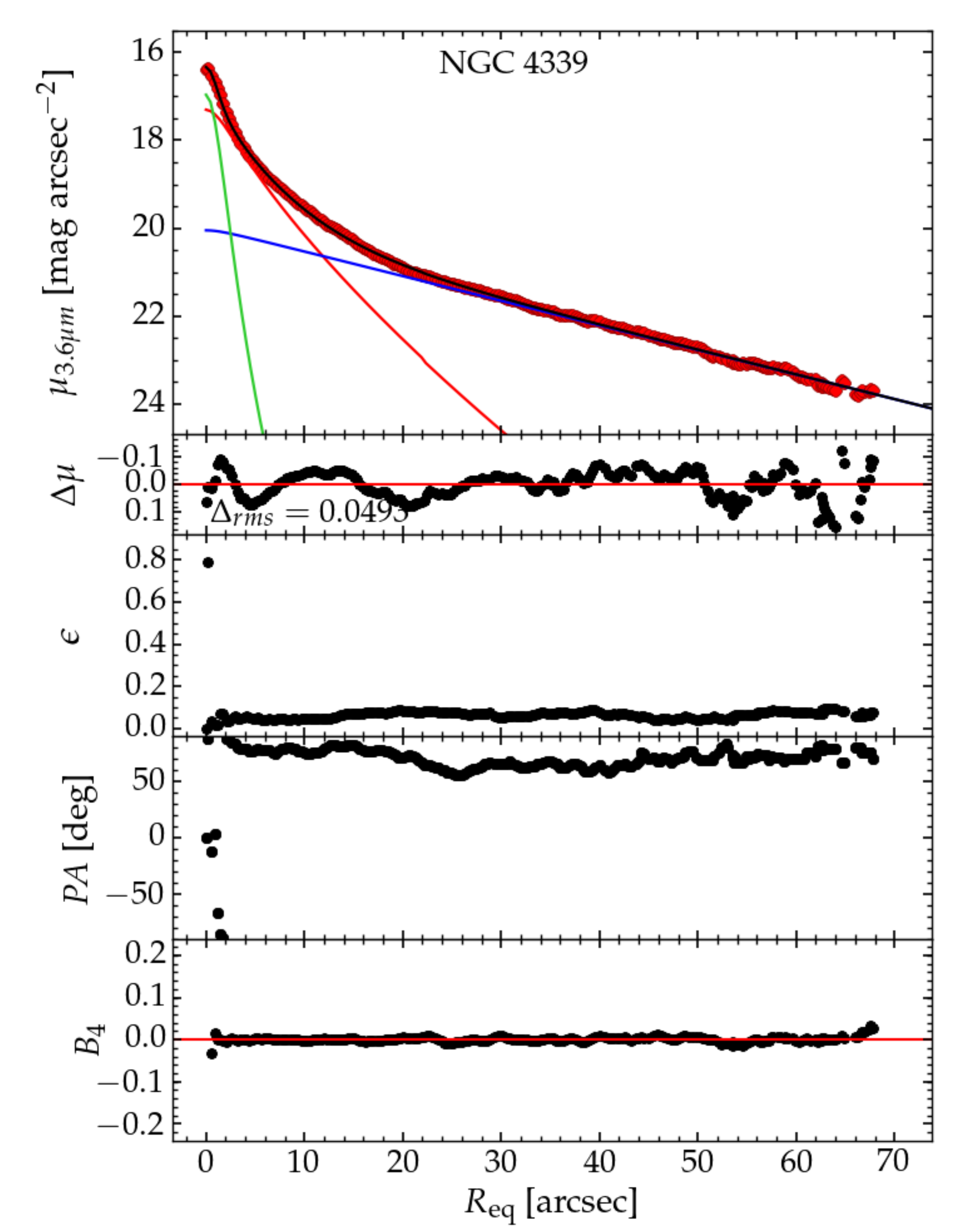}
\caption{NGC 4339: a face-on lenticular galaxy \citep{Halliday:1998}  with a central point source, a S{\'e}rsic bulge (\textcolor{red}{---}), and an exponential disk (\textcolor{blue}{---}).}
\label{NGC 4339}
\end{figure}

\begin{figure}[H]
\includegraphics[clip=true,trim= 1mm 1mm 1mm 1mm,height=12cm,width=0.49\textwidth]{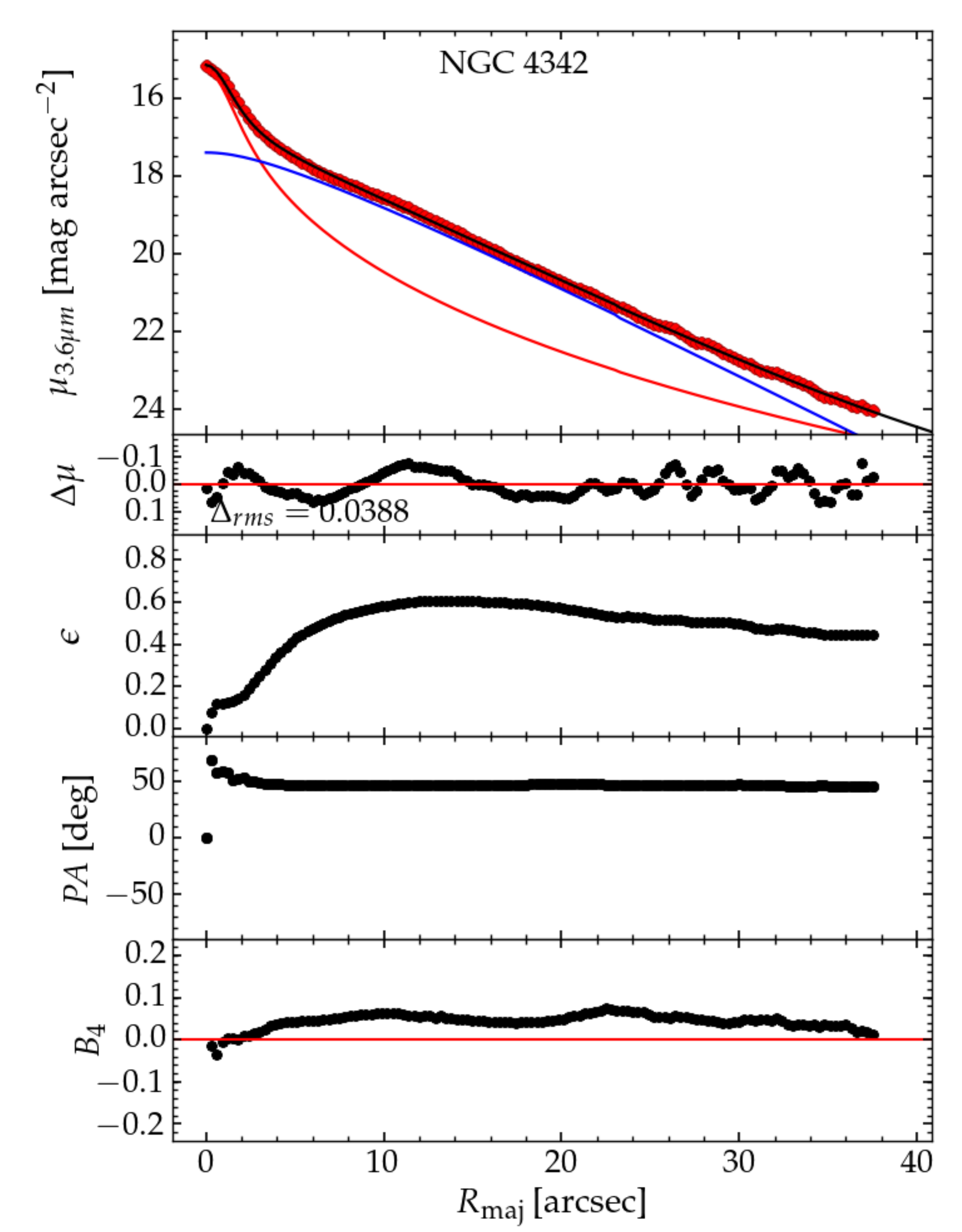}
\includegraphics[clip=true,trim= 1mm 1mm 1mm 1mm,height=12cm,width=0.49\textwidth]{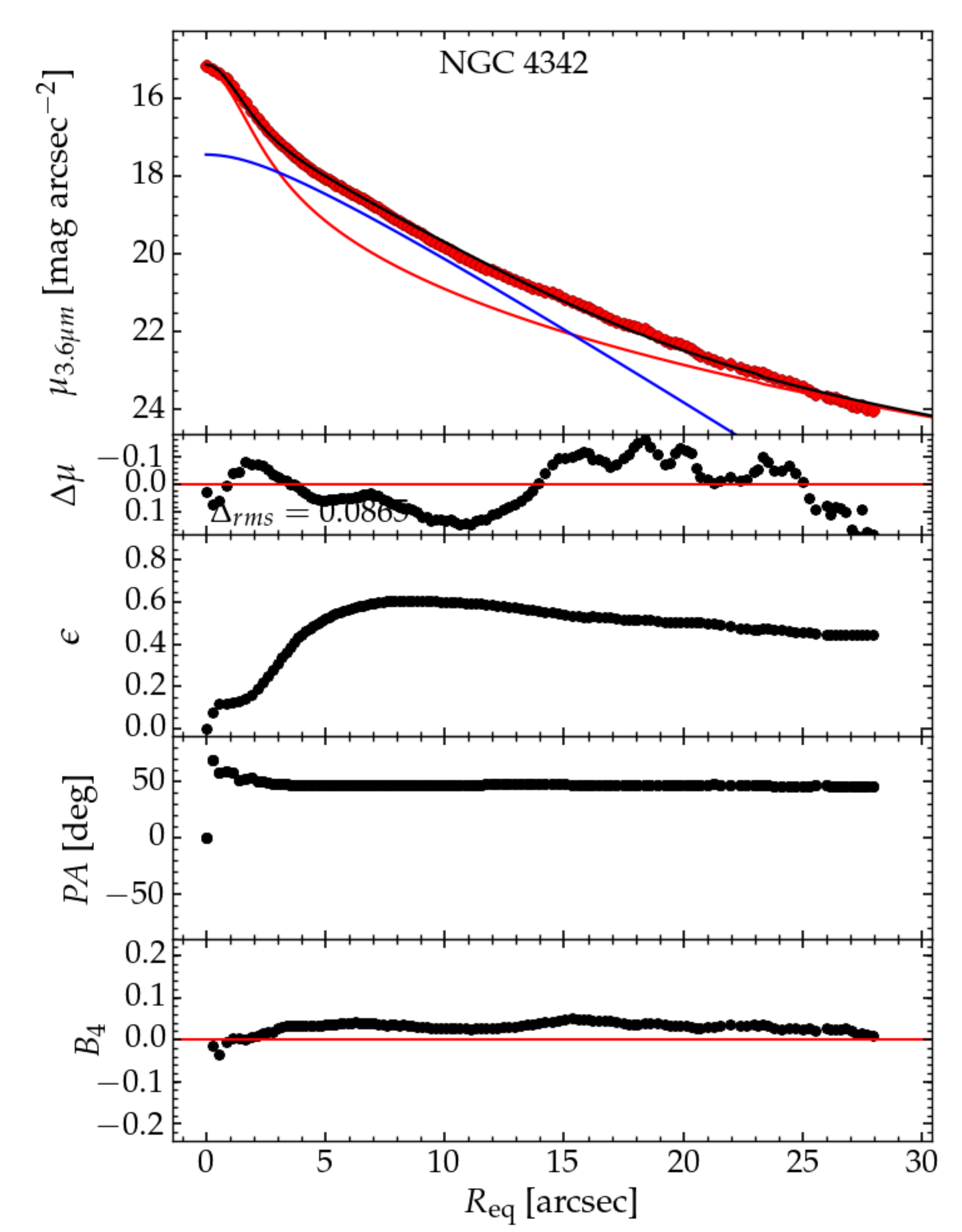}
\caption{NGC 4342: a dwarf ellicular galaxy, with most of its mass tidally stripped by the massive companion galaxy NGC 4365 \citep{Blom:Forbes:2014}. Its light profile has been fit using an extended S{\'e}rsic bulge (\textcolor{red}{---}) and an intermediate-scale inclined disk (\textcolor{blue}{---}), evident from the bump in the ellipticity profile at intermediate radii.  }
\label{NGC 4342}
\end{figure}

\begin{figure}[H]
\includegraphics[clip=true,trim= 1mm 1mm 1mm 1mm,height=12cm,width=0.49\textwidth]{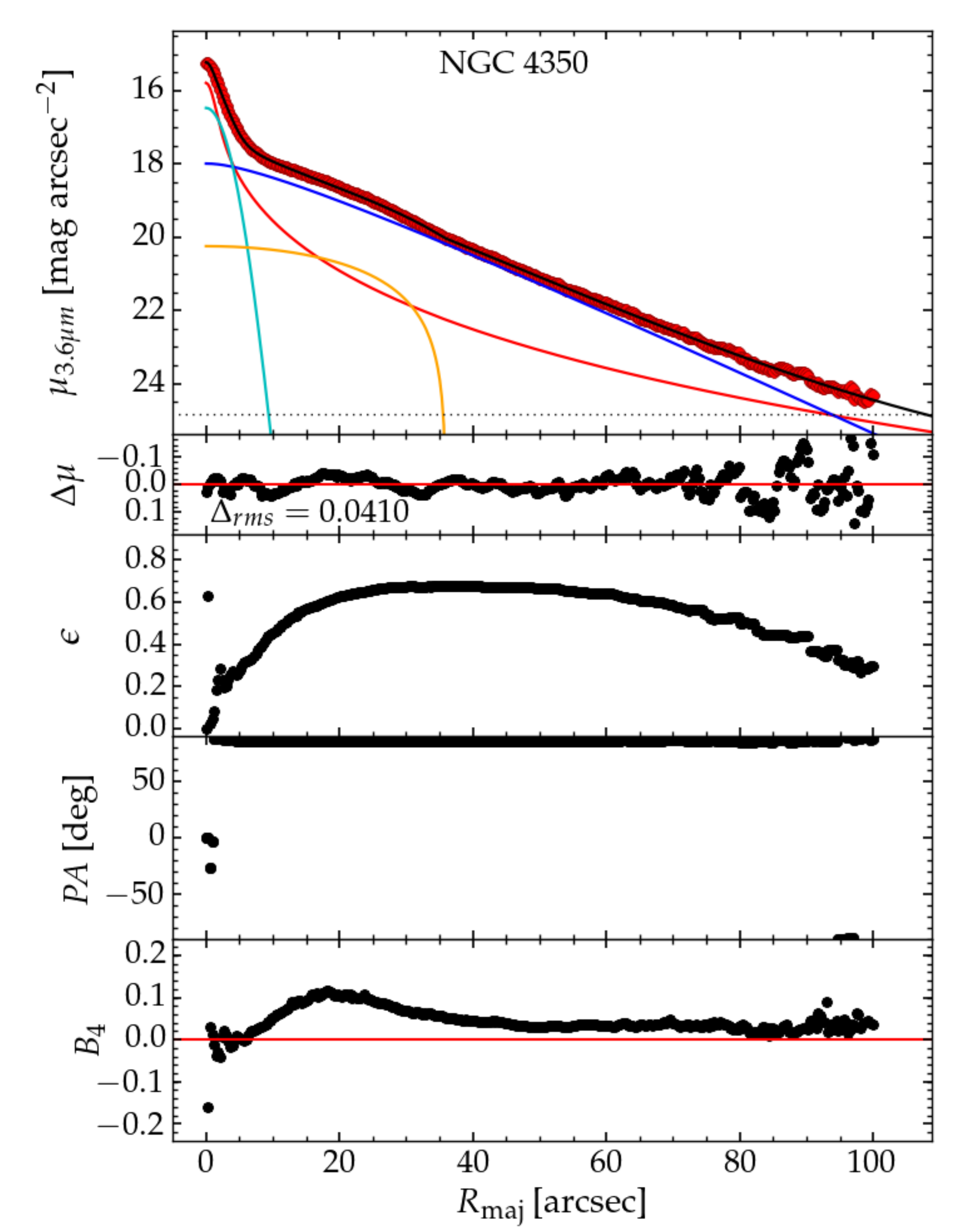}
\includegraphics[clip=true,trim= 1mm 1mm 1mm 1mm,height=12cm,width=0.49\textwidth]{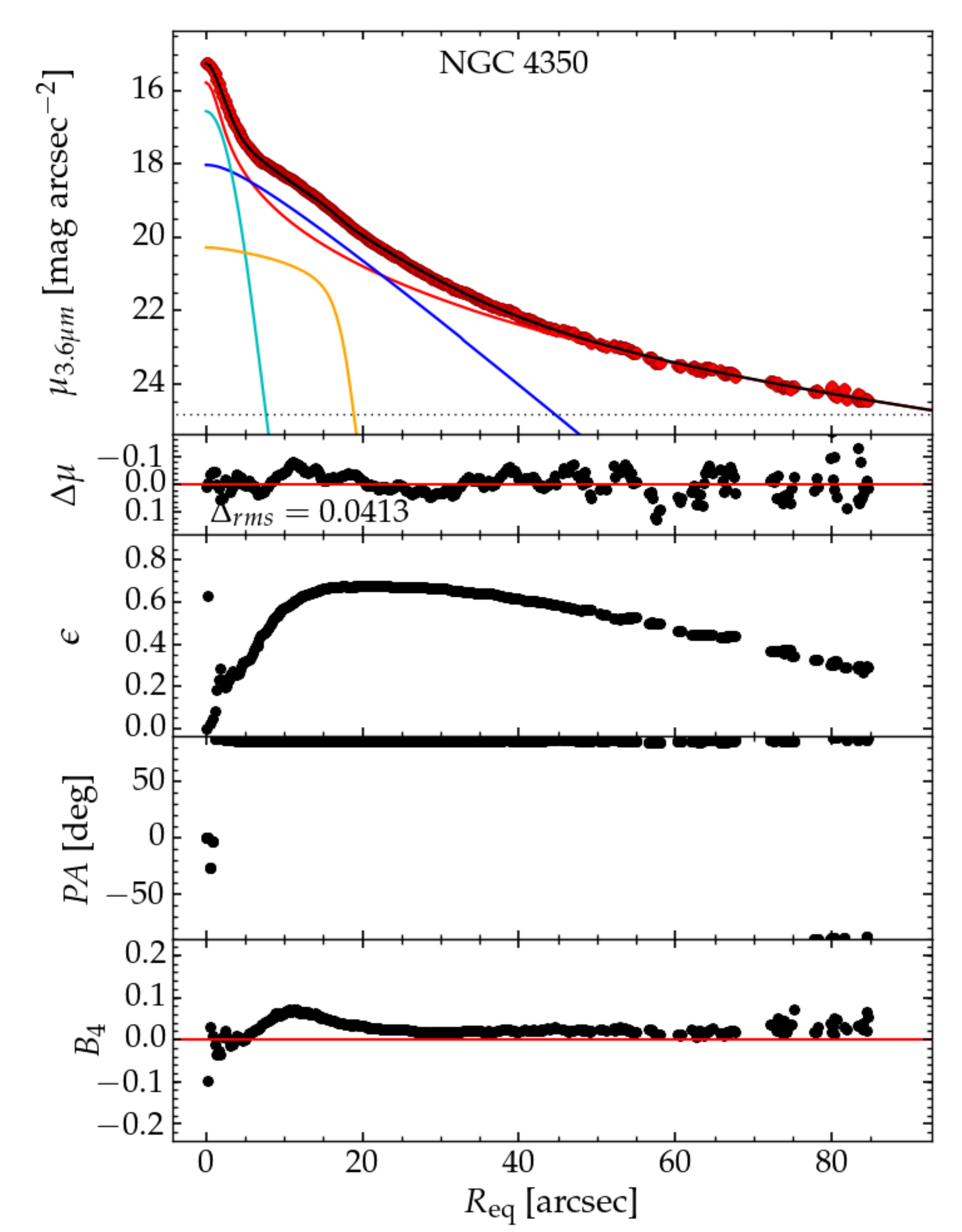}
\caption{NGC 4350: an ellicular (ES) galaxy with a faint bar \citep{Pignatelli:2001}. The bump in the $B_4$ profile at $R_{maj} \approx 20\arcsec$ reflects the combined effect of bar and high inclination of the galaxy. As apparent from the ellipticity profile, the spheroid of NGC 4350, fit using a S{\'e}rsic function (\textcolor{red}{---}), takes over the intermediate-scale disk (\textcolor{blue}{---}), fit using an inclined exponential, at larger radii. The bar component is fit using a Ferrers (\textcolor{orange}{---}) function and the central Gaussian (\textcolor{cyan}{---}) accounts for extra light at the galaxy center \citep{Pignatelli:2001}.}
\label{NGC 4350}
\end{figure}

\begin{figure}[H]
\includegraphics[clip=true,trim= 1mm 1mm 1mm 1mm,height=12cm,width=0.49\textwidth]{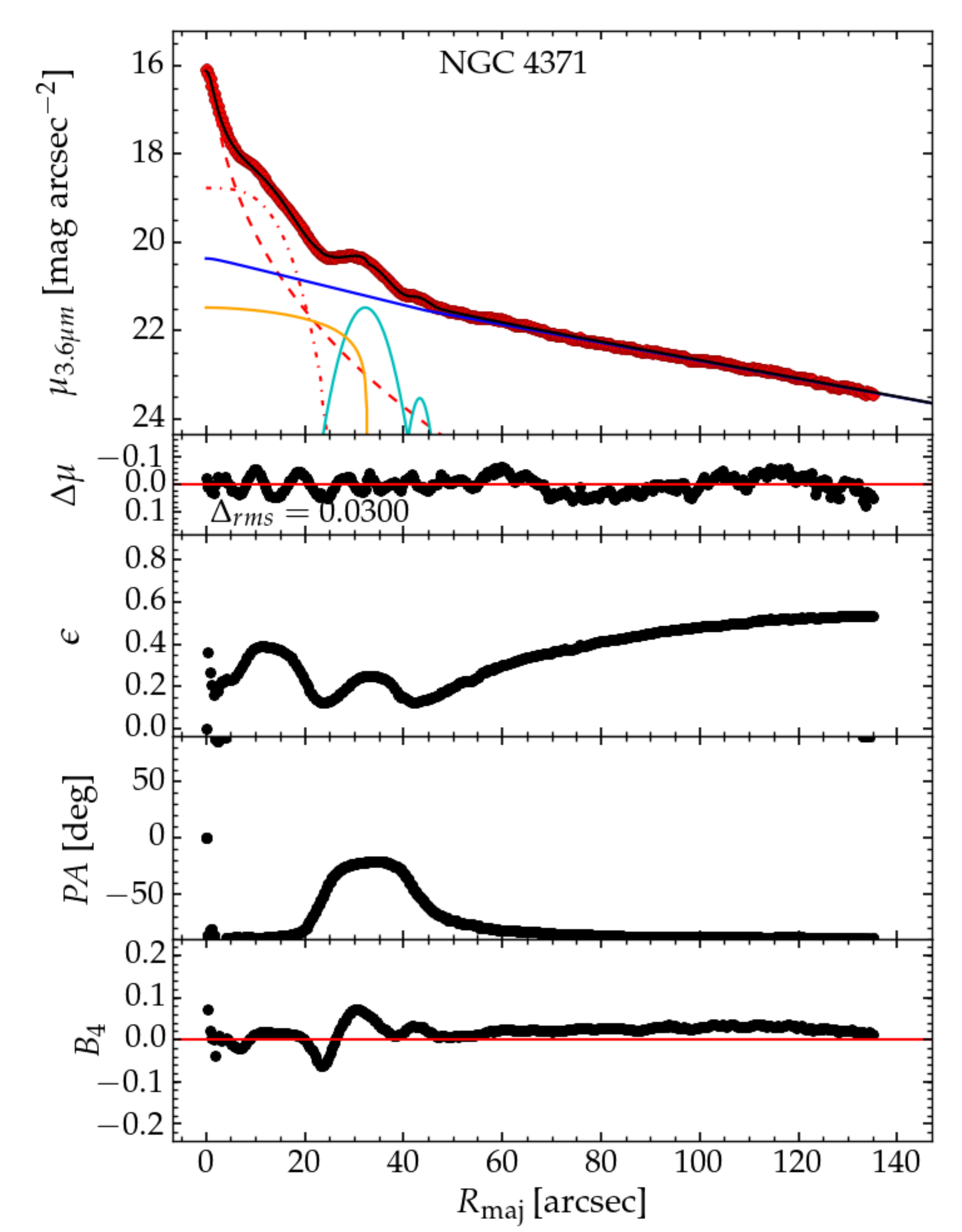}
\includegraphics[clip=true,trim= 1mm 1mm 1mm 1mm,height=12cm,width=0.49\textwidth]{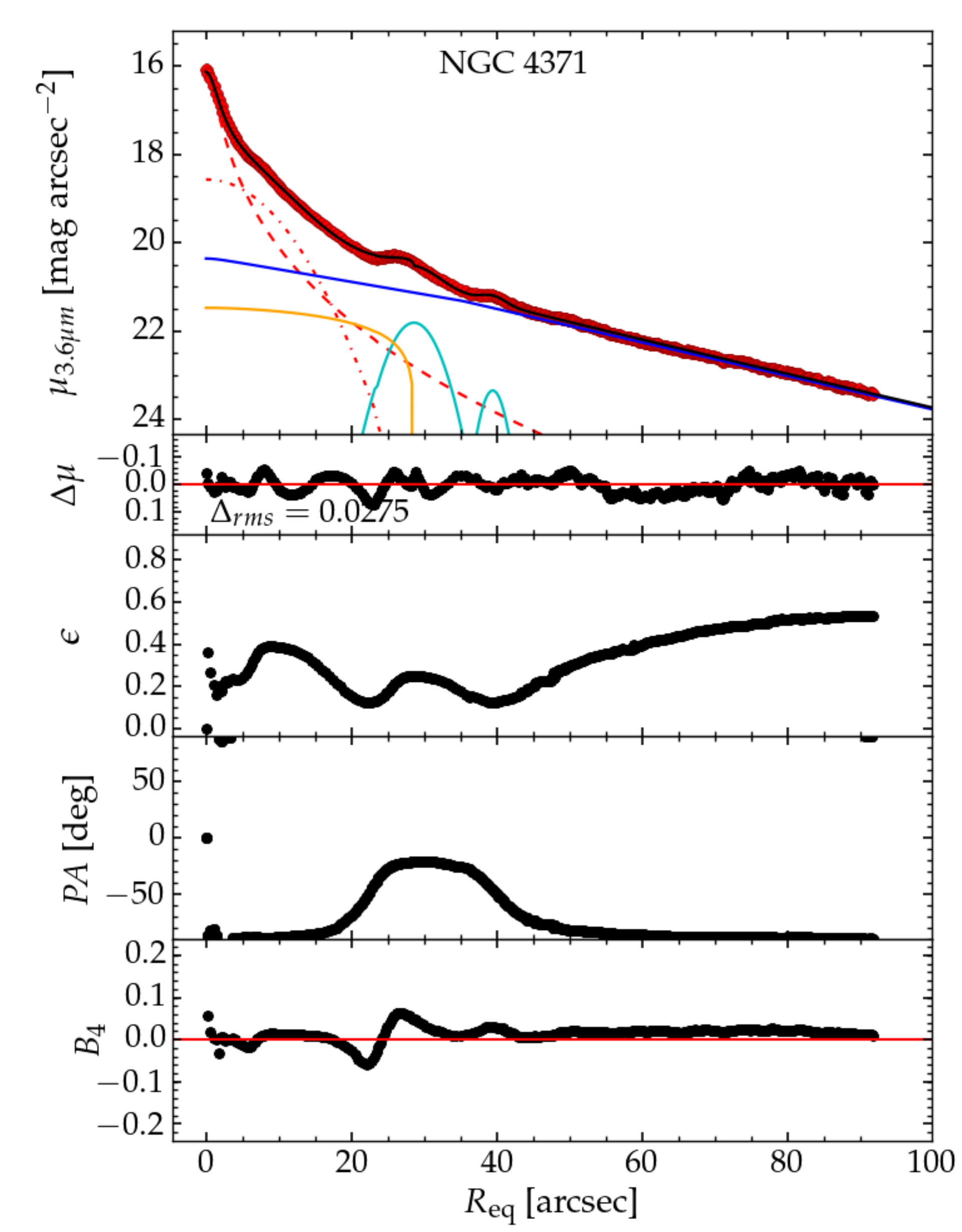}
\caption{NGC 4371: a barred lenticular, SB(r)0, galaxy with a pseudobulge \citep{Erwin:2015} fit here with a S{\'e}rsic function (\textcolor{red}{- - -}) for the bulge, a bar-lens (or pseudobulge) fit using a low S{\'e}rsic index function (\textcolor{red}{ - $\cdot$ - $\cdot$ -}), a bar fit using Ferrers  function (\textcolor{orange}{---}), an ansae at the end of the bar fit using a Gaussian (\textcolor{cyan}{---}), an outer faint ring fit using a low width Gaussian (\textcolor{cyan}{---}), and an extended disk (\textcolor{blue}{---}) truncated at $R_{maj} \approx 44\arcsec$. \citet{Gadotti:2015} call the two parts of the truncated disk as inner disk and (outer) disk. \citet{Erwin:2015} treat the bulge and the (oval-shaped) barlens as a single entity naming it a \enquote{composite bulge}.}
\label{NGC 4371}
\end{figure}

\begin{figure}[H]
\includegraphics[clip=true,trim= 1mm 1mm 1mm 1mm,height=12cm,width=0.49\textwidth]{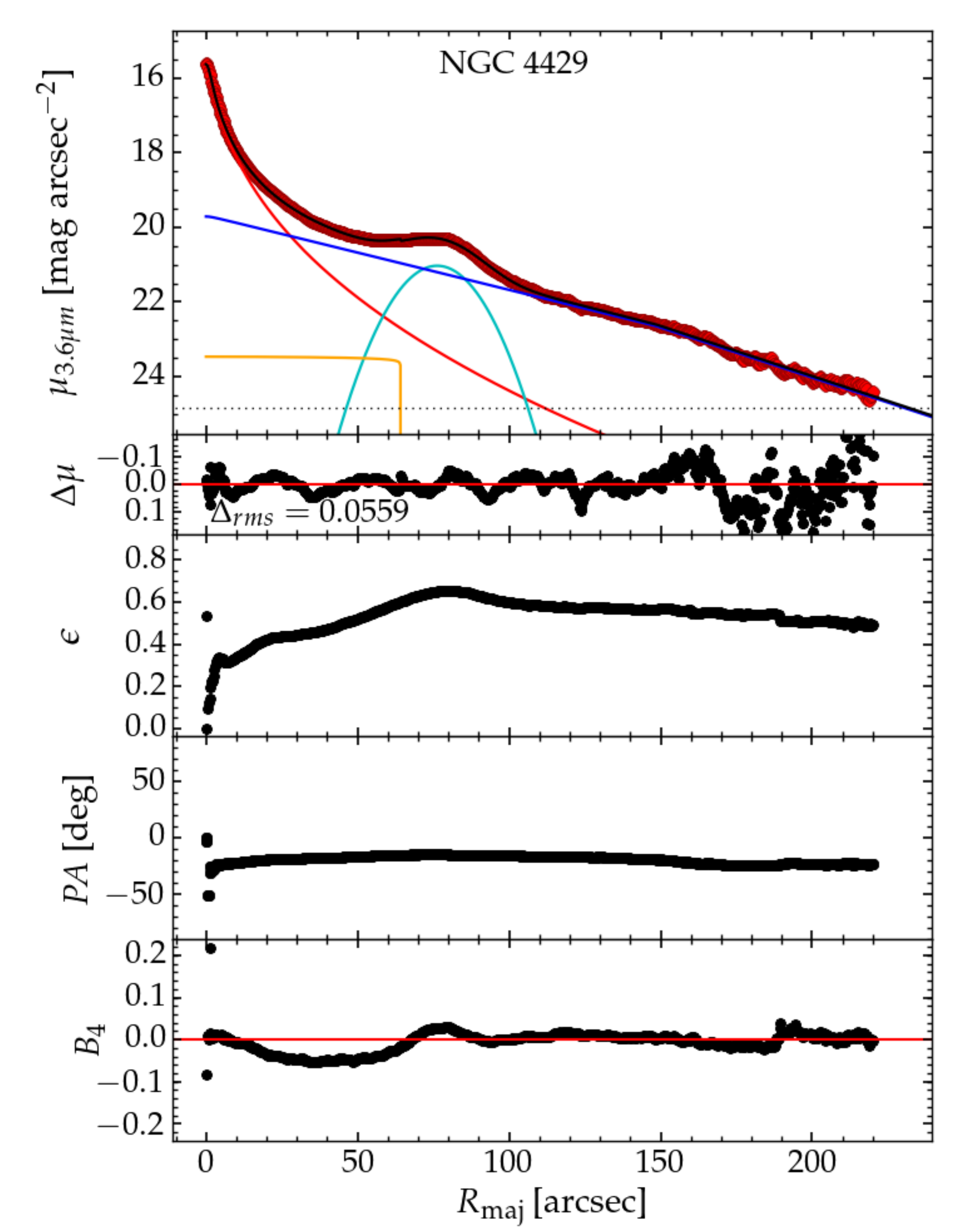}
\includegraphics[clip=true,trim= 1mm 1mm 1mm 1mm,height=12cm,width=0.49\textwidth]{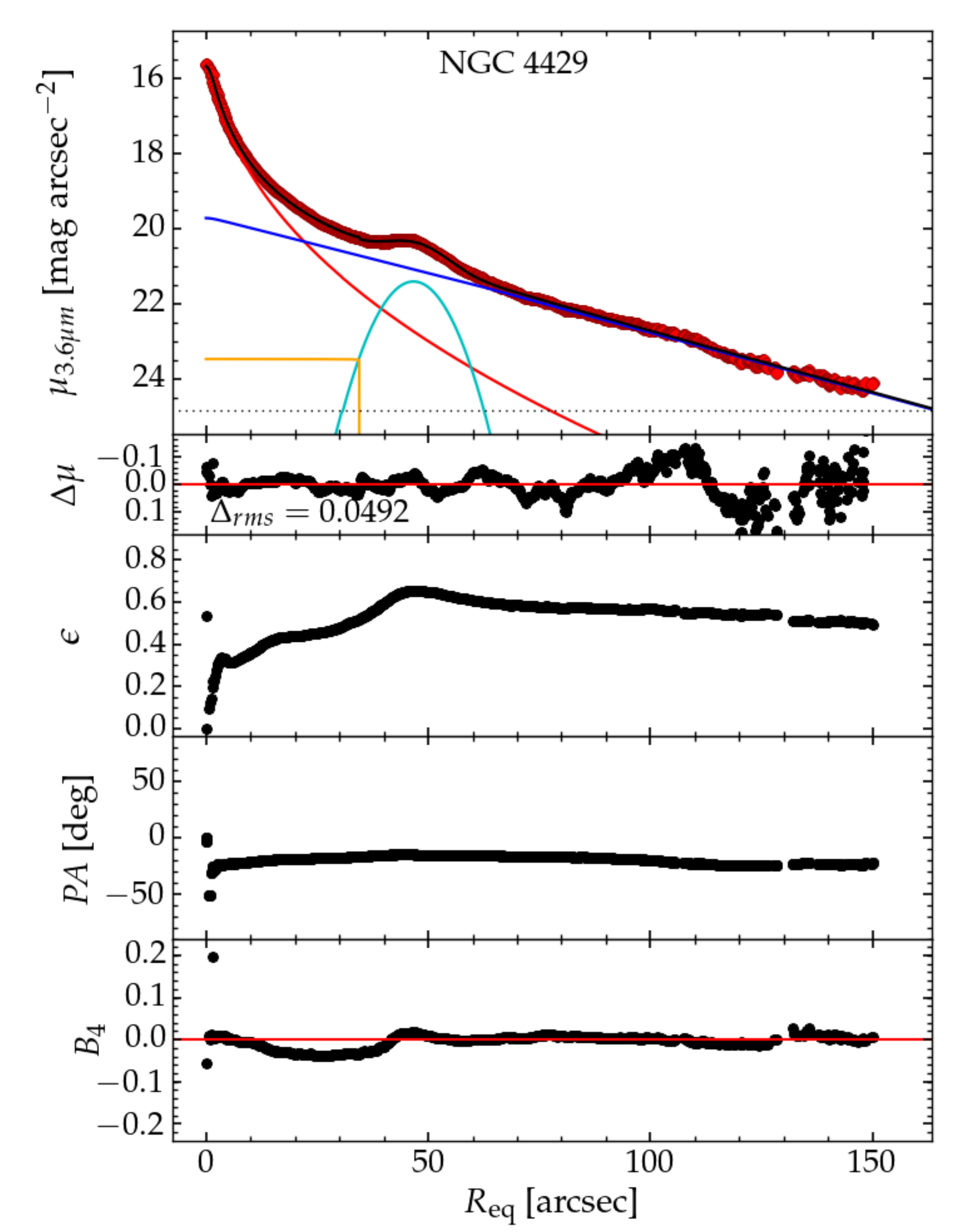}
\caption{NGC 4429: a lenticular galaxy with a boxy (peanut shell)-shaped bulge and a bar \citep{TimothyDavis:2018} fit using a S{\'e}rsic and a Ferrers (\textcolor{orange}{---}) function, respectively. The galaxy has a prominent outer ring at around $R_{maj}\approx 80\arcsec$, fit here using a Gaussian (\textcolor{cyan}{---}), plus a truncated (at around $150\arcsec$ along $R_{maj}$) exponential disk (\textcolor{blue}{---}).}
\label{NGC 4429}
\end{figure}

\begin{figure}[H]
\includegraphics[clip=true,trim= 1mm 1mm 1mm 1mm,height=12cm,width=0.49\textwidth]{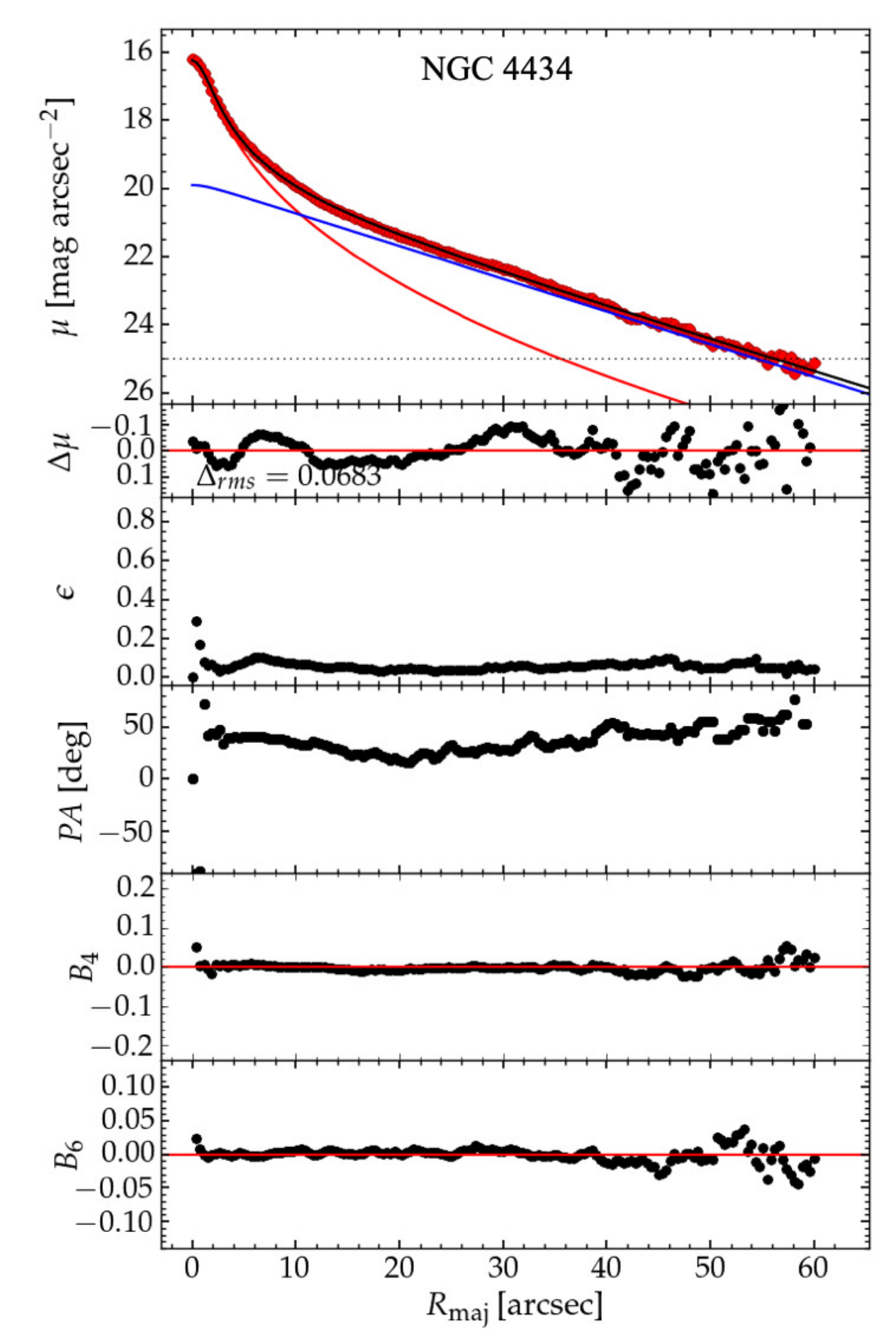}
\includegraphics[clip=true,trim= 1mm 1mm 1mm 1mm,height=12cm,width=0.49\textwidth]{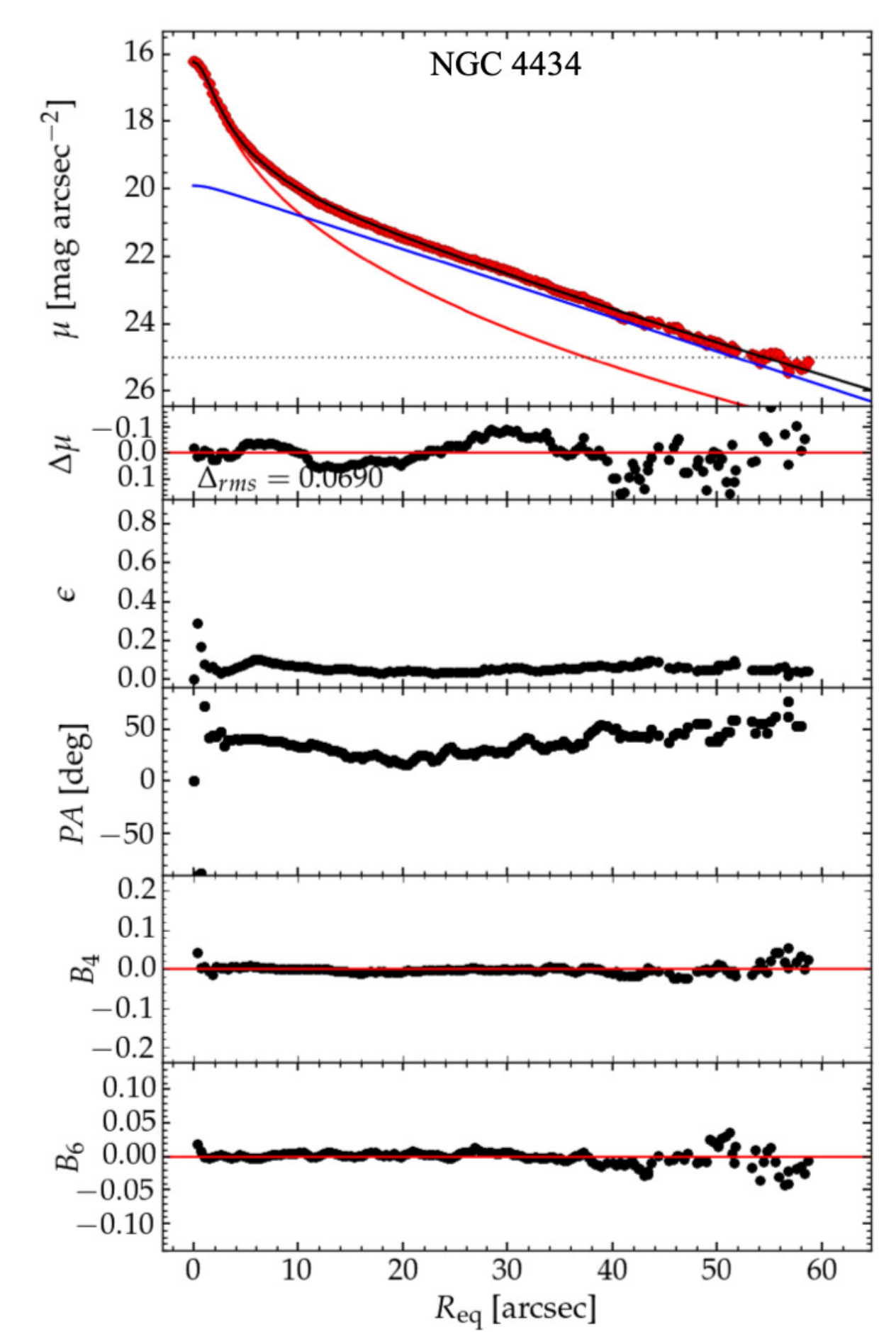}
\caption{NGC 4434: a lenticular galaxy with a S{\'e}rsic bulge (\textcolor{red}{---}) and an exponential disk (\textcolor{blue}{---}).}
\label{NGC 4434}
\end{figure}

\begin{figure}[H]
\includegraphics[clip=true,trim= 1mm 1mm 1mm 1mm,height=12cm,width=0.49\textwidth]{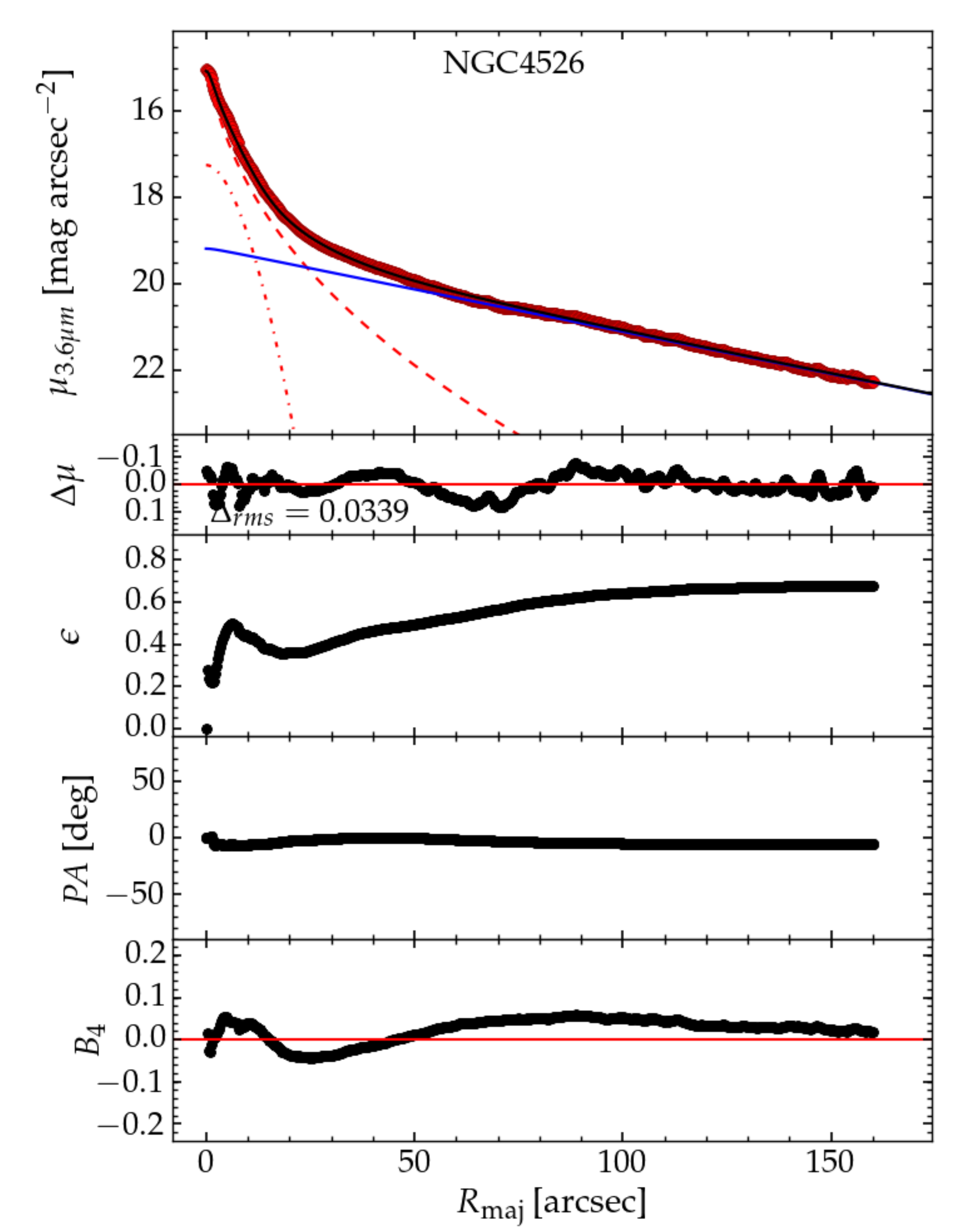}
\includegraphics[clip=true,trim= 1mm 1mm 1mm 1mm,height=12cm,width=0.49\textwidth]{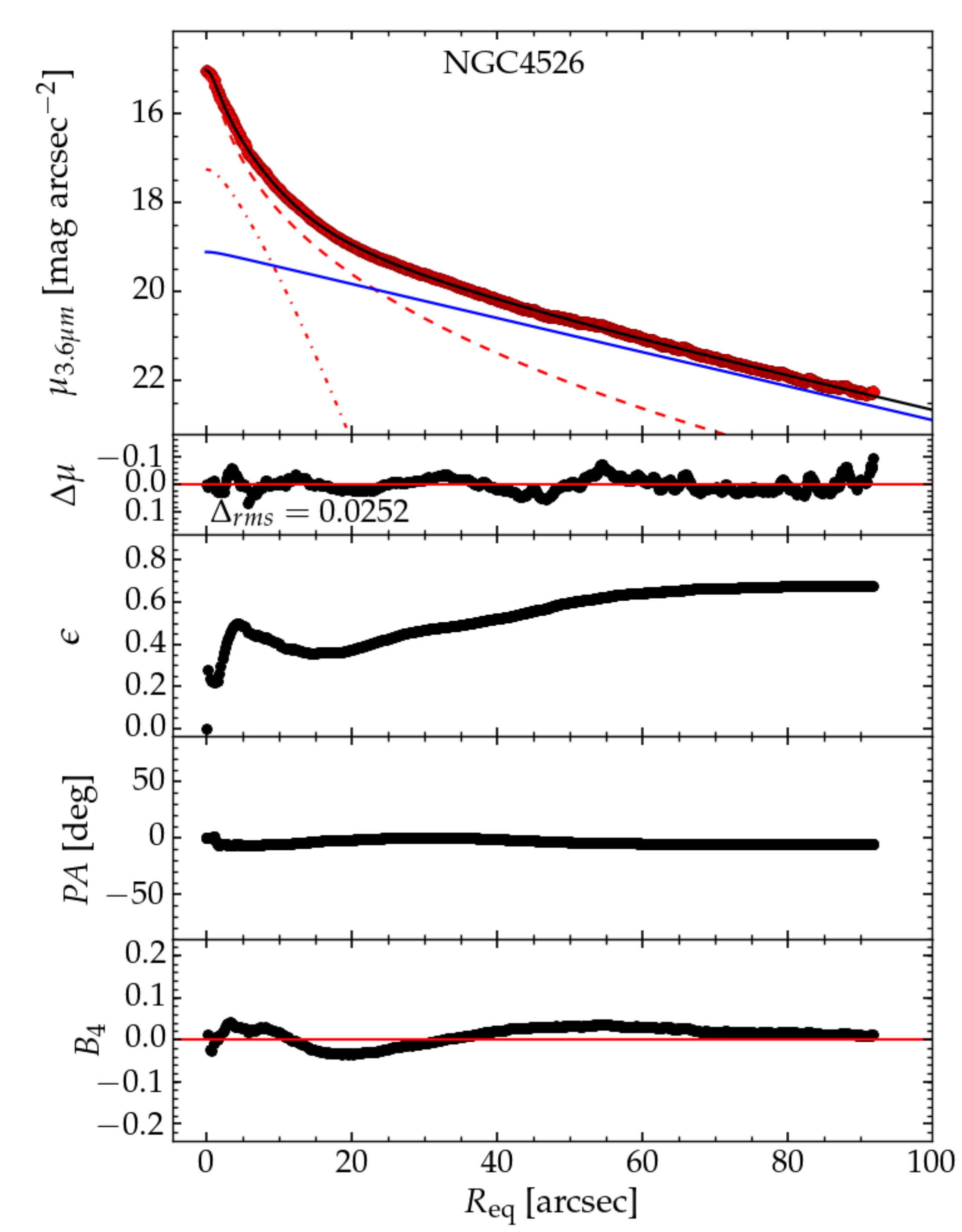}
\caption{NGC 4526: a lenticular galaxy with a S{\'e}rsic bulge (\textcolor{red}{- - -}), an extended exponential disk (\textcolor{blue}{---}), plus a fast rotating nuclear disk \citep{Rubin:1995} extending up to $\sim 20\arcsec$ and causing the bump in the ellipticity and $B_4$ profile. The nuclear disk is fit using a low S{\'e}rsic index function (\textcolor{red}{ - $\cdot$ - $\cdot$ -}). A faint bar, as claimed by \citet{RC3:1991}, could not be clearly seen in the Spitzer image of the galaxy. However, the addition of a weak bar ending at $R_{maj}\approx 40 - 50\arcsec$, coupled with a broken exponential disk with a bend at $R_{maj}\approx 90\arcsec$, might be plausible but would not greatly impact on our bulge parameters.}
\label{NGC 4526}
\end{figure}

\begin{figure}[H]
\includegraphics[clip=true,trim= 1mm 1mm 1mm 1mm,height=12cm,width=0.49\textwidth]{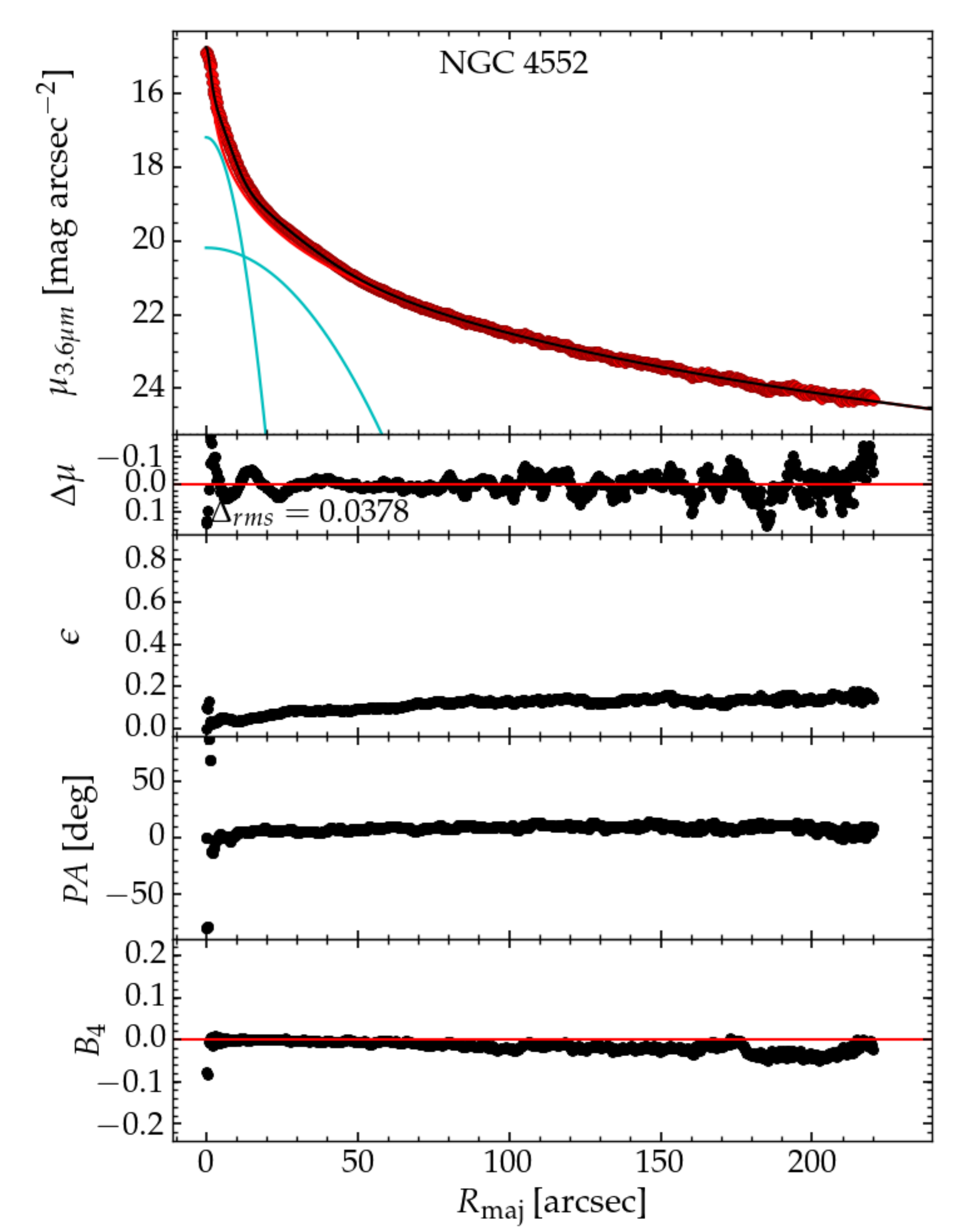}
\includegraphics[clip=true,trim= 1mm 1mm 1mm 1mm,height=12cm,width=0.49\textwidth]{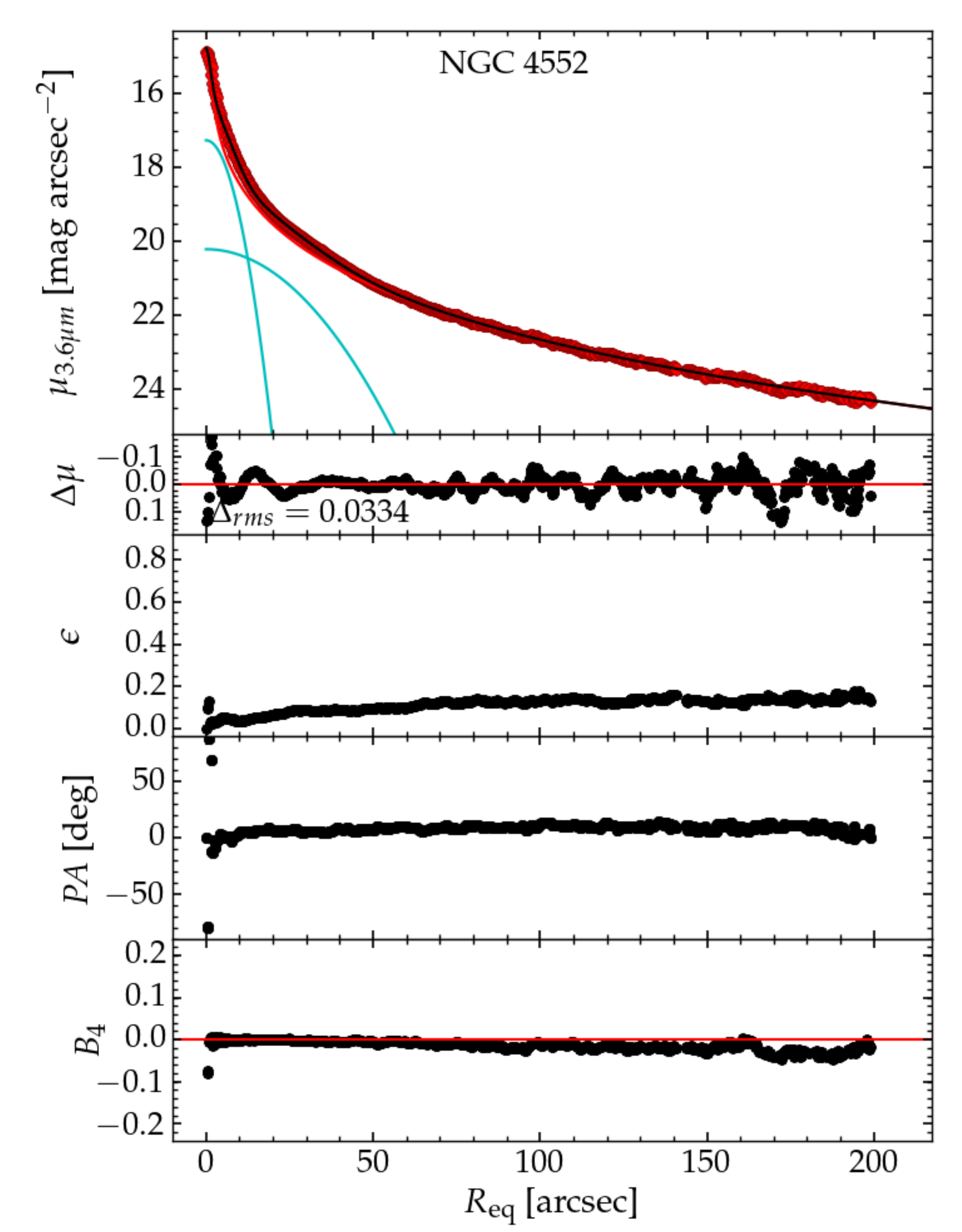}
\caption{NGC 4552: a massive elliptical galaxy with a dust ring at its core \citep{Bonfini:2018} which blocks light in the optical filter and can mimic the depleted core of a core-S{\'e}rsic galaxy, while in near-infrared filters it can mimic a central point source. Hence, we fit a central Gaussian (\textcolor{cyan}{---}) for extra light, a S{\'e}rsic function (\textcolor{red}{---}) for the extended spheroid, and another Gaussian (\textcolor{cyan}{---}) at the bump in the light profile at $R_{maj} \approx 35\arcsec$ which could be due to light from an undigested galaxy. }
\label{NGC 4552}
\end{figure}

\begin{figure}[H]
\includegraphics[clip=true,trim= 1mm 1mm 1mm 1mm,height=12cm,width=0.49\textwidth]{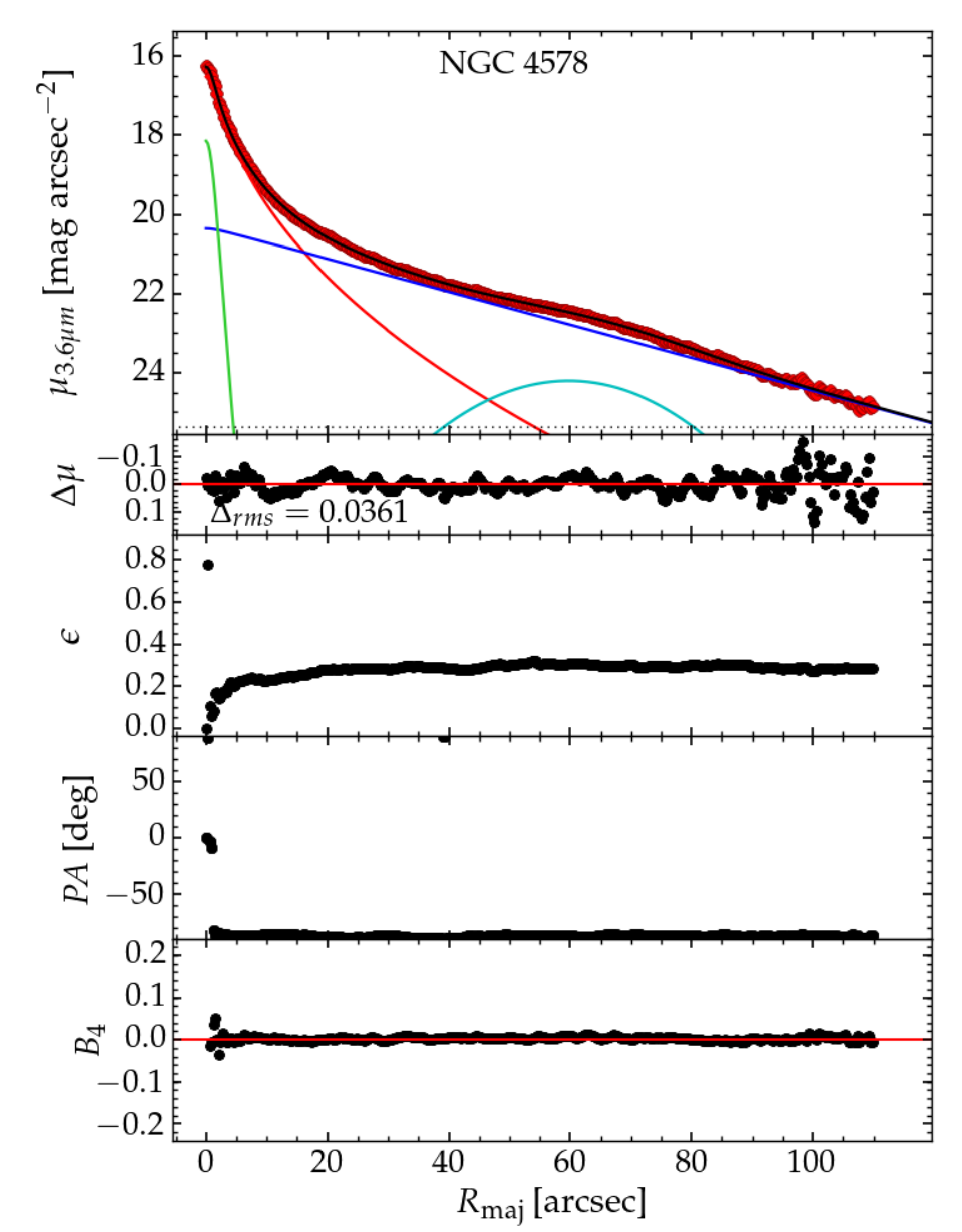}
\includegraphics[clip=true,trim= 1mm 1mm 1mm 1mm,height=12cm,width=0.49\textwidth]{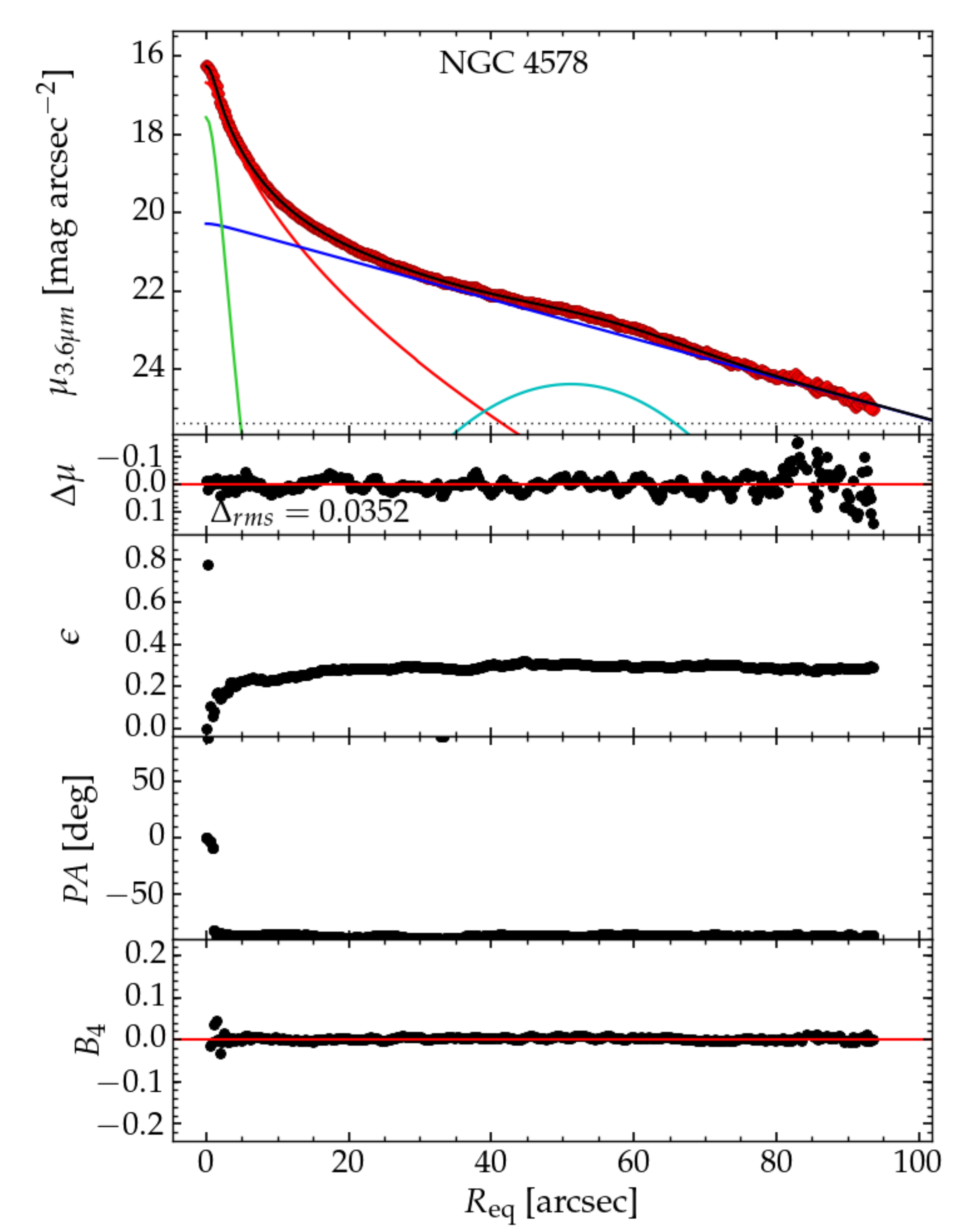}
\caption{NGC 4578: a lenticular galaxy with a central point source (\textcolor{green}{---}), a S{\'e}rsic bulge (\textcolor{red}{---}), an exponential disk (\textcolor{blue}{---}) and a faint ring \citep{RC3:1991} bumping up the light profile at ($R_{maj} \approx 67\arcsec$). }
\label{NGC 4578}
\end{figure}

\begin{figure}[H]
\includegraphics[clip=true,trim= 1mm 1mm 1mm 1mm,height=12cm,width=0.49\textwidth]{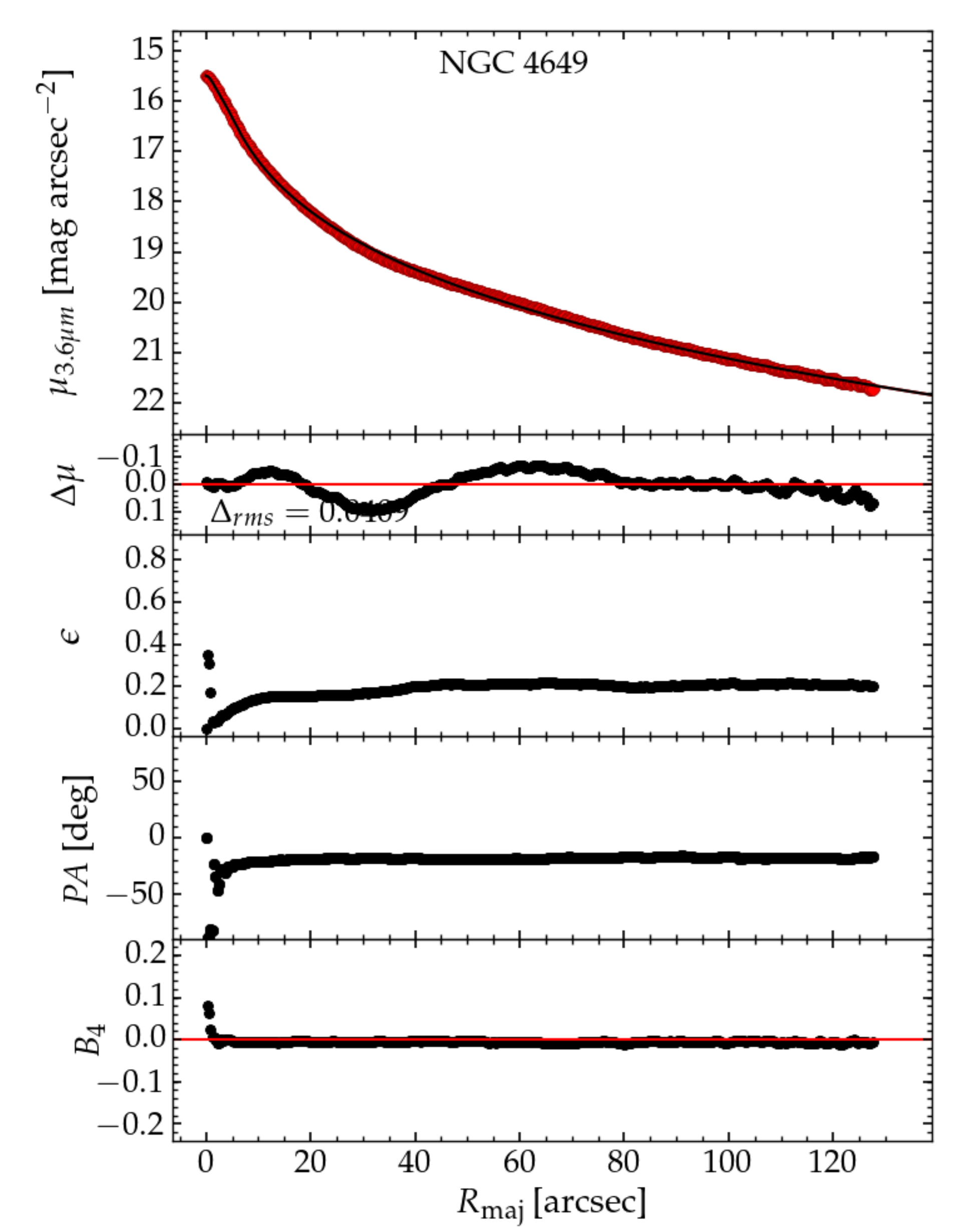}
\includegraphics[clip=true,trim= 1mm 1mm 1mm 1mm,height=12cm,width=0.49\textwidth]{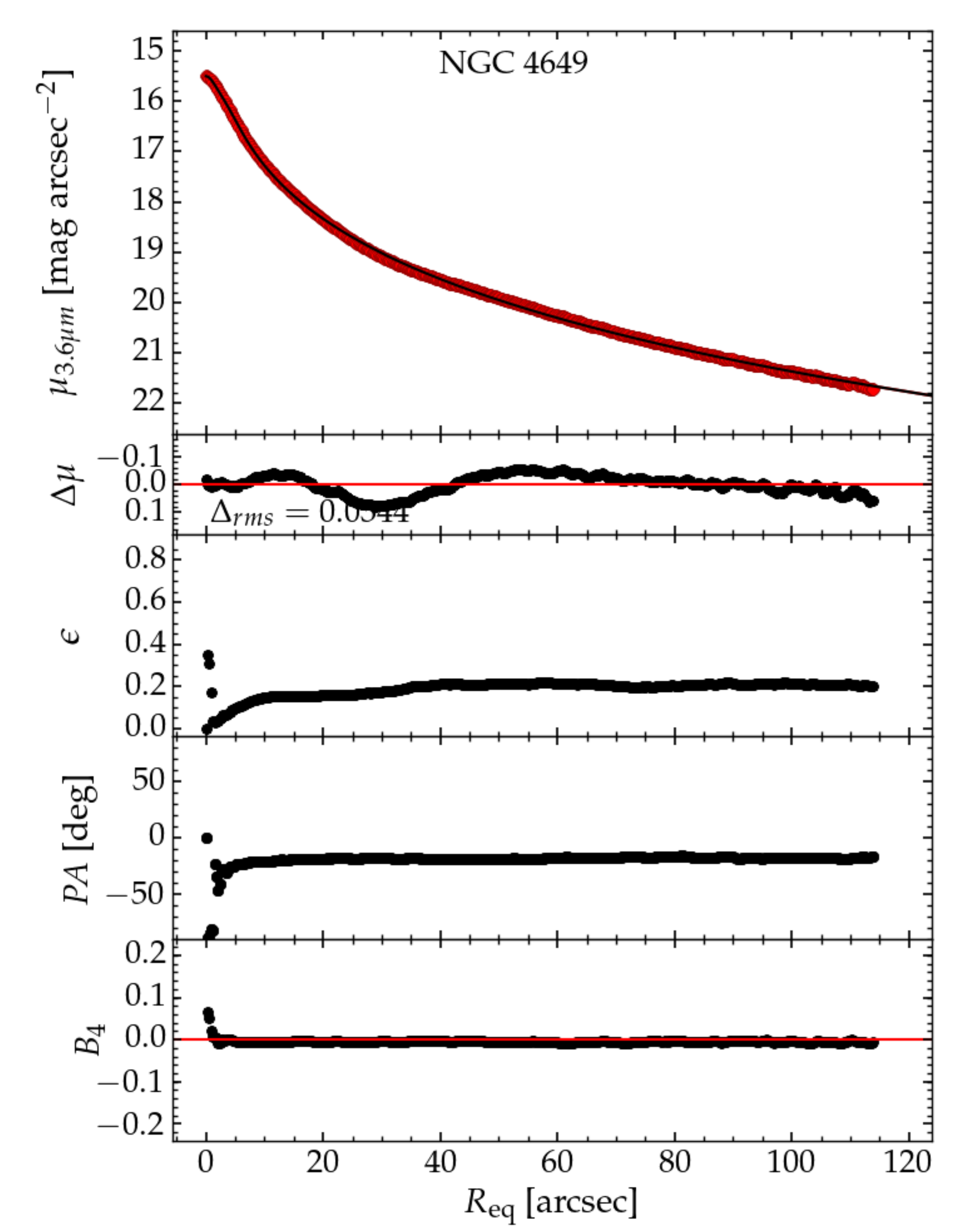}
\caption{NGC 4649: a massive elliptical galaxy with a deficit of light at its core, fit using a core-S{\'e}rsic function (\textcolor{red}{---}).}
\label{NGC 4649}
\end{figure}

\begin{figure}[H]
\includegraphics[clip=true,trim= 1mm 1mm 1mm 1mm,height=12cm,width=0.49\textwidth]{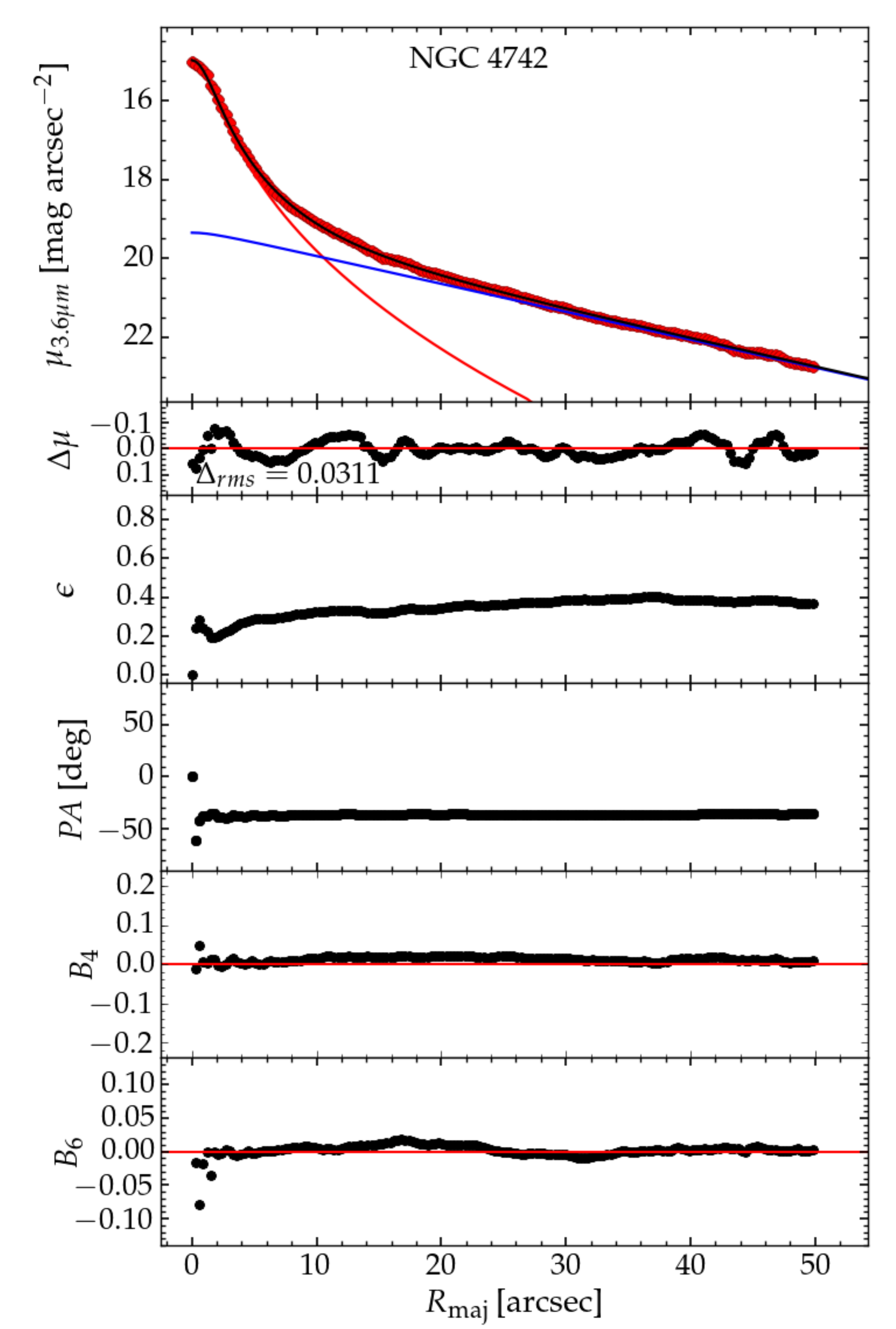}
\includegraphics[clip=true,trim= 1mm 1mm 1mm 1mm,height=12cm,width=0.49\textwidth]{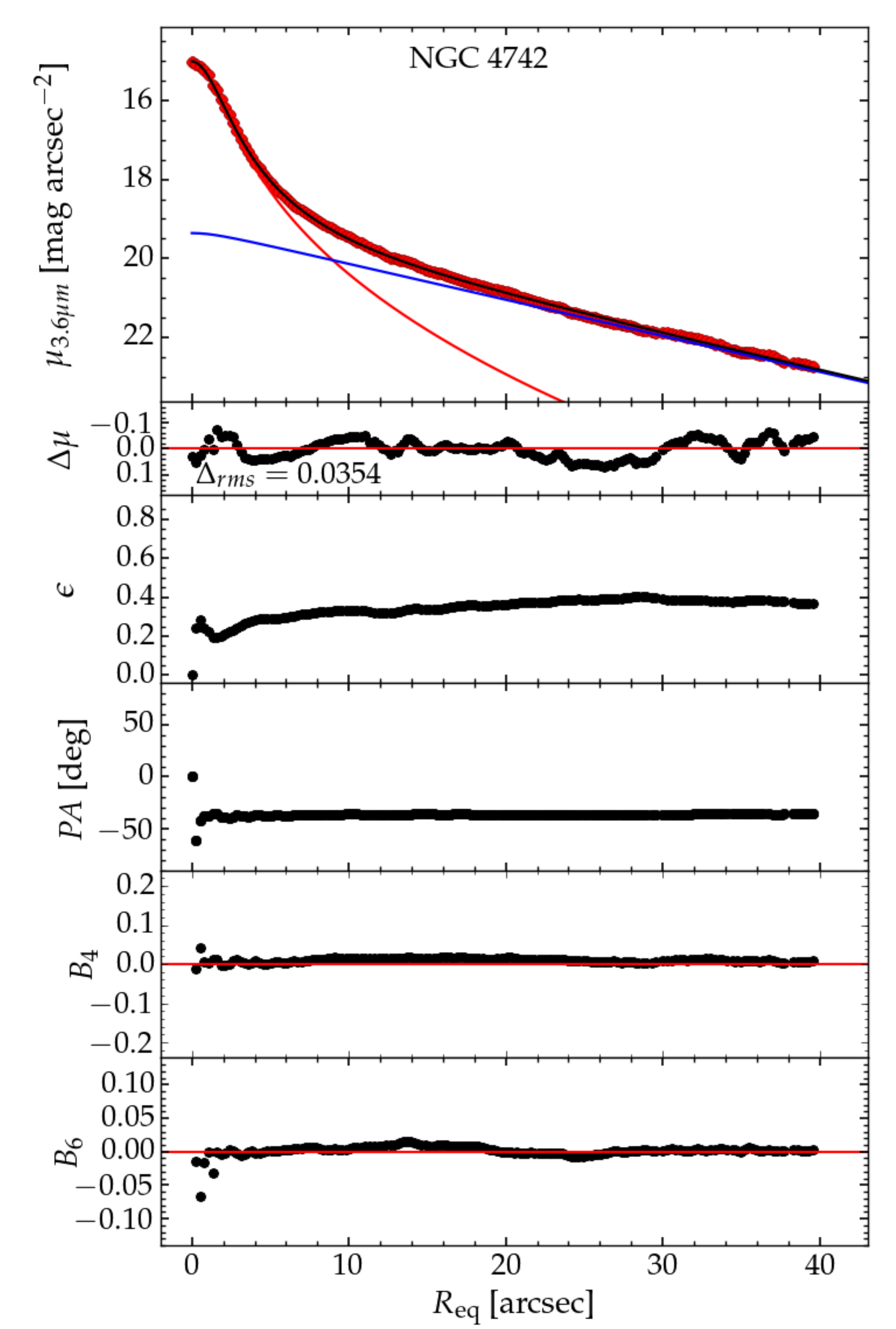}
\caption{NGC 4742: a lenticular galaxy with a S{\'e}rsic bulge (\textcolor{red}{---}) and an exponential disk (\textcolor{blue}{---}).}
\label{NGC 4742}
\end{figure}

\begin{figure}[H]
\includegraphics[clip=true,trim= 1mm 1mm 1mm 1mm,height=12cm,width=0.49\textwidth]{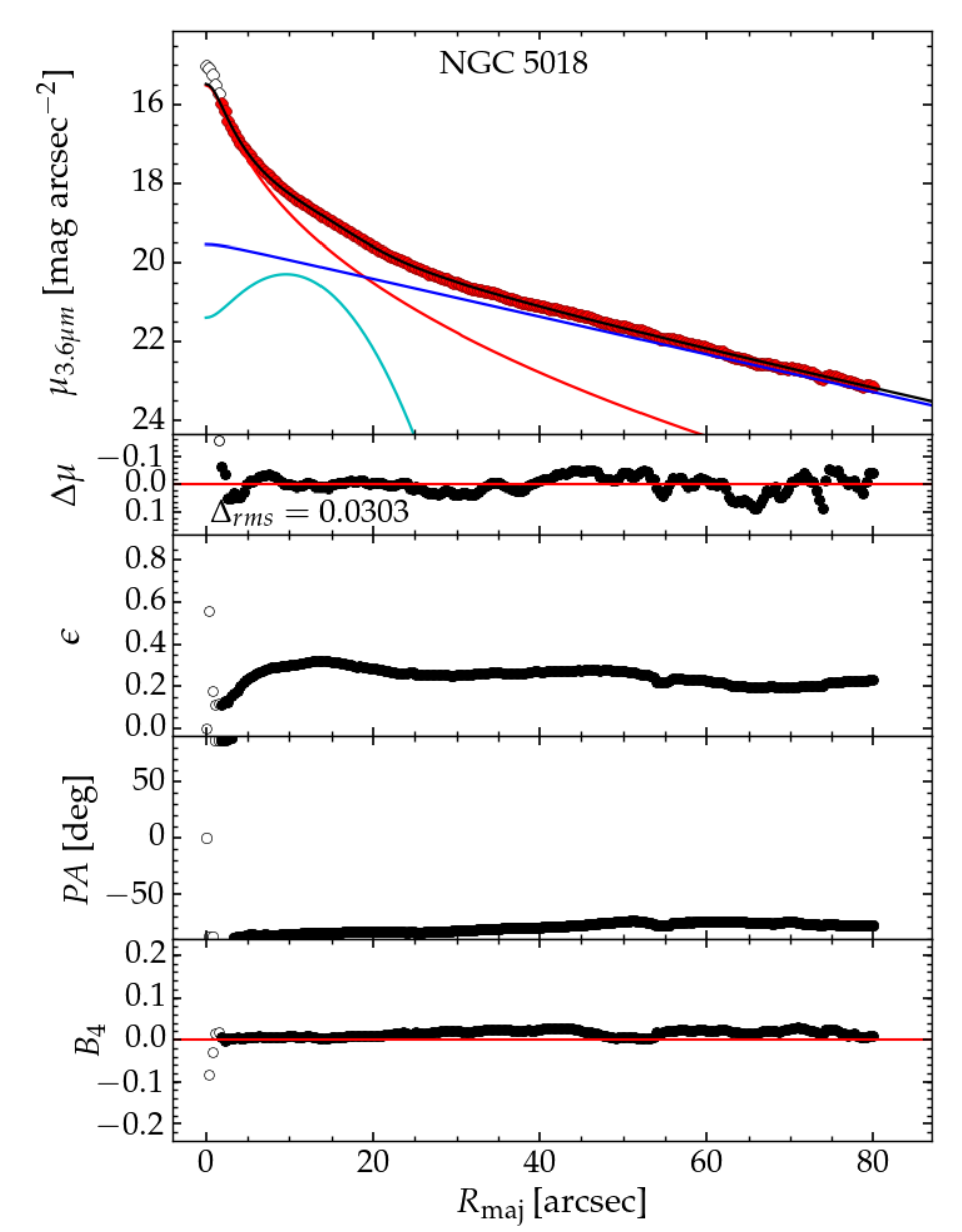}
\includegraphics[clip=true,trim= 1mm 1mm 1mm 1mm,height=12cm,width=0.49\textwidth]{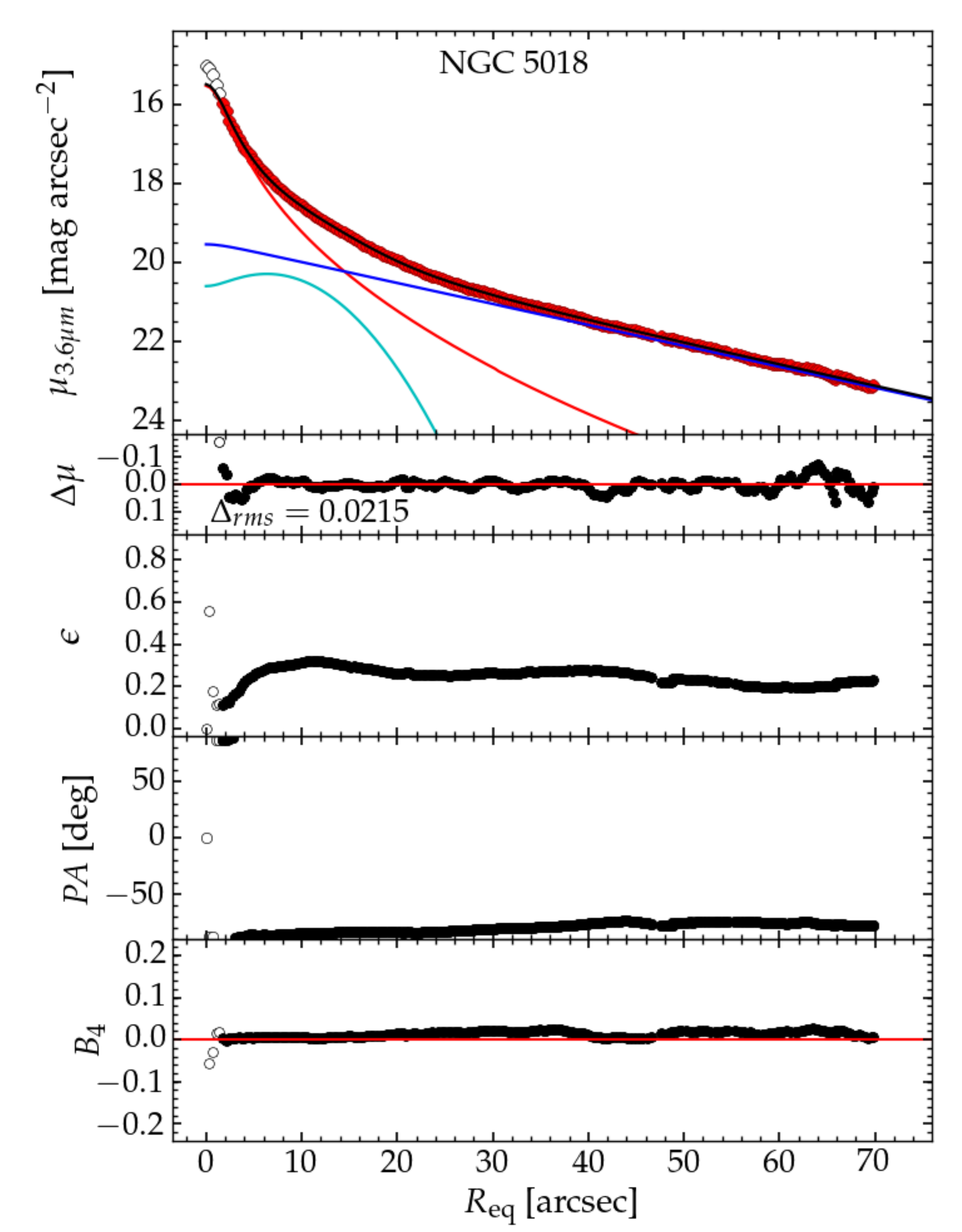}
\caption{NGC 5018: a post-merger remnant \citep{Buson:2004}, lenticular galaxy with an elongated debris tail revealing the previous merger. We have added a Gaussian (\textcolor{cyan}{---}) for the bump in the profile at $R_{maj}\approx 14\arcsec$ --- accounting for the undigested merged galaxy --- along with a S{\'e}rsic bulge (\textcolor{red}{---}), plus an exponential disk (\textcolor{blue}{---}). We excluded the inner data (up to $2\arcsec$) during the fitting.}
\label{NGC 5018}
\end{figure}

\begin{figure}[H]
\includegraphics[clip=true,trim= 1mm 1mm 1mm 1mm,height=12cm,width=0.49\textwidth]{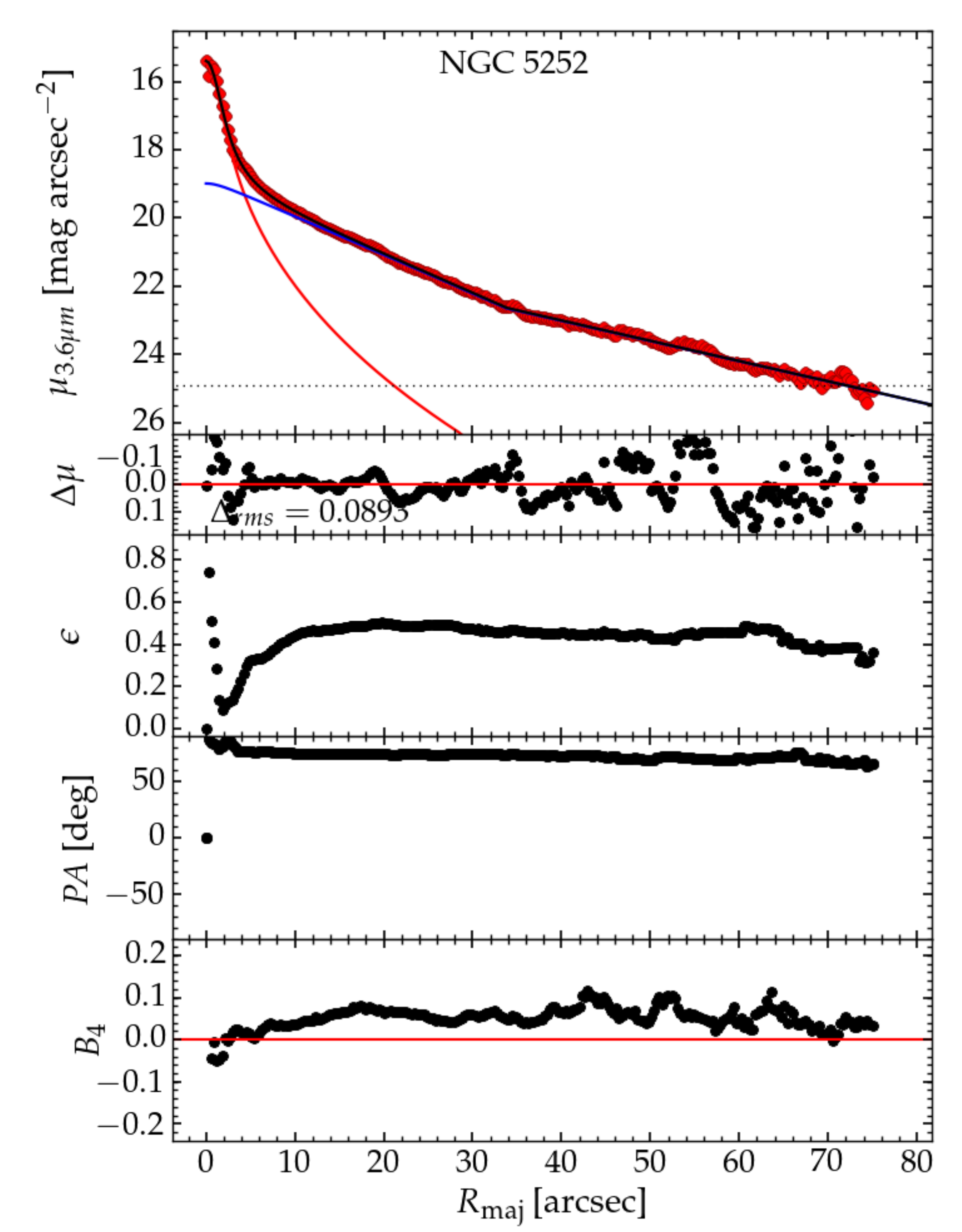}
\includegraphics[clip=true,trim= 1mm 1mm 1mm 1mm,height=12cm,width=0.49\textwidth]{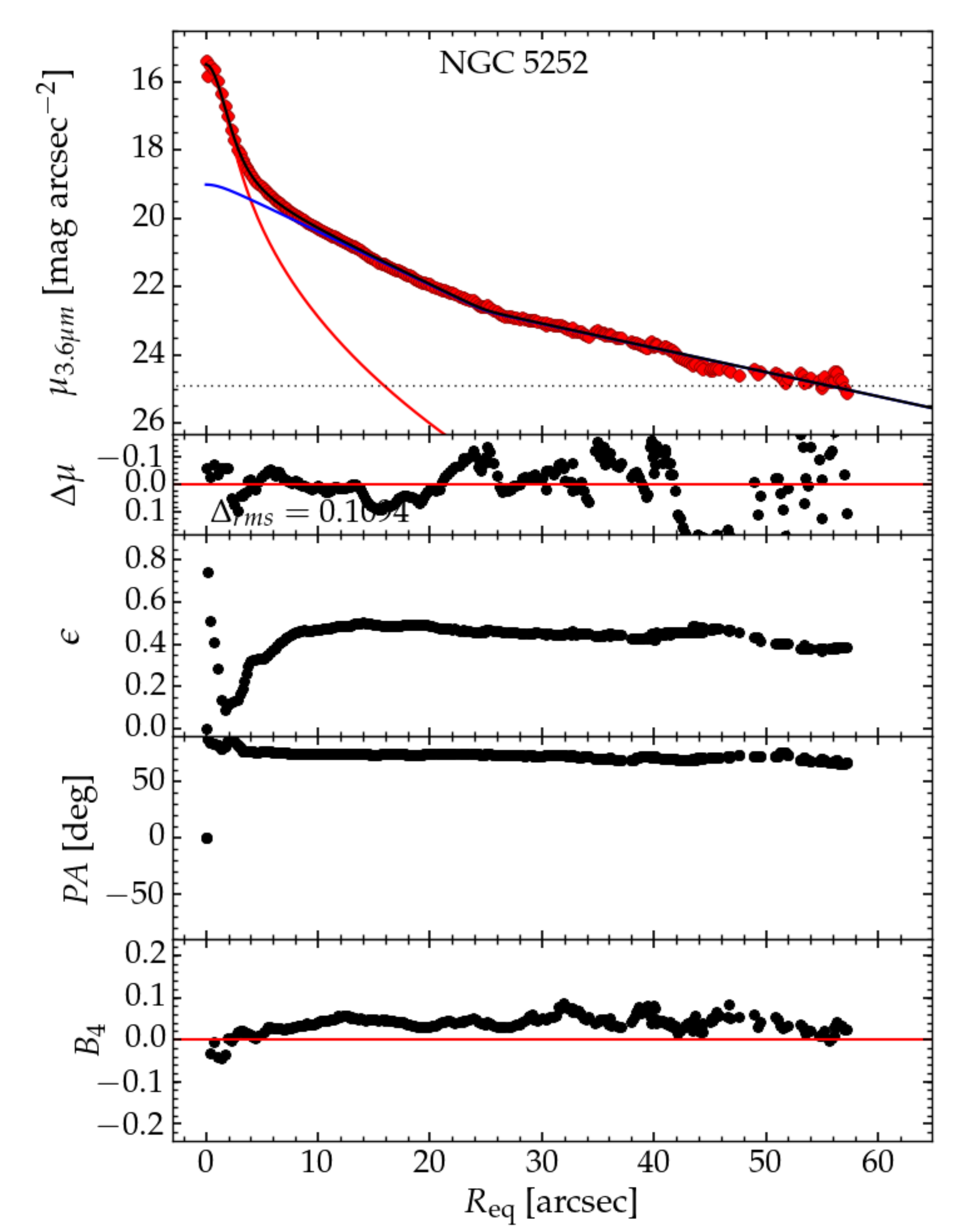}
\caption{NGC 5252:  a lenticular galaxy with a with S{\'e}rsic bulge (\textcolor{red}{---}) and a warped truncated disk (\textcolor{blue}{---}). NGC 5252 hosts a pair of AGNs, one is the central SMBH while the other (at 10 kpc distance from center) is an intermediate mass black hole \citep{yang:2017}. With $R_{e, sph} = 0.672$ kpc, $M_{*,sph}=7.1\times 10^{10} M_{\odot}$,and $M_{*,gal}=2.4\times 10^{11} M_{\odot}$, NGC 5252 is a \enquote{compact massive spheroid}.}
\label{NGC 5252}
\end{figure}

\begin{figure}[H]
\includegraphics[clip=true,trim= 1mm 1mm 1mm 1mm,height=12cm,width=0.49\textwidth]{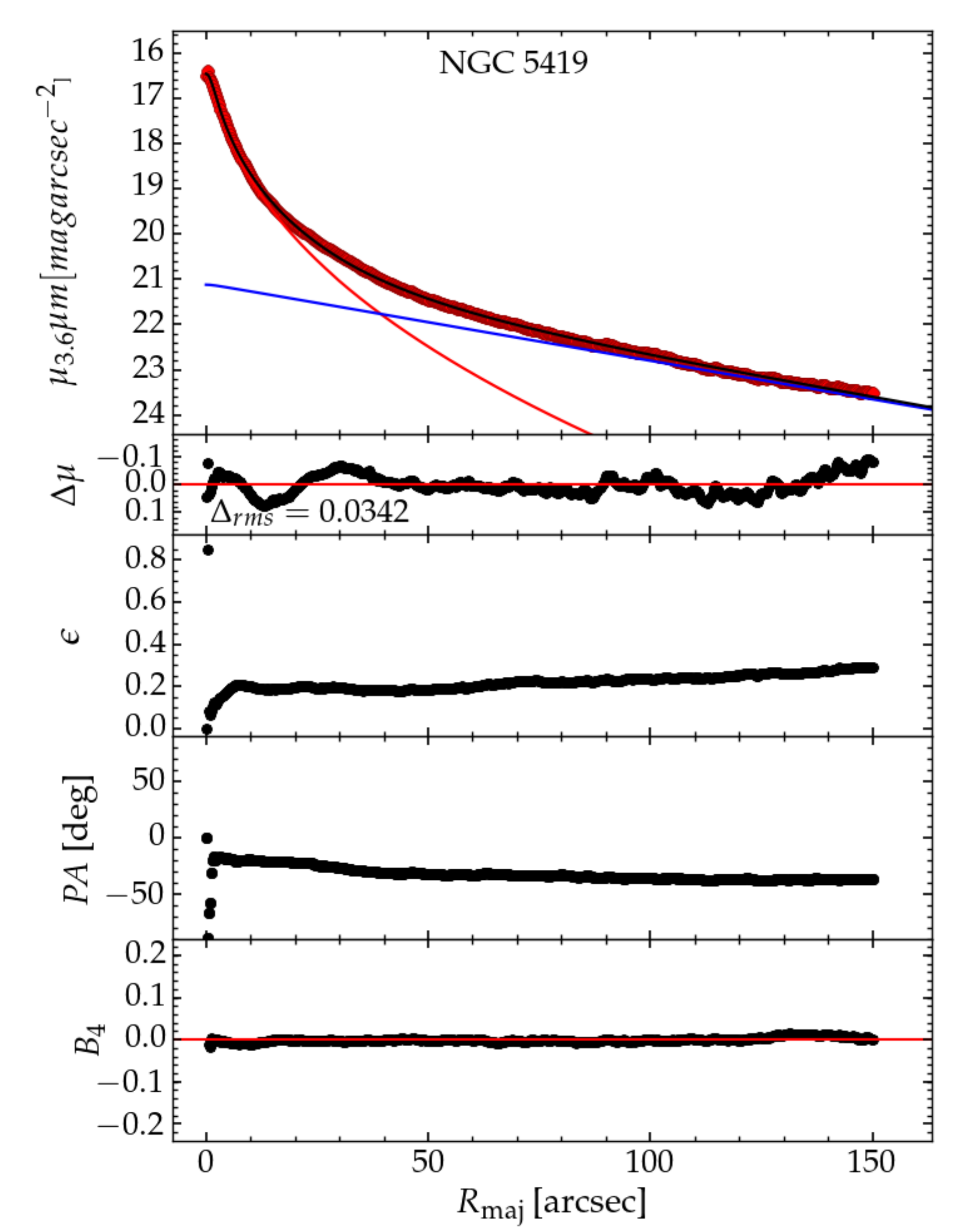}
\includegraphics[clip=true,trim= 1mm 1mm 1mm 1mm,height=12cm,width=0.49\textwidth]{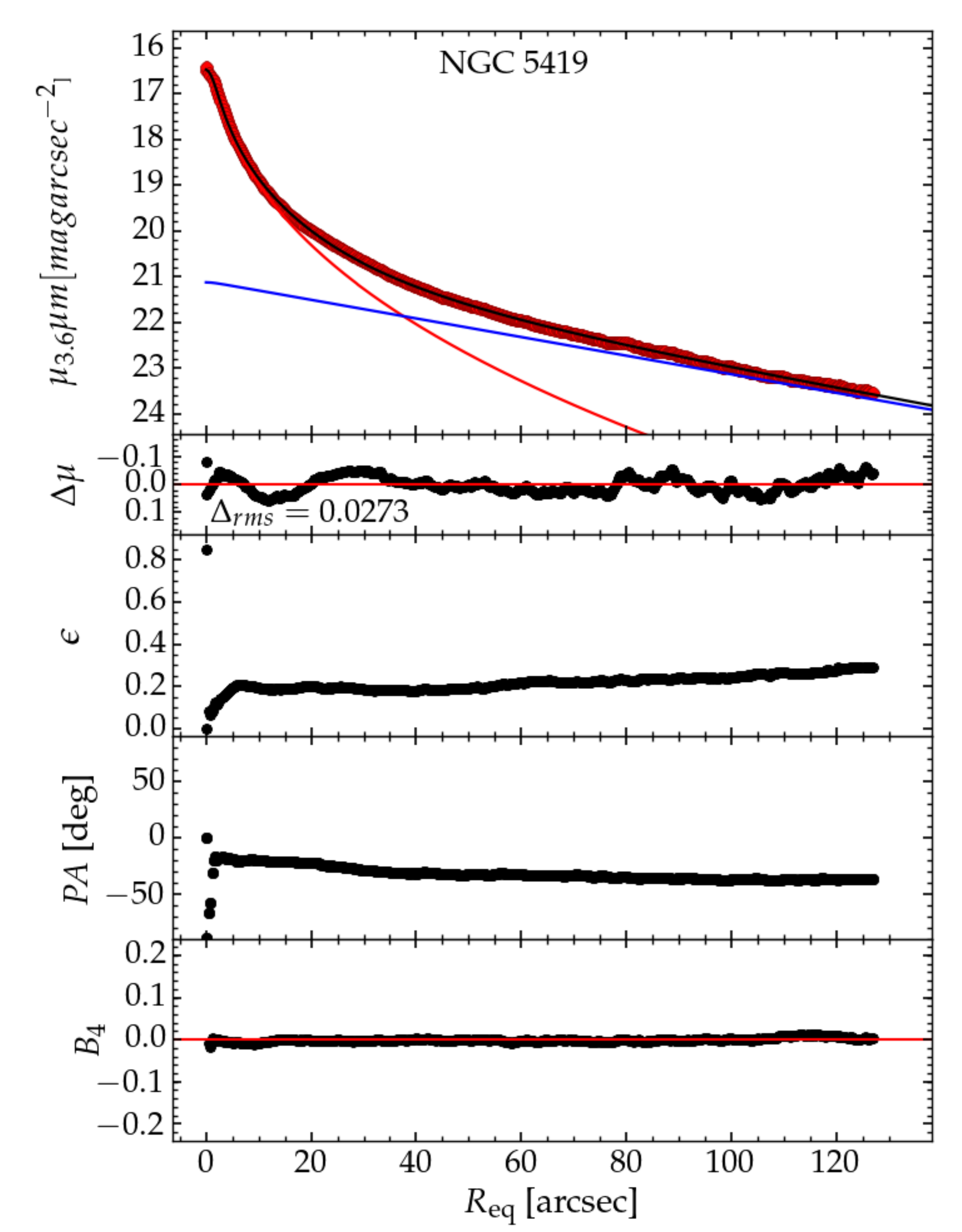}
\caption{NGC 5419: a \enquote{BCG \citep{Coziol:2009}} massive elliptical galaxy with a depleted core \citep{Mazzalay:2016} and an extended stellar halo. Its spheroid is fit using a core-S{\'e}rsic function (\textcolor{red}{---}) and for its halo we use an exponential (\textcolor{blue}{---}) function  \citep{deVaucouleurs:1969, Seigar:2007}. We do not include the (cluster's) halo light as a part of the galaxy's total light.}
\label{NGC 5419}
\end{figure}

\begin{figure}[H]
\includegraphics[clip=true,trim= 1mm 1mm 1mm 1mm,height=12cm,width=0.49\textwidth]{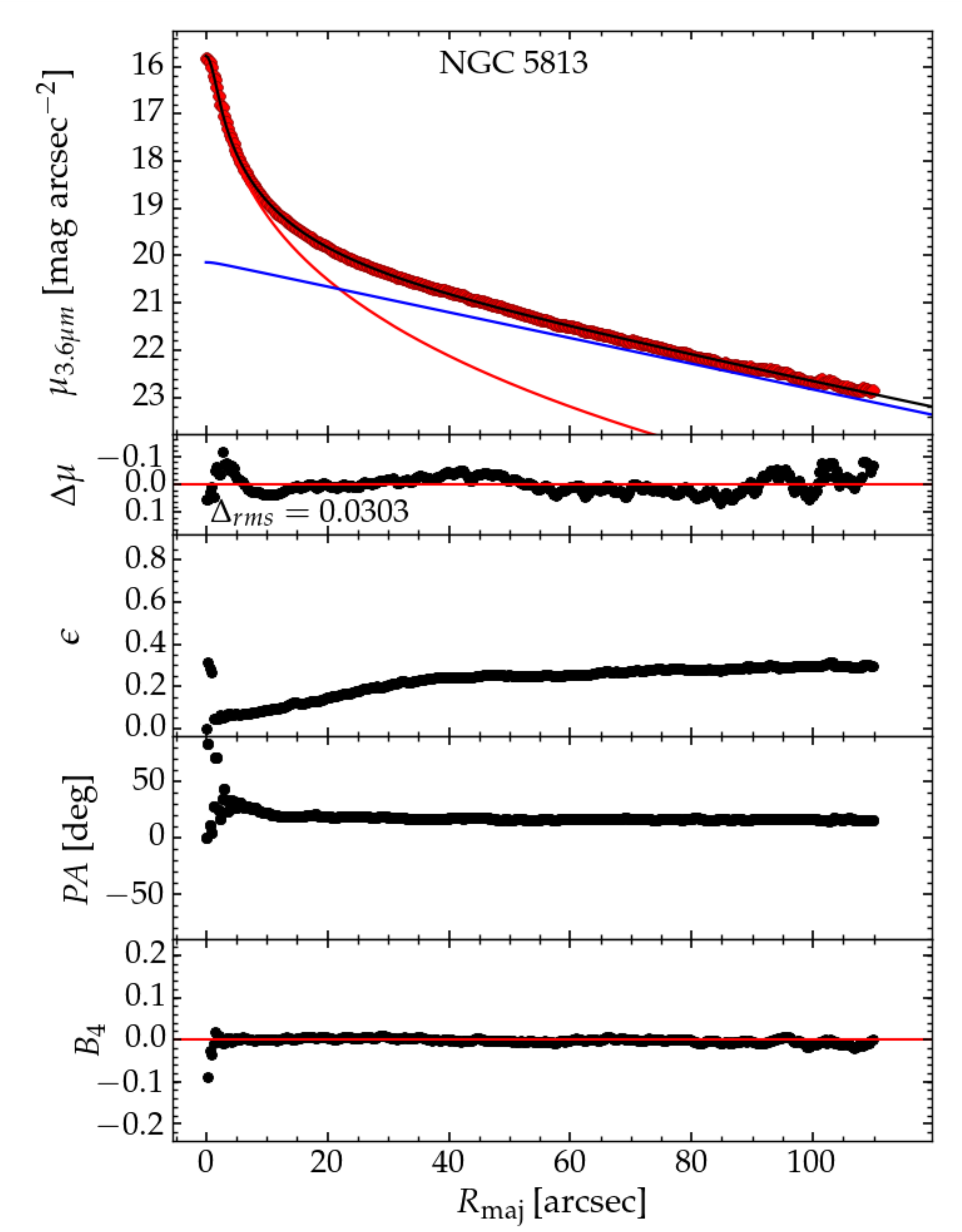}
\includegraphics[clip=true,trim= 1mm 1mm 1mm 1mm,height=12cm,width=0.49\textwidth]{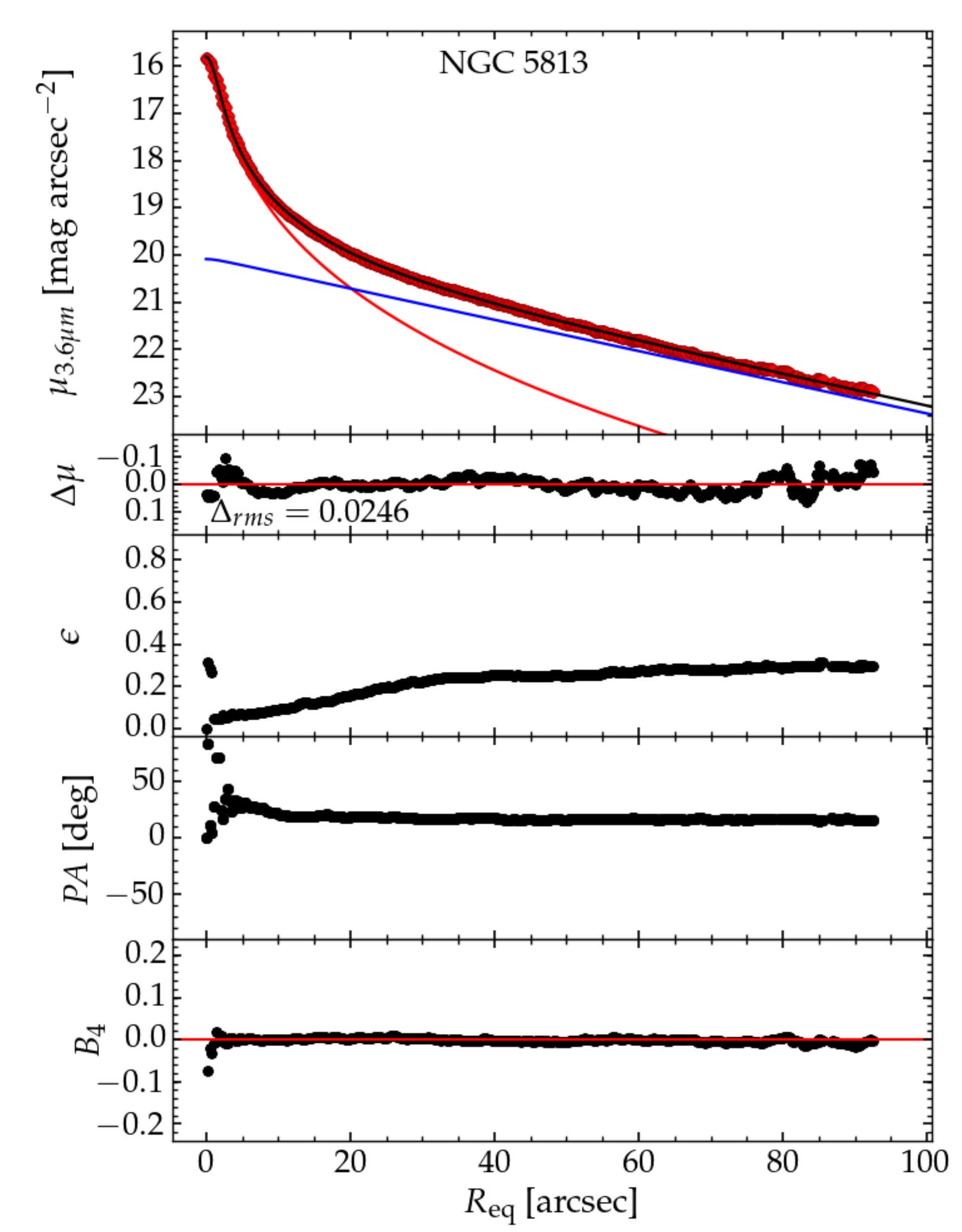}
\caption{NGC 5813: a core-S{\'e}rsic (\textcolor{red}{---}) galaxy \citep{Dullo:2014} with an outer exponential (\textcolor{blue}{---}) disk \citep{Trujillo:2004:A20}. It also has a counter-rotating core \citep{Carter:1993} which could not be resolved.}
\label{NGC 5813}
\end{figure}

\begin{figure}[H]
\includegraphics[clip=true,trim= 1mm 1mm 1mm 1mm,height=12cm,width=0.49\textwidth]{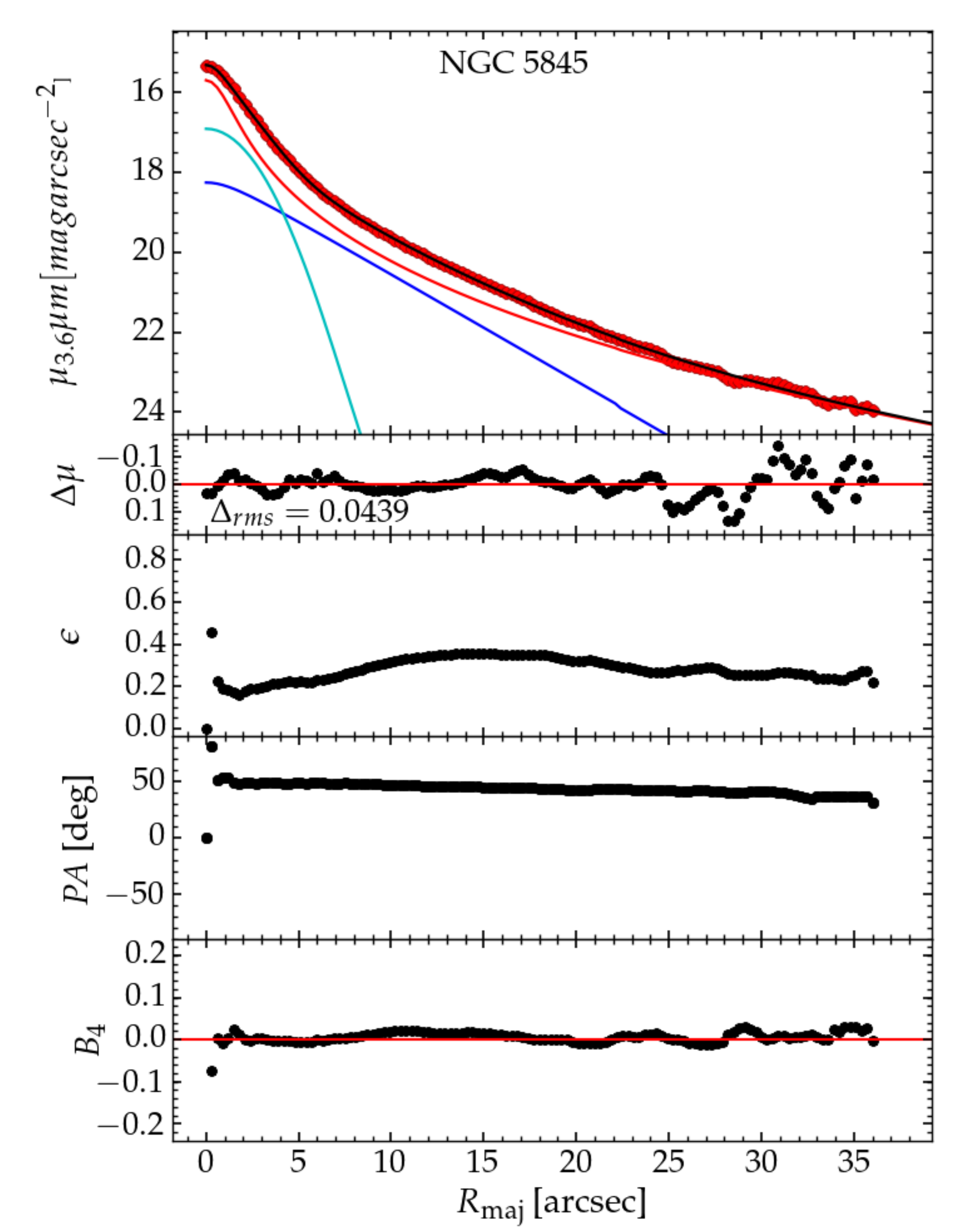}
\includegraphics[clip=true,trim= 1mm 1mm 1mm 1mm,height=12cm,width=0.49\textwidth]{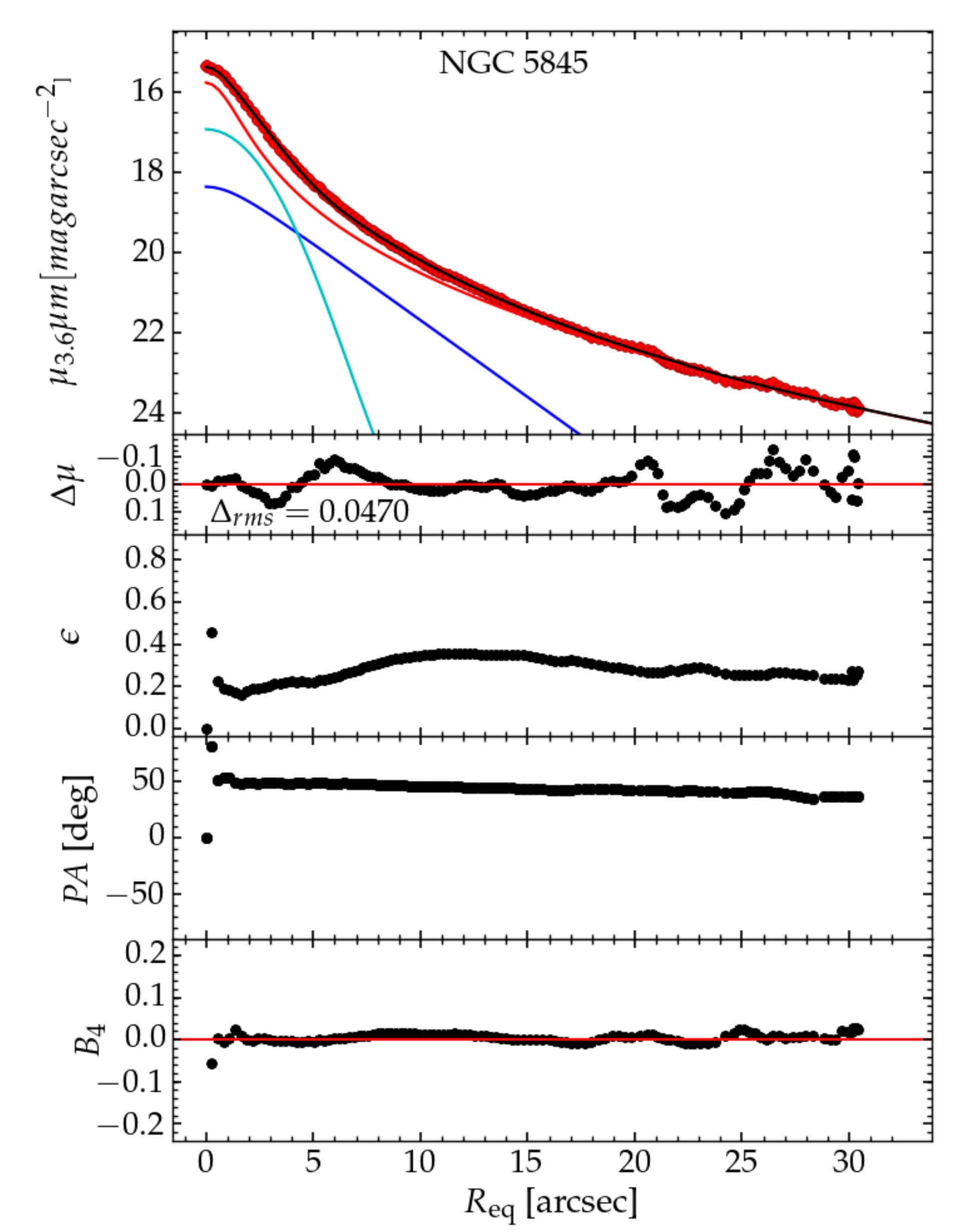}
\caption{NGC 5845: an ellicular galaxy with an extended S{\'e}rsic spheroid (\textcolor{red}{---}) and an intermediate-scale disk (\textcolor{blue}{---}) suggested by the elevation in the ellipticity profile around $R_{maj}\approx 14\arcsec$. \citet{Kormendy:2000} call it a \enquote{Rosetta stone} object which contains a dust disk and a stellar disk. The double peak rotation curve in \citet{Jiang:2012} suggests that there is another inner disk, which we fit with a Gaussian (\textcolor{cyan}{---}). }
\label{NGC 5845}
\end{figure}

\begin{figure}[H]
\includegraphics[clip=true,trim= 1mm 1mm 1mm 1mm,height=12cm,width=0.49\textwidth]{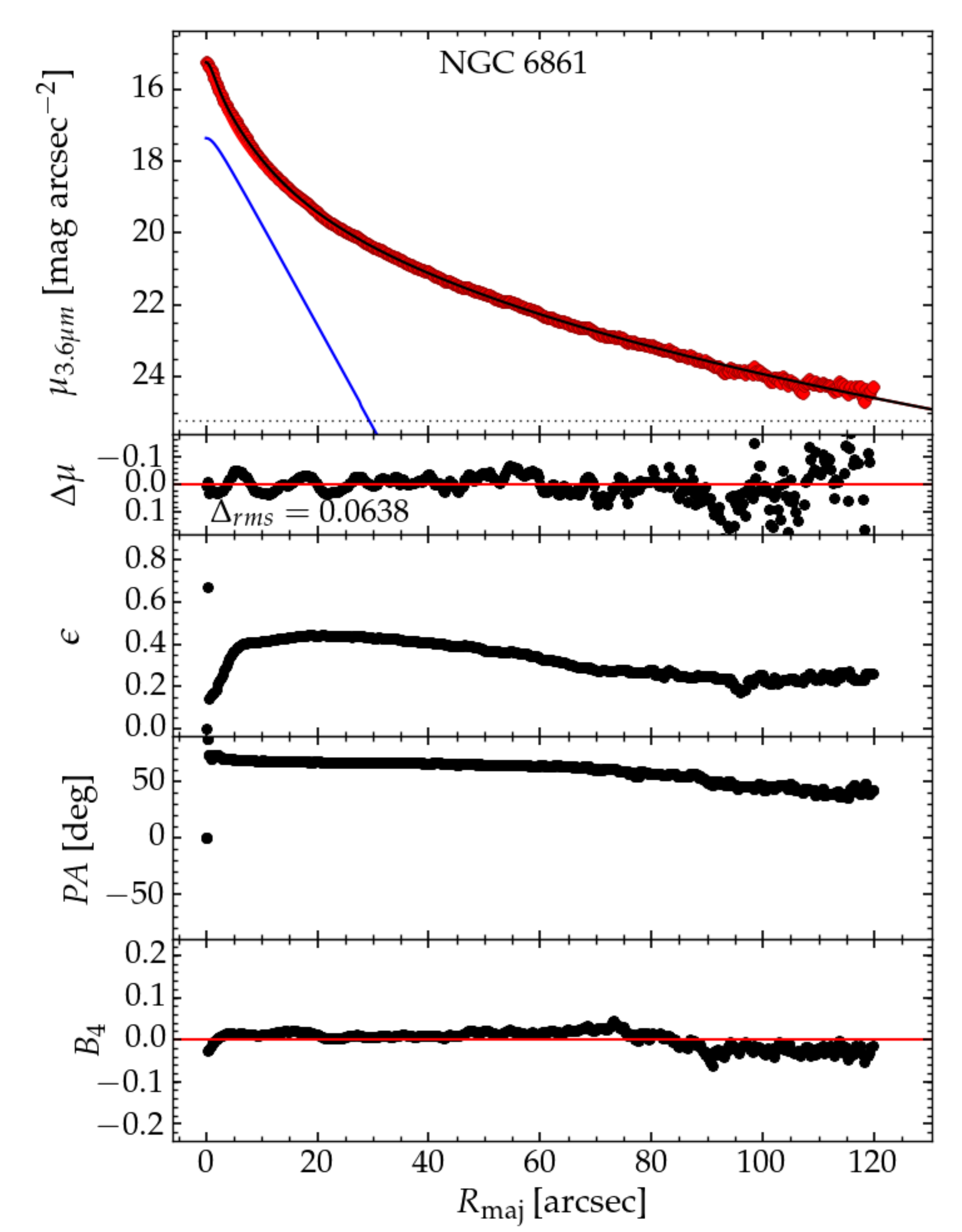}
\includegraphics[clip=true,trim= 1mm 1mm 1mm 1mm,height=12cm,width=0.49\textwidth]{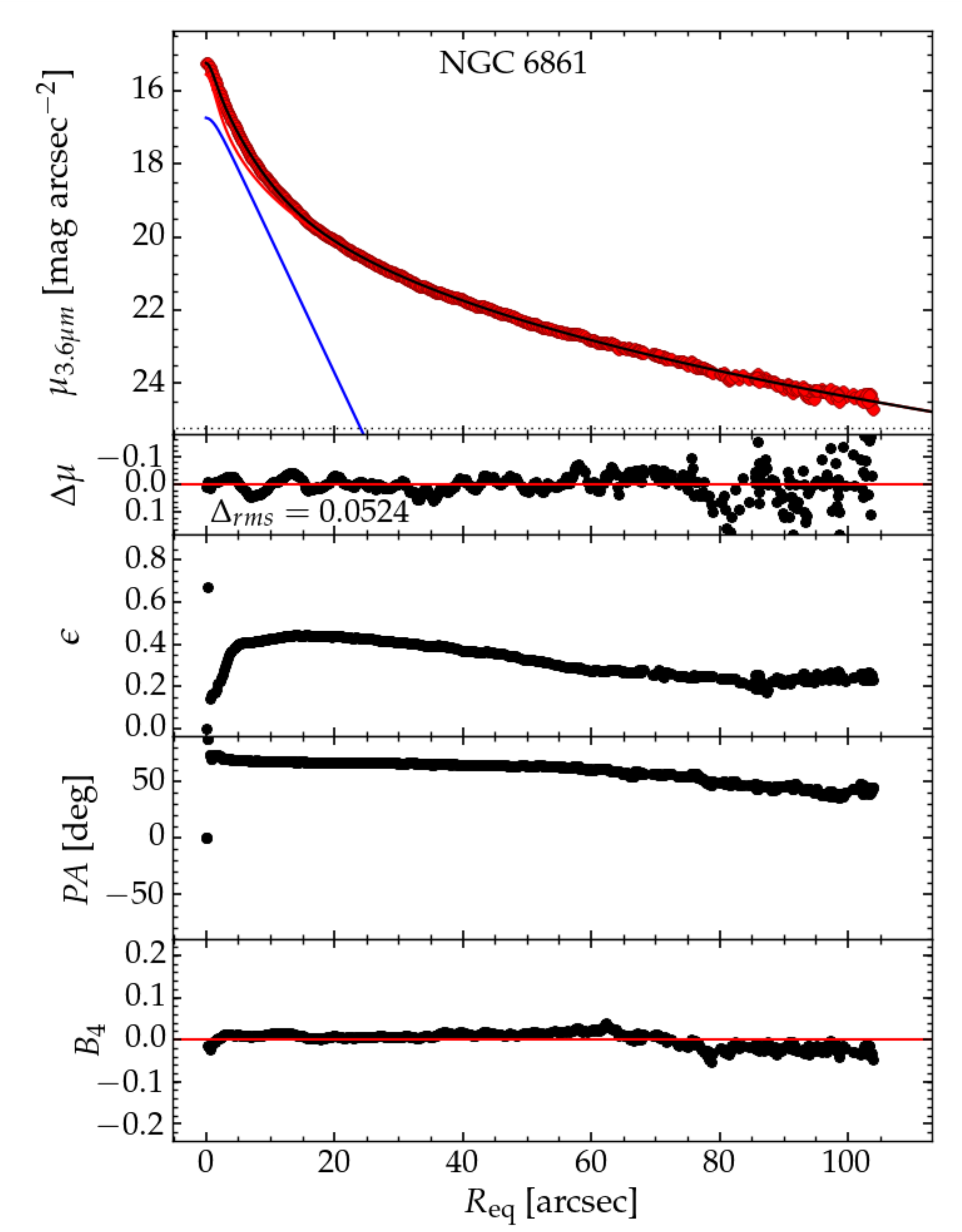}
\caption{NGC 6861: an ellicular (ES) galaxy with an extended S{\'e}rsic bulge (\textcolor{red}{---}) plus an intermediate-scale disk \citep{Rusli:2013, Escudero:2015} fit here using an exponential function (\textcolor{blue}{---}).}
\label{NGC 6861}
\end{figure}

\begin{figure}[H]
\includegraphics[clip=true,trim= 1mm 1mm 1mm 1mm,height=12cm,width=0.49\textwidth]{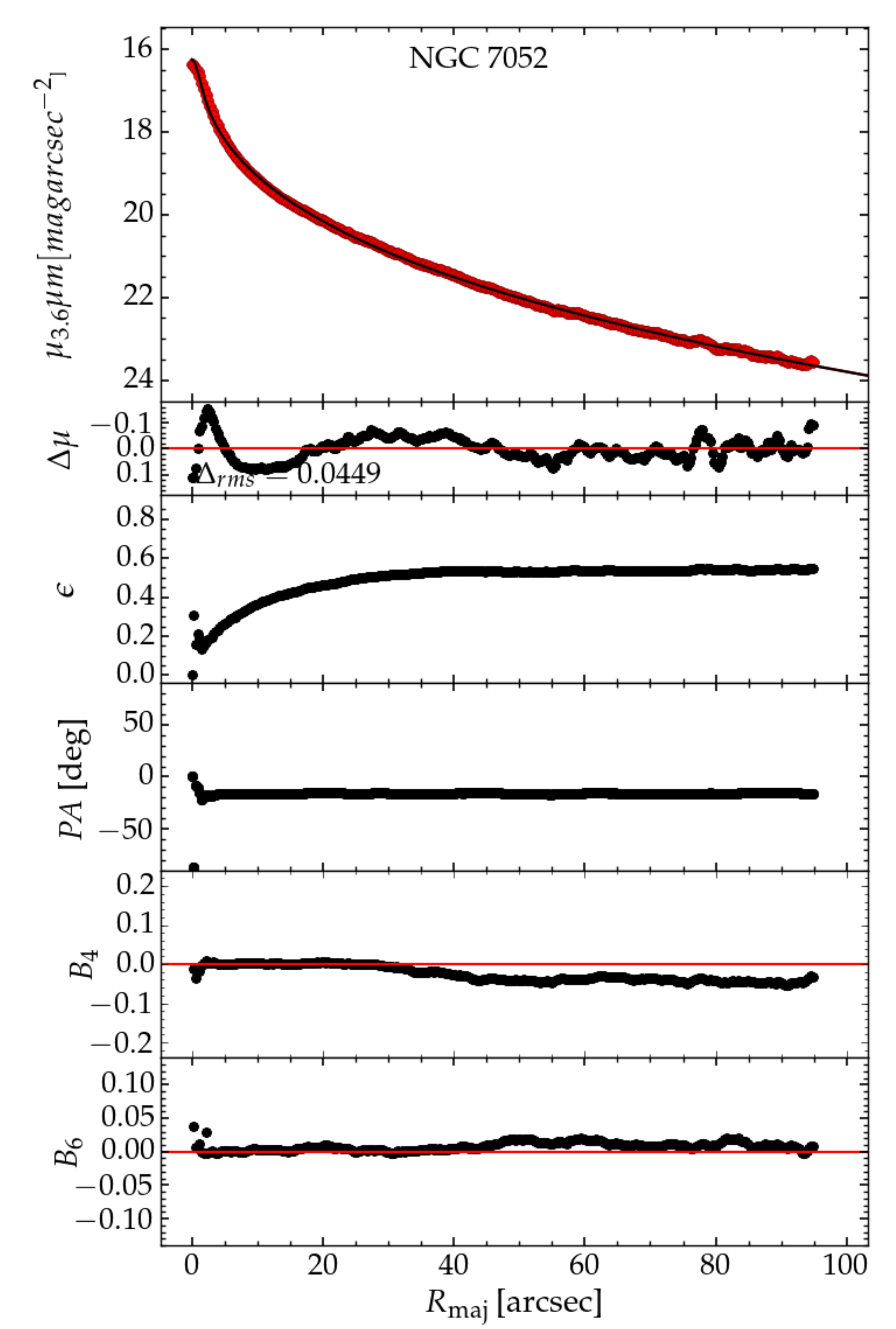}
\includegraphics[clip=true,trim= 1mm 1mm 1mm 1mm,height=12cm,width=0.49\textwidth]{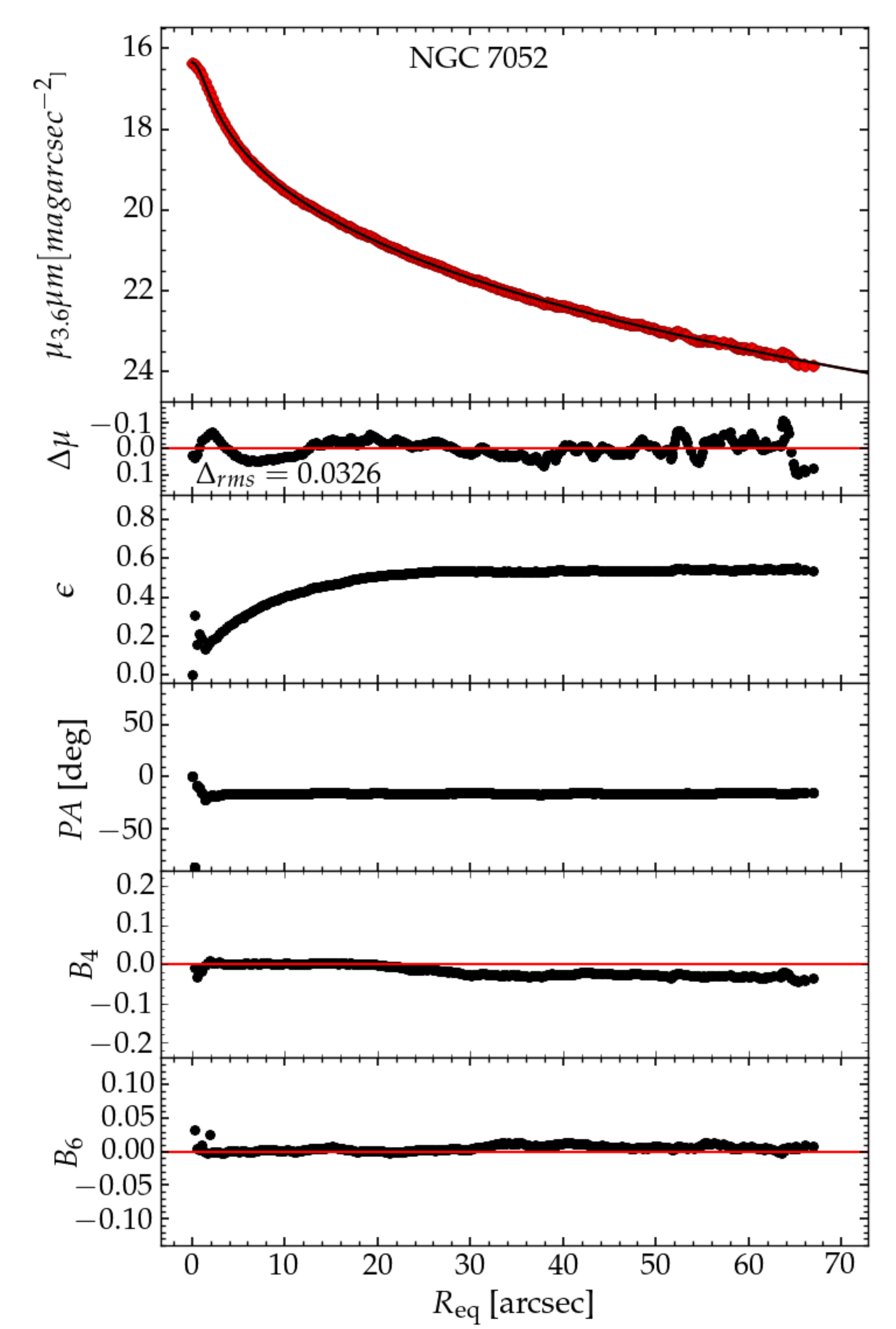}
\caption{NGC 7052: a massive elliptical core-S{\'e}rsic  (\textcolor{red}{---}) galaxy \citep{Quillen:2000}.}
\label{NGC 7052}
\end{figure}

\begin{figure}[H]
\includegraphics[clip=true,trim= 1mm 1mm 1mm 1mm,height=12cm,width=0.49\textwidth]{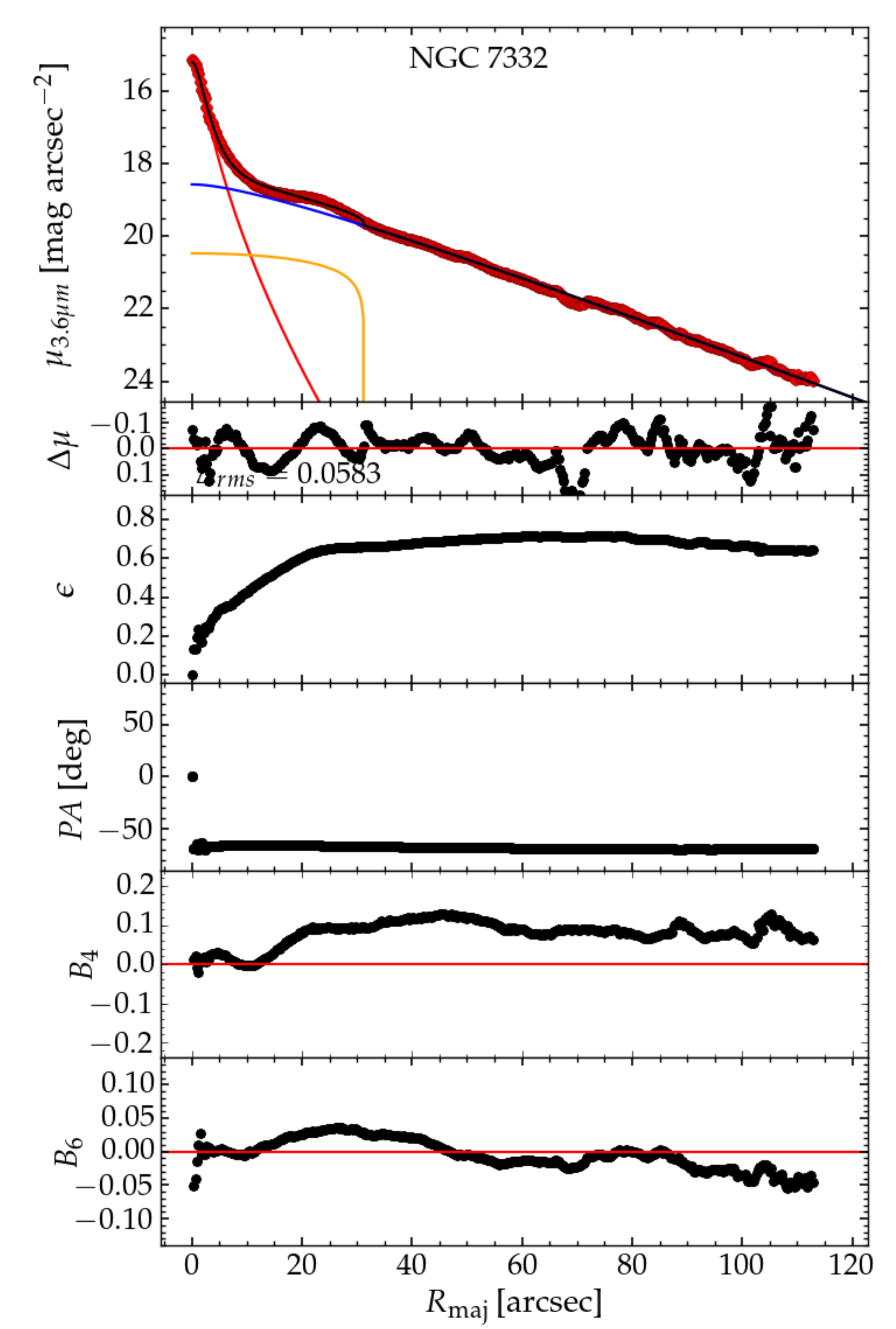}
\includegraphics[clip=true,trim= 1mm 1mm 1mm 1mm,height=12cm,width=0.49\textwidth]{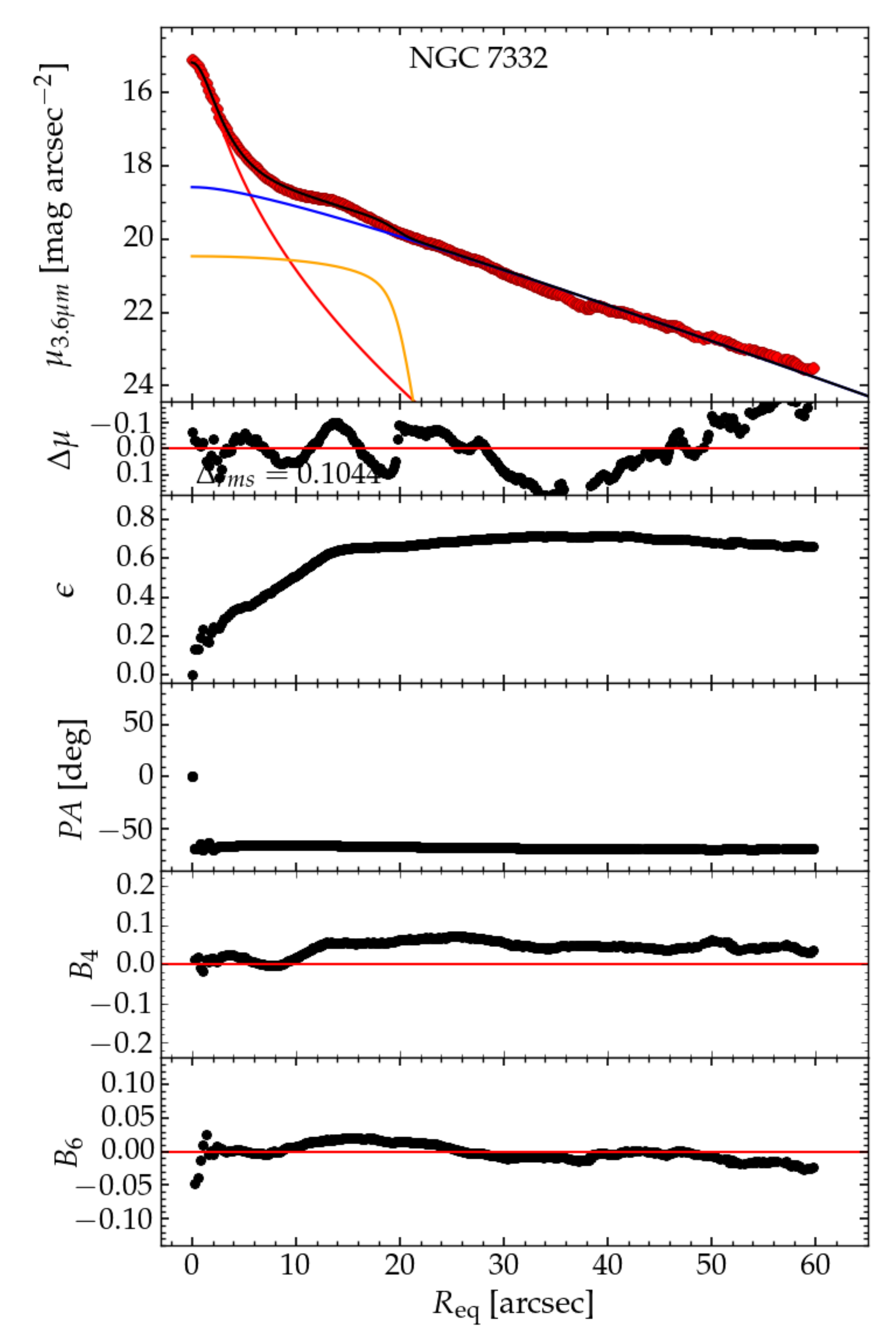}
\caption{NGC 7332: a peculiar (edge-on) lenticular galaxy. It has a S{\'e}rsic bulge (\textcolor{red}{---}), a weak bar \citep{Falcon-Barroso:2004} fit using a Ferrers (\textcolor{orange}{---}) function, and an outer exponential truncated disk (\textcolor{blue}{---}). According to \citet{Falcon-Barroso:2004}, NGC 7332 also has an inner disk but it could not be seen in the Spitzer image.}
\label{NGC 7332}
\end{figure}

\begin{figure}[H]
\includegraphics[clip=true,trim= 1mm 1mm 1mm 1mm,height=12cm,width=0.49\textwidth]{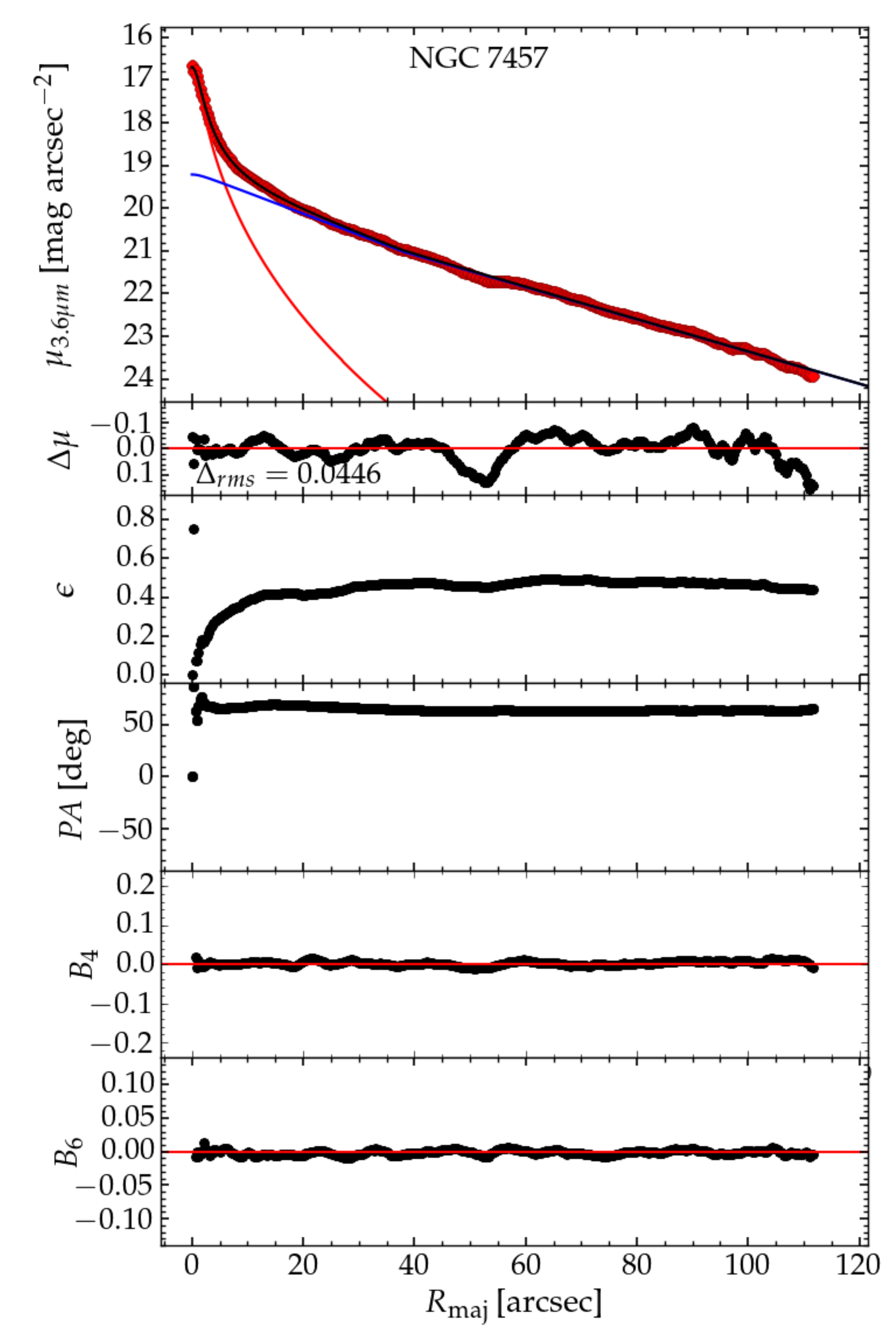}
\includegraphics[clip=true,trim= 1mm 1mm 1mm 1mm,height=12cm,width=0.49\textwidth]{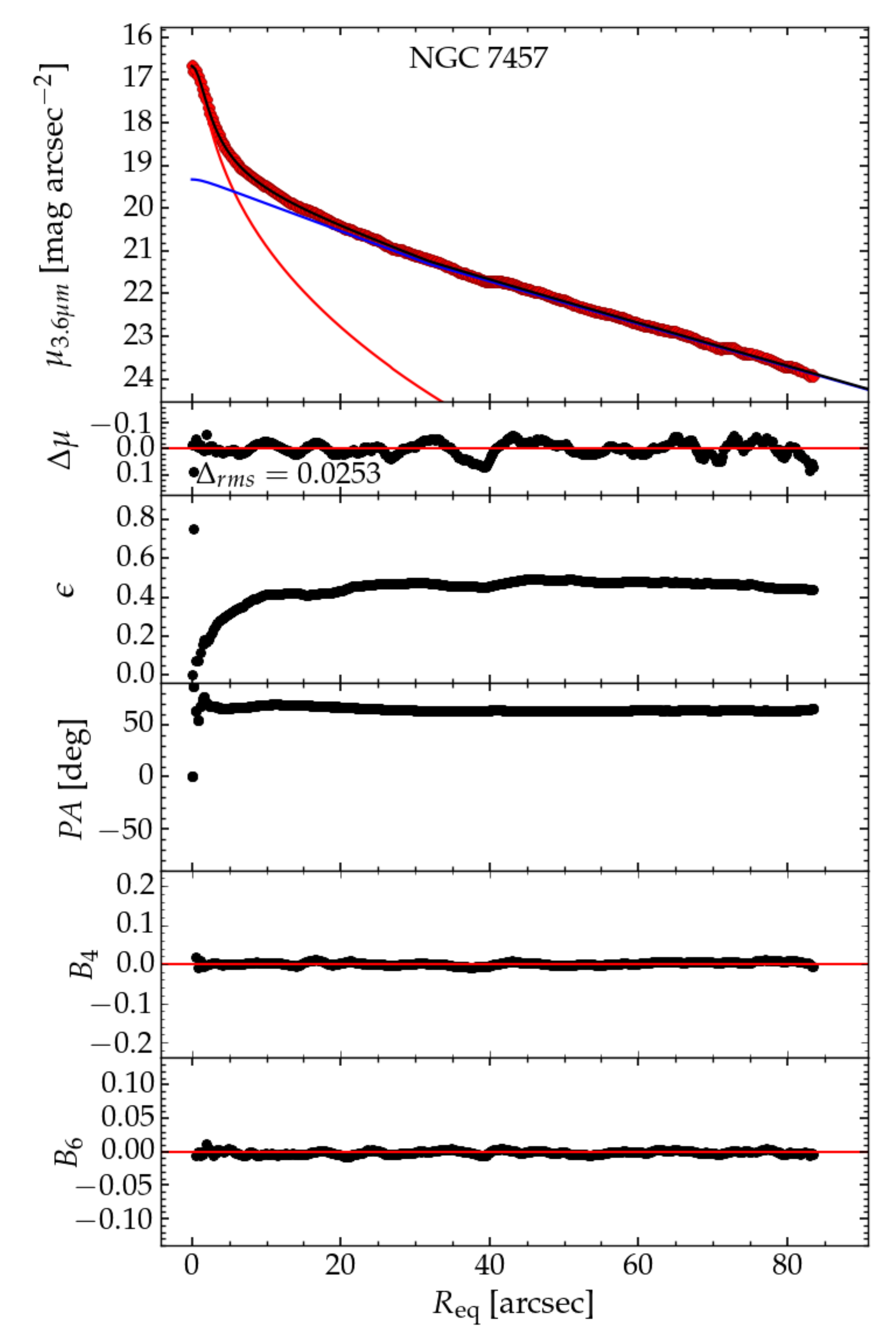}
\caption{NGC 7457: a lenticular galaxy with a S{\'e}rsic bulge (\textcolor{red}{---}) and truncated exponential disk (\textcolor{blue}{---}). }
\label{NGC 7457}
\end{figure}

\subsection{Light profile from $K_{s}$-band images (AB mag)}
\begin{figure}[H]
\includegraphics[clip=true,trim= 1mm 1mm 1mm 1mm,height=12cm,width=0.49\textwidth]{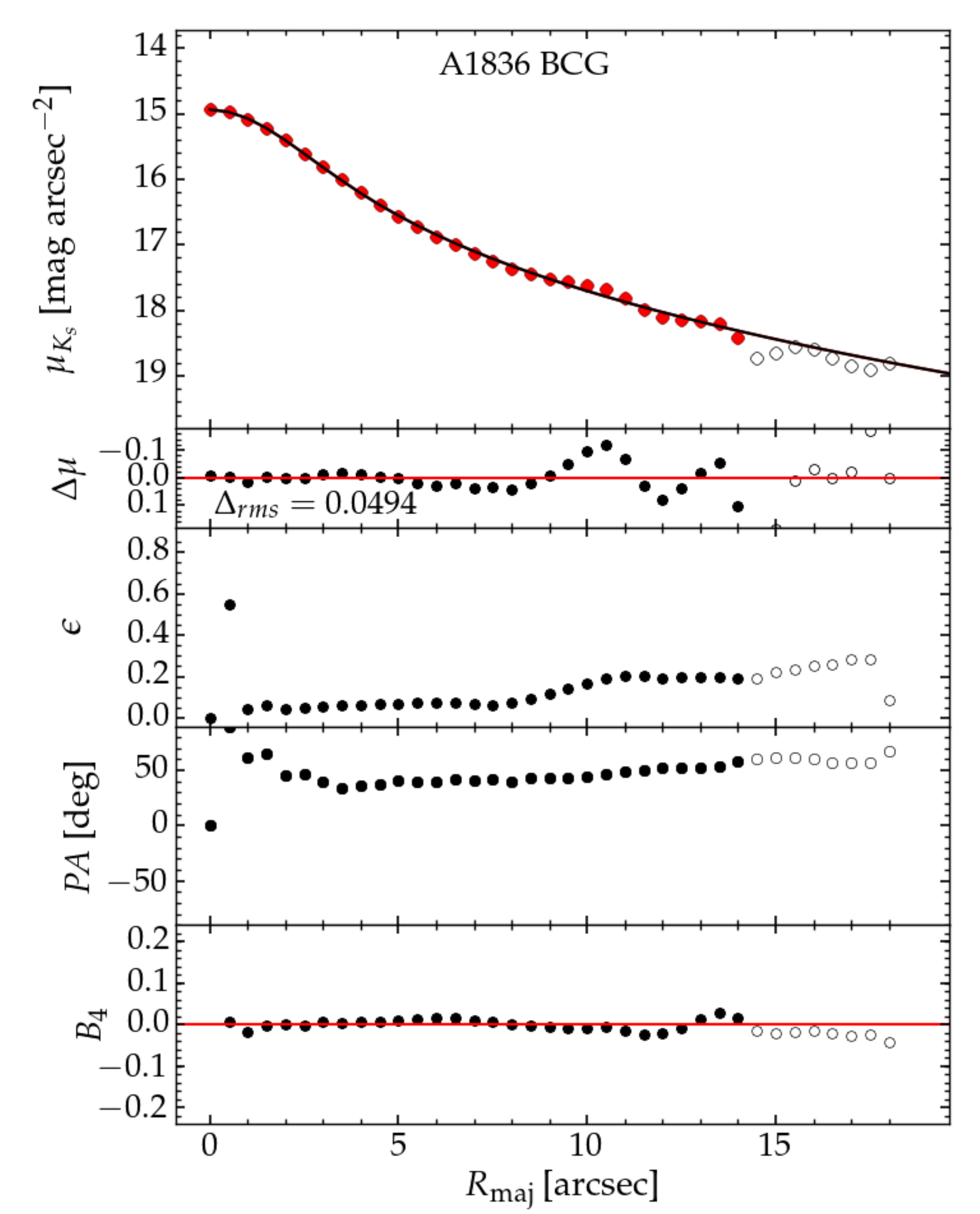}
\includegraphics[clip=true,trim= 1mm 1mm 1mm 1mm,height=12cm,width=0.49\textwidth]{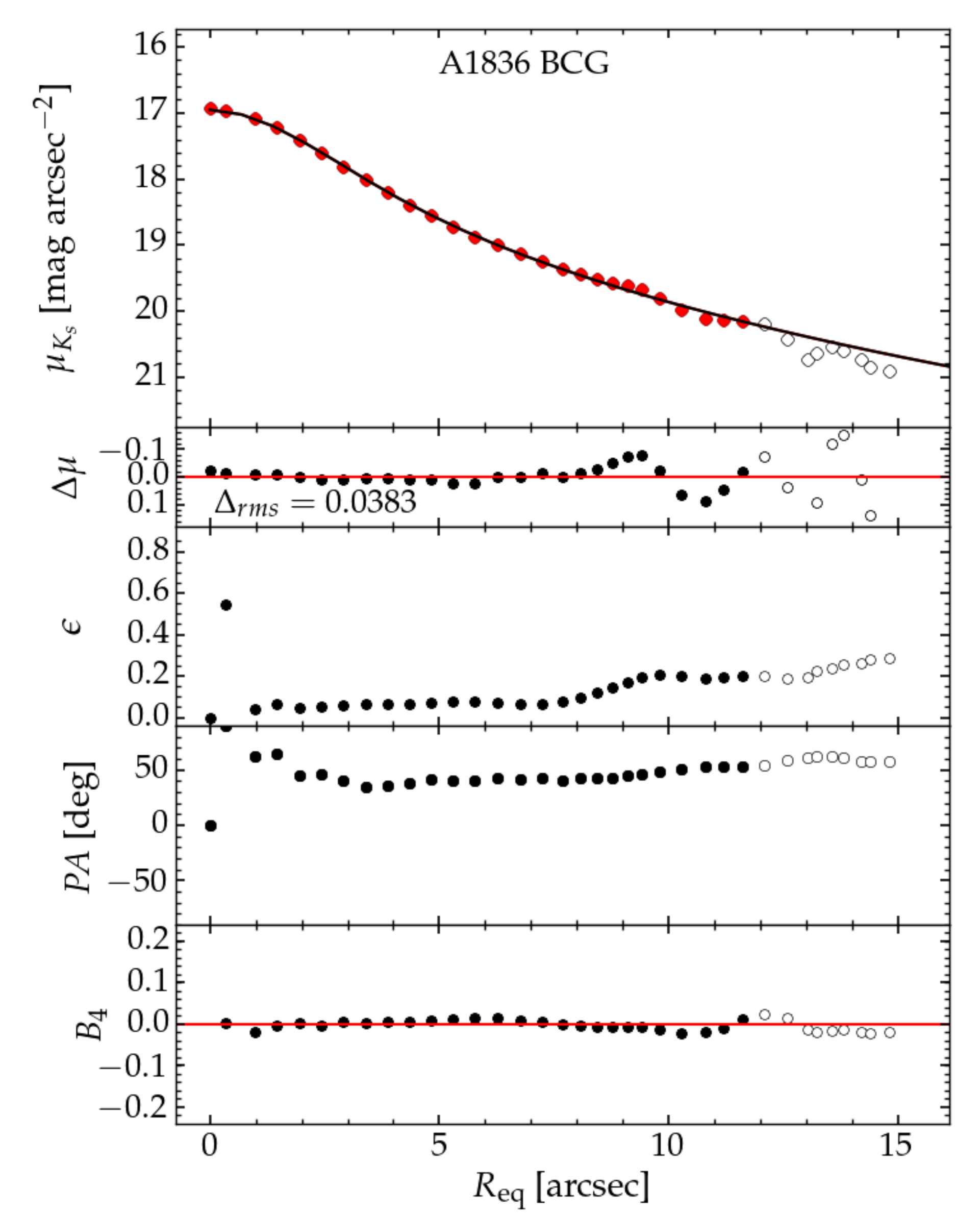}
\caption{A1836 BCG: a massive elliptical Brightest Cluster Galaxy (BCG). Its light profile and very high velocity dispersion suggests that it may have a depleted core, hence we fit its light profile using a core-S{\'e}rsic function (\textcolor{red}{---}). }
\label{A1836}
\end{figure}

\begin{figure}[H]
\includegraphics[clip=true,trim= 1mm 1mm 1mm 1mm,height=12cm,width=0.49\textwidth]{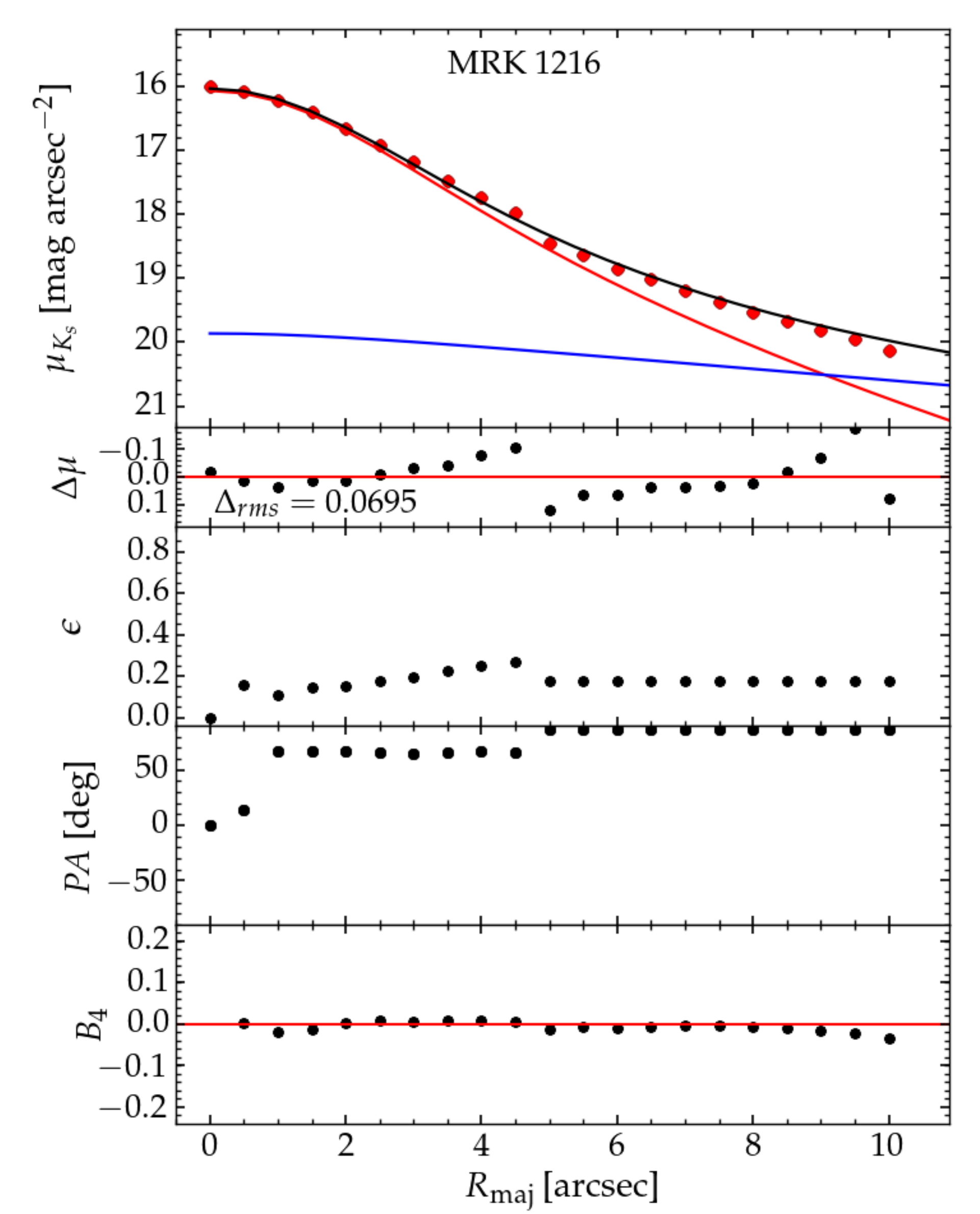}
\includegraphics[clip=true,trim= 1mm 1mm 1mm 1mm,height=12cm,width=0.49\textwidth]{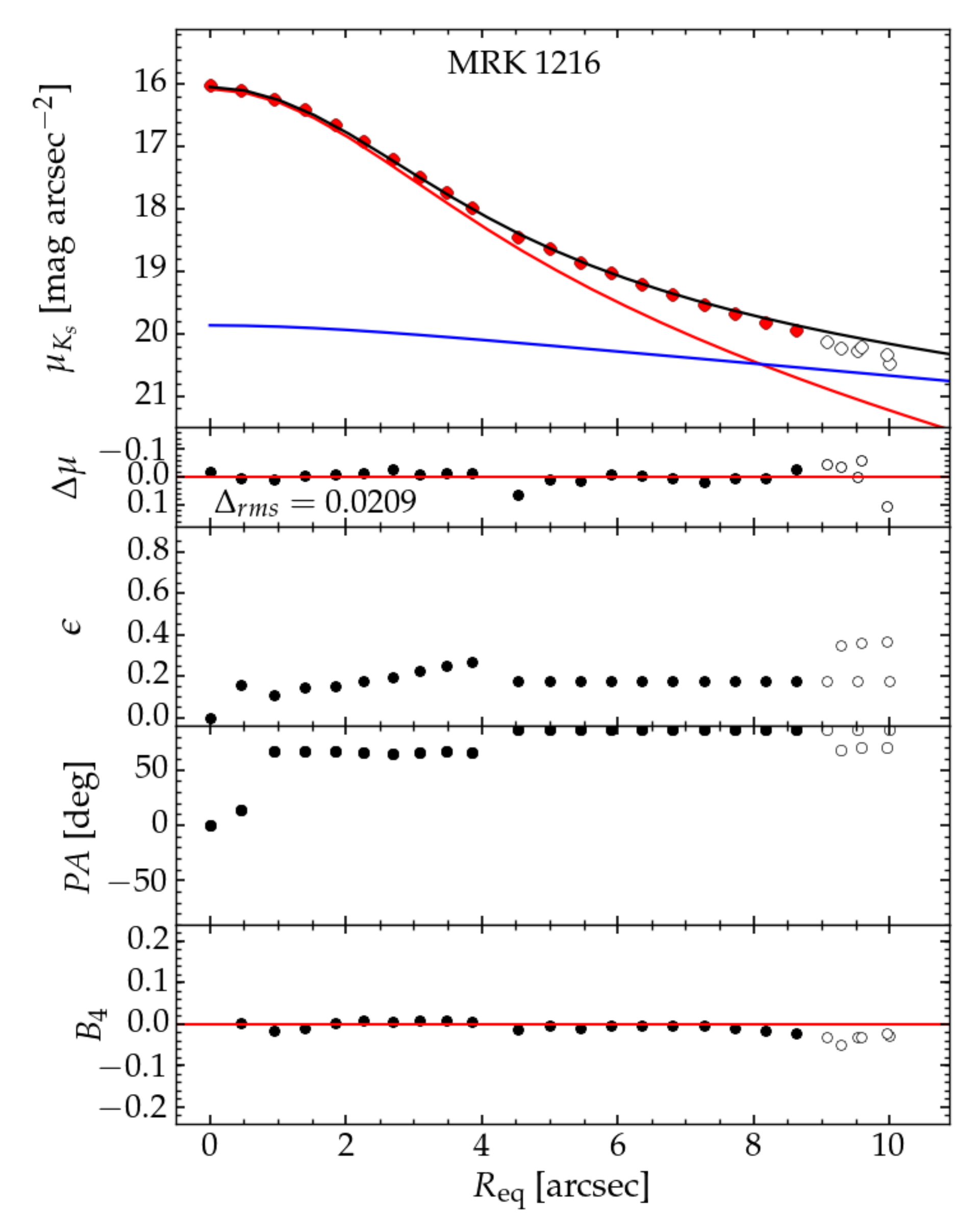}
\caption{MRK 1216: an ellicular galaxy with a S{\'e}rsic bulge (\textcolor{red}{---}) and with a flat exponential model (\textcolor{blue}{---}) fit here to the stellar halo \citep{Yildirim:2015}.  With limited radial extent, and mediocre spatial resolution, our surface brightness profile does not enable a detailed decomposition. Comparison with \citet{Savorgnan:Graham:ES:2016} suggests that we may be in error with this galaxy. However it does not stand out as unusual in our diagrams involving $M_{*,sph}$. }
\label{MRK 1216}
\end{figure}

\begin{figure}[H]
\includegraphics[clip=true,trim= 1mm 1mm 1mm 1mm,height=12cm,width=0.49\textwidth]{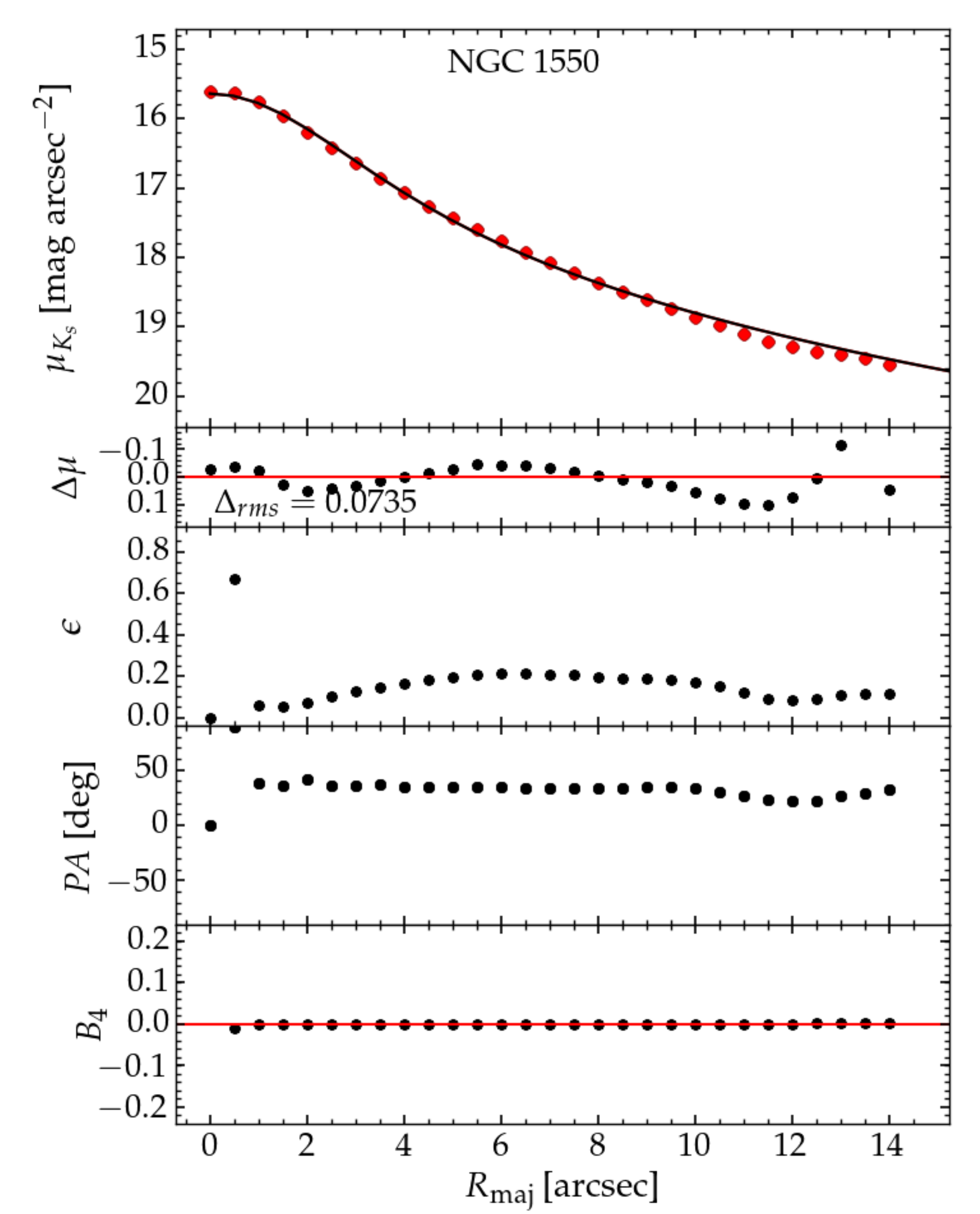}
\includegraphics[clip=true,trim= 1mm 1mm 1mm 1mm,height=12cm,width=0.49\textwidth]{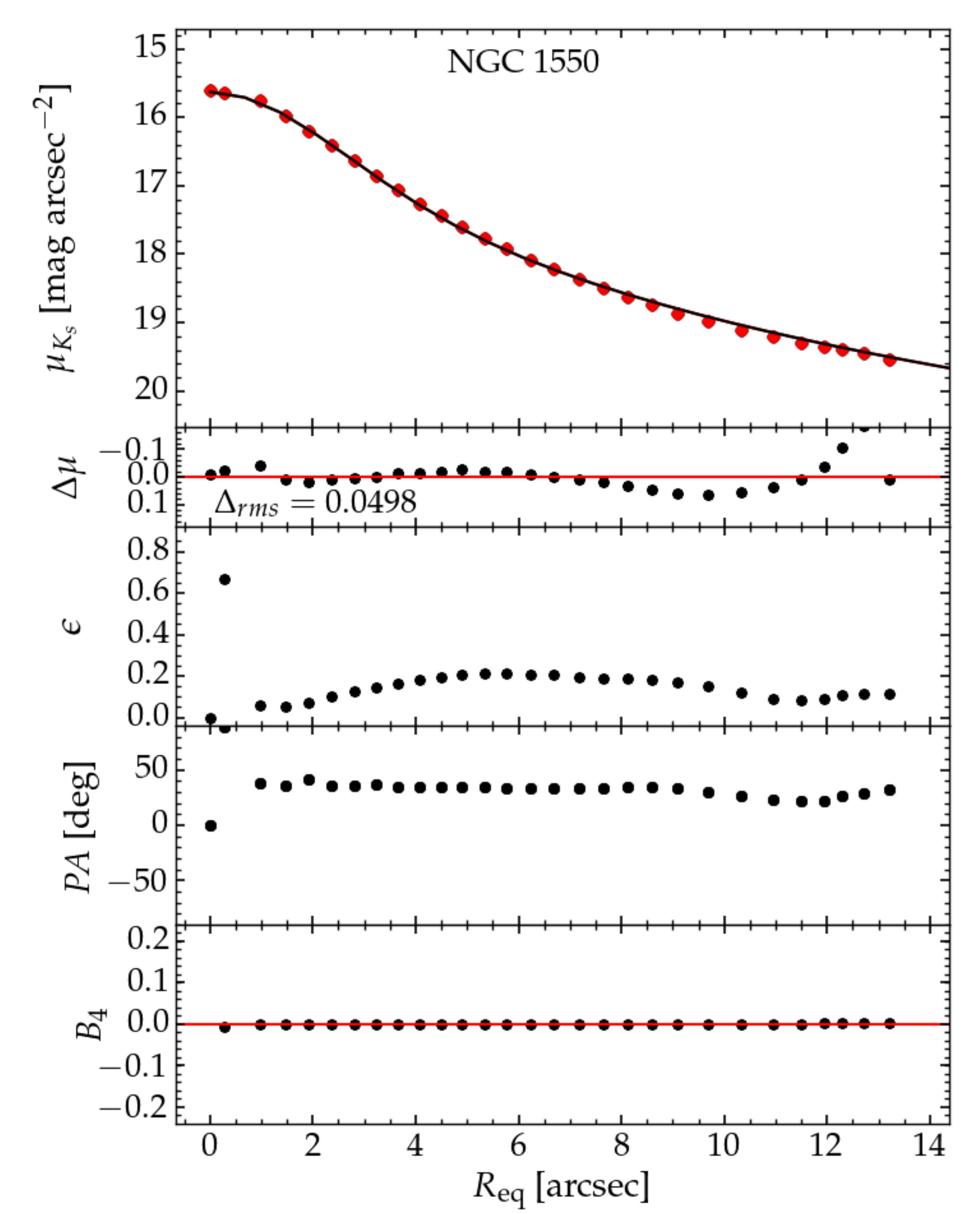}
\caption{NGC 1550: an elliptical galaxy with a depleted core \citep{Rusli:core:2013}, fit using a core-S{\'e}rsic function (\textcolor{red}{---}).}
\label{NGC 1550}
\end{figure}

\begin{figure}[H]
\includegraphics[clip=true,trim= 1mm 1mm 1mm 1mm,height=12cm,width=0.49\textwidth]{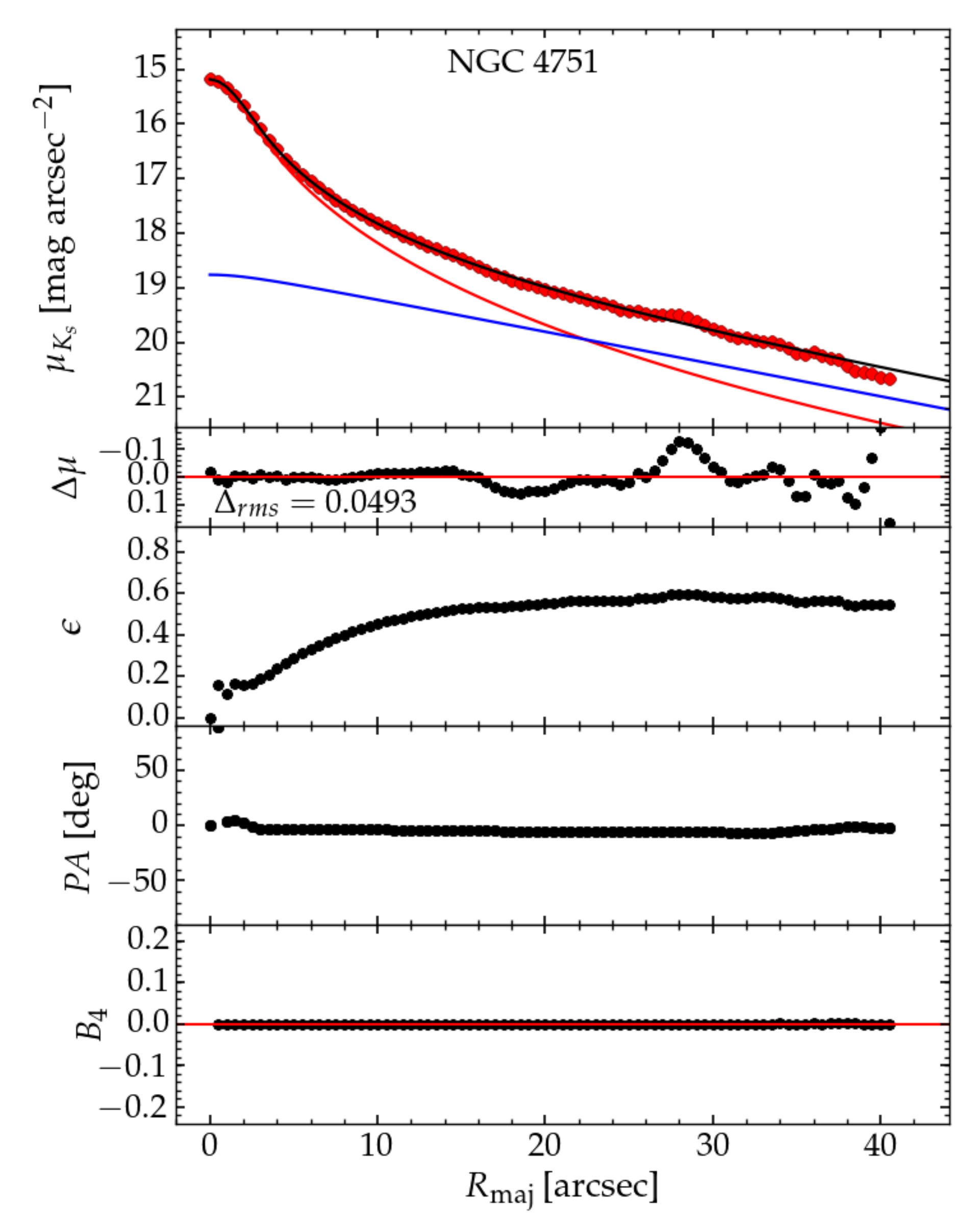}
\includegraphics[clip=true,trim= 1mm 1mm 1mm 1mm,height=12cm,width=0.49\textwidth]{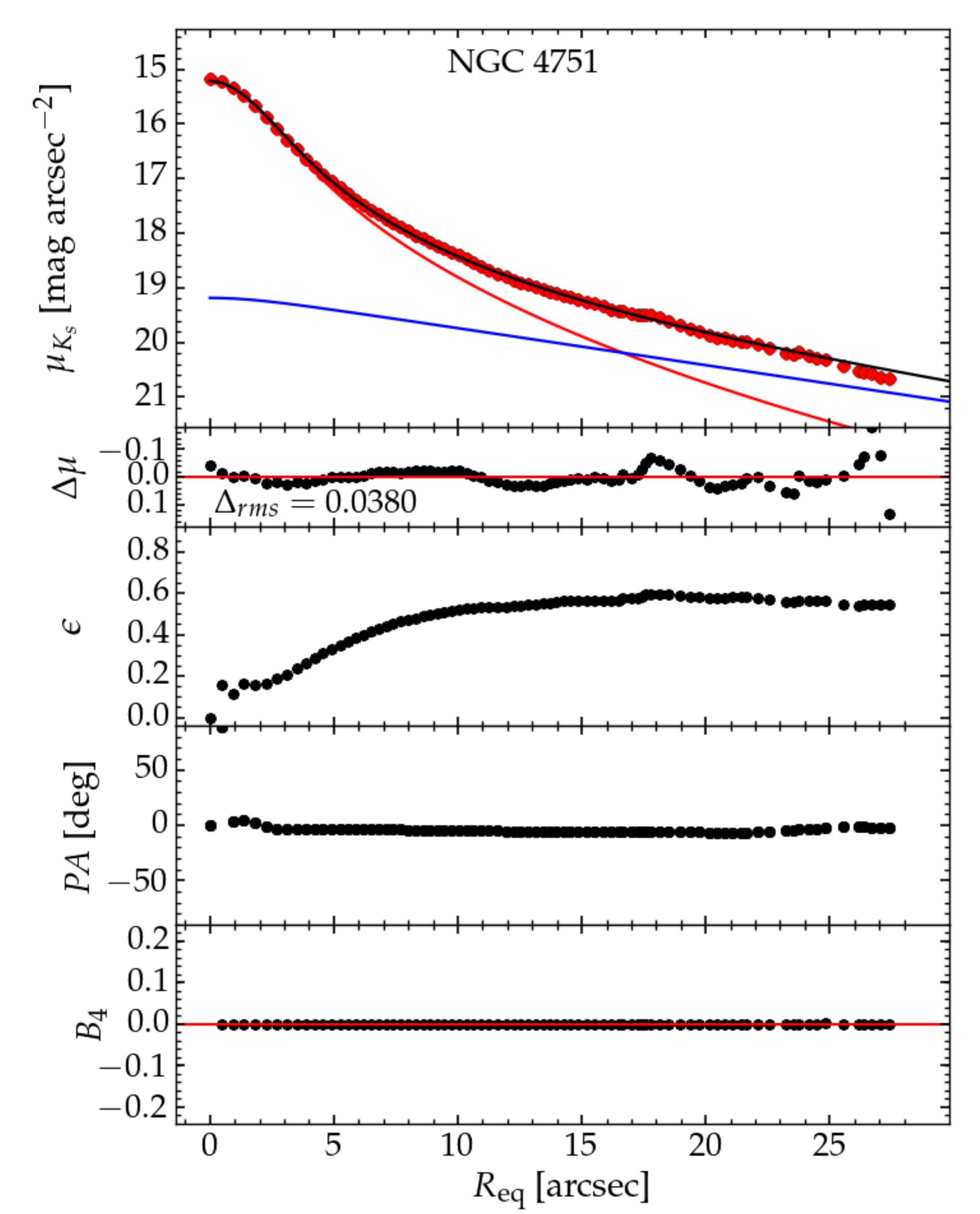}
\caption{NGC 4751: a lenticular galaxy with a very high velocity dispersion and $M_{BH}$ which suggest it may have a depleted core. Hence, we fit a core-S{\'e}rsic function (\textcolor{red}{---}) to its spheroid, plus an extended exponential disk (\textcolor{blue}{---}). }
\label{NGC 4751}
\end{figure}

\begin{figure}[H]
\includegraphics[clip=true,trim= 1mm 1mm 1mm 1mm,height=12cm,width=0.49\textwidth]{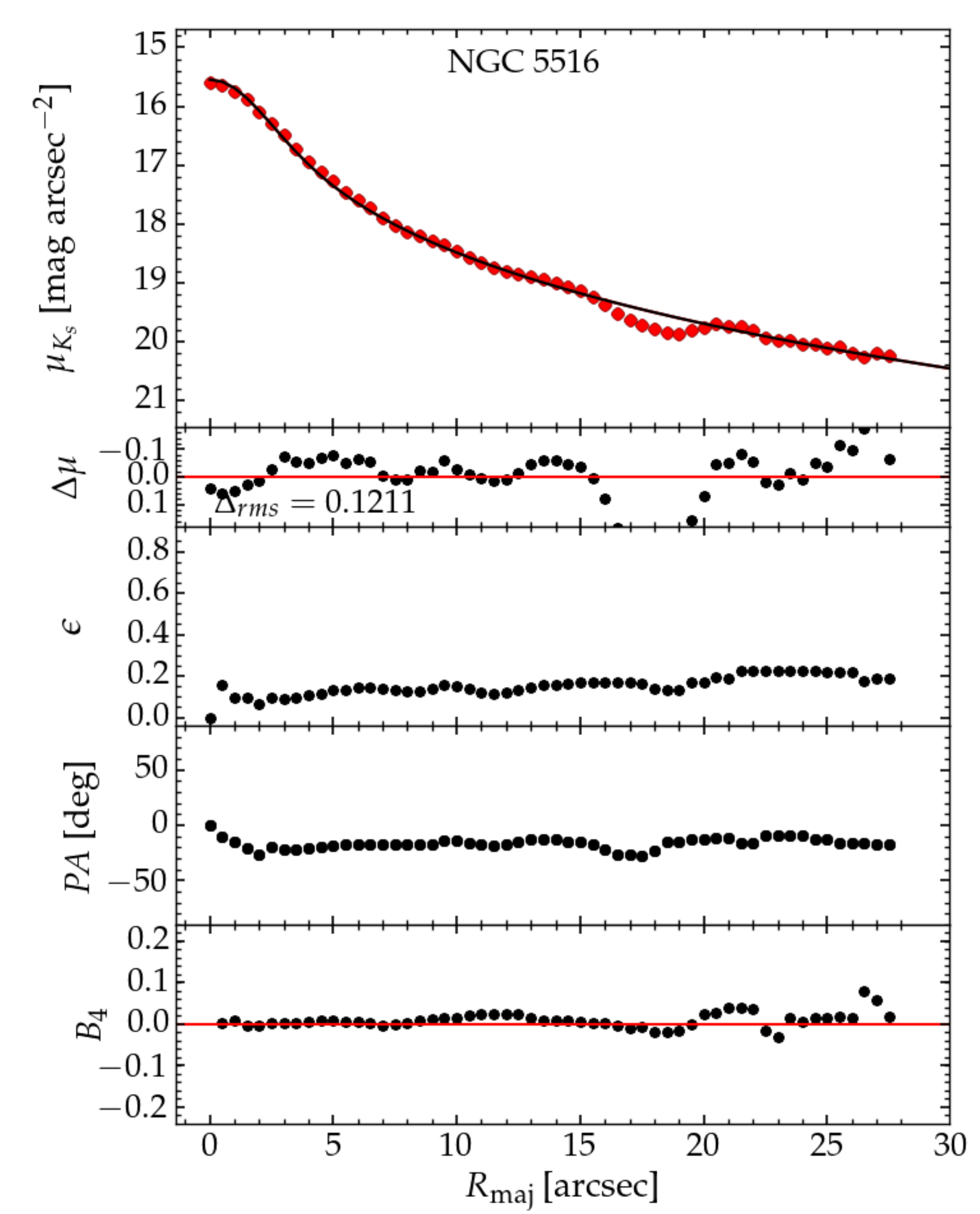}
\includegraphics[clip=true,trim= 1mm 1mm 1mm 1mm,height=12cm,width=0.49\textwidth]{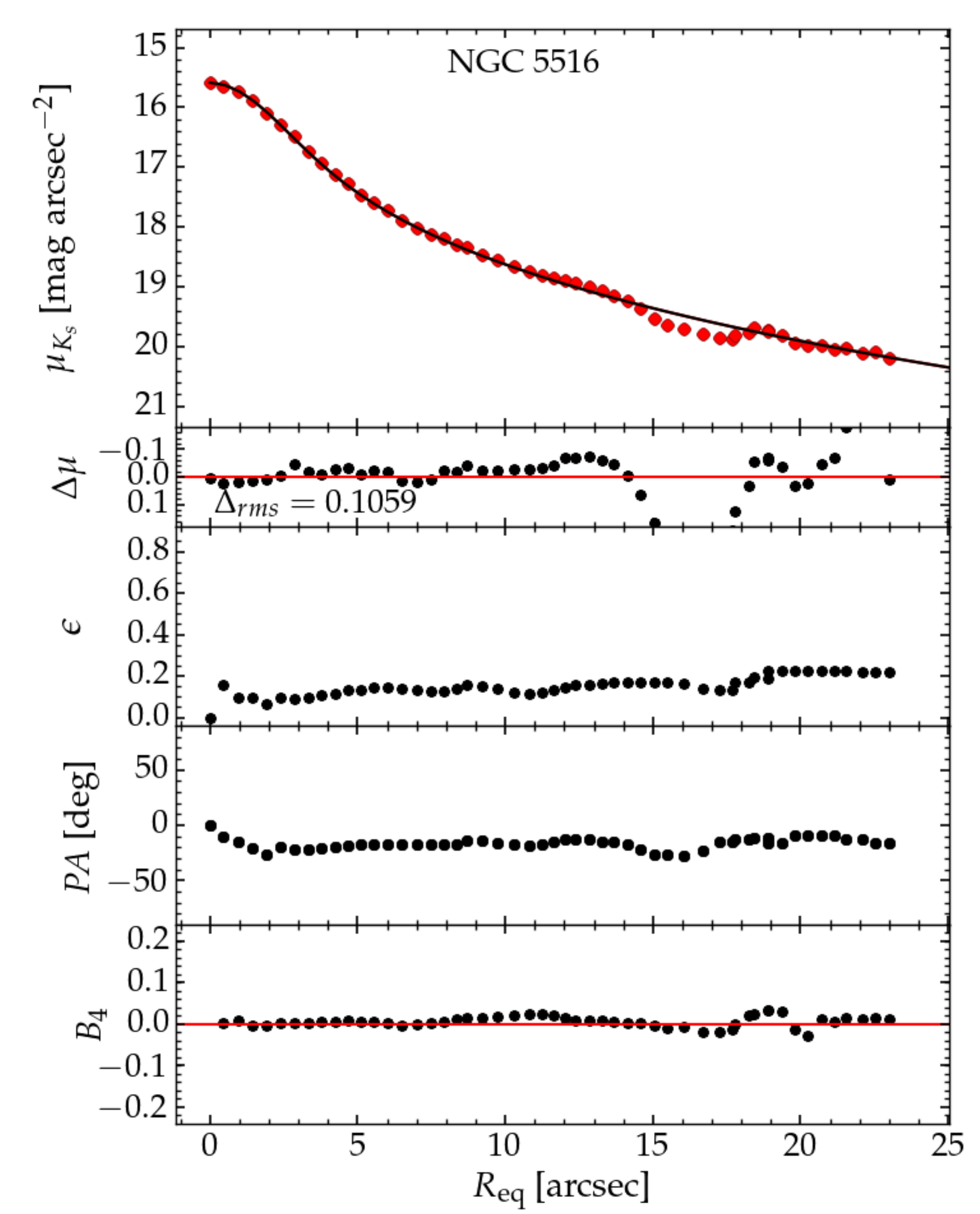}
\caption{NGC 5516: an elliptical galaxy \citep{Rusli:core:2013} fit using a core-S{\'e}rsic function (\textcolor{red}{---}).}
\label{NGC 5516}
\end{figure}

\begin{figure}[H]
\includegraphics[clip=true,trim= 1mm 1mm 1mm 1mm,height=12cm,width=0.49\textwidth]{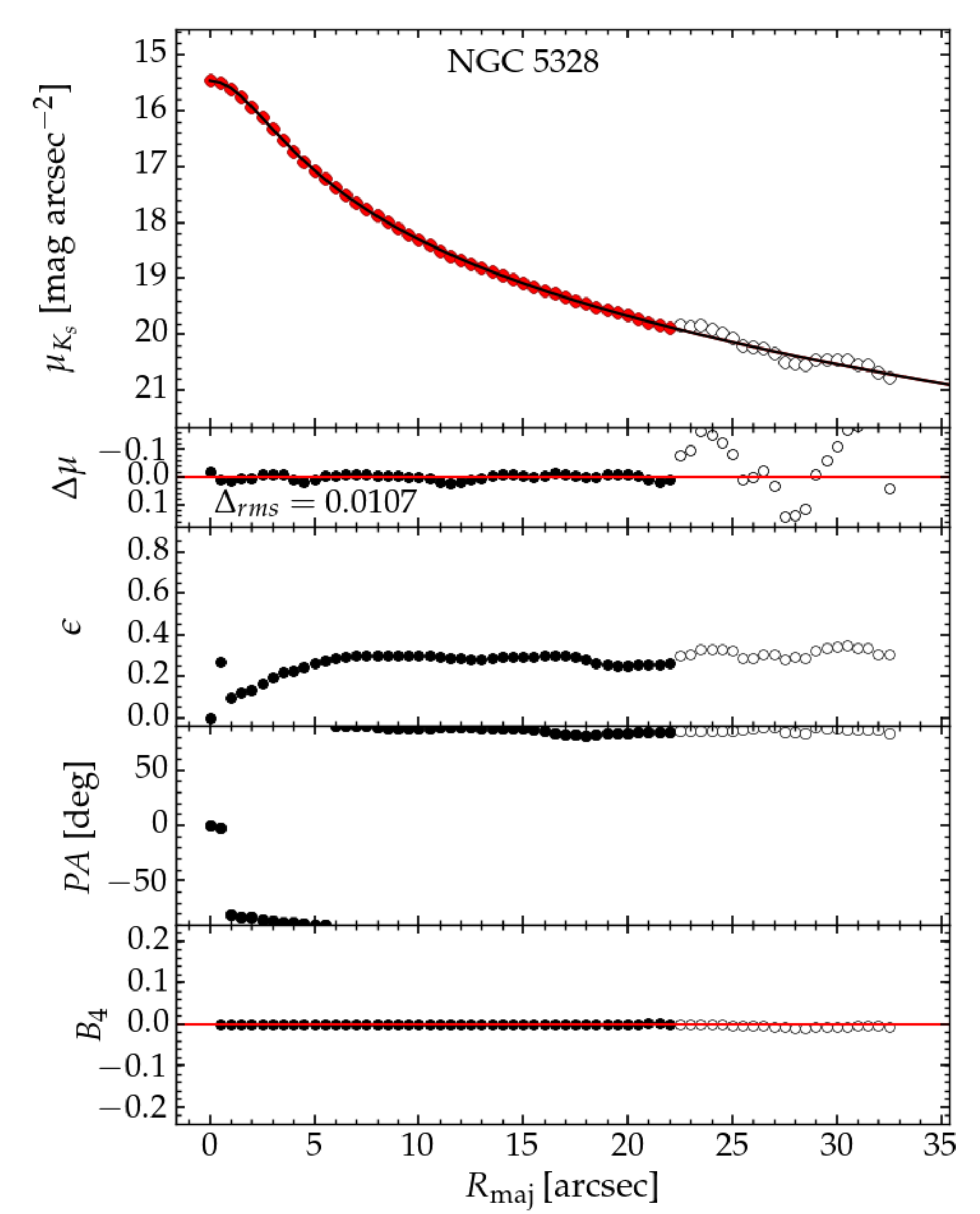}
\includegraphics[clip=true,trim= 1mm 1mm 1mm 1mm,height=12cm,width=0.49\textwidth]{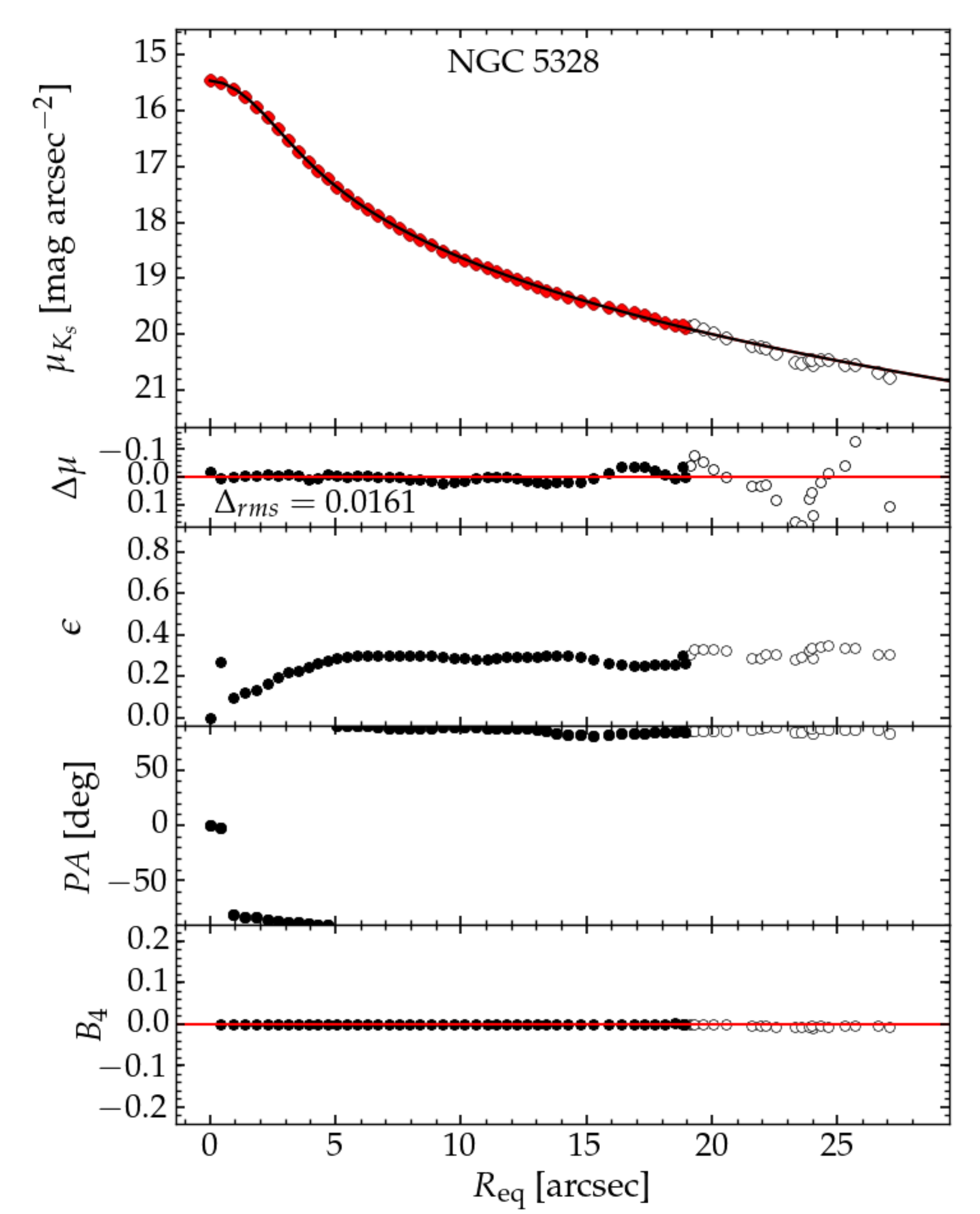}
\caption{NGC 5328: a massive elliptical core-S{\'e}rsic  (\textcolor{red}{---}) galaxy \citep{Rusli:core:2013}.}
\label{NGC 5328}
\end{figure}

\subsection{Light profile from SDSS $r^{\prime}$-band images (AB mag)}

\begin{figure}[H]
\includegraphics[clip=true,trim= 1mm 1mm 1mm 1mm,height=12cm,width=0.49\textwidth]{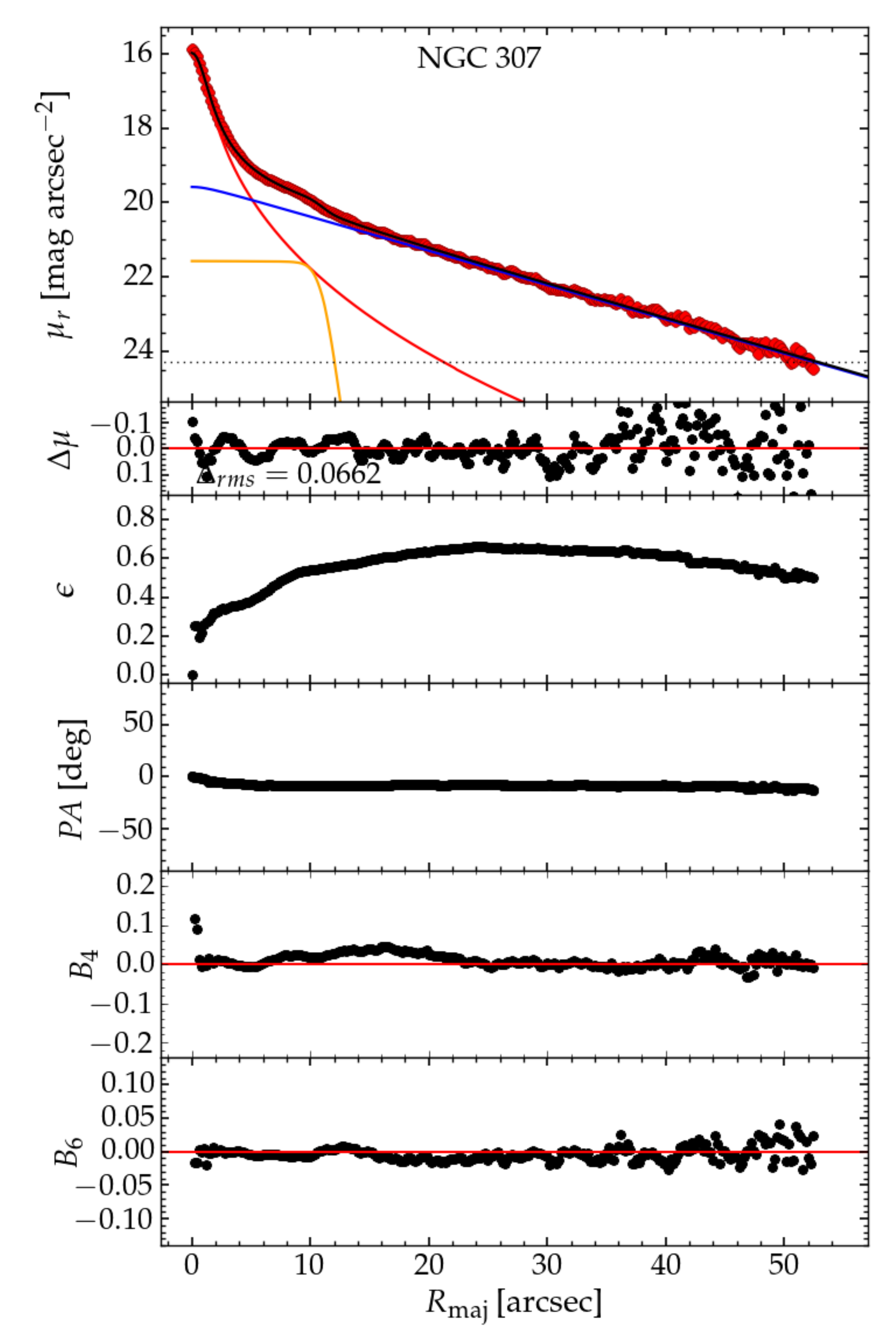}
\includegraphics[clip=true,trim= 1mm 1mm 1mm 1mm,height=12cm,width=0.49\textwidth]{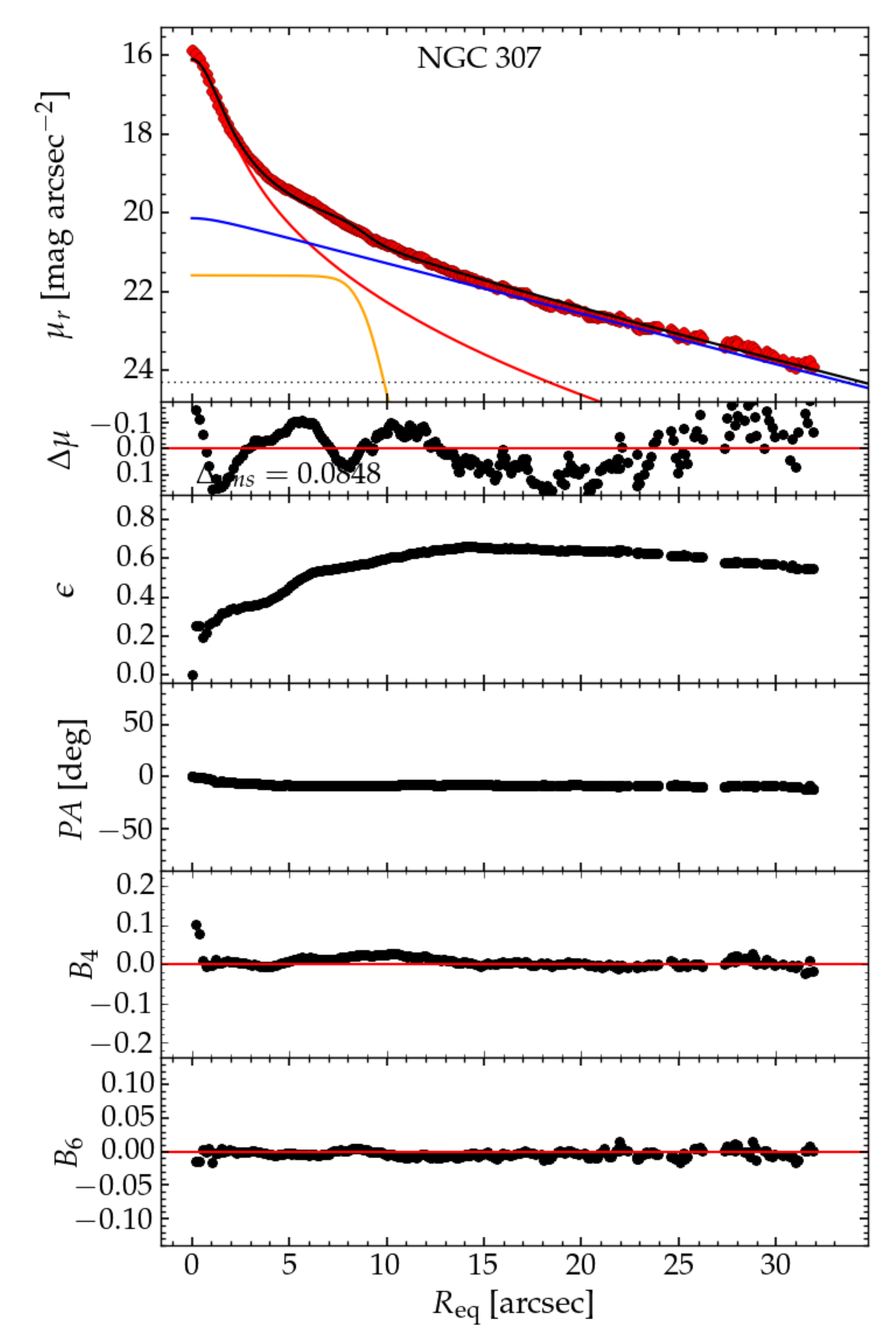}
\caption{NGC 307: a lenticular galaxy with a weak bar \citep{Erwin:2018} fit using a Ferrers (\textcolor{orange}{---}) function, along with a S{\'e}rsic bulge (\textcolor{red}{---}), and an exponential disk (\textcolor{blue}{---}).}
\label{NGC 0307}
\end{figure}

\begin{figure}[H]
\includegraphics[clip=true,trim= 1mm 1mm 1mm 1mm,height=12cm,width=0.49\textwidth]{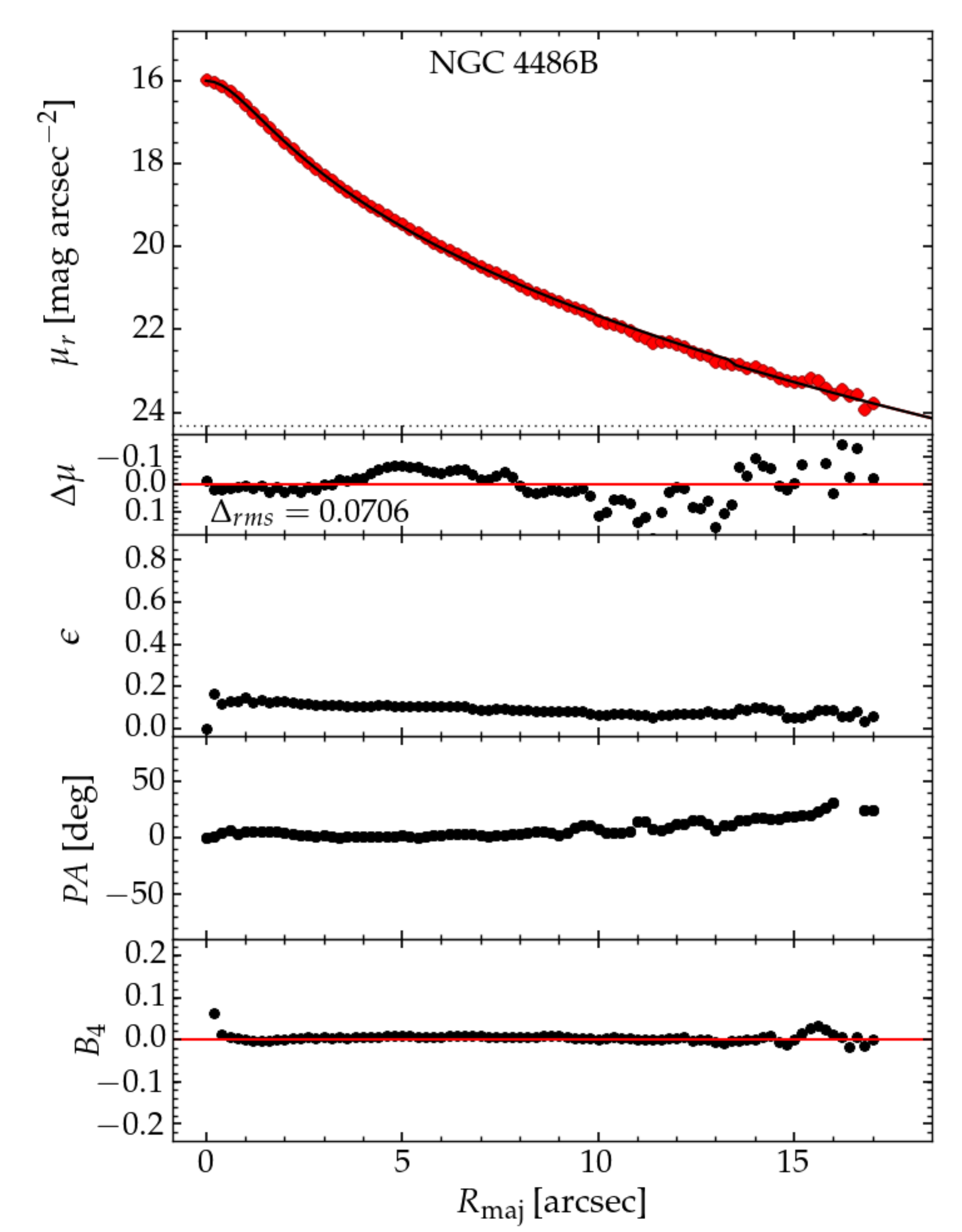}
\includegraphics[clip=true,trim= 1mm 1mm 1mm 1mm,height=12cm,width=0.49\textwidth]{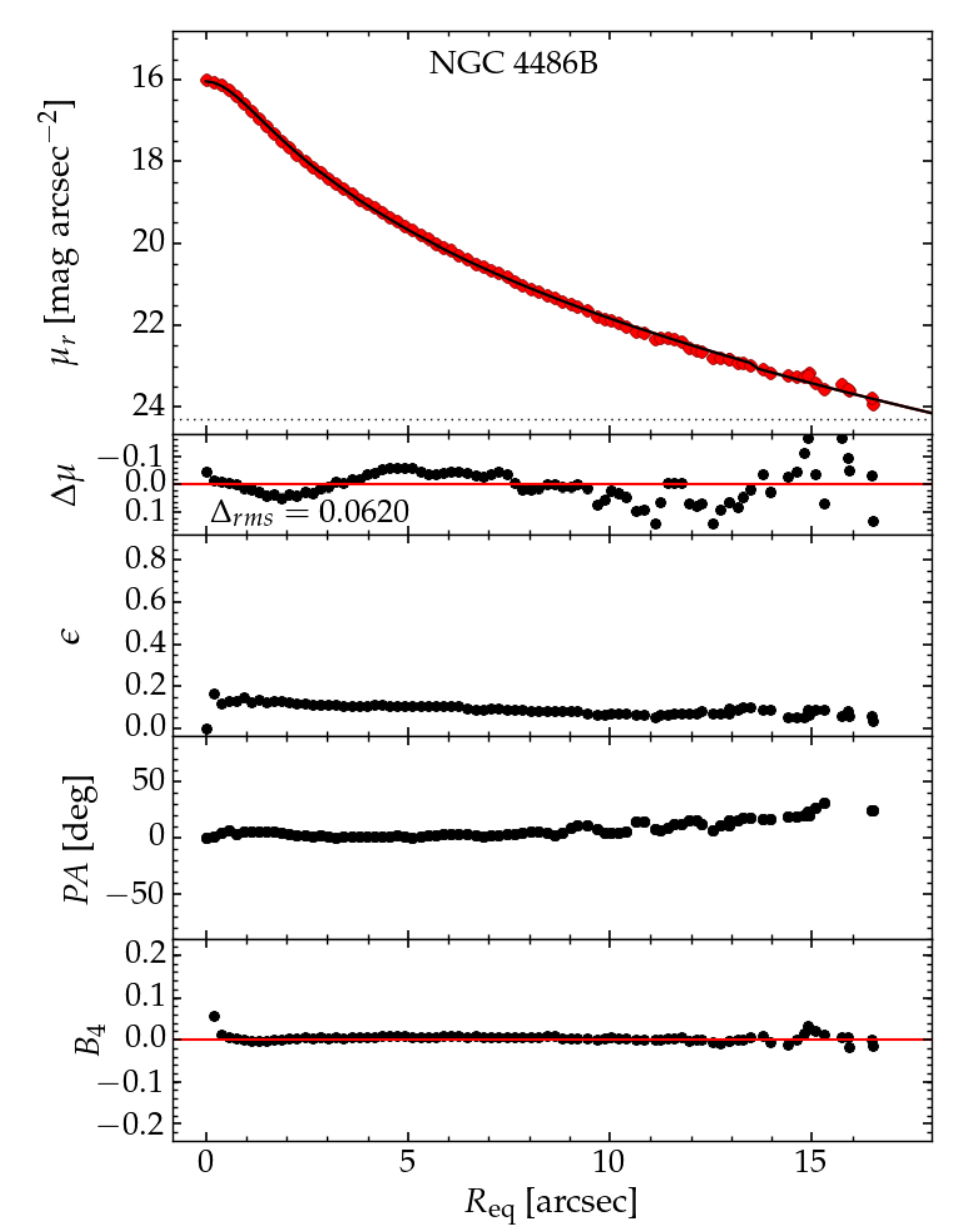}
\caption{NGC 4486B: a \enquote{compact elliptical} galaxy fit with a S{\'e}rsic bulge (\textcolor{red}{---}). Most of its mass is stripped off due to the gravitational interaction with the massive companion galaxy NGC 4486.}
\label{NGC 4486B}
\end{figure}

\begin{figure}[H]
\includegraphics[clip=true,trim= 1mm 1mm 1mm 1mm,height=12cm,width=0.49\textwidth]{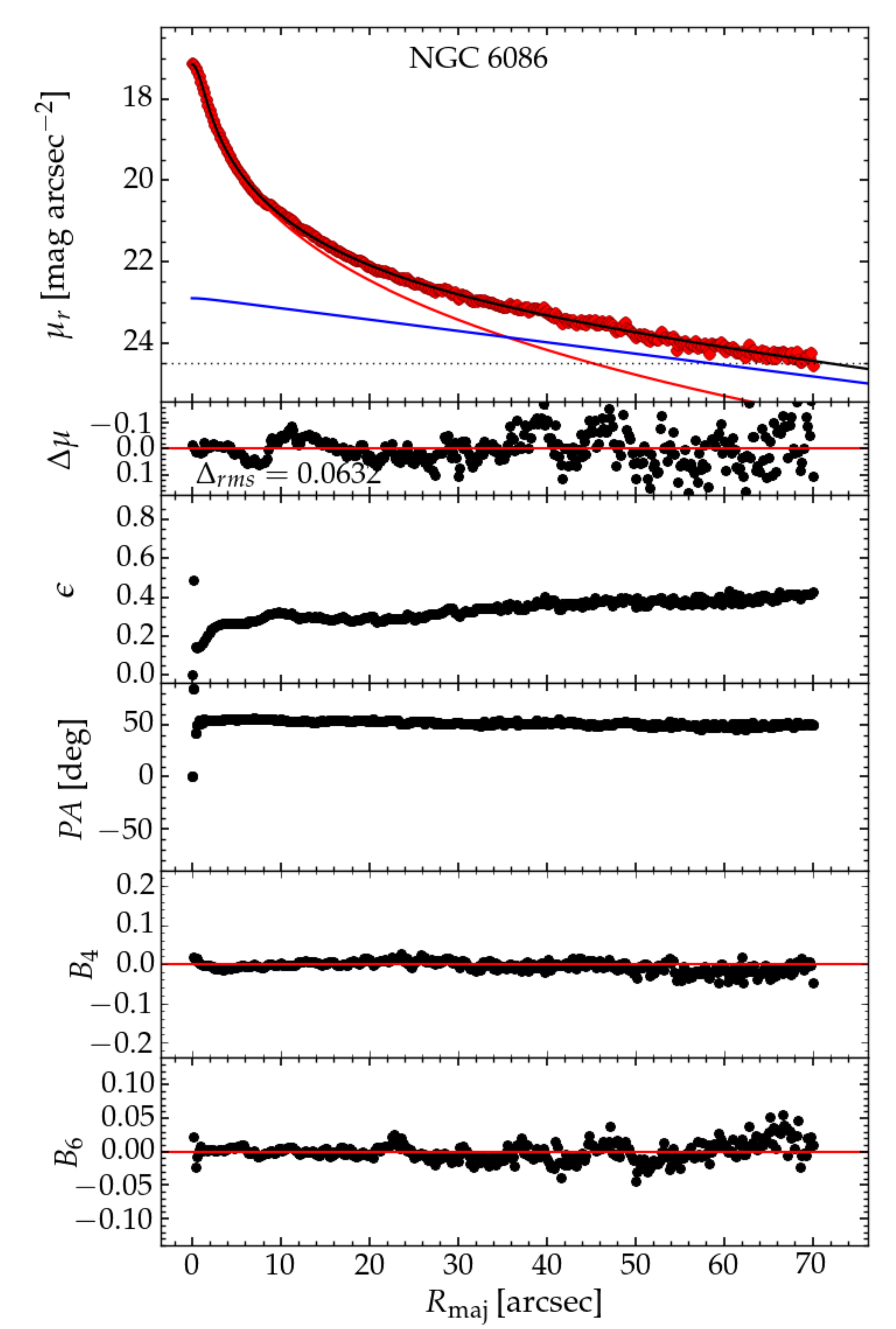}
\includegraphics[clip=true,trim= 1mm 1mm 1mm 1mm,height=12cm,width=0.49\textwidth]{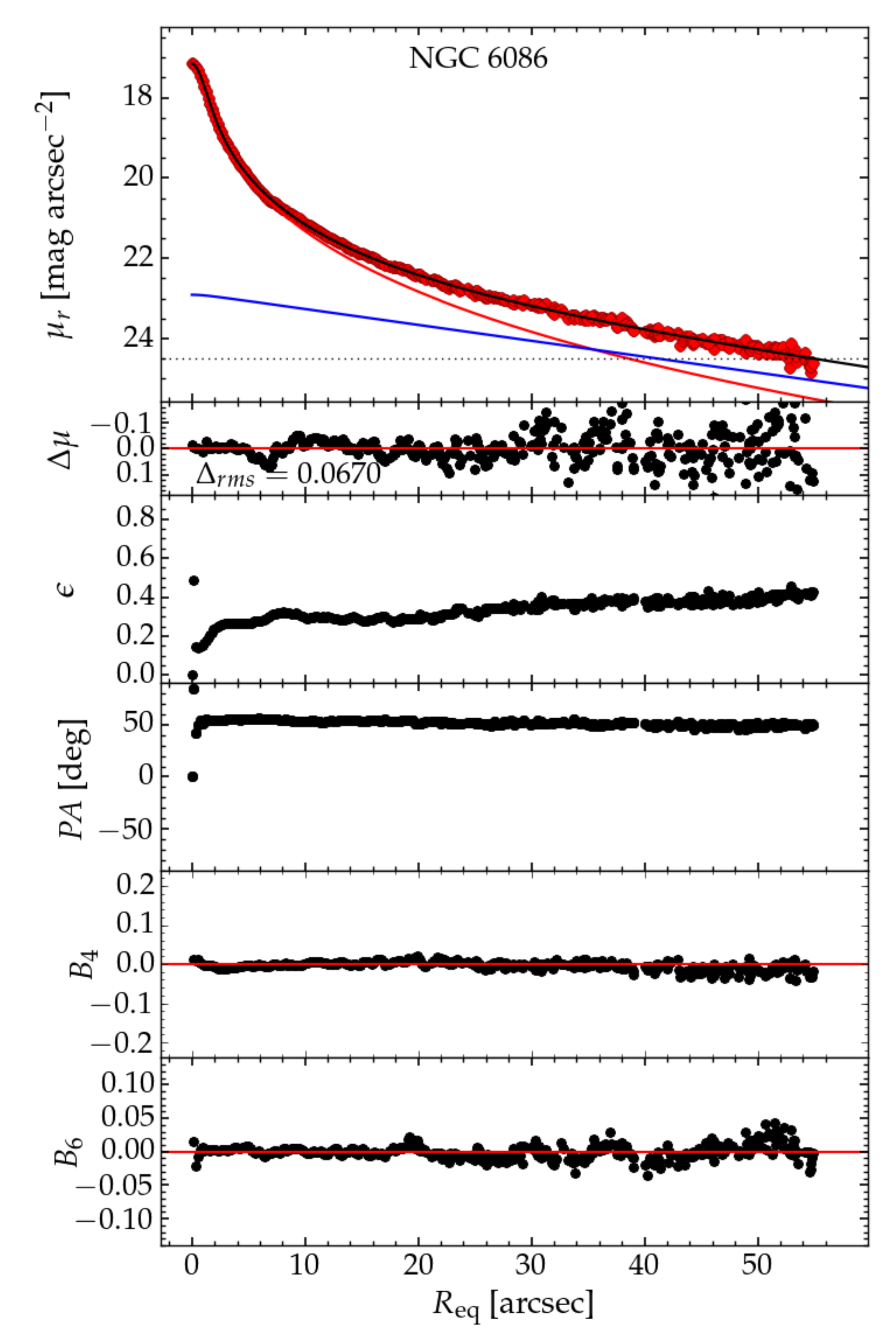}
\caption{NGC 6086: a massive elliptical BCG, with a depleted core \citep{Laine:2003} fit using a core-S{\'e}rsic function (\textcolor{red}{---}) plus an extended halo fit using an exponential  (\textcolor{blue}{---}) function \citep{deVaucouleurs:1969, Seigar:2007}. According to \citet{Carter:1999}, NGC 6086 has a counter-rotating core but a rather slow rotation at the outer radii. The total galaxy light does not include the light from  the (cluster) halo. }
\label{NGC 6086}
\end{figure}

\clearpage

\end{document}